\documentclass[letter]{aa} 

\usepackage{graphicx}
\usepackage{txfonts}
\begin{document}

\title{Dust correction factors over $0 < z < 3$ in massive star-forming galaxies from stacking in \emph{Herschel}\thanks{{\it Herschel} is an ESA space observatory with science instruments provided by European-led Principal Investigator consortia and with important participation from NASA.}}
\titlerunning{Dust correction factors over $0 < z < 3$ from stacking in \emph{Herschel}}

\author{
I. Oteo \inst{1,2}\fnmsep\thanks{mail to: ivanoteogomez@gmail.com}
          }

\institute{Institute for Astronomy, University of Edinburgh, Royal Observatory, Blackford Hill, Edinburgh EH9 3HJ
\and
European Southern Observatory, Karl-Schwarzschild-Str. 2, 85748 Garching, Germany
}

   \date{Received --,  --; accepted --, --}

 
  \abstract{In this work we use stacking analysis in \emph{Herschel} PACS to study the accuracy of several dust-correction factors typically employed to estimate total star-formation rate (SFR) of high-redshift star-forming (SF) galaxies. We also analyze what stacking suggests about the relation between SFR and stellar mass and the redshift evolution of the specific SFR (${\rm sSFR} = {\rm SFR} / {\rm M_*}$). We find that the dust properties of massive SF galaxies evolve with redshift, being galaxies at $z \sim 2-3$ more attenuated than at $z \sim 1$ for a given UV continuum slope and stellar mass. As a consequence, a single IRX-$\beta$ relation can not be used to recover the total SFR of massive SF galaxies at $0 \lesssim z \lesssim 3$. This might have implications for higher redshift studies, where a single IRX-$\beta$ relation derived for local starburst is usually assumed to be valid. However, we find that the local relation between dust attenuation and stellar mass is valid at least up to $z \sim 1$, although deviations are found for higher redshift galaxies where only $\log{\left( M_* / M_\odot \right)} > 10.25-10.50$ galaxies are detected through stacking. This, therefore, does not rule out the possibility that the local dust-mass relation can be valid for less massive SF galaxies at $z \sim 2-3$. The SED fitting procedure with stellar population templates gives over-estimated values (about 0.3--0.5 dex in $\log{\rm SFR}$) of the dust-corrected SFR at all redshifts studied here. We find that the slope of the main-sequence of star formation is less steep than previously found in massive galaxies with $\log{\left( M_* / M_\odot \right)} \geq 10$, and the redshift evolution of the sSFR reported in previous works in massive is well recovered.}
\maketitle
%

\section{Introduction}\label{intro}


In order to estimate dust attenuation and total SFRs of SF galaxies at different redshifts several recipes have been traditionally employed in the literature when FIR information is not available. These include the IRX-$\beta$ relation derived for local starburst (SB) \citep{Meurer1999} or using properties derived using UV/optical/NIR SED fits that do not include FIR data. However, there is increasing evidence that the dust properties of massive SF galaxies might have evolved with redshift and that local relations cannot be applied to high-redshift or special kinds of galaxies \citep{Goldader2002,Oteo2013A&A...554L...3O}. 


Most previous that analyze dust correction factors in SF galaxies are based on direct detection in \emph{Herschel} \citep{Pilbratt2010} bands \citep{Buat2010,Wuyts2011_SED,Oteo2014MNRAS.439.1337O}. In fact, the most accurate values of dust attenuation and total SFR are those derived with individual detections in the FIR. Unfortunately, for a given redshift slice, only a small fraction of galaxies those with the highest total SFRs can be individually detected by \emph{Herschel} even in the deepest surveys \citep{Magdis2010LBGs,Rigopoulou2010,Heinis2013,Oteo2013MNRAS.435..158O,Oteo2013A&A...554L...3O,Oteo2014MNRAS.439.1337O}. Furthermore, a significant population of the high-redshift galaxies detected by \emph{Herschel} are starburst rather than normal SF galaxies \citep[see for example][]{Rodighiero2014arXiv1406.1189R}. Therefore, if we want to study less extreme sources, we need to rely on stacking analysis. In this work we study the accuracy of several dust-correction factors (IRX-$\beta$ relation, dust-mass correlation, and SED fits) at different redshifts from stacking analysis in \emph{Herschel} PACS-160 $\mu$m \citep{Poglitsch2010}. We also explore what stacking tells us about the relation between SFR and stellar mass in massive SF galaxies and about the redshift evolution of the specific SFR (${\rm sSFR}$). This paper is organized as follows: Section \ref{selection_sources} explains the source selection and the methodology employed in this work. The main results are presented in Section \ref{results_work}. Finally, we summarize our main conclusions in Section \ref{concluuuuu}. Throughout the paper, we assume a flat universe with $(\Omega_m, \Omega_\Lambda, h_0)=(0.3, 0.7, 0.7)$, and all magnitudes are listed in the AB system \citep{Oke1983}.

\section{Source selection and methodology}\label{selection_sources}

We focus our work on the COSMOS field \citep{Scoville2007} and select all the rest-frame UV-selected sources whose photometric redshifts (taken from \cite{Ilbert2013A&A...556A..55I}) are within $0.02 \leq z \leq 4.0$. We avoid AGN contamination by discarding all galaxies detected in X-rays \citep{Elvis2009}. In this work we are interested in SF galaxies. Different ways of selecting SF galaxies have been proposed in the literature, mostly based on single and double colors \citep{Williams2009,Rodighiero2010A&A...518L..25R}. However, a single color cut might produce the loss of SF galaxies with red rest-frame $U-V$ colors \cite[see figure 11 in][]{Williams2009}. Also, red SF galaxies might occupy the green valley of even the red sequence of galaxies \citep{Oteo2014MNRAS.439.1337O}. Since SF galaxies are galaxies that form stars, we select SF galaxies basing on their SFR. In this way, we consider only galaxies whose ${\rm SFR_{UV}}$ (uncorrected from dust attenuation) is at least 1 $M_\odot \, {\rm yr}^{-1}$. This is a conservative approach, since a high percentage of low-redshift galaxies have SFR higher than that threshold \citep{Elbaz2007}, and SF galaxies with ${\rm SFR} \geq 1 M_\odot \, {\rm yr}^{-1}$ can be detected at the highest redshifts studied in this work with the depth of the data we are using. We have also compared with results adopting a threshold ${\rm SFR_{UV}} \geq 5 M_{\odot} \, {\rm yr}^{-1}$ at $z \geq 1$. Note that since the above limit is before any dust correction, the actual SFR would be even higher.

The UV-to-NIR SEDs are fitted with \cite{Bruzual2003} templates to obtain the stellar masses, UV continuum slope ($\beta$), and rest-frame UV luminosities ($L_{\rm UV}$) of our studied galaxies \citep[see for example][]{Oteo2013MNRAS.433.2706O}. Optical and NIR photometric information is also taken from \cite{Ilbert2013A&A...556A..55I}, while GALEX data are taken from \cite{Zamojski2007} for galaxies at $z \leq 1$. In this step, \cite{Bruzual2003} templates associated to constant SFR and $Z=0.2 Z_{\odot}$ metallicity are considered. Dust attenuation is included with the \cite{Calzetti2000} law. We also include intergalactic medium absorption adopting the prescription of \cite{Madau1995}. The SED fits are carried out with the Zurich Extragalactic Bayesian Redshift Analyzer \citep[ZEBRA,][]{Feldmann2006} code. 

Since most of the galaxies are undetected by \emph{Herschel} we must rely on an stacking analysis to estimate the FIR emission of UV-selected galaxies. To this aim we use the IAS stacking library \citep{Bethermin2010A&A...512A..78B} and focus on the PACS-160 $\mu$m band. We have checked that stacking in PACS-100 $\mu$m gives similar results \citep[see also][]{Rodighiero2010A&A...518L..25R}. We stack in the residual images and median stacked image are considered. It should be noted that only \emph{Herschel}-undetected galaxies are considered in the stacking analysis. The stacked fluxes are obtained with a PSF-fitting procedure and their uncertainties are calculated with bootstrap. We studied the effect of clustering effect \citep{Bethermin2012A&A...542A..58B} by comparing the PSF of PACS-160 $\mu$m with the actual radial profile of each stacked detection. This is similar to the method used in \cite{Heinis2013} (see also Method C in \citealt{Bethermin2012A&A...542A..58B}). Comparing the profile of the stacked detections and the profile of the PSF of the PACS-160 $\mu$m observations, we do not find additional broadening of the stacked emission due to clustering of the input catalog and, consequently, we do not include any extra correction \citep{Magnelli2014A&A...561A..86M}. Radial profiles and stacked images are shown in Appendix \ref{app_plots} for stacks as a function of the UV continuum slope and stellar mass (see later in the text). The PACS-160 $\mu$m are converted to total IR luminosities ($L_{\rm IR}$) with single band extrapolations using the \cite{Chary2001} templates. This step provides good estimation of $L_{\rm IR}$ since PACS-160 $\mu$m samples the dust emission peak at the redshift range of the galaxies studied. The stacked total SFR are obtained from luminosities with the \cite{Kennicutt1998} relations and adopting ${\rm SFR_{total}} = {\rm SFR_{UV}} + {\rm SFR_{IR}}$. Dust attenuation is obtained from the ratio between $L_{\rm IR}$ and $L_{\rm UV}$ luminosities assuming the \cite{Buat2005} calibration. We will stack as a function of UV continuum slope, stellar mass, and dust attenuation. We only consider in this study stacked detections with $f_{\rm stack} \geq 3 \sigma_{\rm stack}$, where $\sigma_{\rm stack}$ is the uncertainty in each stacked flux. With the PACS data used in this work, stacked detection are only recovered for massive galaxies, typically with $\log{\left( M_*/M_\odot \right)} \gtrsim 10$. Therefore, we will only deal with a massive population os SF galaxies. Furthermore, despite we consider galaxies up to $z \sim 4$, we only recover stacked detections for galaxies at $z \leq 3$. Tables \ref{hola_tabla} and \ref{hola_tabla_masa} summarize the stacked fluxes, rest-frame UV and total IR luminosities, and number of sources in each redshift and UV slope bin for the stacked detection as a function of UV continuum slope and stellar mass. It should be noted that we do not employ individual detections in our study since with the depth of \emph{Herschel} observations in COSMOS, only the most-extreme sources are detected, and they might not be normal SF galaxies, but likely have a SB nature. In any case, the detection rate in \emph{Herschel} bands is low and this does not change the main conclusions of our work.

\begin{figure}
\centering
\includegraphics[width=0.48\textwidth]{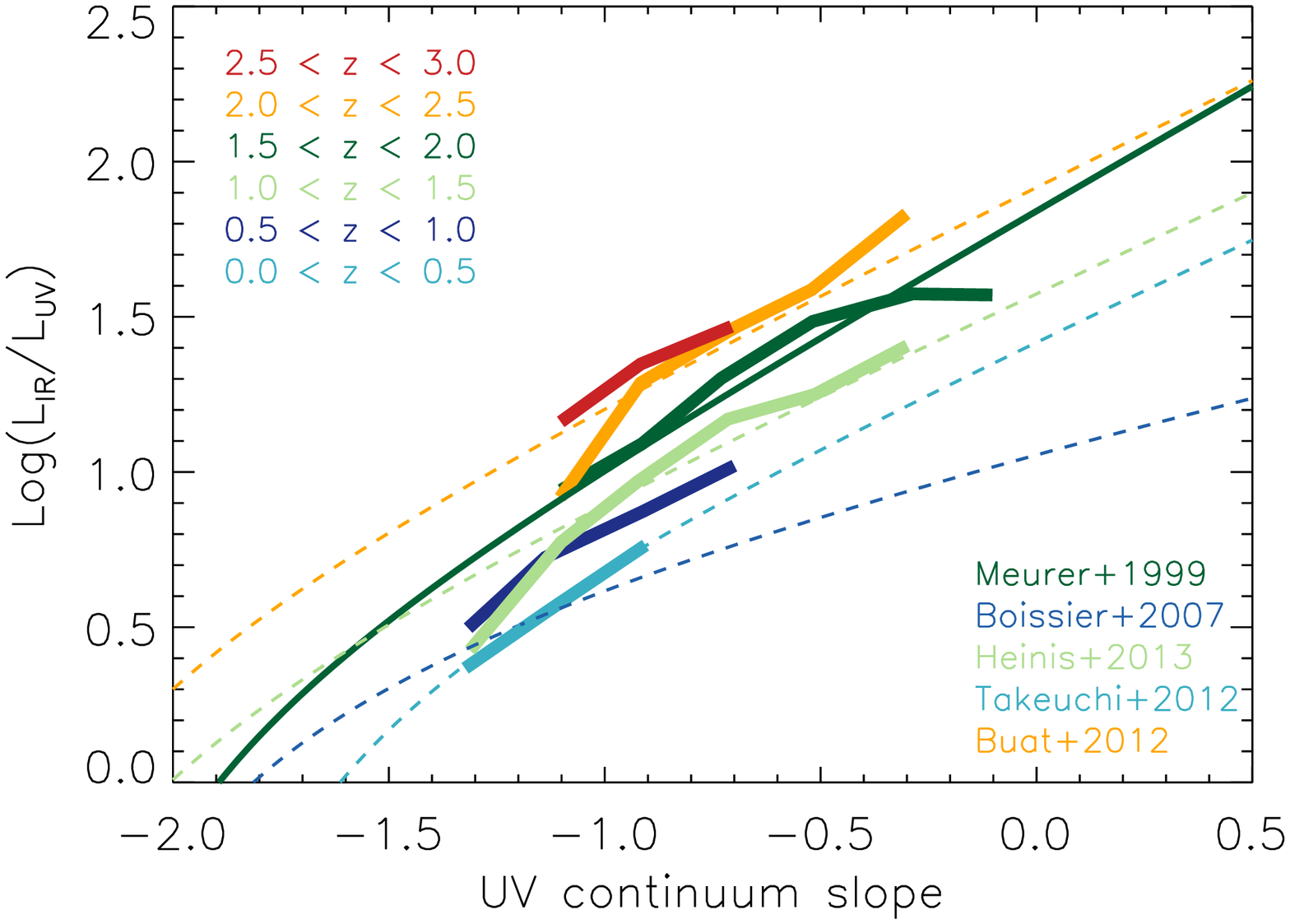}
\caption{Relation between the total IR and rest-frame UV luminosities (a proxy for dust attenuation) agains the UV continuum slope as seen by our stacking analysis over $0.02 \lesssim z \lesssim 3.0$, with the color-code shown in the top-left legend. We also represent the IRX-$\beta$ relations reported in previous works \citep{Meurer1999, Boissier2007, Heinis2013, Takeuchi2012, Buat2012}, as indicated in the bottom-right legend. Their color code is related to the data they fits best to.
              }
\label{IRXbeta}
\end{figure}

\section{Results}\label{results_work}

Figure \ref{IRXbeta} shows the relation between dust attenuation (parametrized by the $L_{\rm IR} / L_{\rm UV}$ ratio) against the UV continuum slope for our SF galaxies at different redshifts. As expected, redder galaxies are more attenuated. There is a clear trend with redshift for galaxies with $\beta \geq -1.0$: for a given UV continuum slope, galaxies at $z \sim 2-3$ are more attenuated than those at $z \sim 1$. This result is similar to the ones found for individual detections in PACS/SPIRE \citep{Oteo2013A&A...554L...3O,Oteo2013MNRAS.435..158O,Oteo2014MNRAS.439.1337O}, but now with a staking analysis and, therefore, being able to probe less dusty and IR-fainter galaxies at each epoch. At $z \lesssim 1$, where we can detect bluer galaxies, dust attenuation has not significantly changed with redshift for $\beta < -1.0$ SF galaxies. This might indicate that dust attenuation evolved only in massive, red SF galaxies, although deeper data would be needed to confirm this.

These results confirm the evolution of the dust properties of massive galaxies over $0 < z < 3$. As a consequence of this evolutionary trend, there is not a single IRX-$\beta$ relation that can be applied at all redshifts to accurately recover the dust attenuation or total SFR in massive SF galaxies. At low redshift, stacked points are in agreement with the \cite{Takeuchi2012} relation, but the \cite{Buat2012} relation should be applied at $z \sim 2-3$, where galaxies were more attenuation for a given UV continuum slope. We recall that at $z \sim 1.5$ our stacked points are in great agreement with the \cite{Heinis2013} relation, that was obtained with galaxies around that redshift through stacking in SPIRE bands.

As shown in Figure \ref{IRXbeta}, even with stacking analysis in \emph{Herschel}, we do not recover stacked detections for galaxies with $\beta \lesssim -1.5$ at any redshift \citep[see also][]{Heinis2013}, nor with $\beta < -1.1$ at $z \geq 1.5$. This represents a problem for testing the accuracy of the IRX-$\beta$ relations at recovering dust attenuation. This is because the SF galaxies selected in UV, optical, or NIR surveys have actually UV continuum slopes $\beta \lesssim -1.25$ in most cases \citep{Oteo2013A&A...554L...3O,Oteo2013MNRAS.433.2706O,Oteo2014MNRAS.439.1337O}. Therefore, applying a single IRX-$\beta$ relation should be done with care, because it has been shown here to evolve with redshift for massive, red ($\beta \geq -1.1$) SF galaxies but we do not know its behavior at lower masses and bluer UV slopes at different redshifts.

\begin{figure}
\centering
\includegraphics[width=0.48\textwidth]{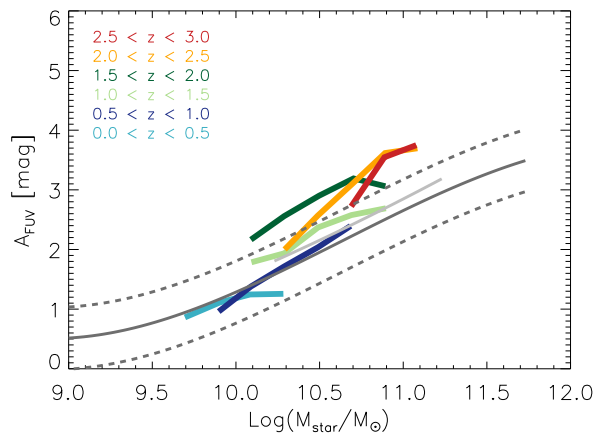}
\caption{Relation between dust attenuation and stellar mass for our stacked galaxies over $0.02 \lesssim z \lesssim 3.0$, with the color-code shown in the top-left legend. Dust attenuation has been derived with the $L_{\rm IR} / L_{\rm UV}$ ratio and the \cite{Buat2005} calibration. For a reference, we represent the local relation as reported in \cite{Sobral2012MNRAS.420.1926S} with a solid dark grey curve and the relation at $z \sim 1.5$ presented in \citet{Heinis2014MNRAS.437.1268H} with a light grey solid line.
              }
\label{dustmass}
\end{figure}

Figure \ref{dustmass} represents the dust attenuation in the FUV band (rest-frame 1500 \AA) as a function of stellar mass for our SF galaxies at different redshifts. More massive galaxies are more attenuated, as obtained in several previous works \citep{Ibar2013MNRAS.434.3218I,Garn2010MNRAS.409..421G,Heinis2014MNRAS.437.1268H}. A trend with redshift is also present, being massive ($\log{\left( M_* / M_\odot \right)} \geq10$) galaxies at $z \sim 2-3$ more attenuated for a given stellar mass than at $z \lesssim 1$. The \cite{Heinis2014MNRAS.437.1268H} is well recovered at their same redshift, $z \sim 1.5$. Since only galaxies with $\log{\left( M_* / M_\odot \right)} \geq 10$ have stacked detections at $z \geq 1$, we can only confirm the evolution in that mass range. At lower redshift, where we can detect less massive galaxies, we find that the relation between dust attenuation and stellar mass is in very good agreement to the one found for local galaxies \citep{Garn2010MNRAS.409..421G}. This might indicate that, as it might occur for blue galaxies, the dust attenuation in less massive galaxies does not change significantly with redshift. This is what it was actually obtained in \cite{Sobral2012MNRAS.420.1926S} for H$\alpha$ emitters at $z \sim 1.47$ and also with stacking analysis in HAEs at the same redshift in \cite{Ibar2013MNRAS.434.3218I}. This result would be quite important for recovering the dust attenuation or total SFR in $\log{\left( M_* / M_\odot \right)} \lesssim 10$ SF galaxies when FIR information is not available. 

\begin{figure}
\centering
\includegraphics[width=0.45\textwidth]{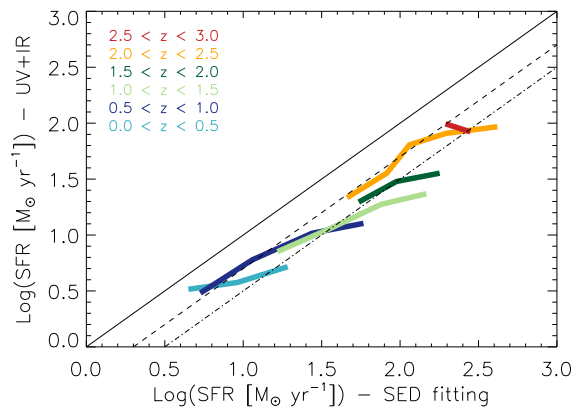}
\caption{Relation between the dust-corrected total SFRs derived with stacking analysis in PACS and from a SED fitting method assuming a constant SFR and fixed metallicity $Z = 0.2 Z_\odot$. The one-to-one relation is shown with a black solid line, whereas deviations of -0.3 and -0.5 dex are indicated with dashed and dotted-dashed lines, respectively.
              }
\label{sssssssssssss}
\end{figure}

The power of the SED fitting method for recovering the total SFR in massive galaxies is shown in Figure \ref{sssssssssssss}. It can be seen that, at any redshift, the SED-derived total SFR is systematically higher than those obtained from stacking analysis. The over-estimation is about 0.3--0.5 dex at all redshifts, despite a little bit higher for the galaxies with the highest SED-derived SFRs. At high redshift and for FIR-bright galaxies individually detected with PACS, the SED-derived total SFR is an under-estimation of the much more accurate SFR derived with direct UV and IR data \citep{Wuyts2011_SED,Oteo2013A&A...554L...3O}. However, the stacking analysis presented here indicates that the SED-fitting method actually over-estimates the total SFR at $0 < z < 3$ when galaxies with lower $L_{\rm IR}$ can be studied.


\begin{figure}
\centering
\includegraphics[width=0.45\textwidth]{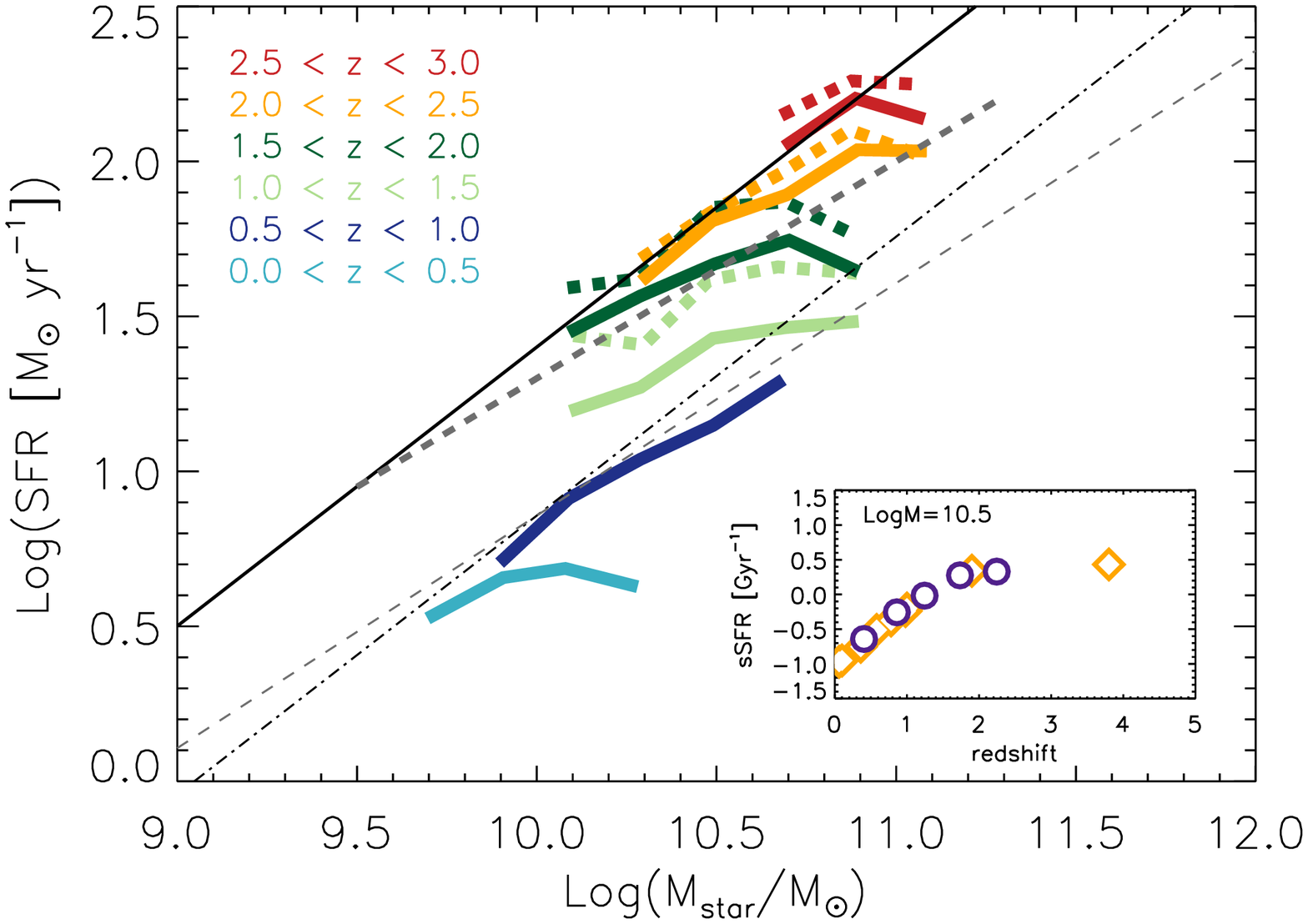}
\caption{Relation between ${\rm SFR} = {\rm SFR_{UV}} + {SFR_{IR}}$ and stellar mass for massive galaxies as revealed by our stacking analysis. We represent with dashed curves the points when the ${\rm SFR_{UV}}$ is limited to values higher than 5 $M_\odot \, {\rm yr}^{-1}$. The \citet{Elbaz2007} (black dotted-dashed line), \citet{Daddi2007} (black solid line), and \citet{Heinis2014MNRAS.437.1268H} (thick grey dashed line) MS are included. We also show the MS with a slope of 0.75 and the same normalization of \citet{Daddi2007} to reflect the \citet{Rodighiero2010A&A...518L..25R} results at $z \leq 1$. We show in the bottom-right inset plot the evolution of the ${sSFR} = {SFR} / {M_*}$ for galaxies with $\log{\left( M_* / M_\odot \right)} \sim 10.5$. We also include the \citet{Dutton2010MNRAS.405.1690D} compilation with orange open squares.
              }
\label{smsmsmsmsmsmsmsmsms}
\end{figure}

The relation between SFR and stellar mass for our stacked massive galaxies is shown in Figure \ref{smsmsmsmsmsmsmsmsms}. Most previous works agree that there is a relation between SFR and stellar mass for normal SF galaxies, the so-called main sequence \citep[MS,][]{Daddi2007,Elbaz2007}. Despite there are many evidence that the MS exists, there is no consensus about its slope at different redshifts. This is mainly due to the different methods that can be employed to derive the stellar mass and total SFR of SF galaxies. Our stacking analysis reveals that the slope of the MS for massive ($\log{\left( M_* / M_\odot \right)} \geq 10$) SF galaxies is less stepper than previously reported \citep{Elbaz2007,Daddi2007}, mainly at our highest redshifts. We cannot detect less massive galaxies with stacking under the depths of the PACS images used here. Thus, it might be possible that there is a break of the MS in massive galaxies, being the slope steeper for galaxies with lower stellar masses and less steep for the most massive galaxies at each redshift. Deeper FIR data would be needed to confirm this. The results shown in \cite{Heinis2014MNRAS.437.1268H} also indicate that the slope of the MS might be lower for massive galaxies. They fit a linear relation to their points, although the points tend to follow the \cite{Daddi2007} relation for their lowest massive galaxies and then they flattens at higher stellar masses. This flattening would be in agreement with a break in the SFR-mass relation and it is consistent with our results at the same redshift and over the same stellar mass range. At $z \sim 1$, our results are in agreement with \cite{Rodighiero2010A&A...518L..25R}. At higher redshifts, \cite{Rodighiero2010A&A...518L..25R} obtained higher slopes although their galaxies are selected by having a blue optical color and therefore might be biased towards bluer galaxies than those studied here, that cover a wider range of optical colors. It might be argued that the flattening of the MS at the highest redshifts is due to the presence of quiescent galaxies. Despite we have selected our galaxies for being SF due to their SFR, we have repeated the stacking when increasing the ${\rm SFR_{UV}}$ threshold to 5 $M_\odot \, {\rm yr}^{-1}$ for galaxies at $z \geq 1$ (again, before any dust correction, being the real SFR even higher) to include only more active SF galaxies. It can be seen in Figure \ref{smsmsmsmsmsmsmsmsms} that even for these more active SF galaxies the slope of the MS is lower than previously reported, suggesting that the flattening of the MS in massive galaxies is not due to the presence of quiescent galaxies.

The redshift evolution of the specific SFR (${\rm sSFR} = {\rm SFR} / {\rm M_\odot}$) determined through stacking is in agreement with previous compilations obtained with different methodologies, increasing about one order of magnitude between $z \sim 0$ and $z \sim 2$. This is shown in the inset plot of Figure \ref{smsmsmsmsmsmsmsmsms} where we compare the sSFR obtained with stacking with the \cite{Dutton2010MNRAS.405.1690D} compilation at the same redshift range for galaxies with $\log{\left( M_* / M_\odot \right)} \sim 10.5$.

%

\section{Conclusions}\label{concluuuuu}

By using a stacking analysis in \emph{Herschel} bands, we have studied the accuracy of several dust-correction factors traditionally employed to recover the total SFR of high-redshift star-forming (SF) galaxies. Our main conclusions are:

   \begin{enumerate}
   
	\item We find that the dust attenuation in massive ($\log{\left( M_* / M_\odot \right)} > 10$) SF galaxies is higher at $z \sim 2-3$ than at $z \sim 1$ for a given UV continuum slope and stellar mass. This is consistent what it was previously found with individual \emph{Herschel} detections. At $z \lesssim 1$, where stacking is able to detected less massive, we do not find significant evolution of dust attenuation for a given stellar mass with respect to galaxies in the local universe.
	
	\item There is not a single IRX-$\beta$ relation that can be applied to accurately recover the dust attenuation or total SFR of massive SF galaxies at any redshifts. Low-redshift galaxies are better parametrized by the \cite{Takeuchi2012} IRX-$\beta$ relation, but the \cite{Buat2012} one is more appropriate at $z \sim 2$, when dust attenuation was higher for a given UV slope. This might have implications in high-redshift studies, where a single IRX-$\beta$ relation is normally assumed to correct for dust attenuation. The SED-derived dust attenuation gives over-estimated total SFRs at all redshifts studied here.
      
      \item Despite the dust attenuation of massive galaxies is higher at $z \sim 2-3$ than at $z \sim 1$, our stacking analysis indicates that the local relation between dust attenuation and stellar mass is valid up to $z \sim 1$. Since we do not recover stacked detections at $z \geq 1.5$ for $\log{\left( M_* / M_\odot \right)} \lesssim 10$ galaxies we cannot confirm the non evolution of the local dust-mass relation in low-mass galaxies at higher redshift reported in previous works.
            
      \item We obtain that the slope of the MS of massive SF galaxies is lower than previously reported. Since we do not recover stacked detection for galaxies with $\log{\left( M_* / M_\odot \right)} \lesssim 10$, we cannot distinguish if this is also true for less massive galaxies or there is a break in the MS for massive galaxies. The redshift evolution of the sSFR is in agreement with previous findings, increasing in about one order of magnitude from $z \sim 0.02$ to $z \sim 2$.
      
   \end{enumerate}

\begin{acknowledgements}
      We acknowledge the anonymous referee for his/her report, which has improved the presentations of the results. IO acknowledges support from the European Research Council (ERC) in the form of Advanced Grant, {\sc cosmicism}. 
      
\end{acknowledgements}


\appendix

\onecolumn

\section{Additional plots and table}\label{app_plots}


\begin{figure}
\centering
\includegraphics[width=0.19\textwidth]{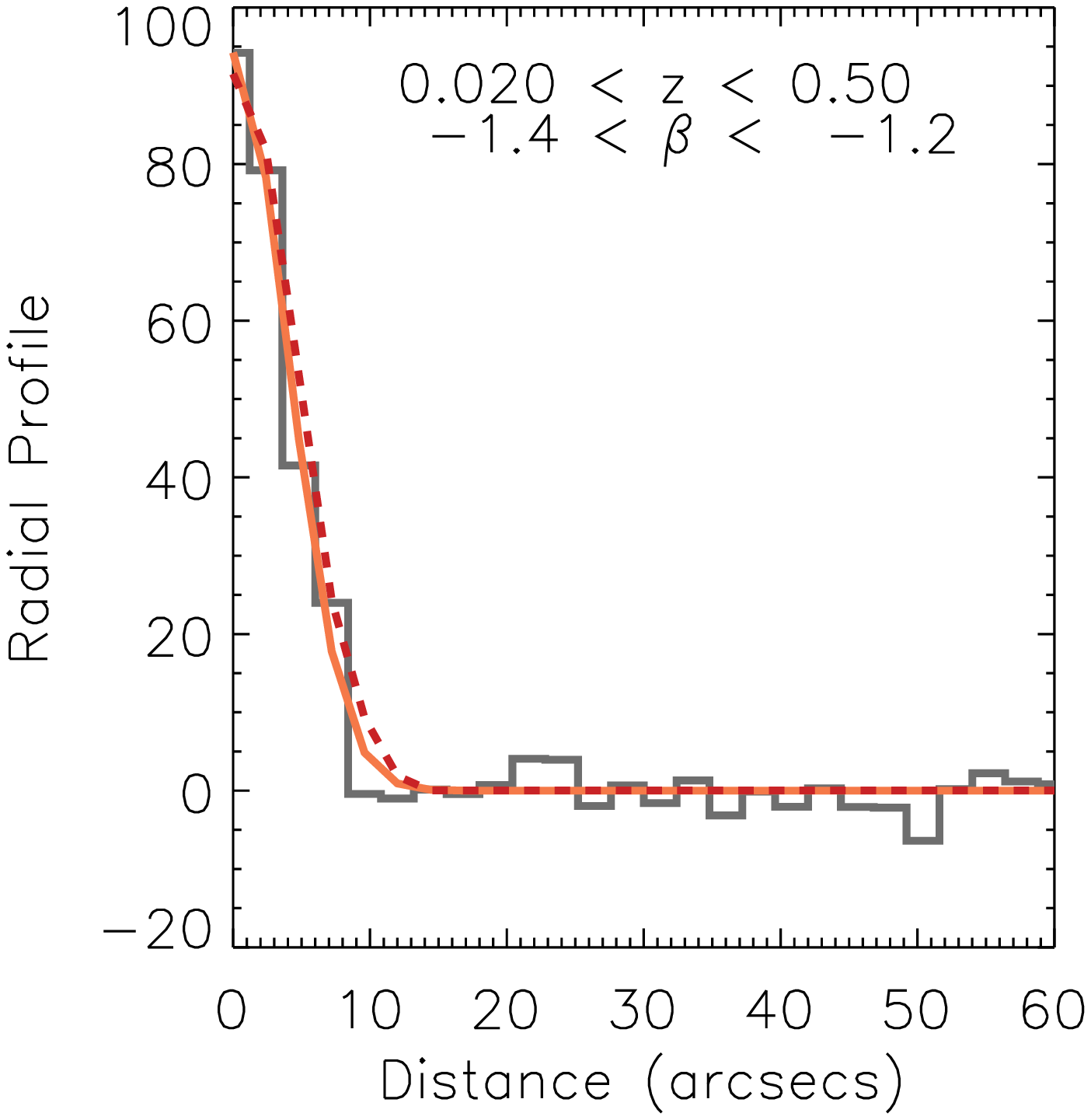}
\includegraphics[width=0.19\textwidth]{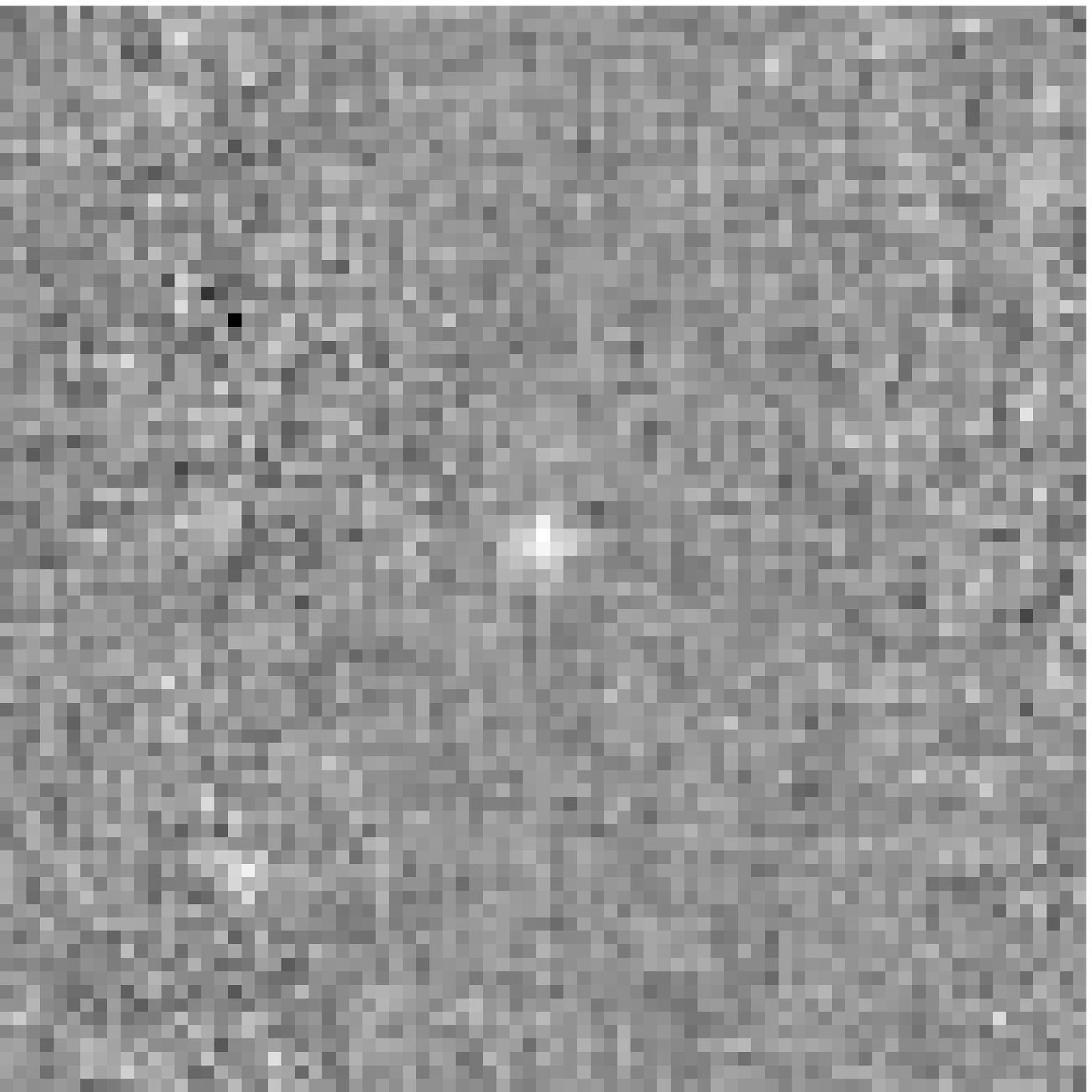}
\includegraphics[width=0.19\textwidth]{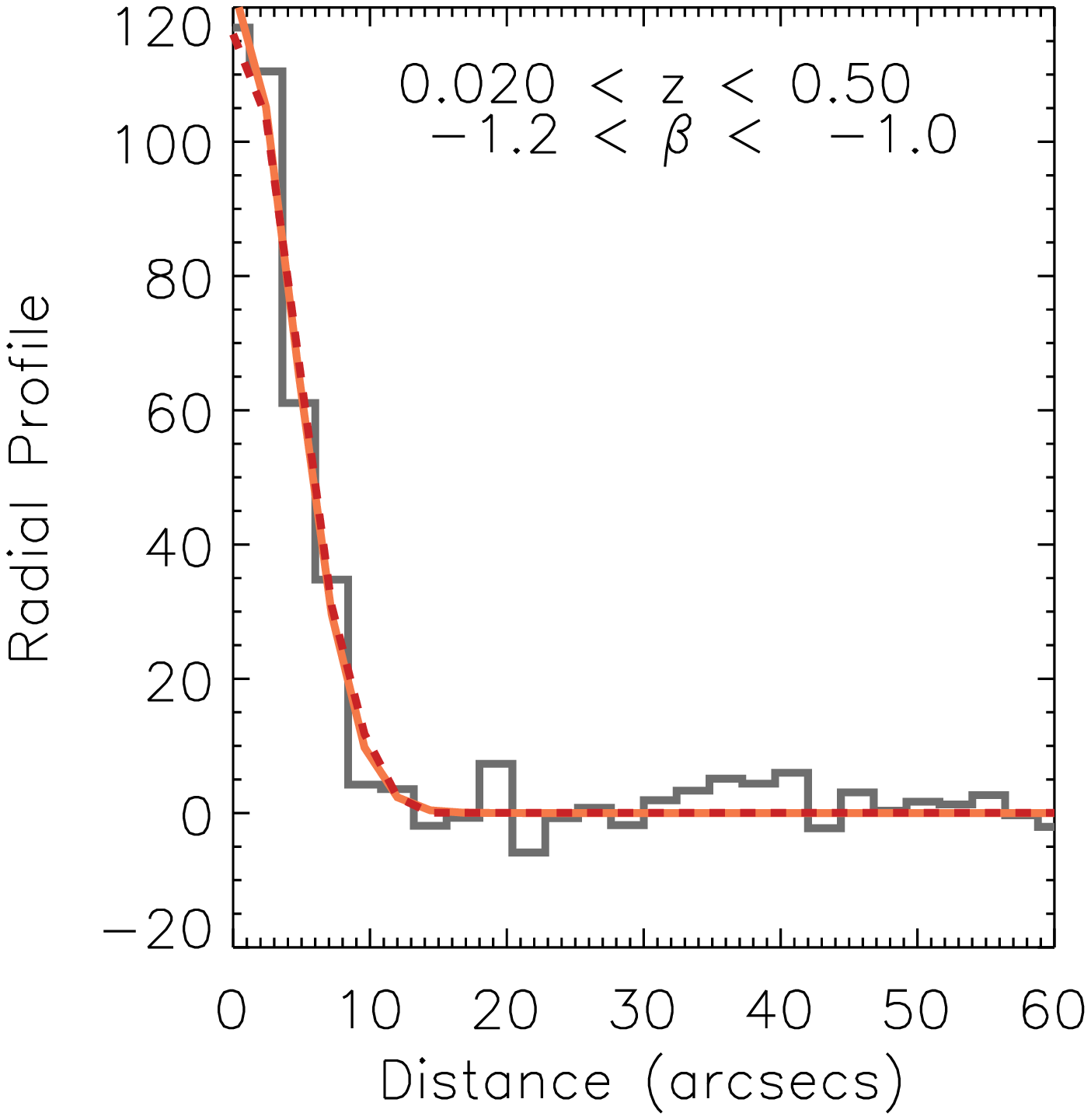}
\includegraphics[width=0.19\textwidth]{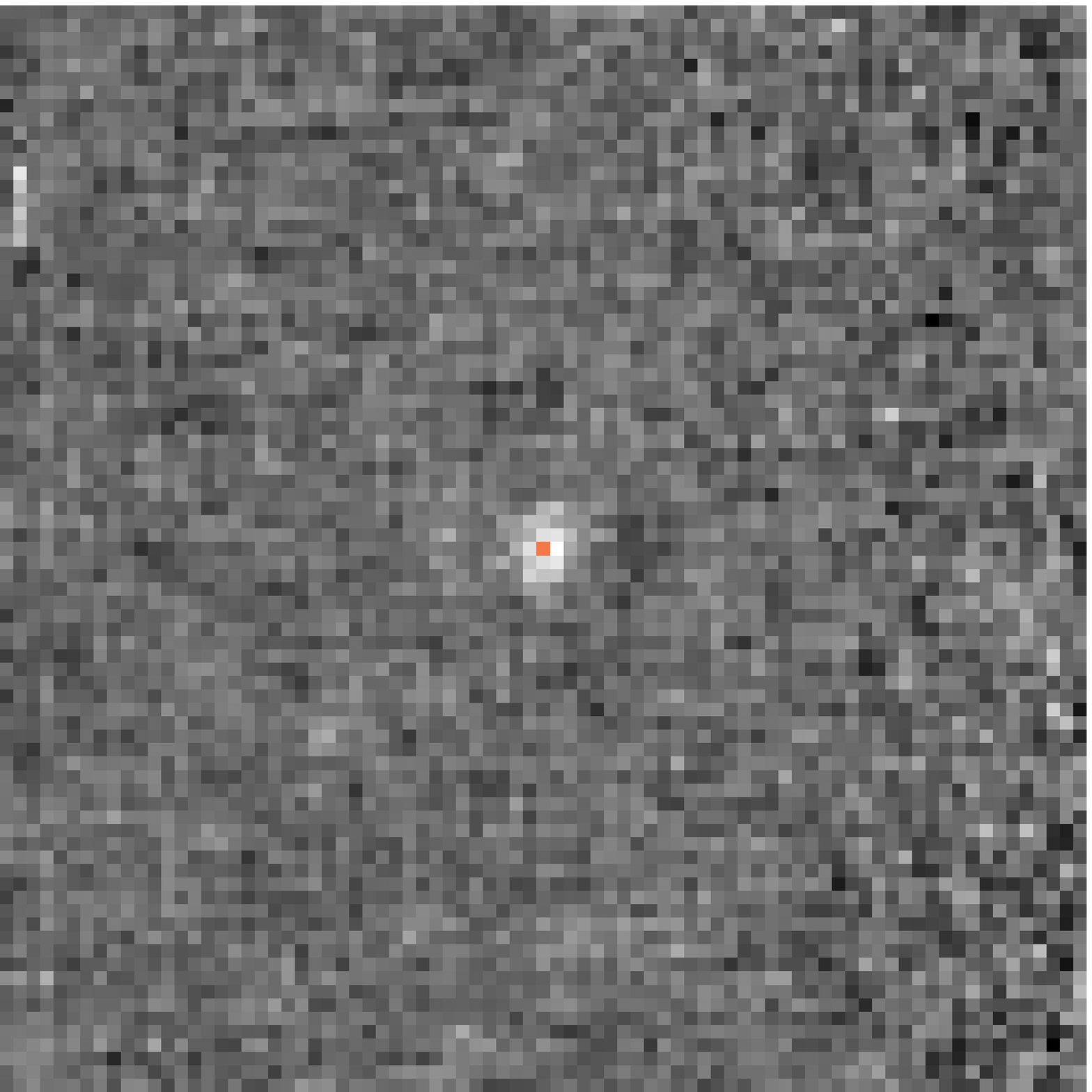}\\
\includegraphics[width=0.19\textwidth]{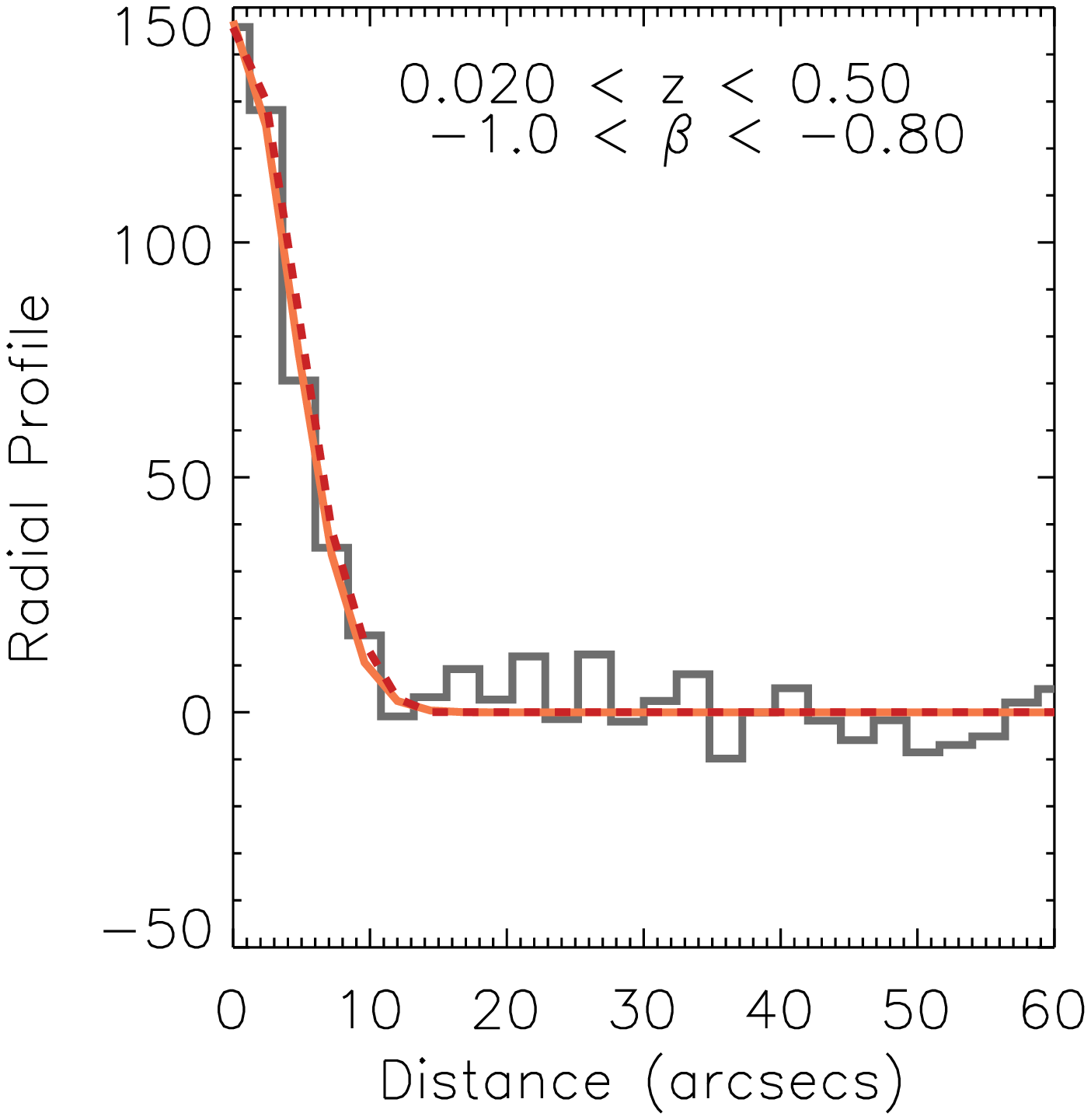}
\includegraphics[width=0.19\textwidth]{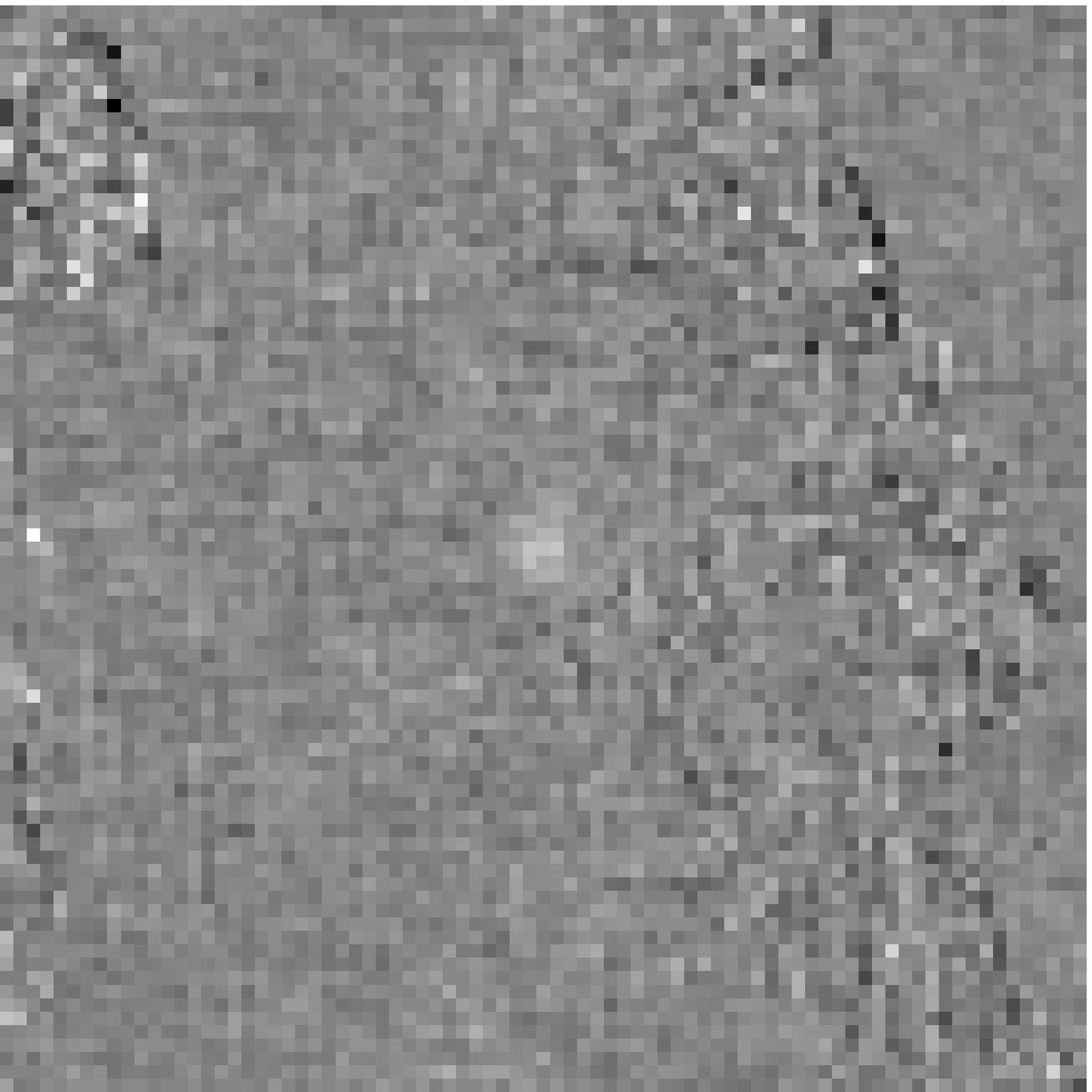}
\includegraphics[width=0.19\textwidth]{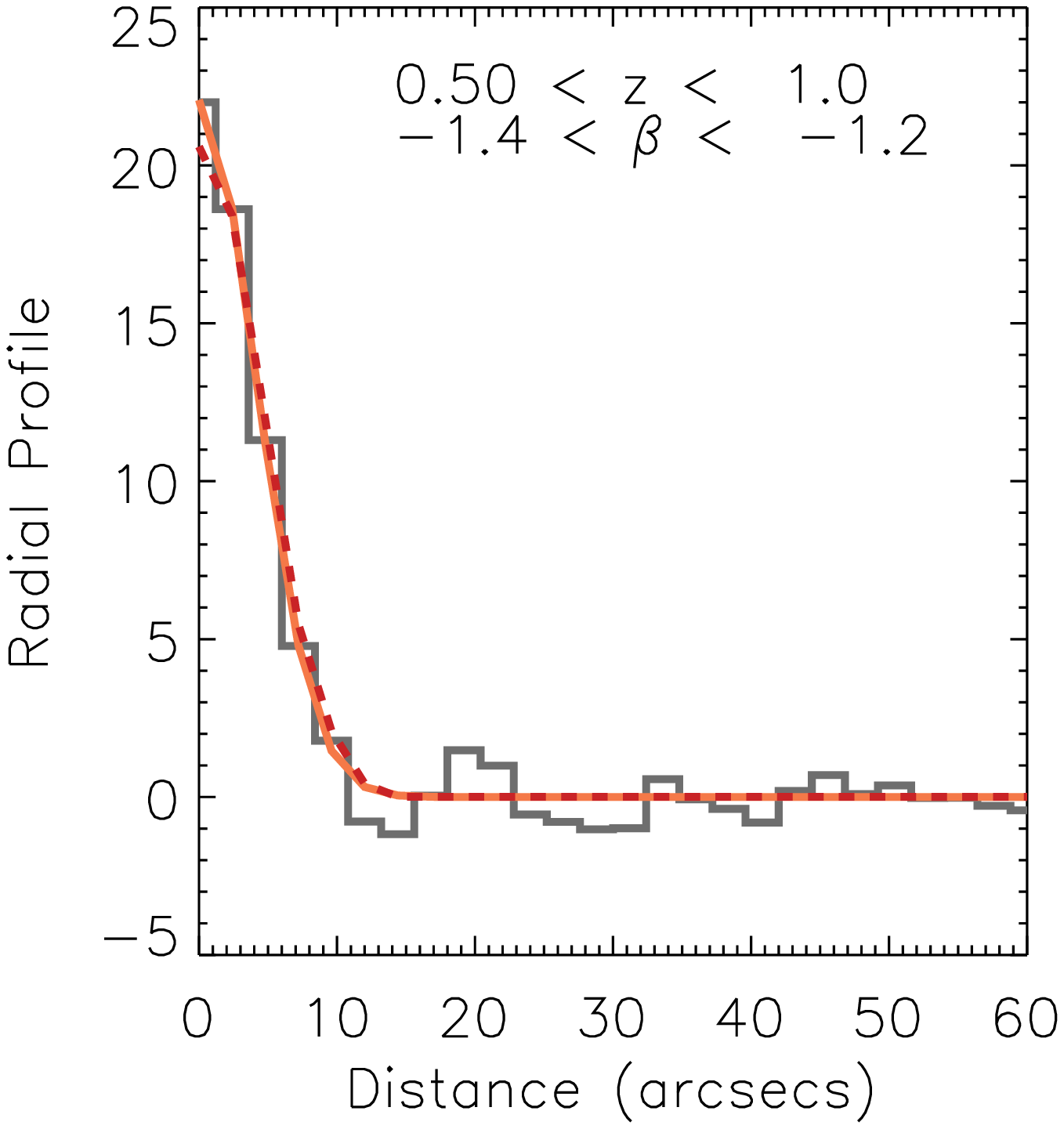}
\includegraphics[width=0.19\textwidth]{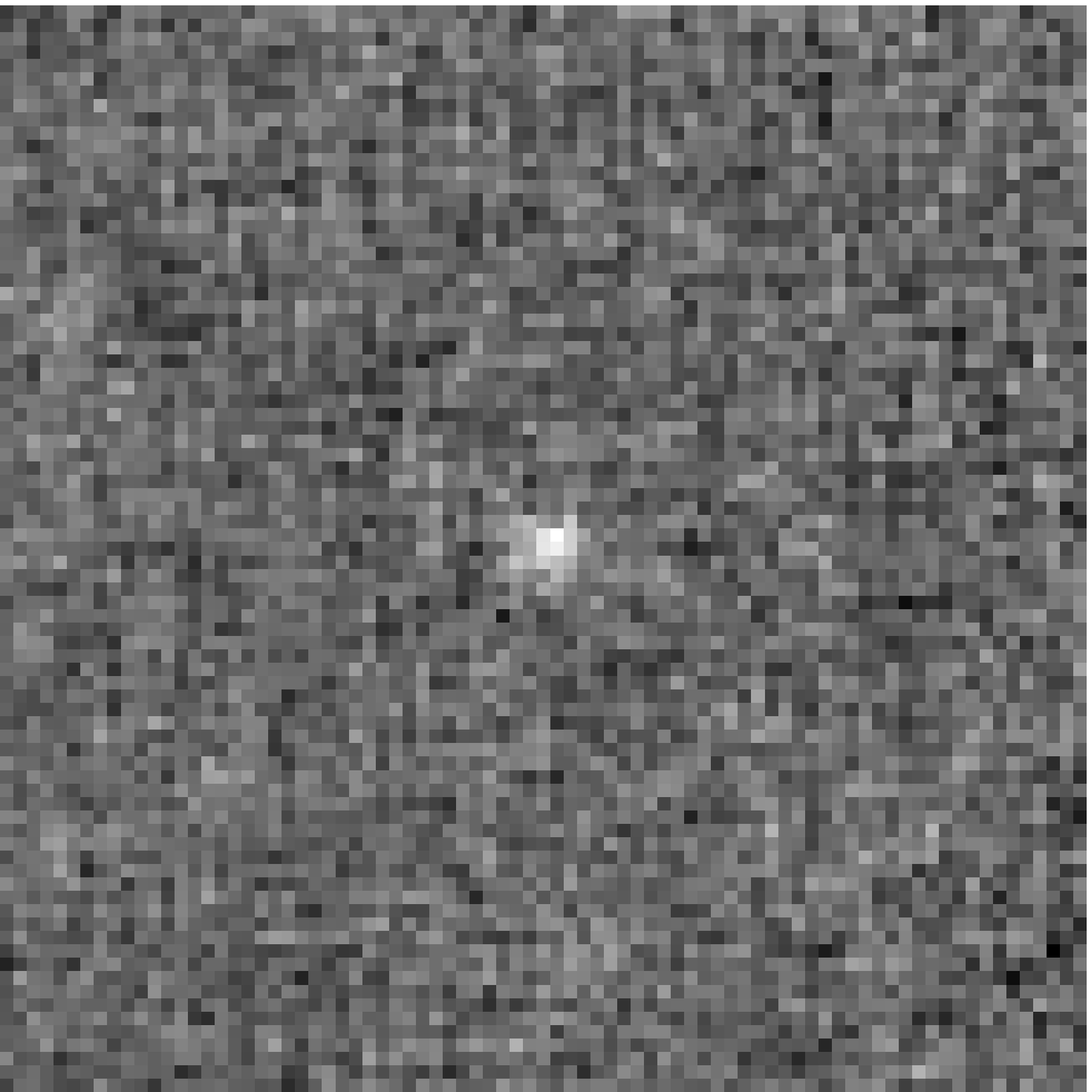}\\
\includegraphics[width=0.19\textwidth]{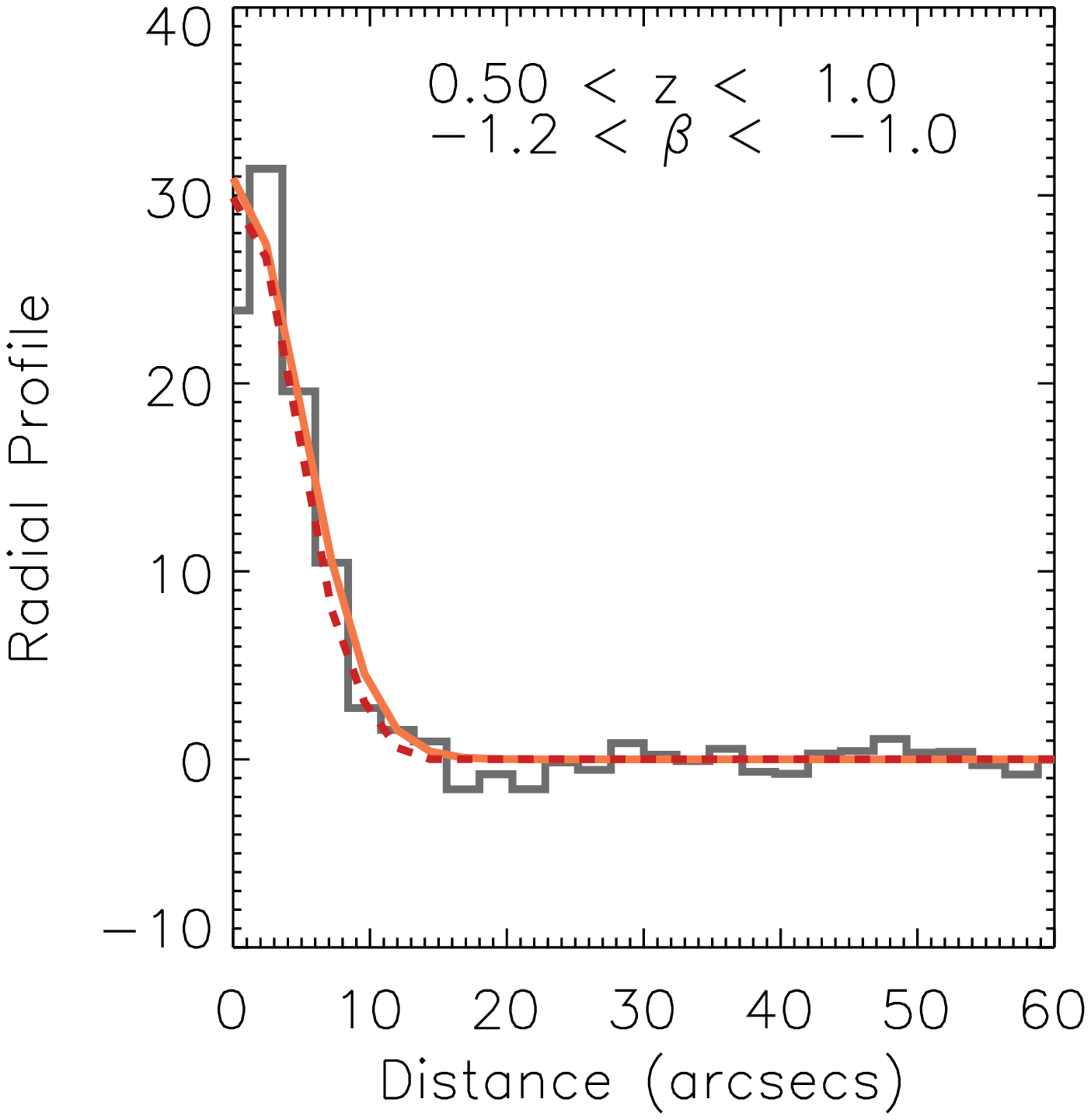}
\includegraphics[width=0.19\textwidth]{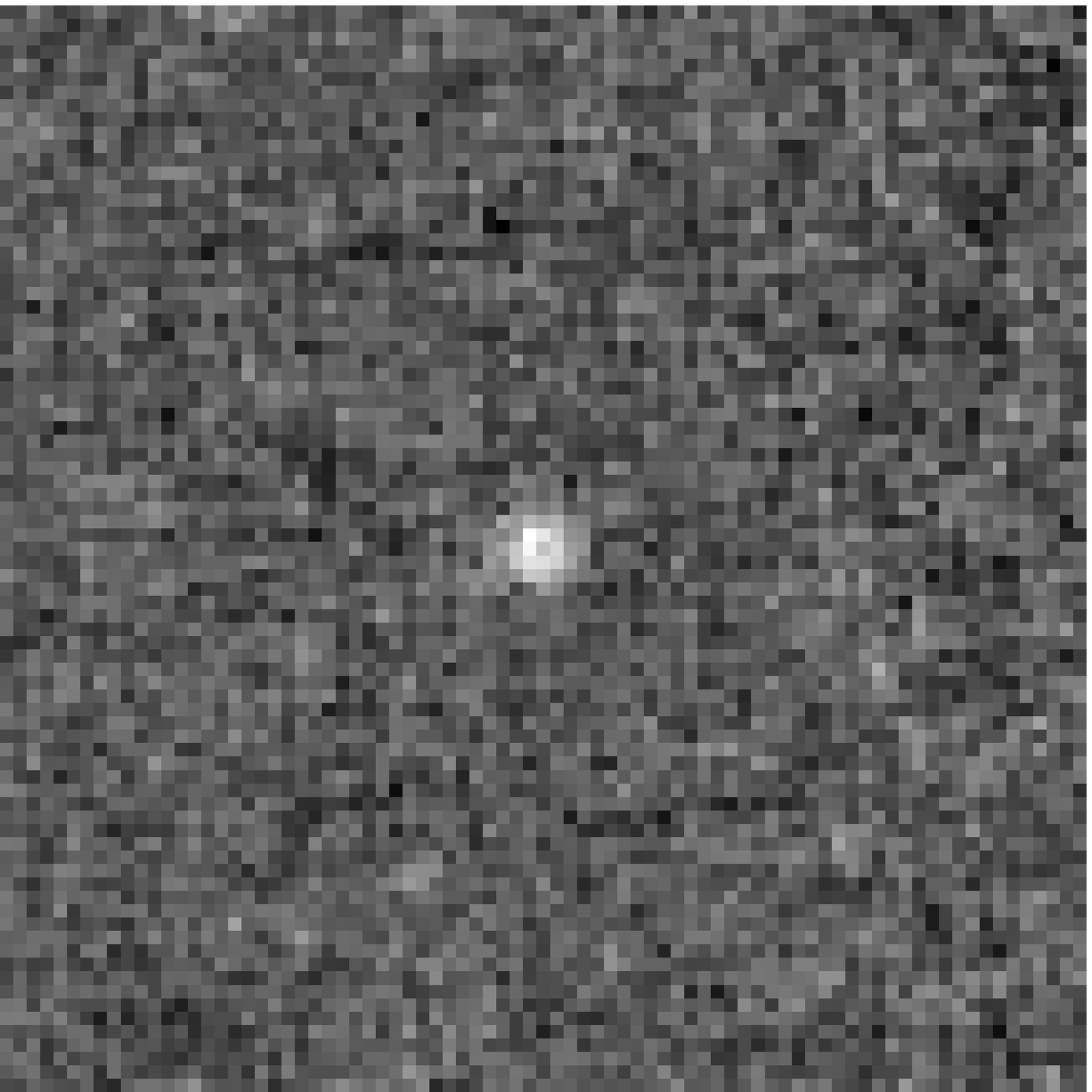}
\includegraphics[width=0.19\textwidth]{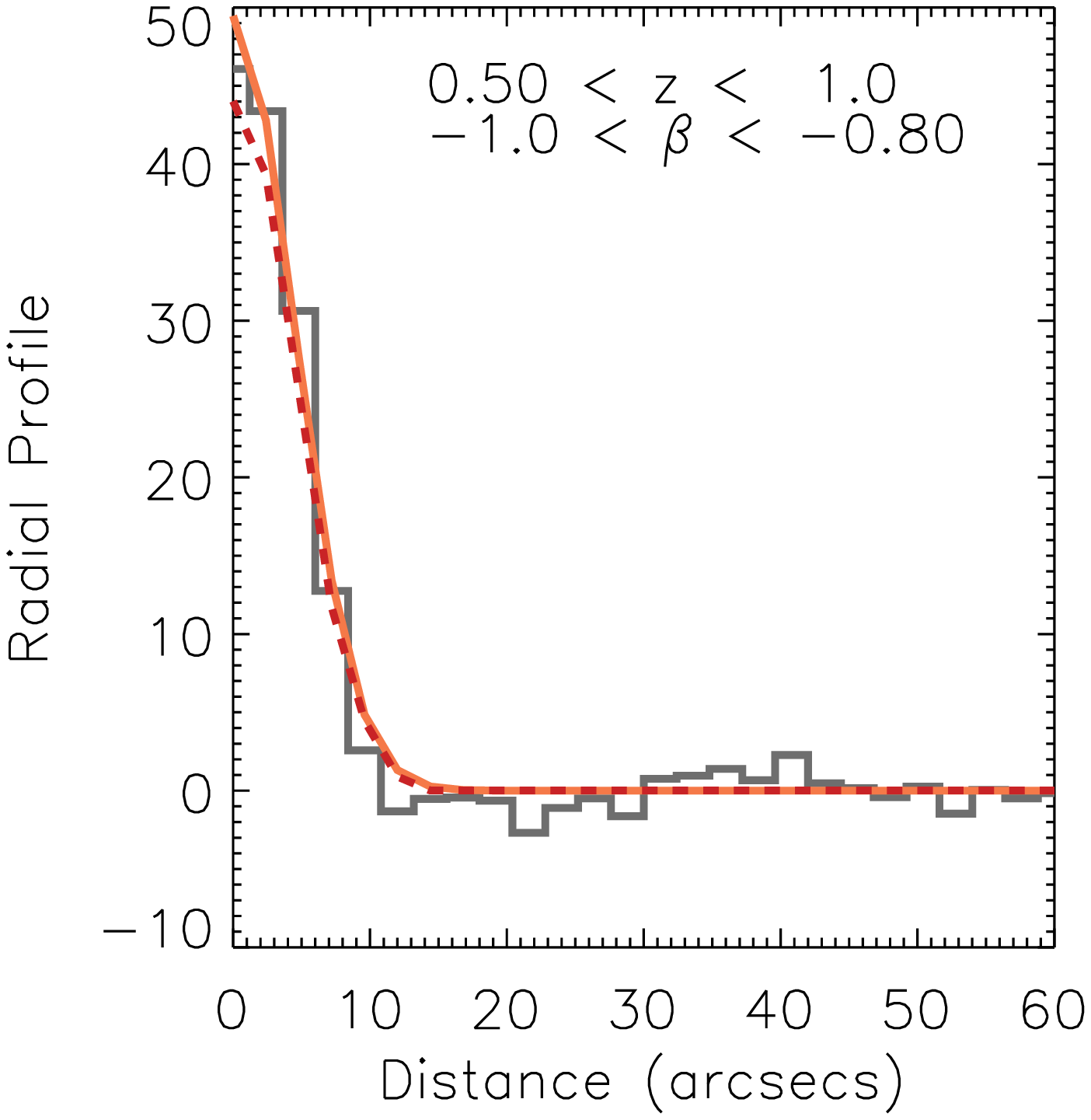}
\includegraphics[width=0.19\textwidth]{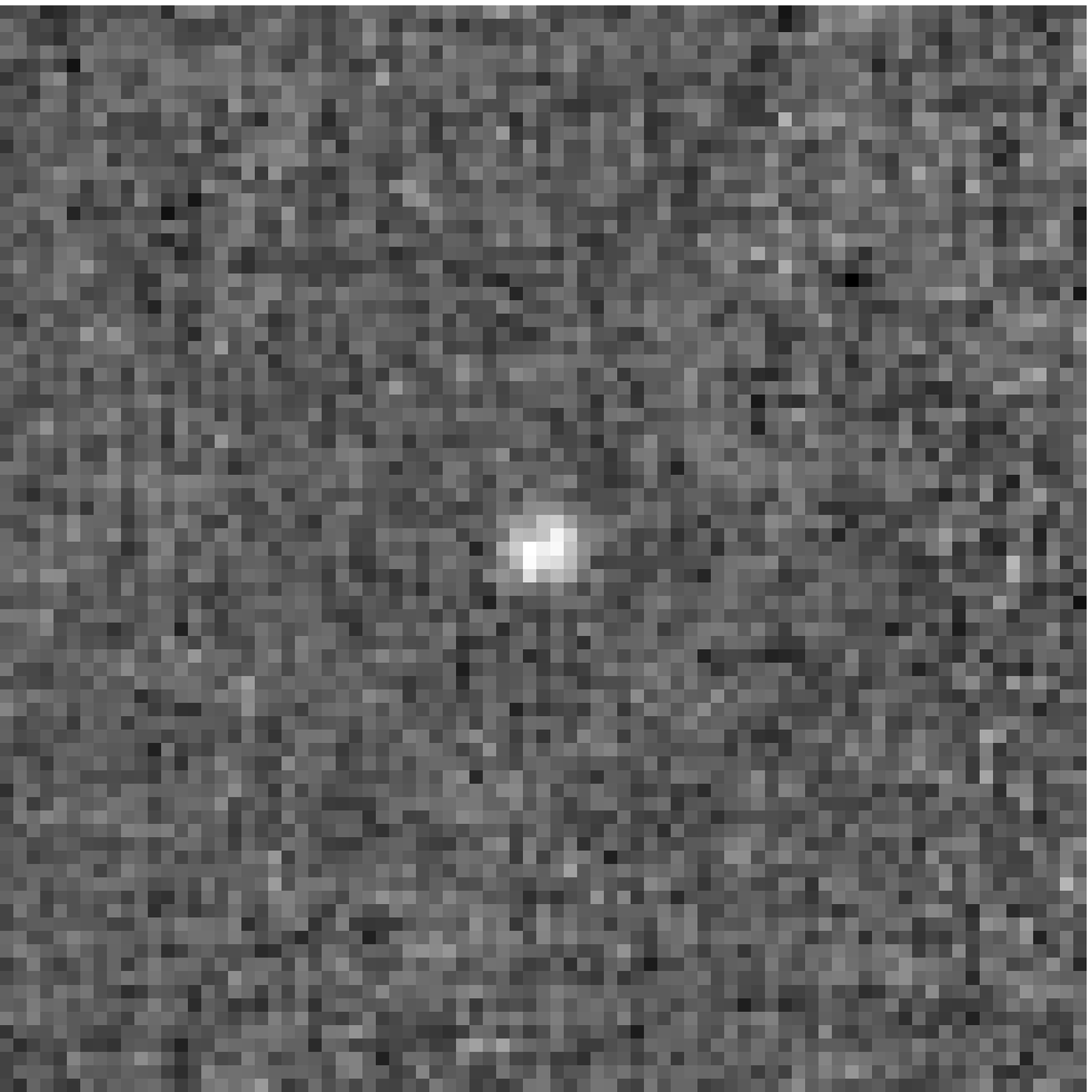}\\
\includegraphics[width=0.19\textwidth]{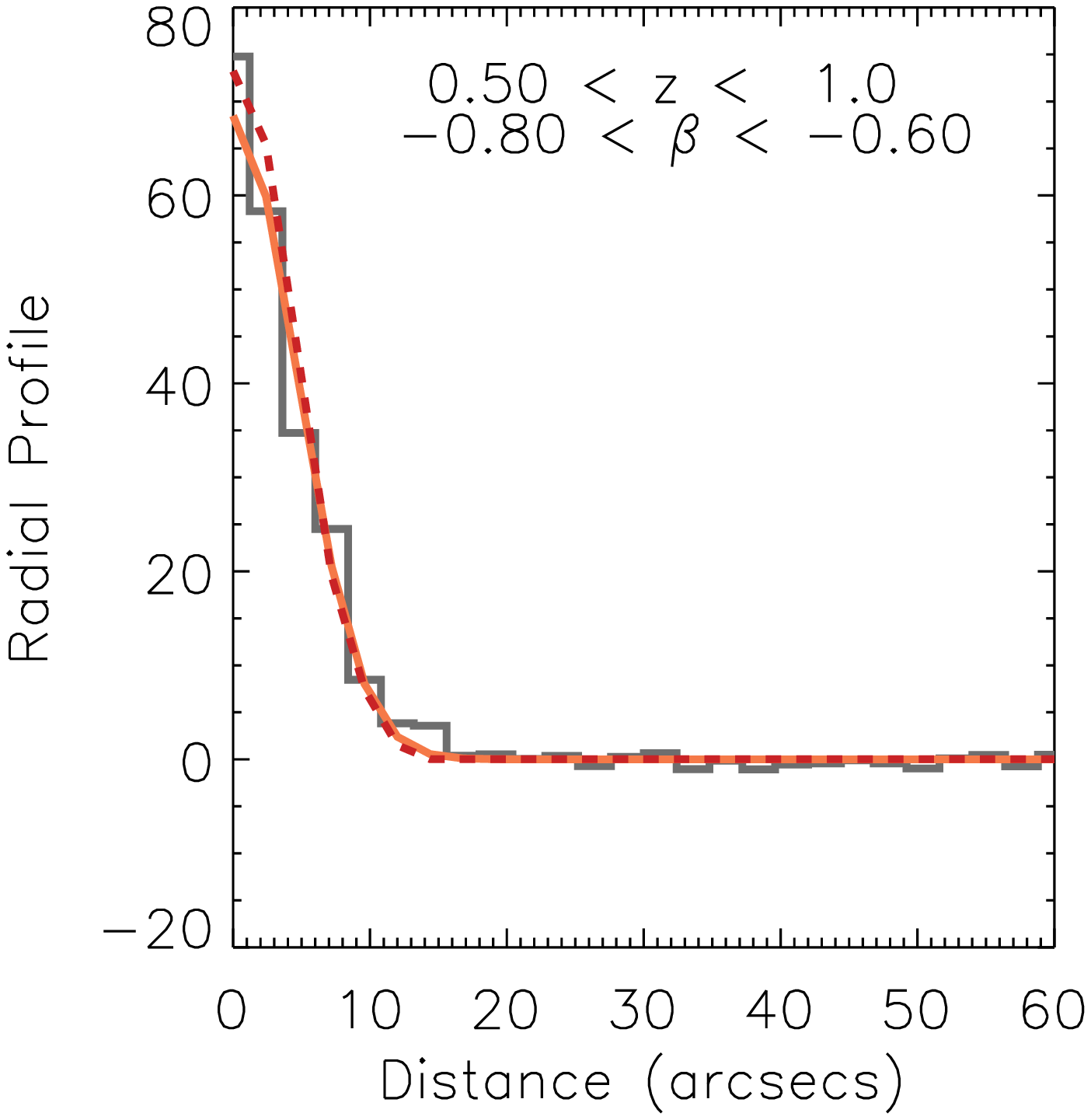}
\includegraphics[width=0.19\textwidth]{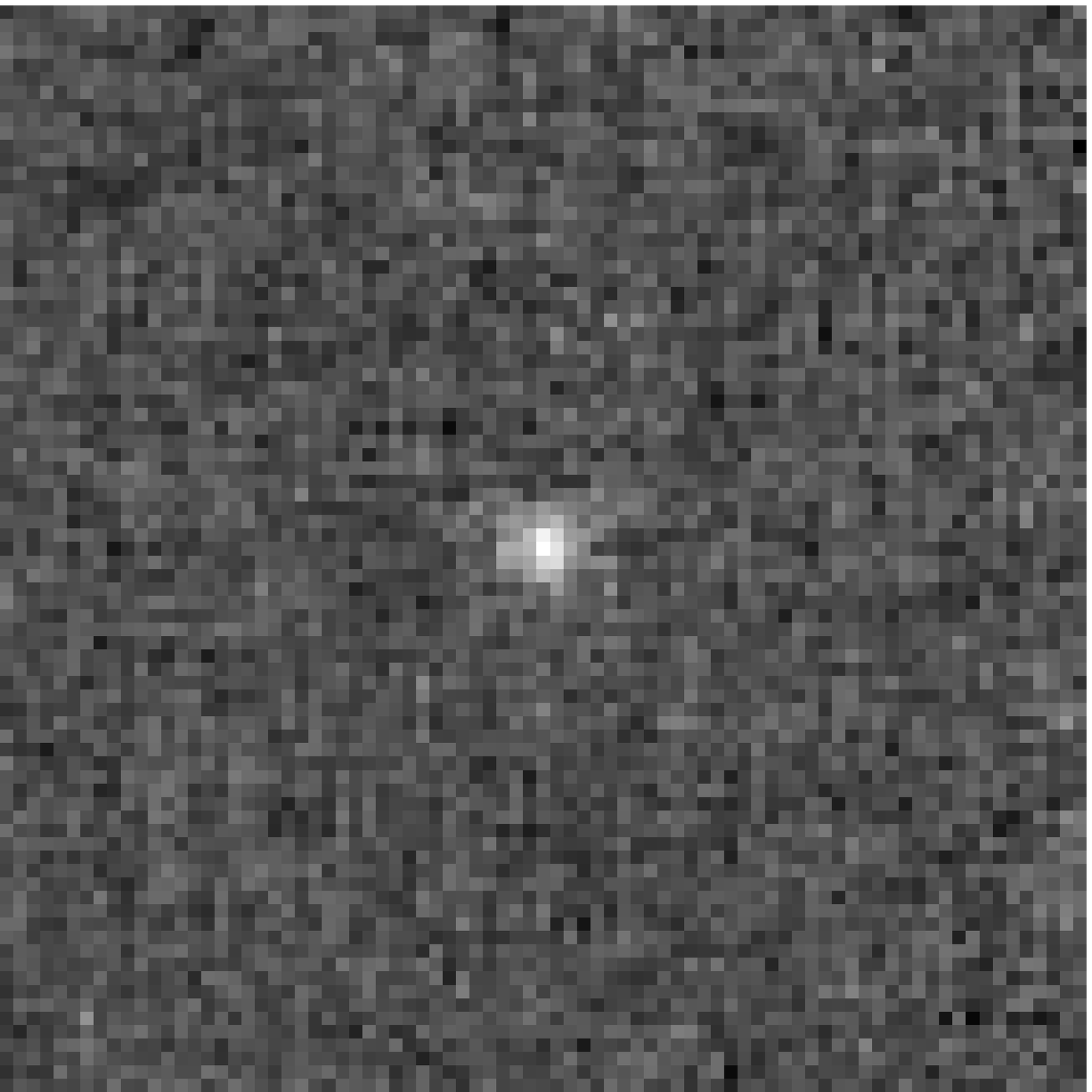}
\includegraphics[width=0.19\textwidth]{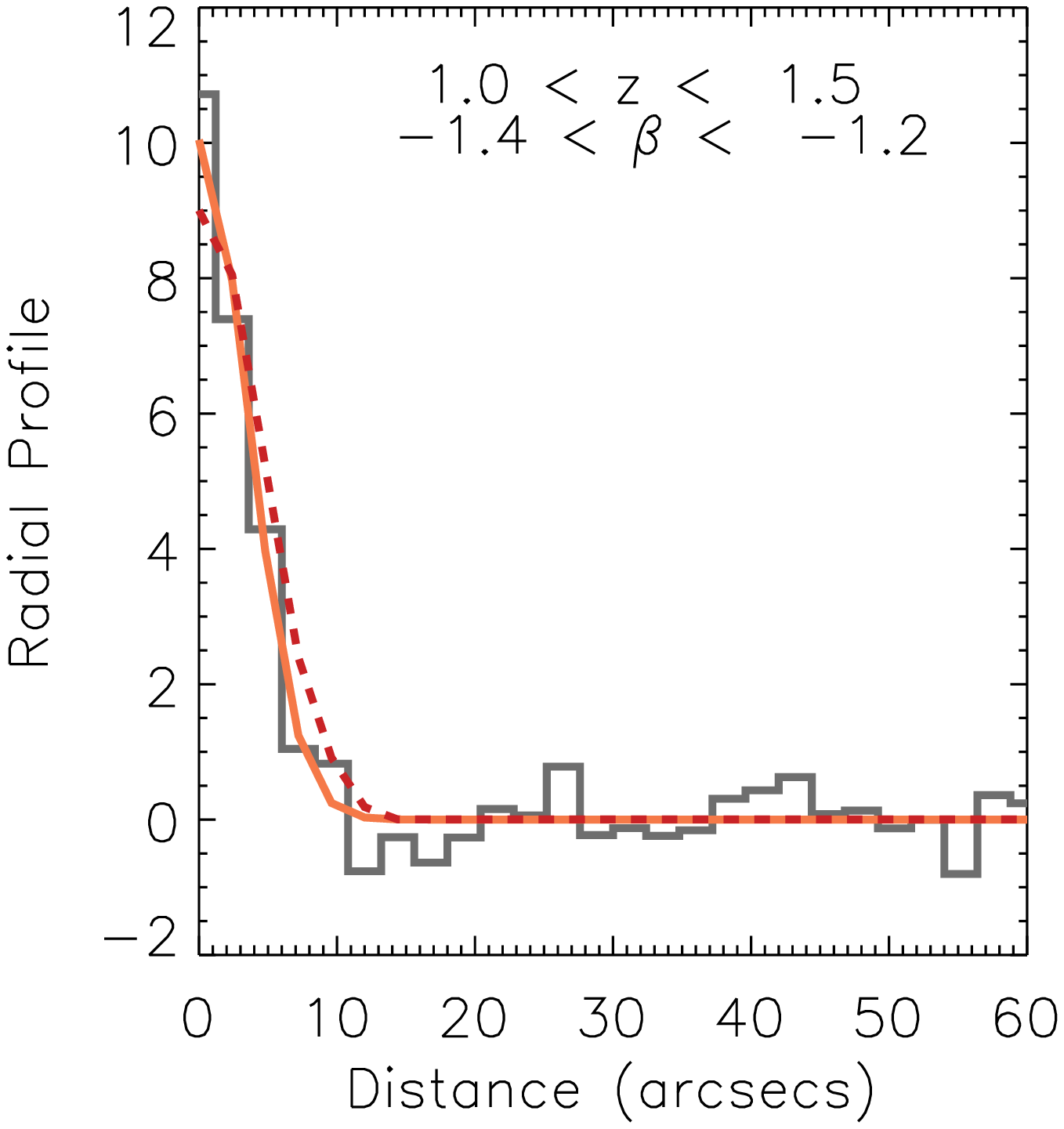}
\includegraphics[width=0.19\textwidth]{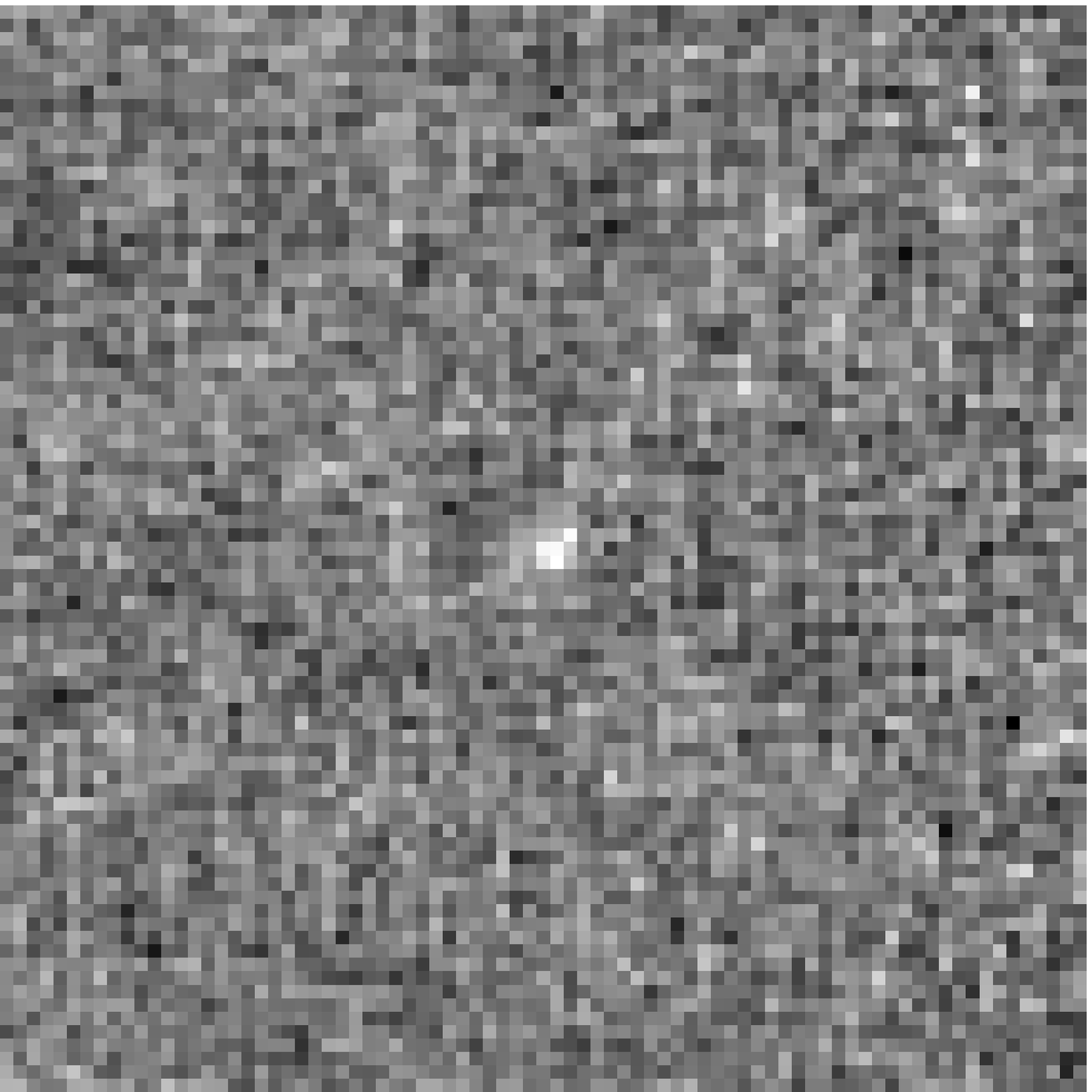}\\
\includegraphics[width=0.19\textwidth]{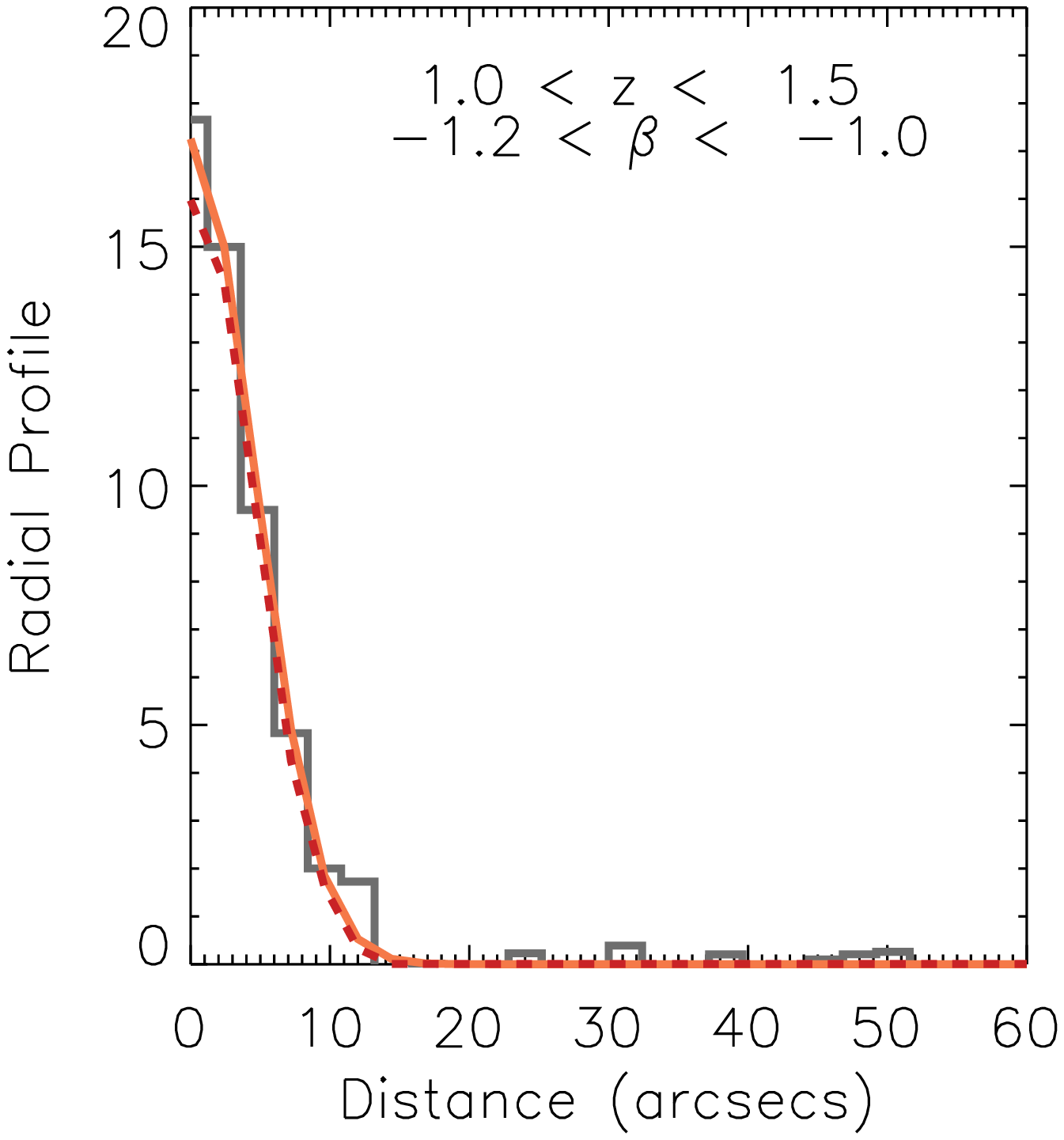}
\includegraphics[width=0.19\textwidth]{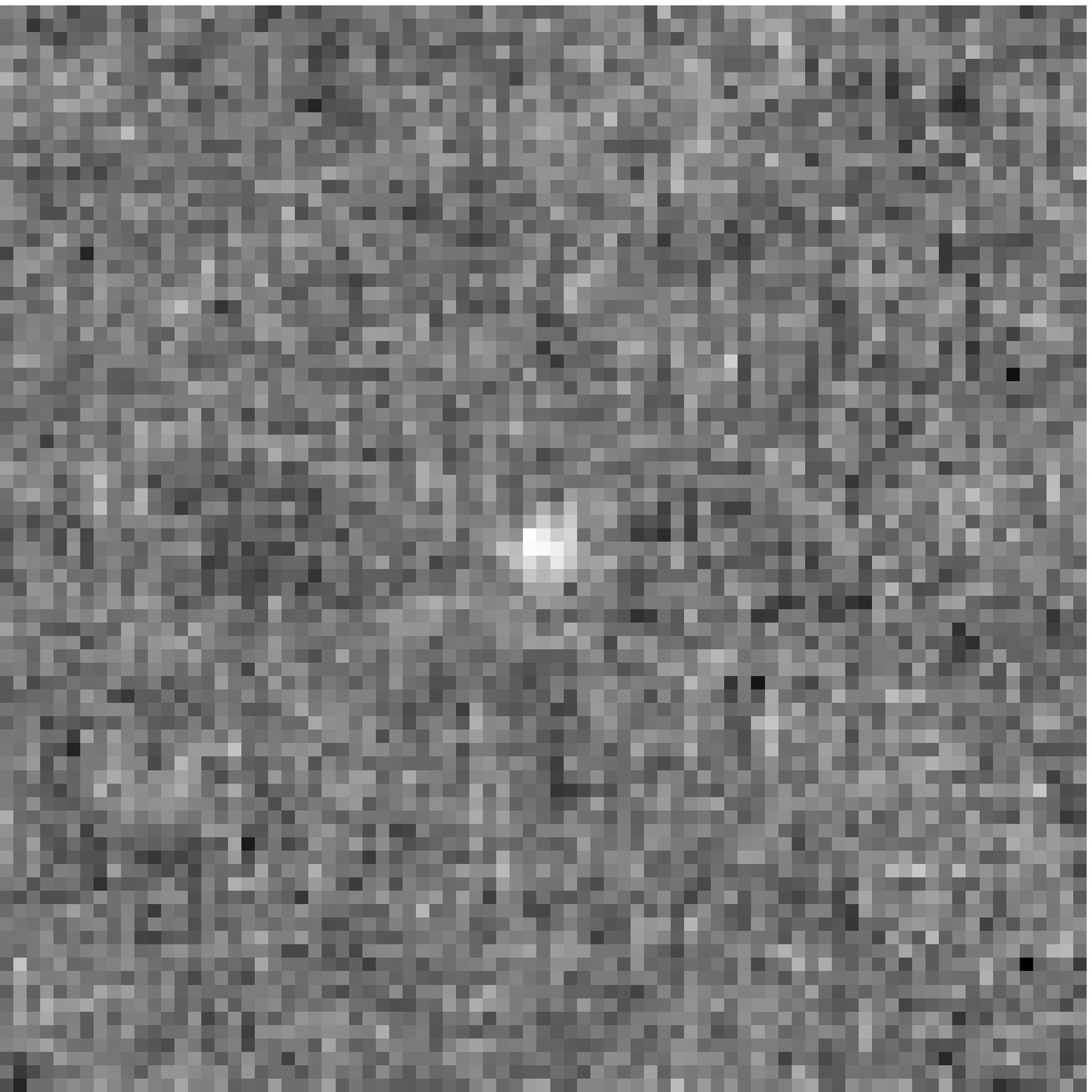}
\includegraphics[width=0.19\textwidth]{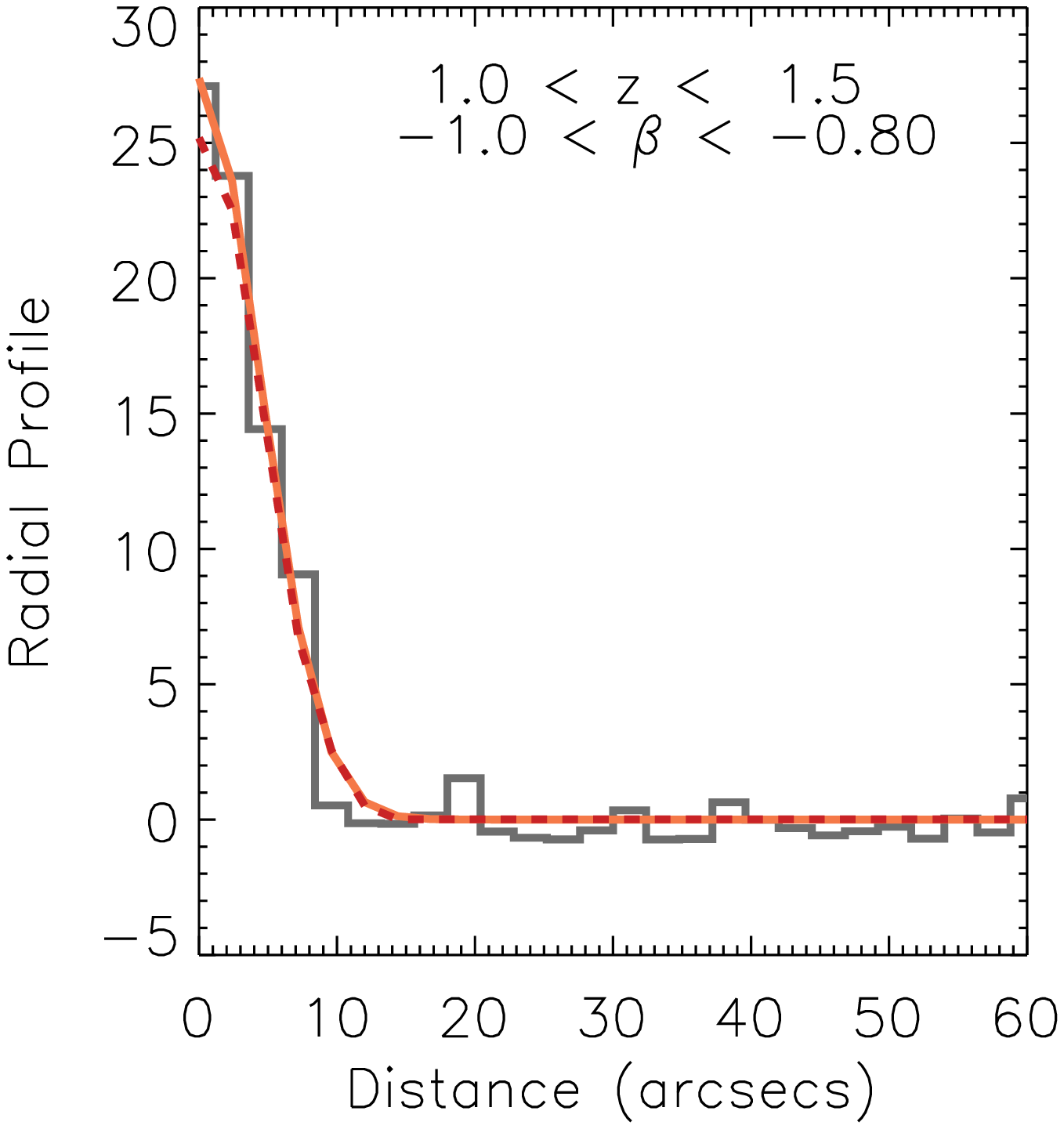}
\includegraphics[width=0.19\textwidth]{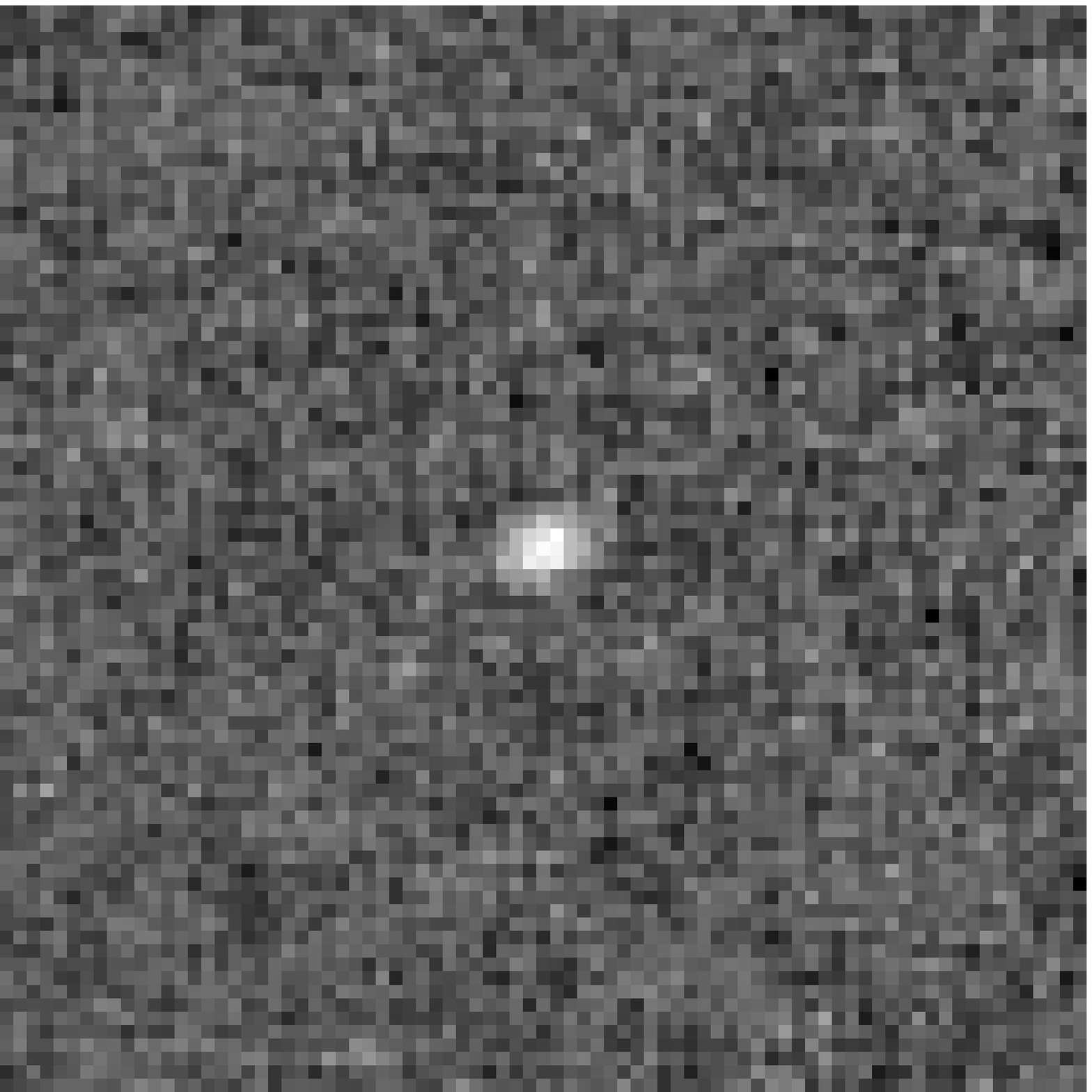}\\
\includegraphics[width=0.19\textwidth]{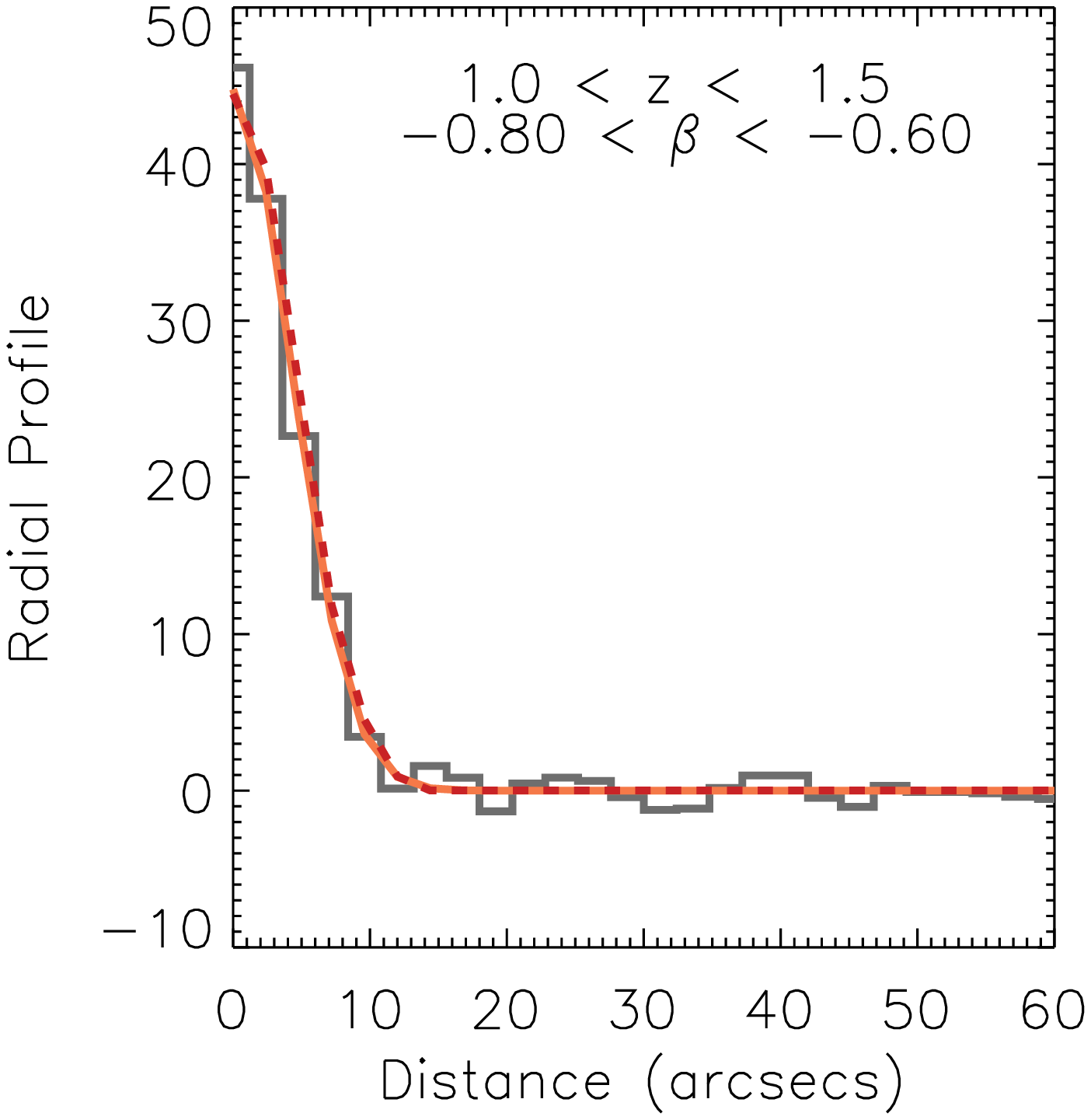}
\includegraphics[width=0.19\textwidth]{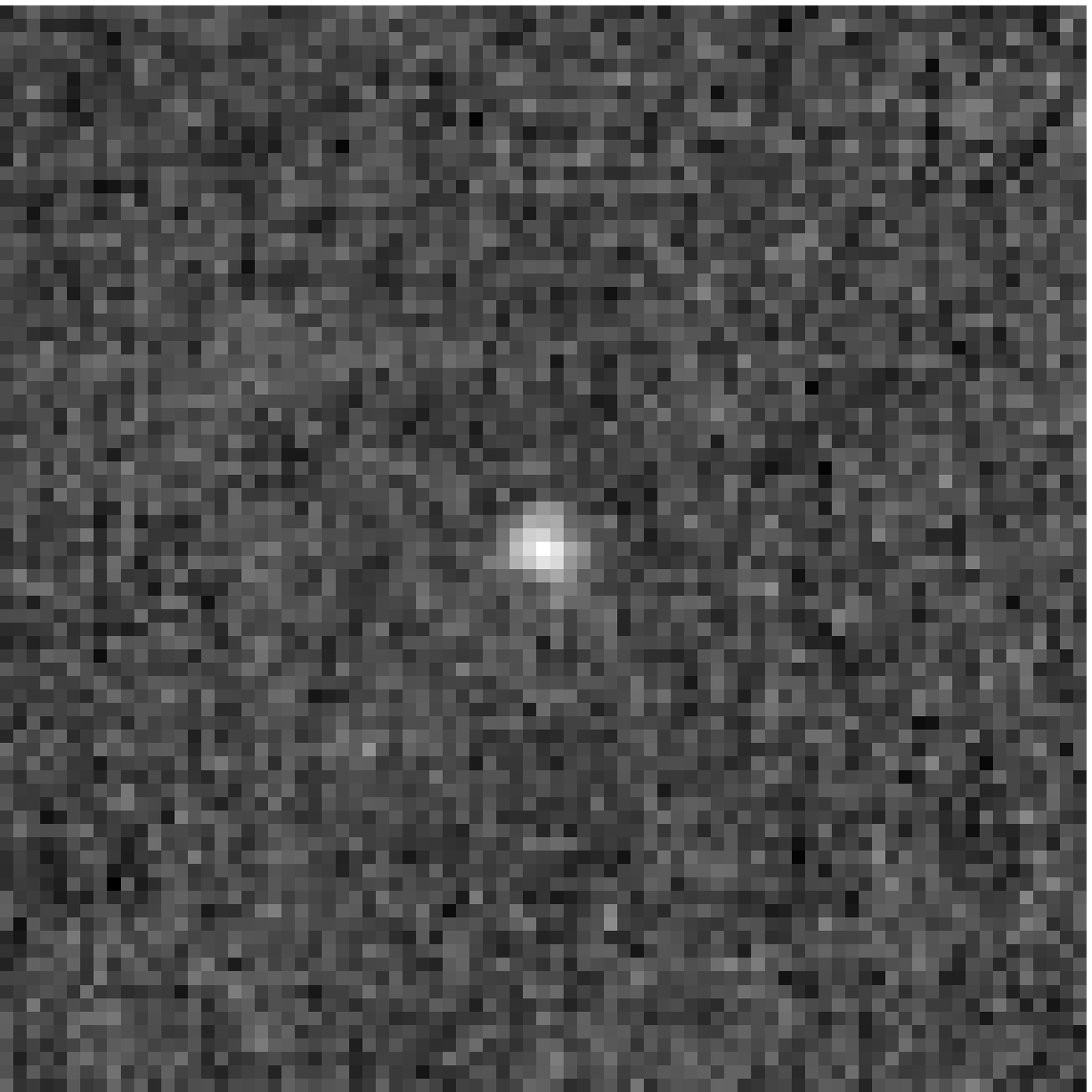}
\includegraphics[width=0.19\textwidth]{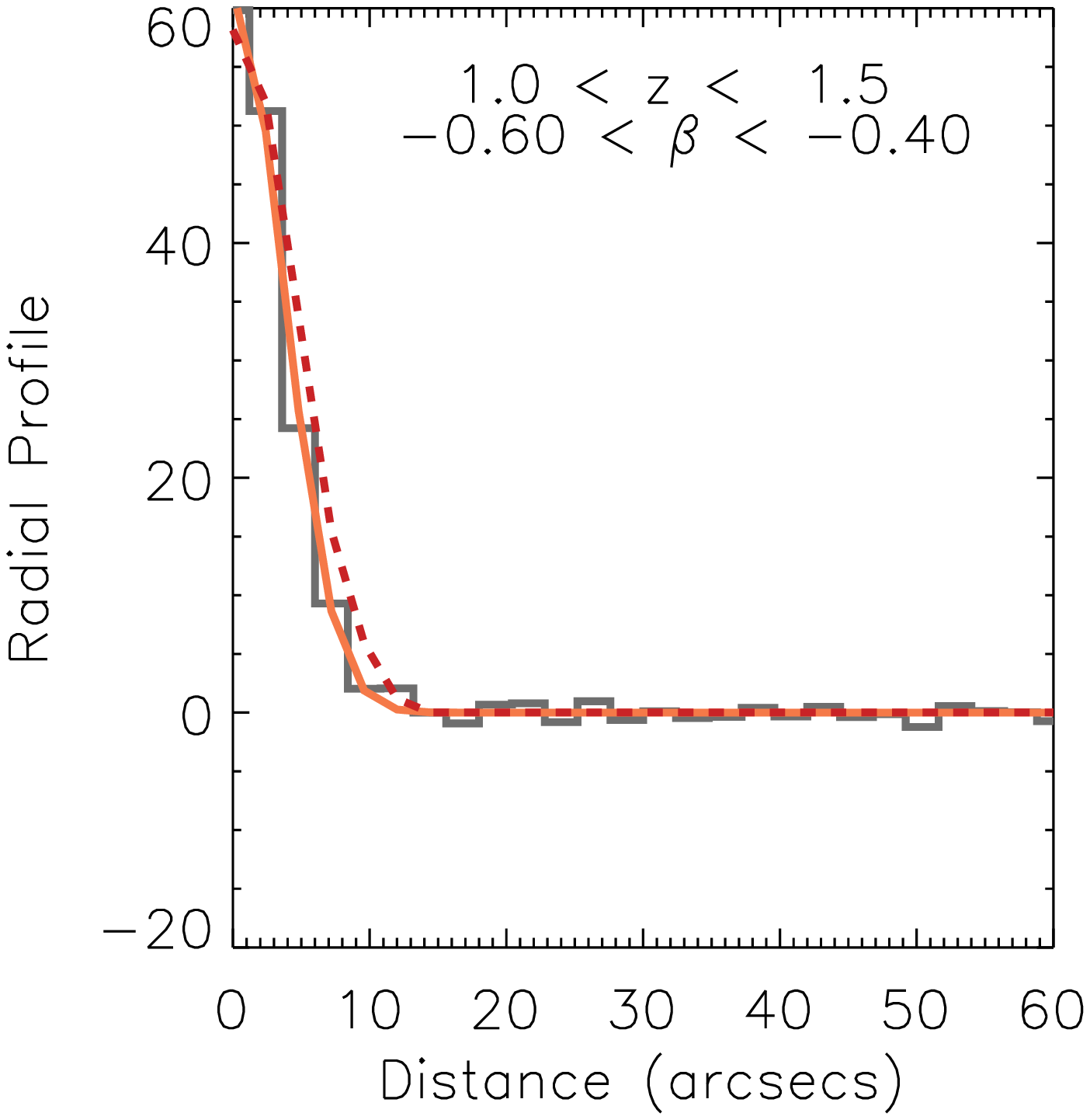}
\includegraphics[width=0.19\textwidth]{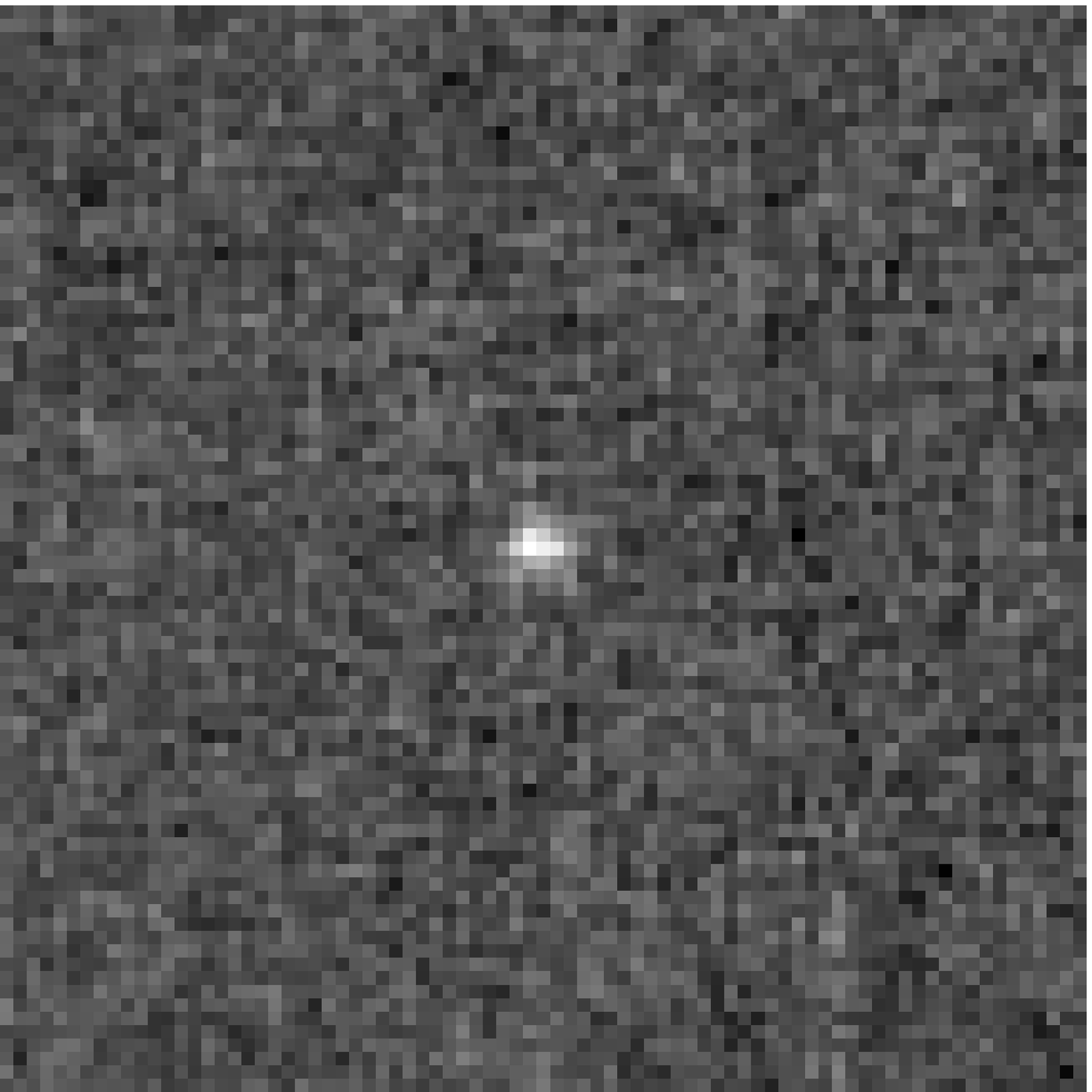}\\

\caption{Radial profiles and associated stacked images when stacking as a function of the UV continuum slope. Redshift and UV continuum bins are indicated in each case. Orange curves are the best-fitted gaussian to the radial profile, while red dashed curves represent the shape of the PACS-160 $\mu$m PSF.
              }
\label{images1}
\end{figure}

\begin{figure}
\centering
\includegraphics[width=0.19\textwidth]{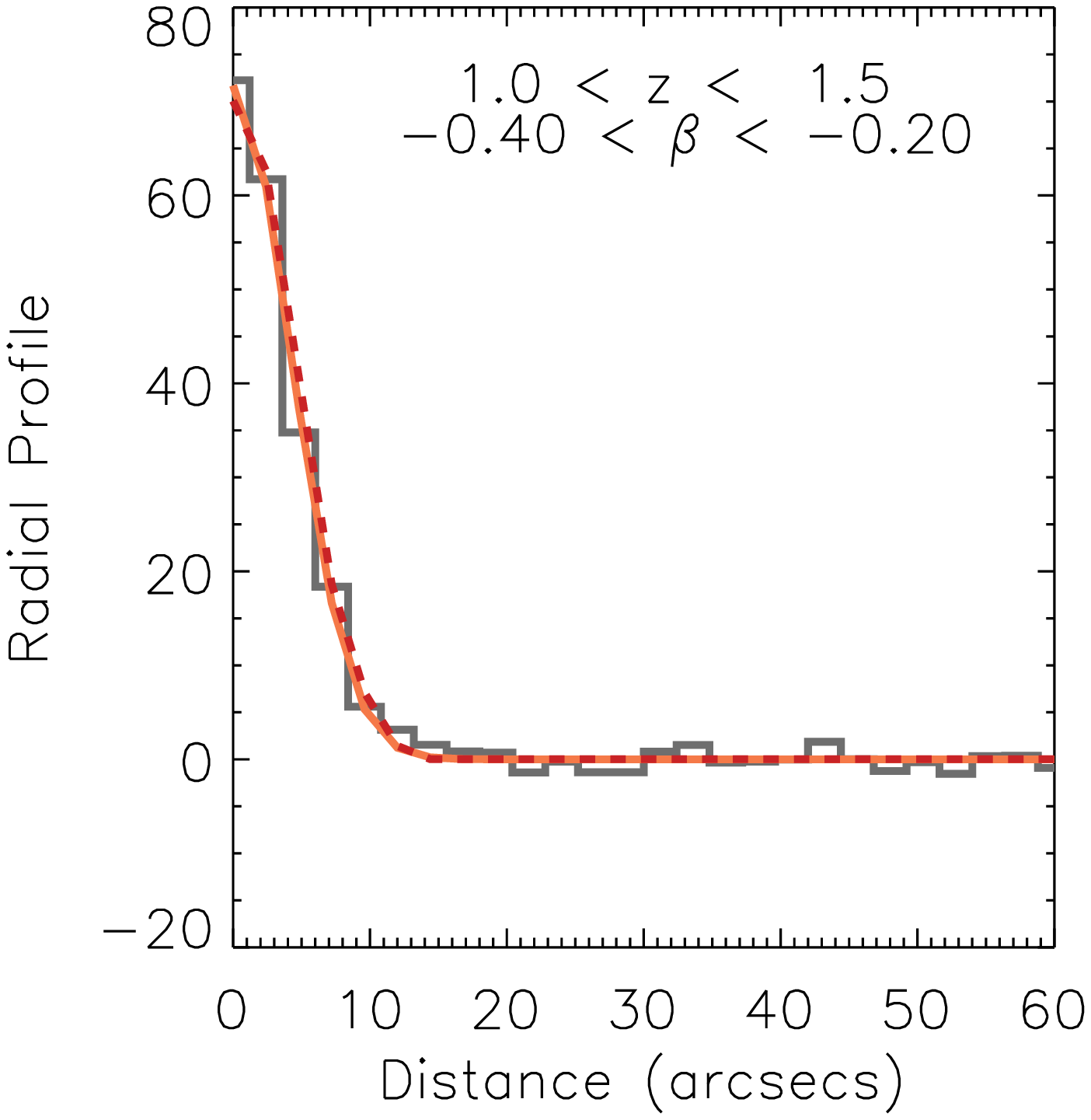}
\includegraphics[width=0.19\textwidth]{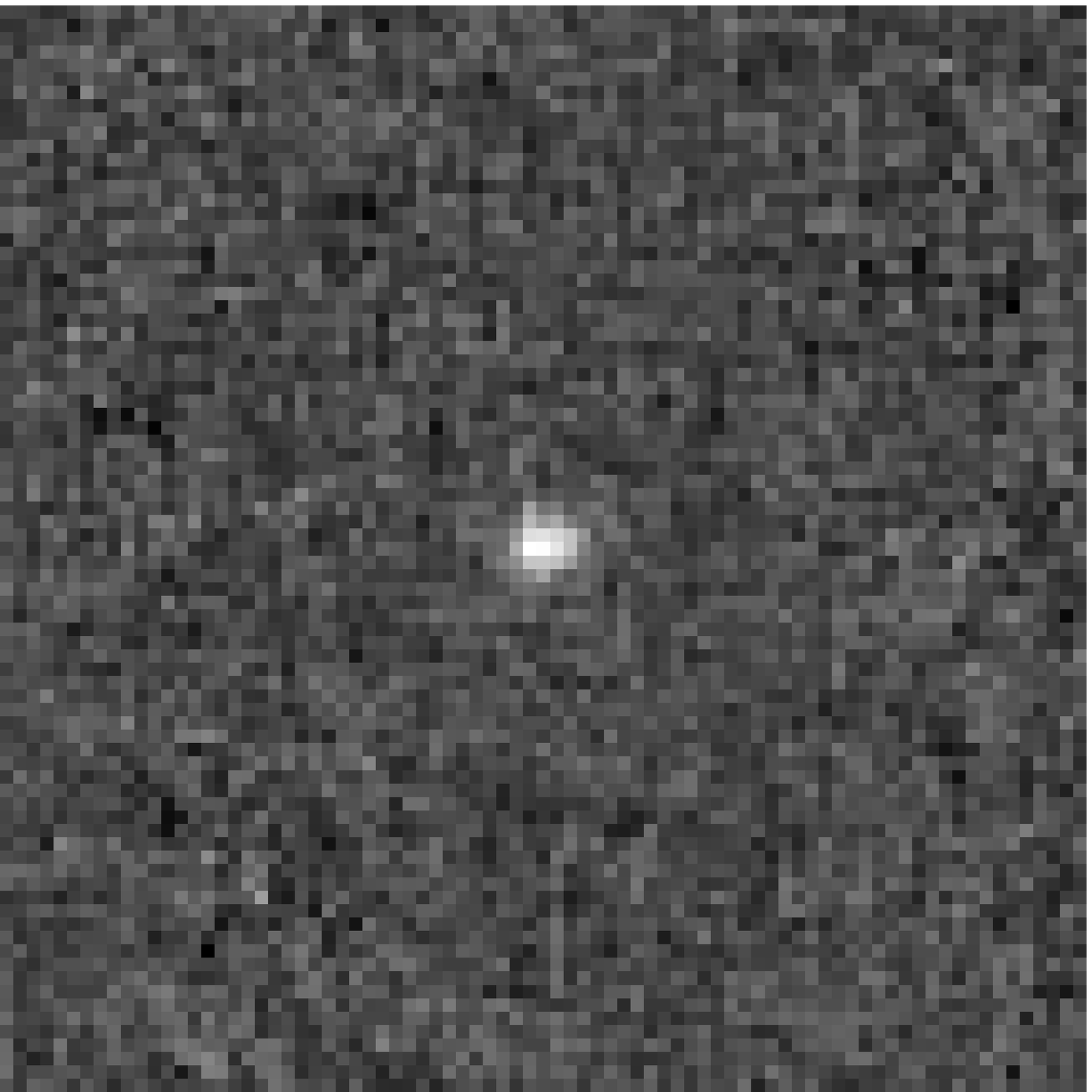}
\includegraphics[width=0.19\textwidth]{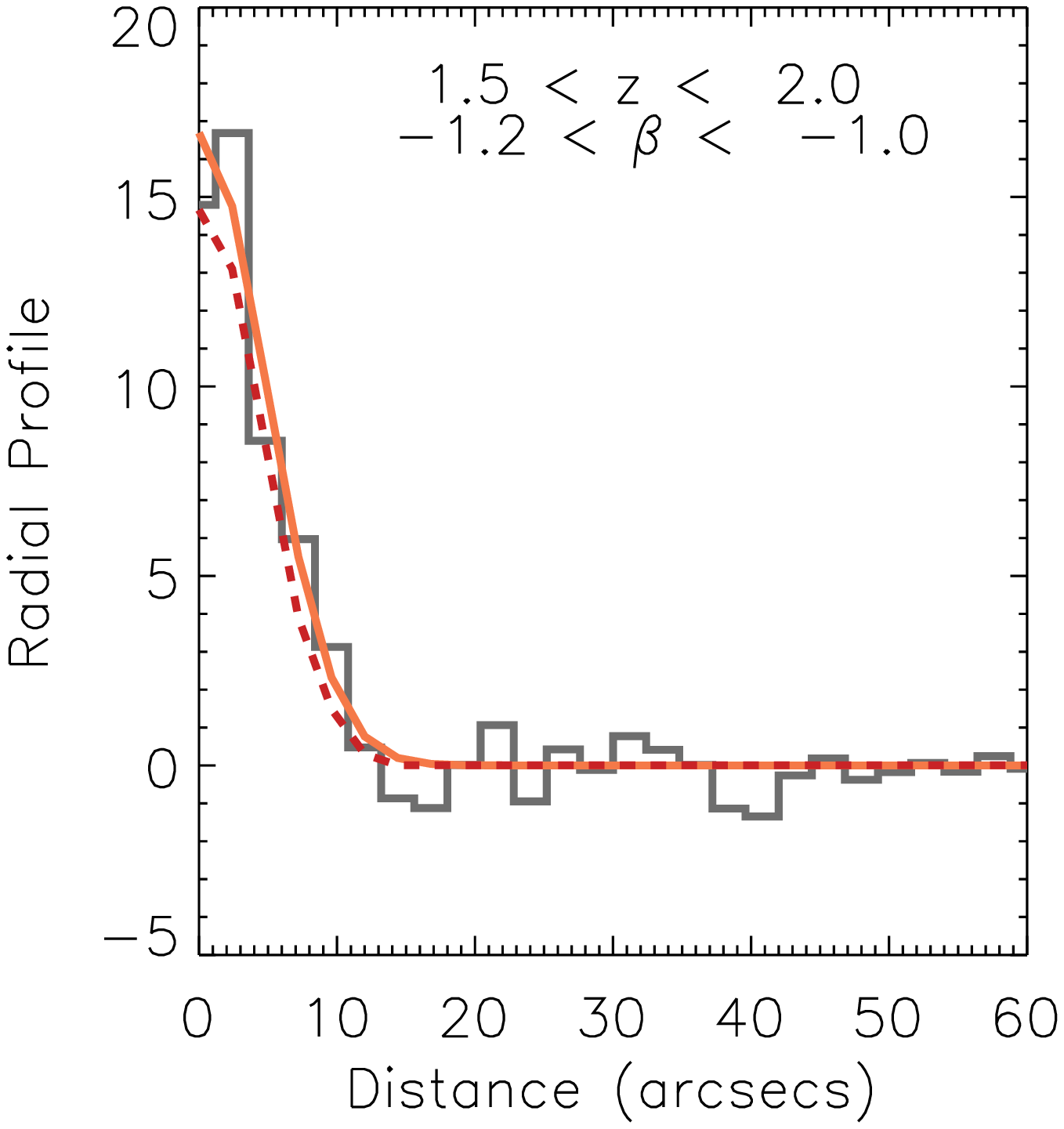}
\includegraphics[width=0.19\textwidth]{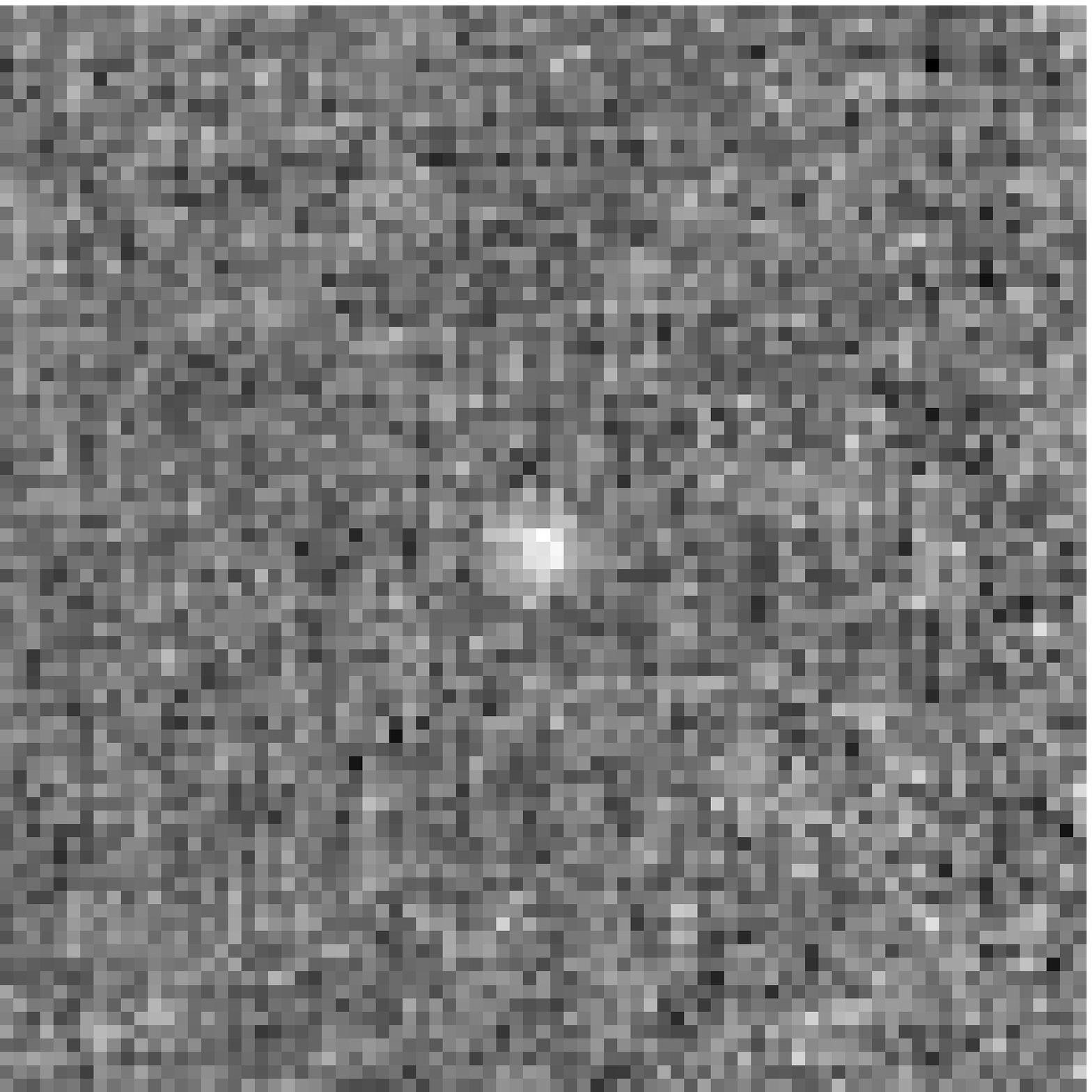}\\
\includegraphics[width=0.19\textwidth]{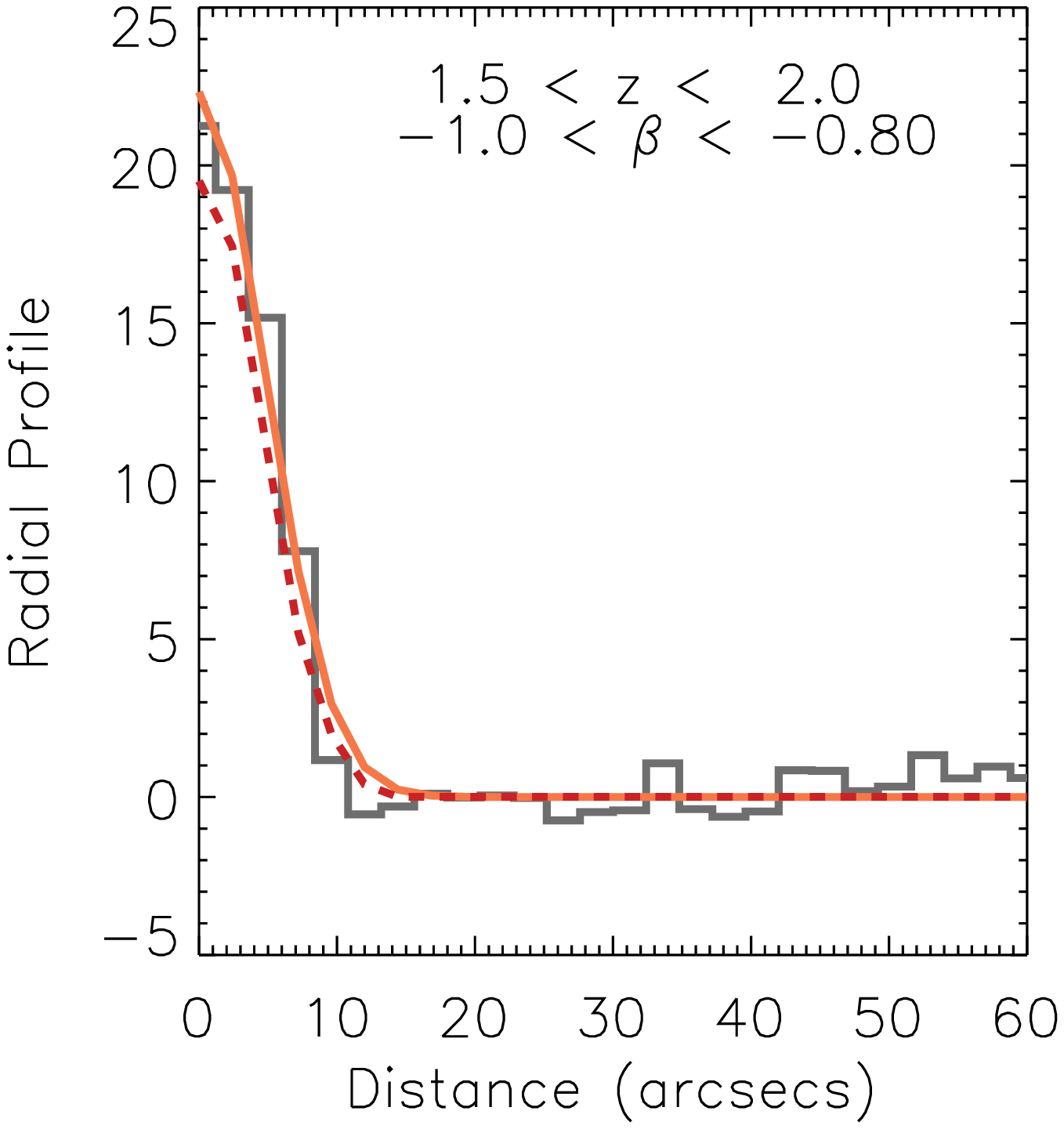}
\includegraphics[width=0.19\textwidth]{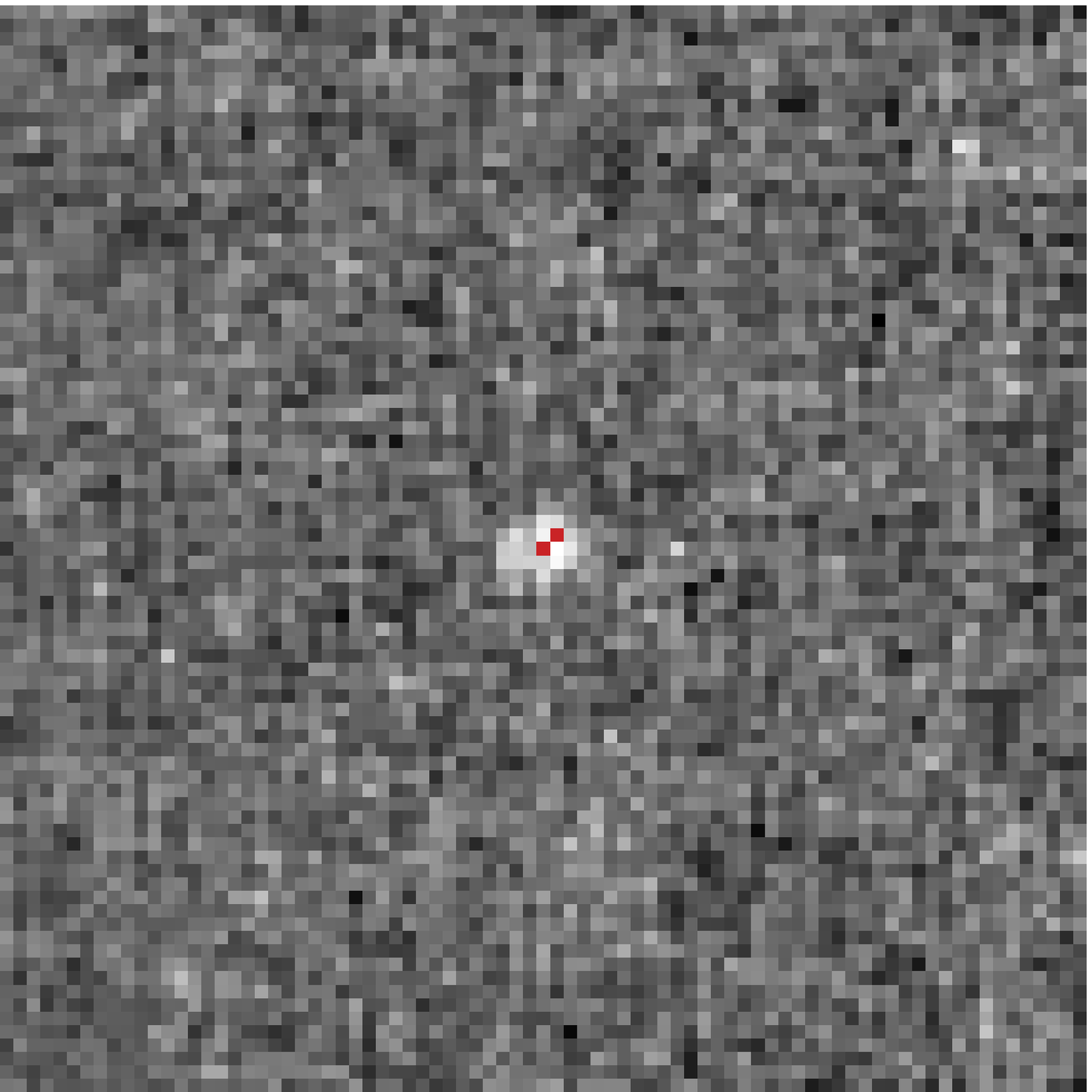}
\includegraphics[width=0.19\textwidth]{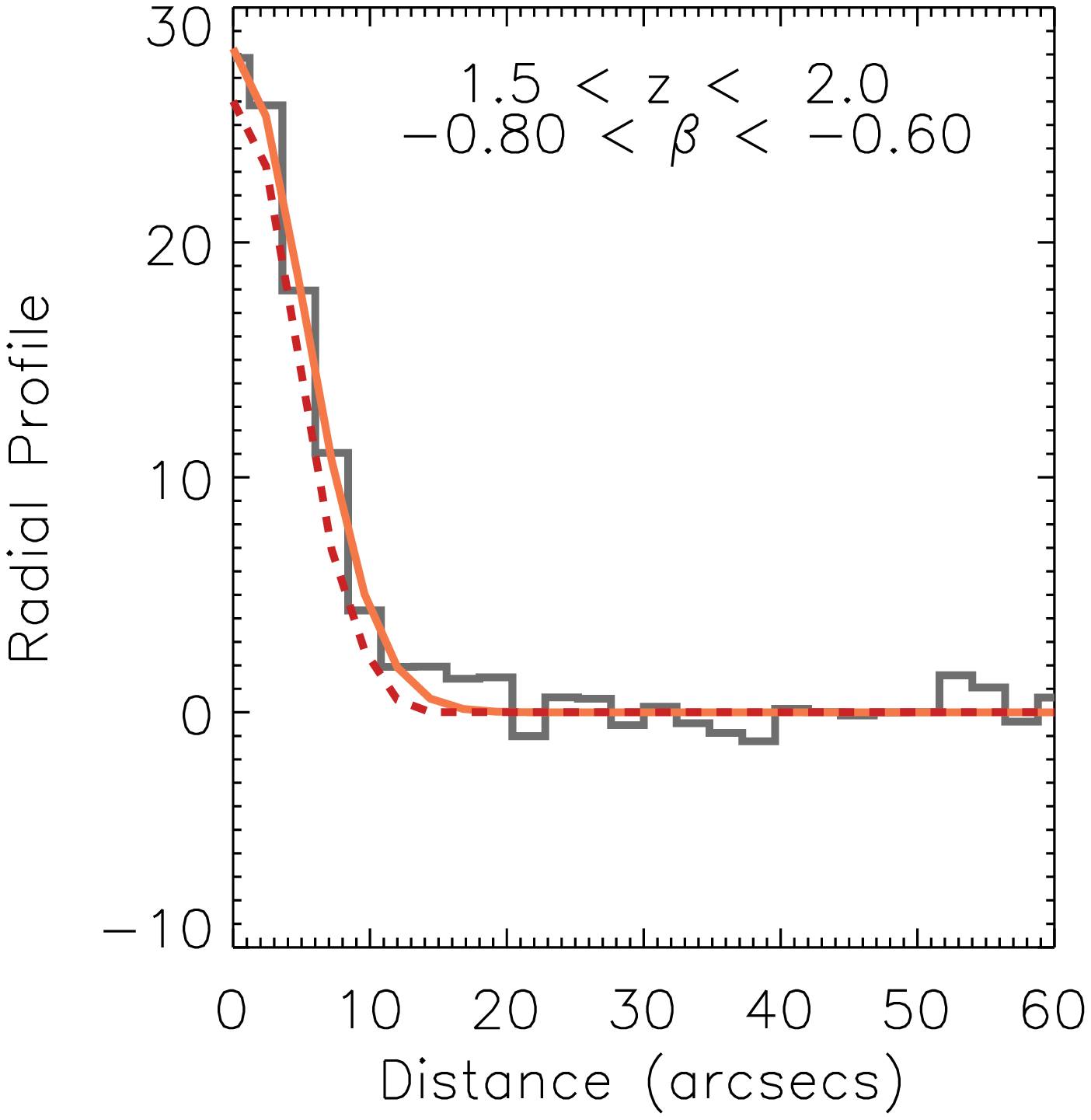}
\includegraphics[width=0.19\textwidth]{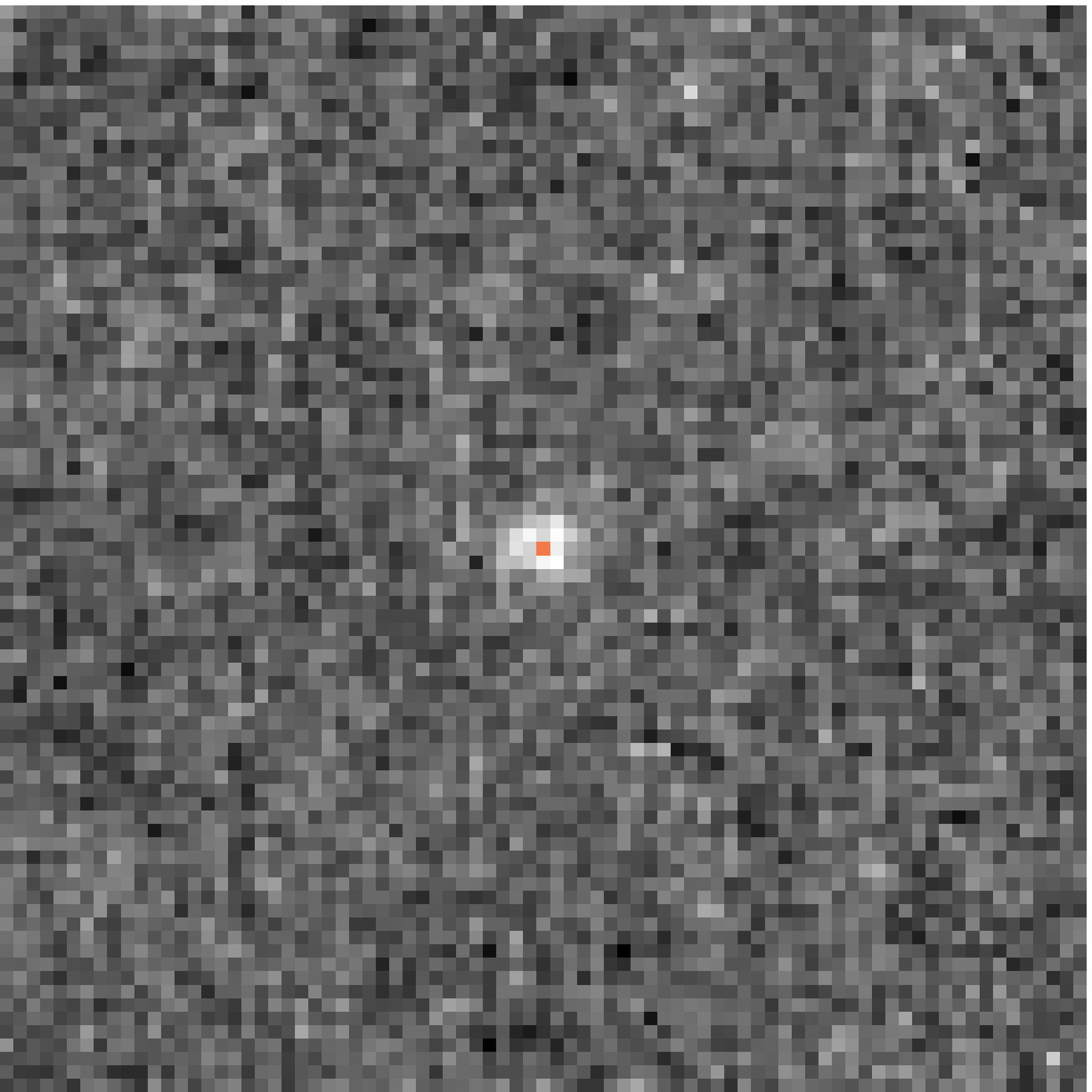}\\
\includegraphics[width=0.19\textwidth]{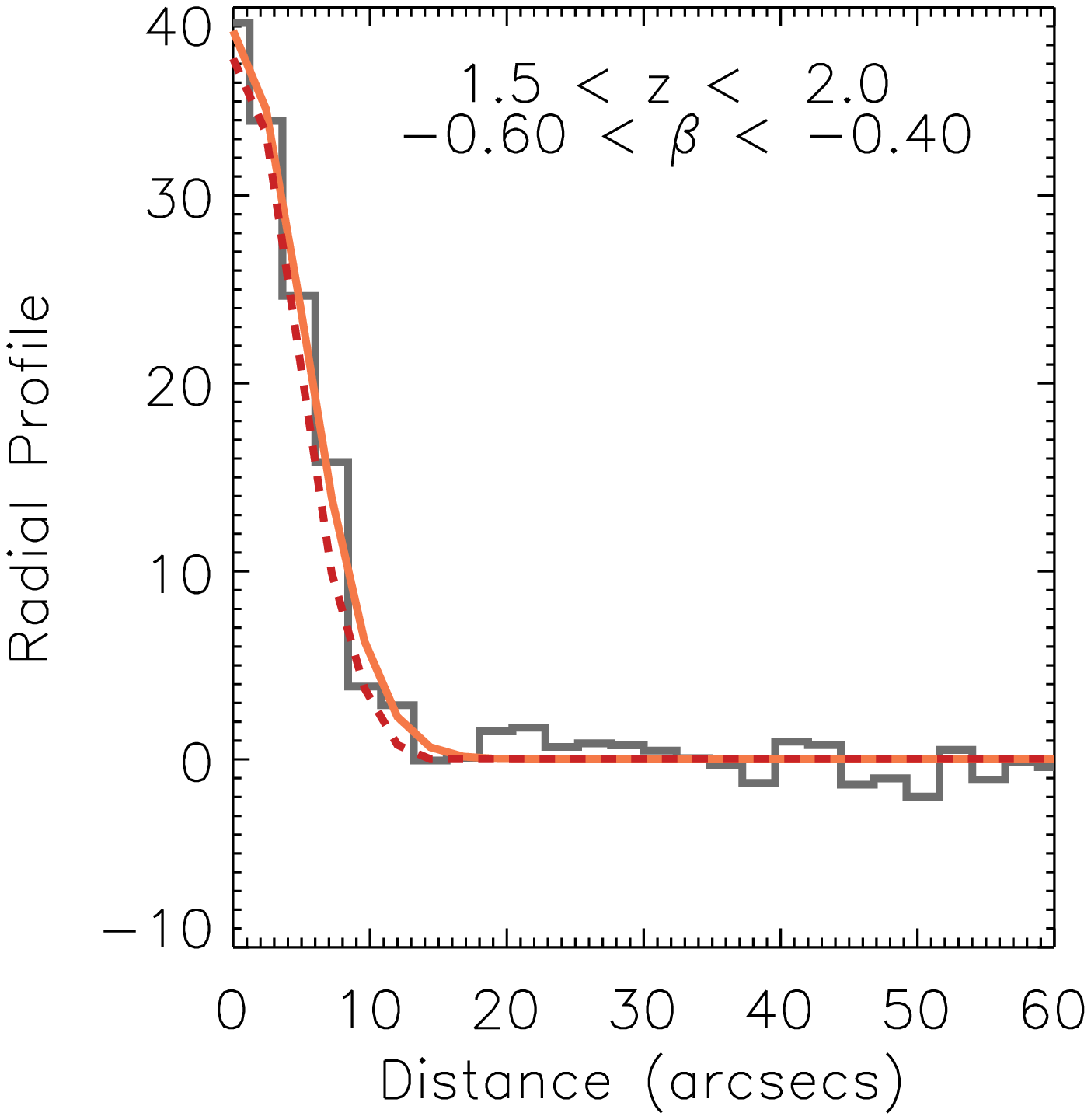}
\includegraphics[width=0.19\textwidth]{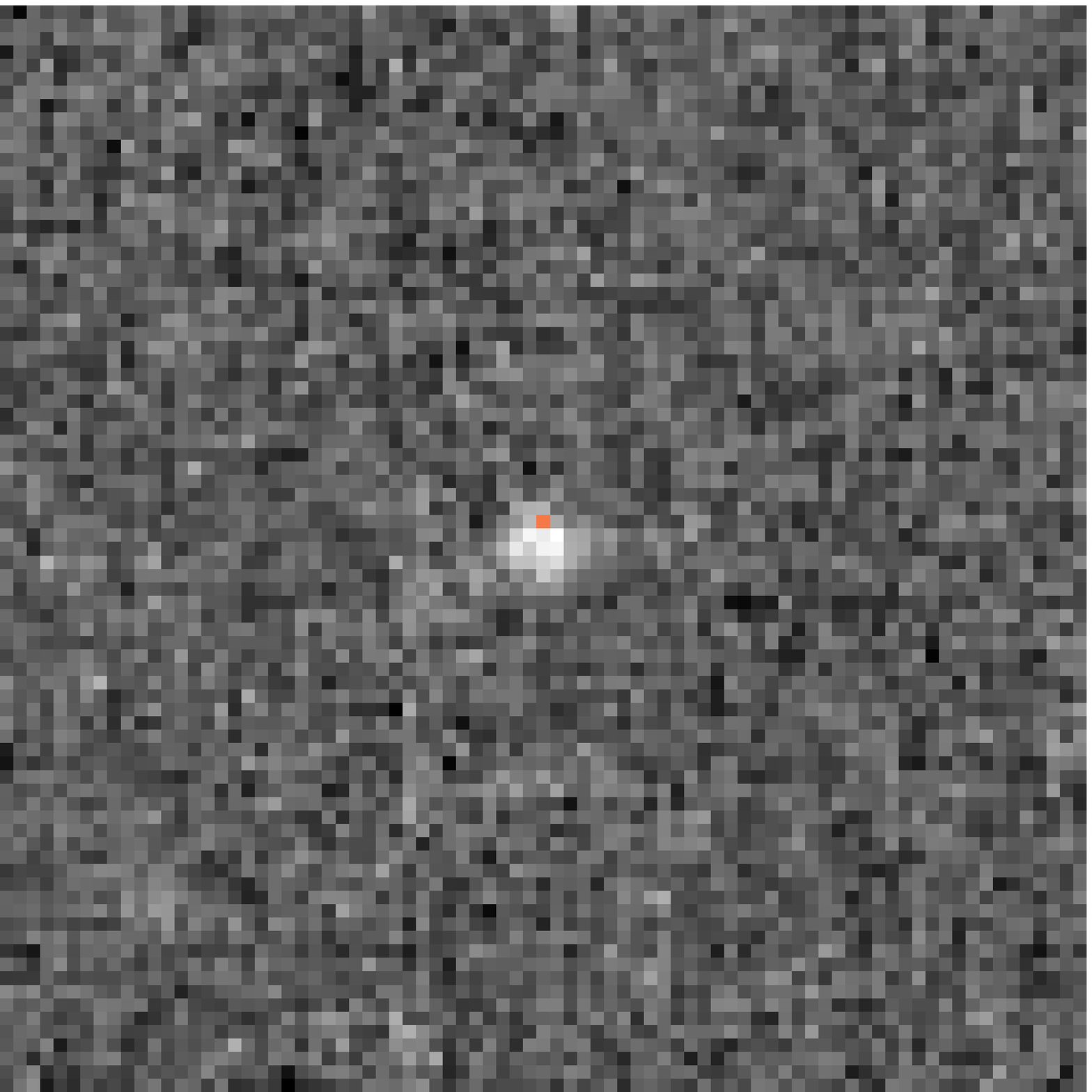}
\includegraphics[width=0.19\textwidth]{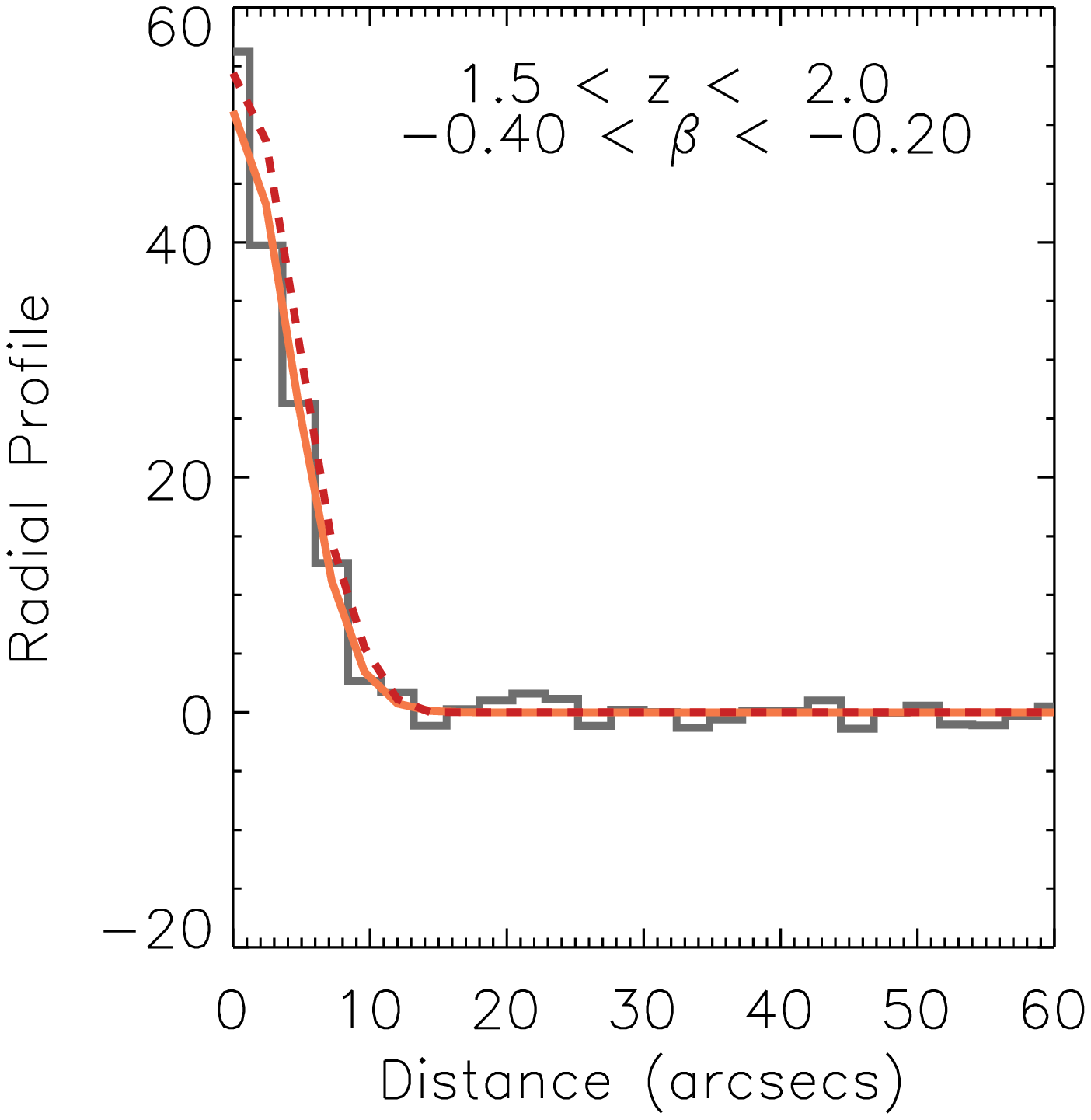}
\includegraphics[width=0.19\textwidth]{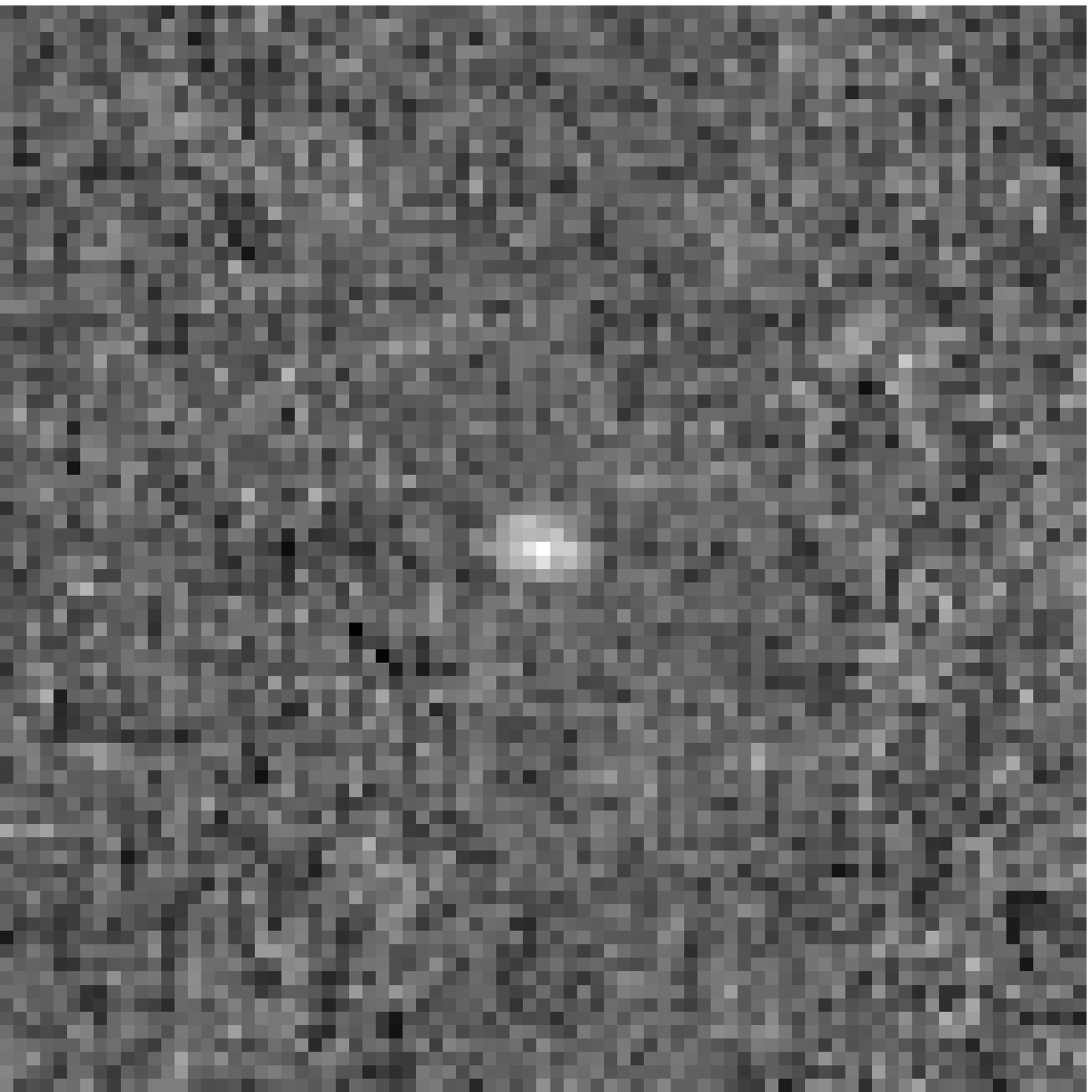}\\
\includegraphics[width=0.19\textwidth]{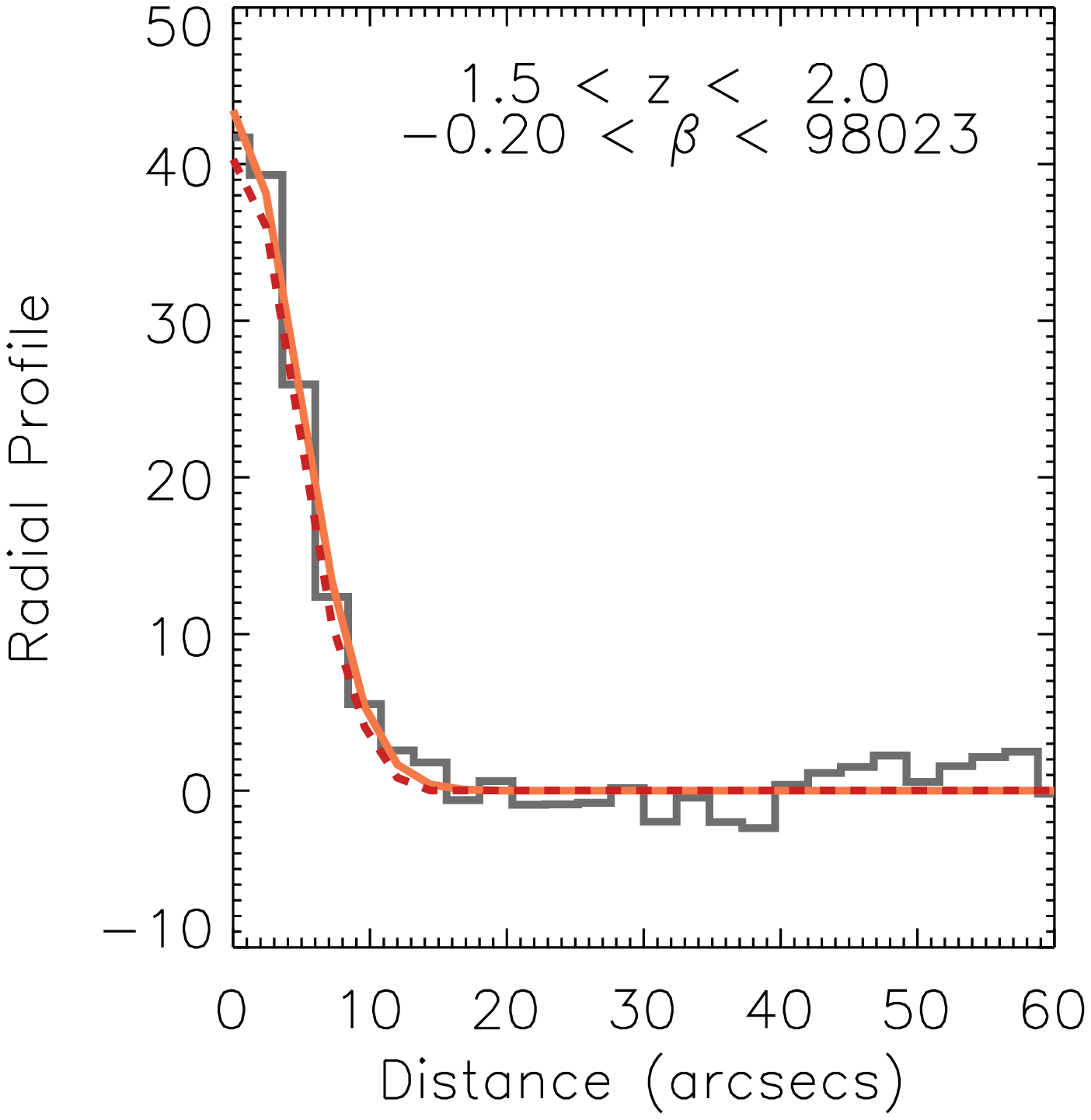}
\includegraphics[width=0.19\textwidth]{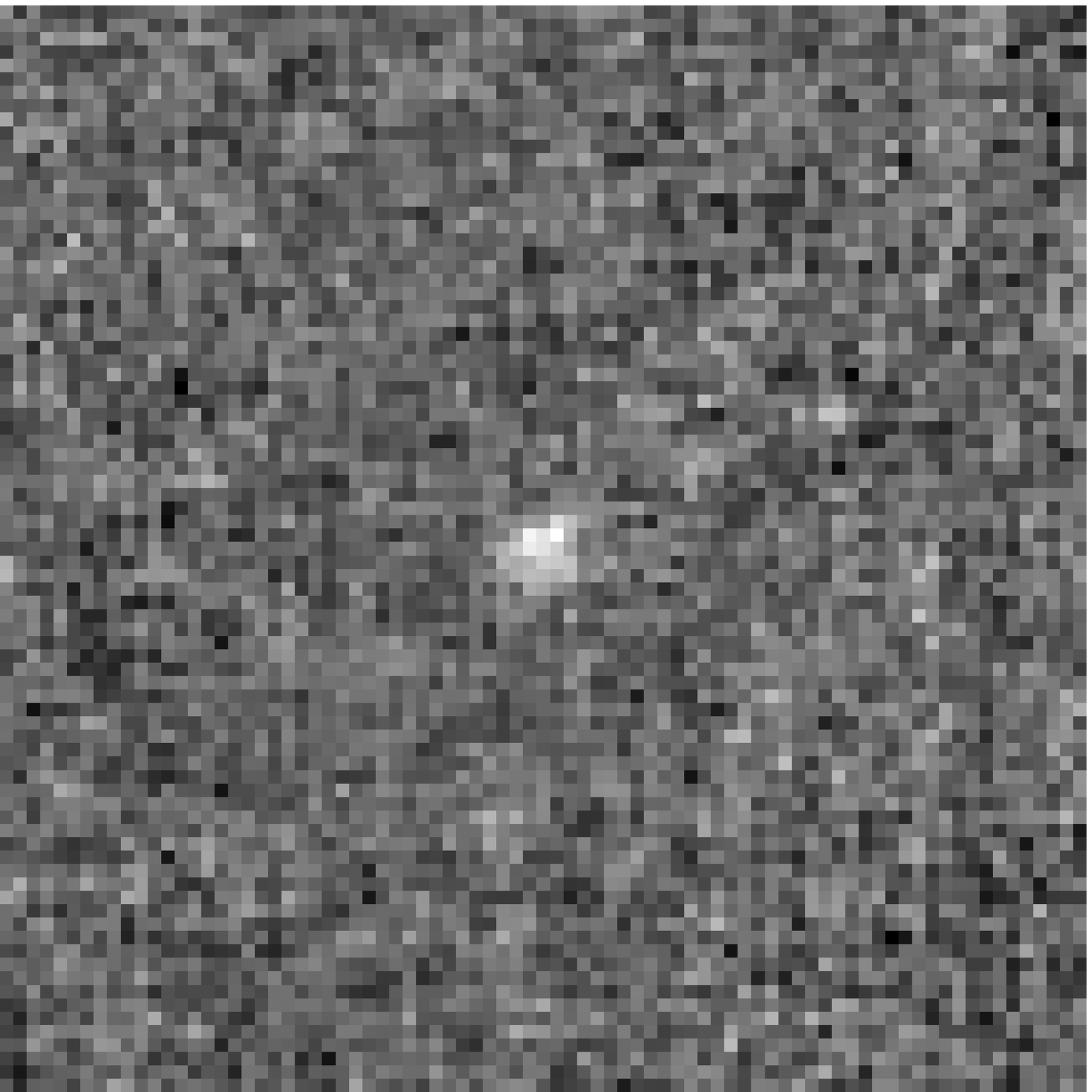}
\includegraphics[width=0.19\textwidth]{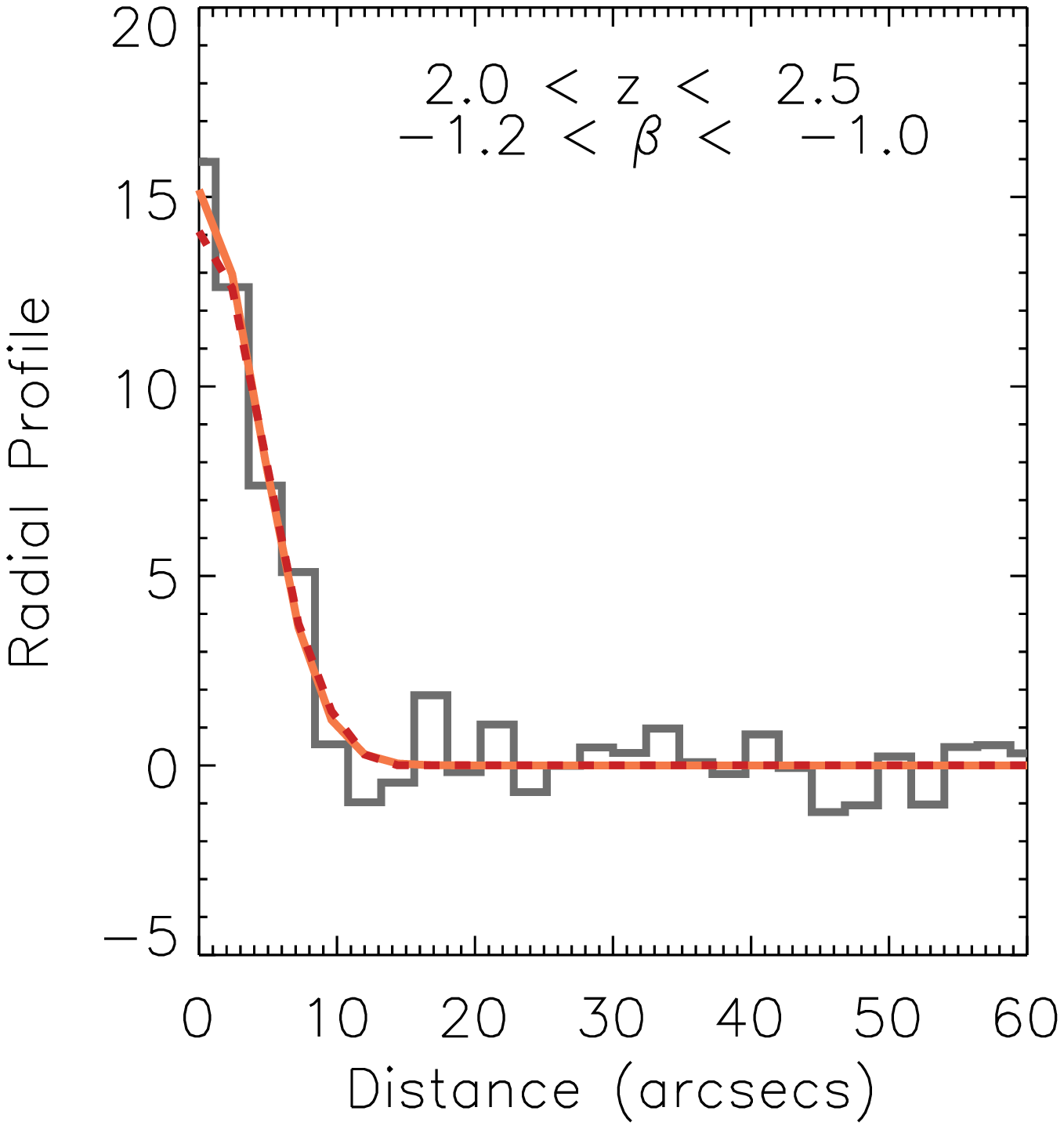}
\includegraphics[width=0.19\textwidth]{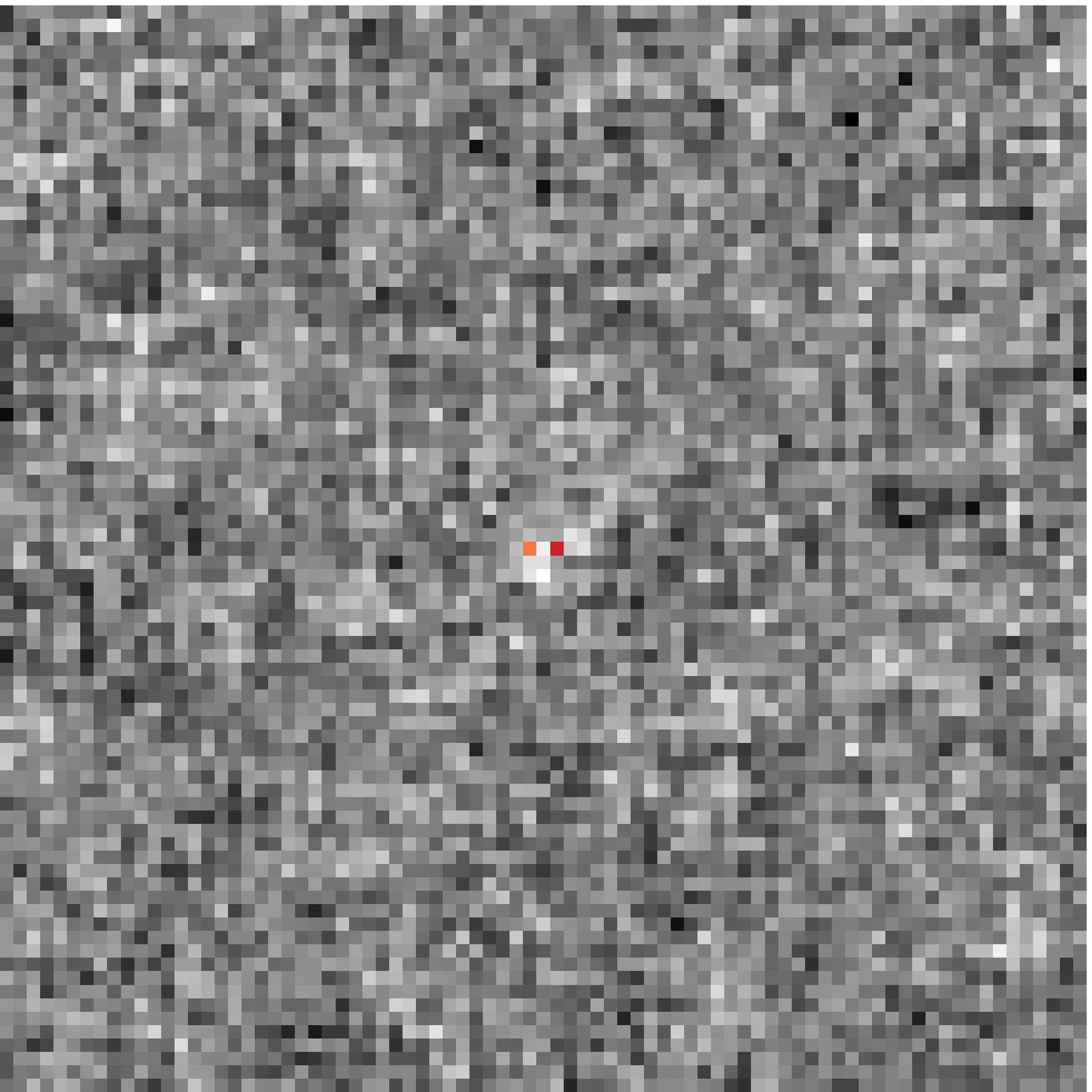}\\
\includegraphics[width=0.19\textwidth]{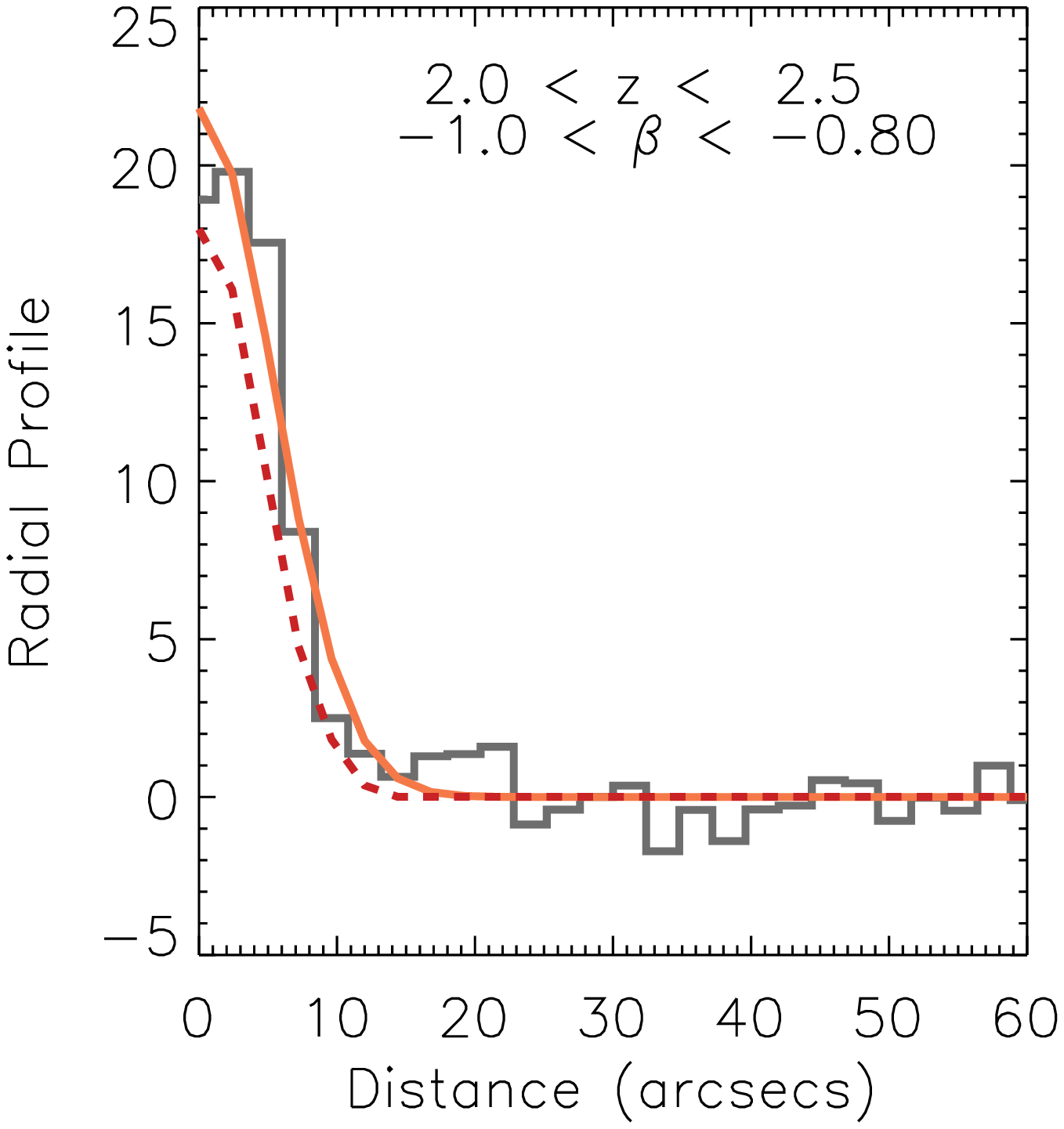}
\includegraphics[width=0.19\textwidth]{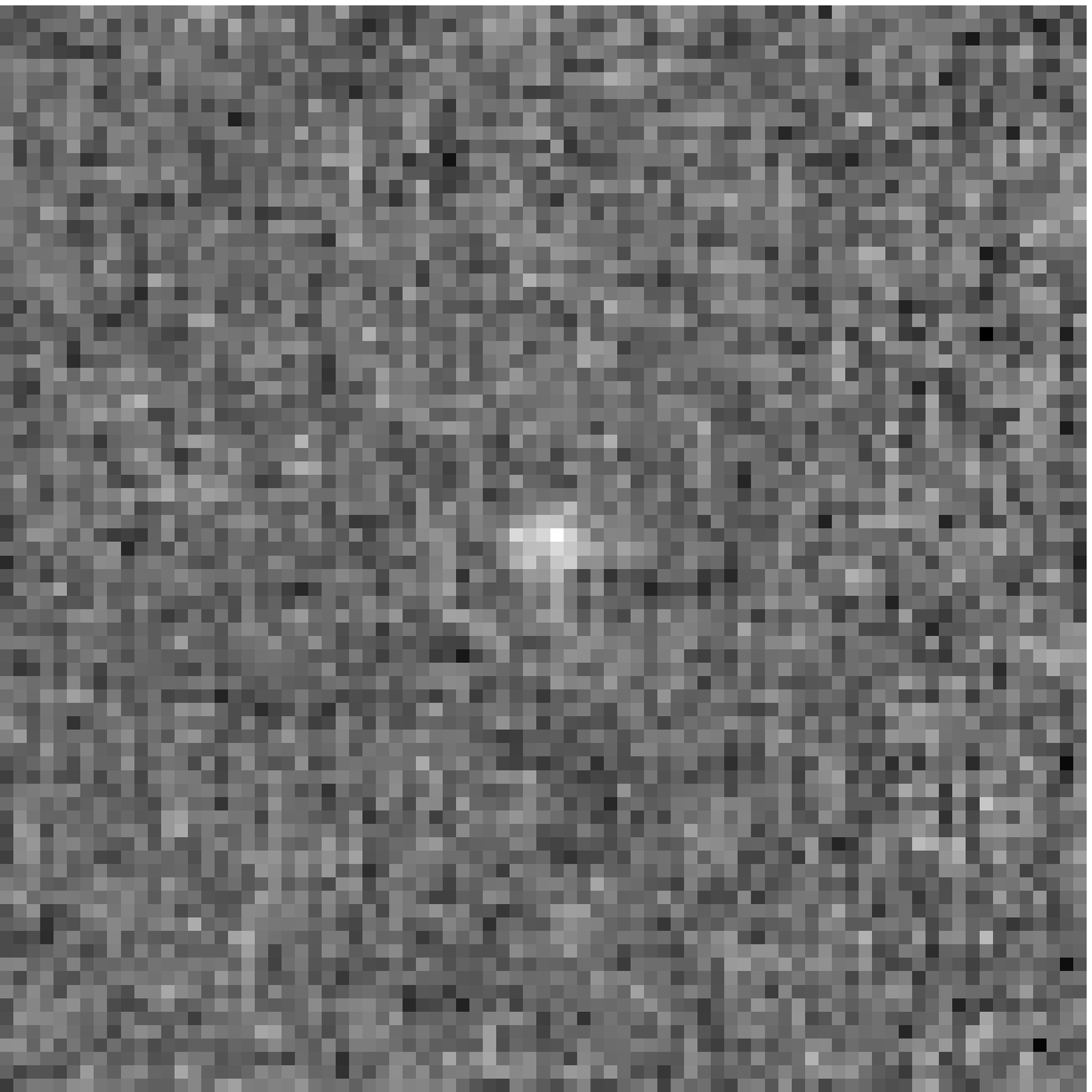}
\includegraphics[width=0.19\textwidth]{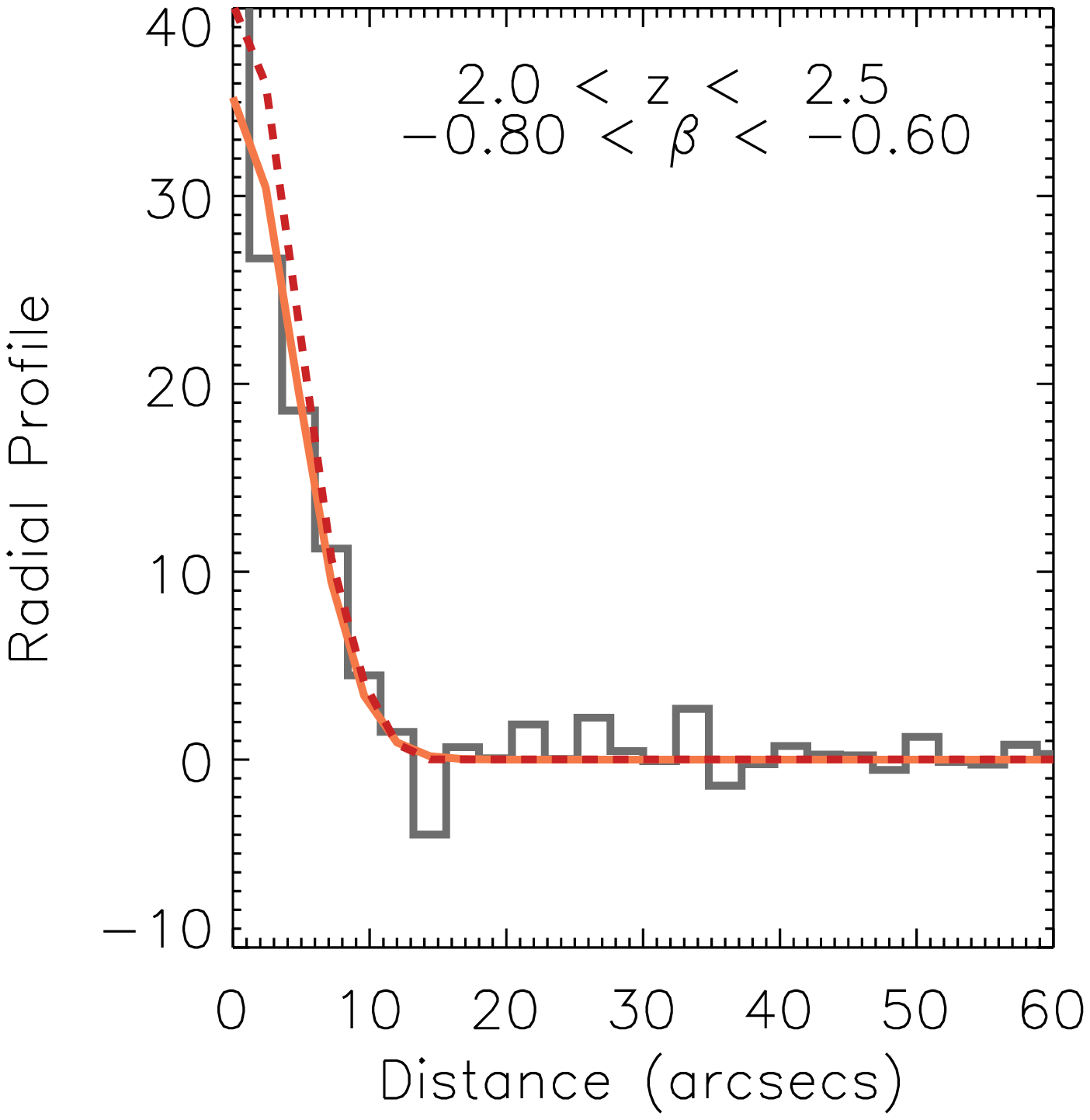}
\includegraphics[width=0.19\textwidth]{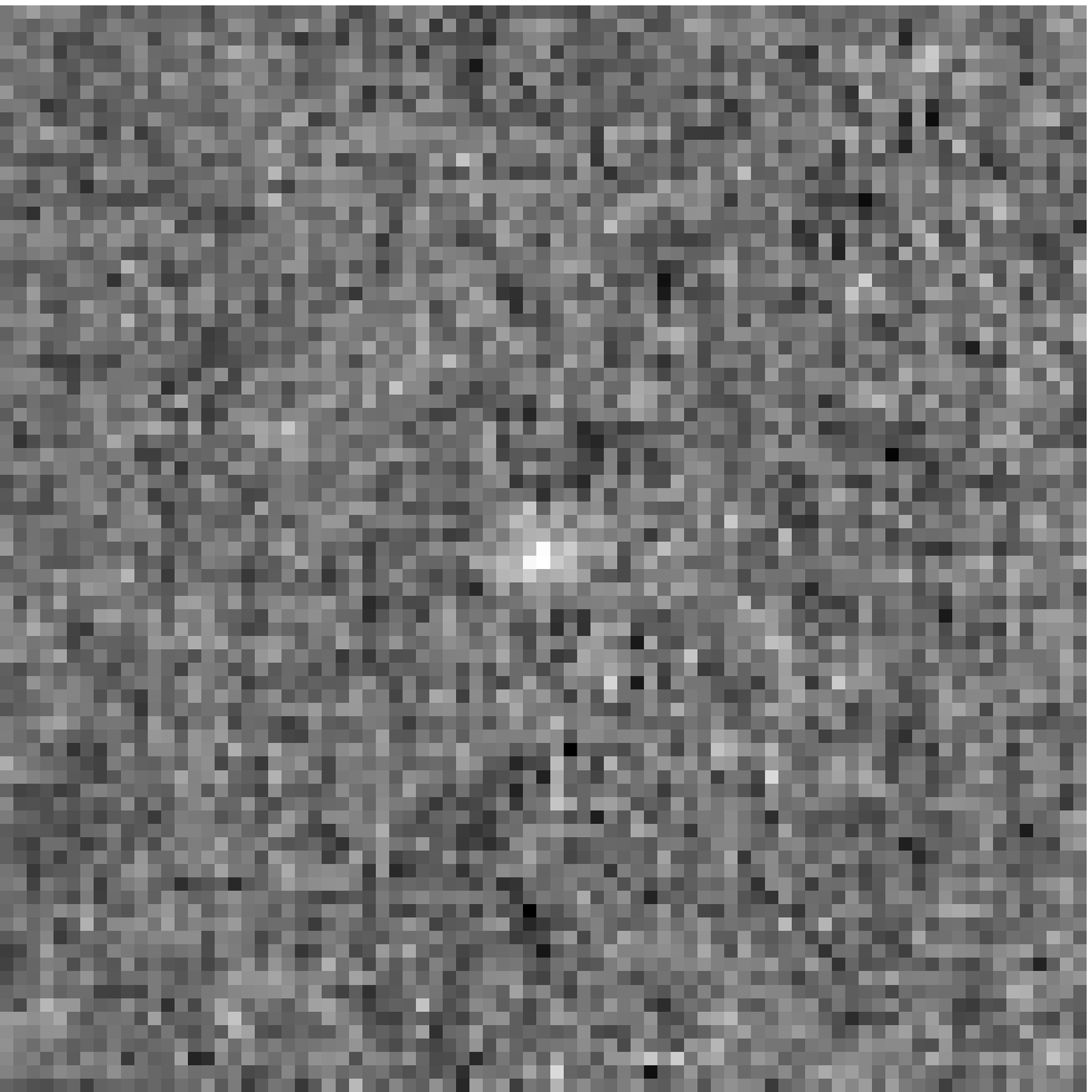}\\
\includegraphics[width=0.19\textwidth]{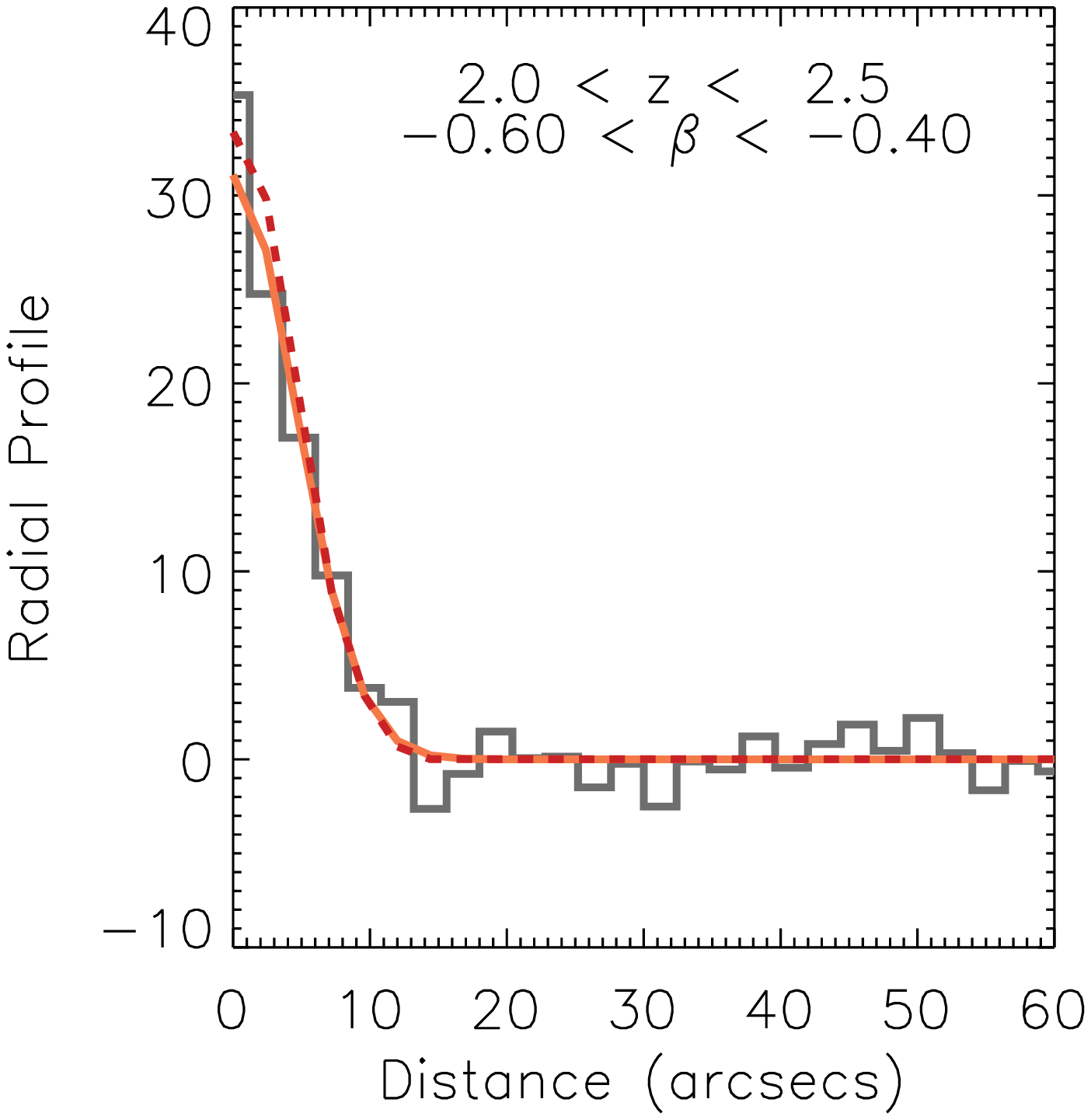}
\includegraphics[width=0.19\textwidth]{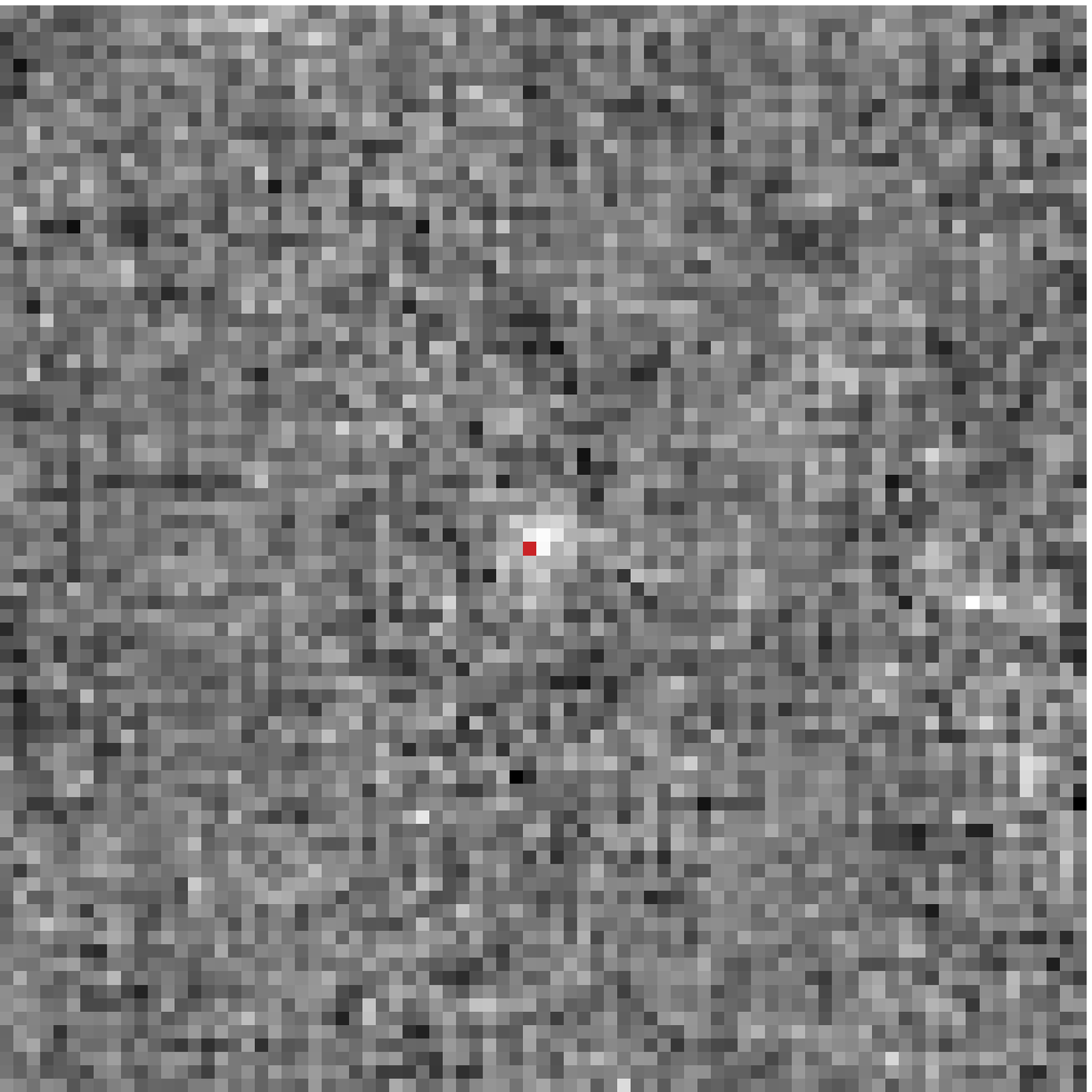}
\includegraphics[width=0.19\textwidth]{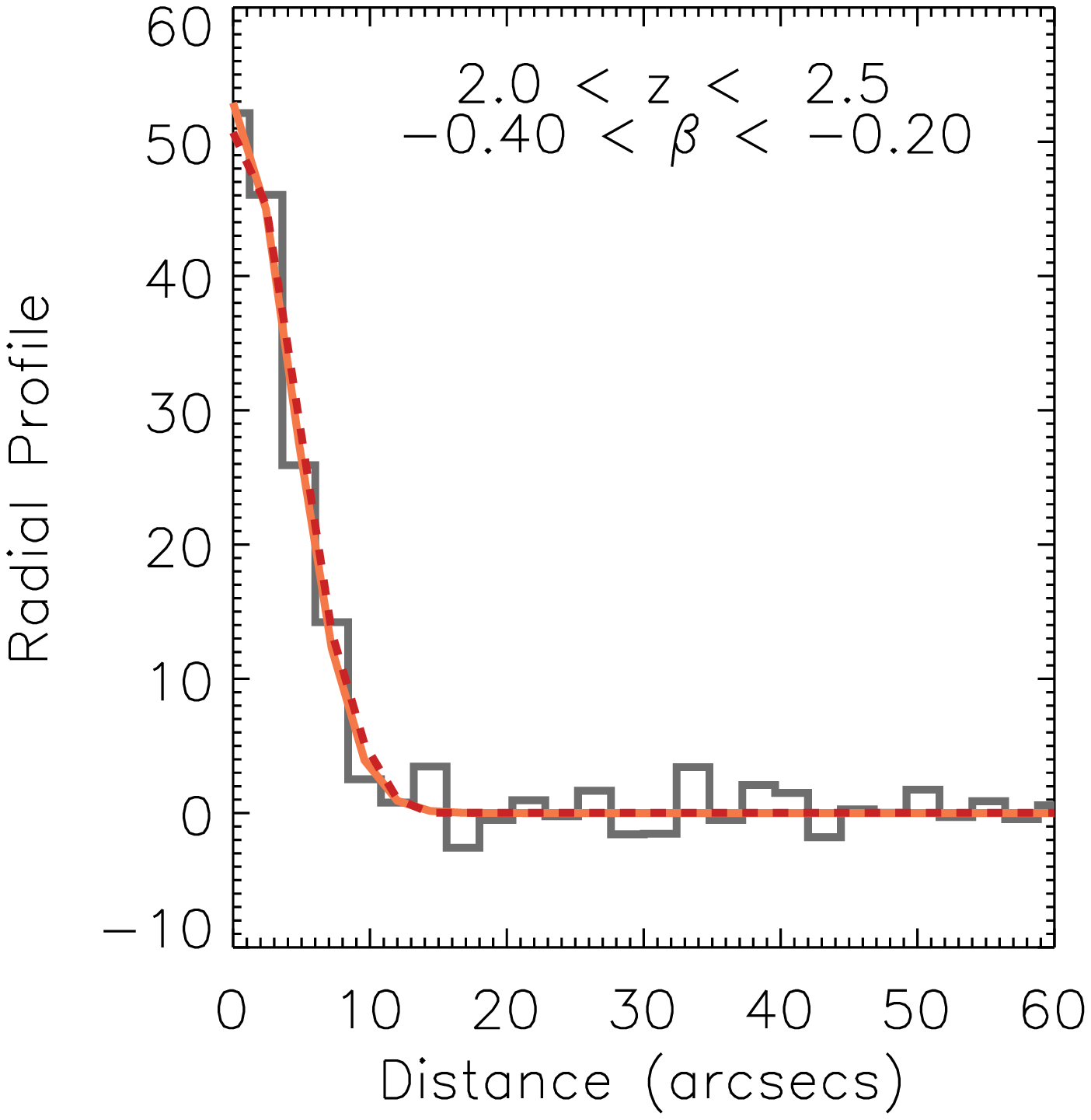}
\includegraphics[width=0.19\textwidth]{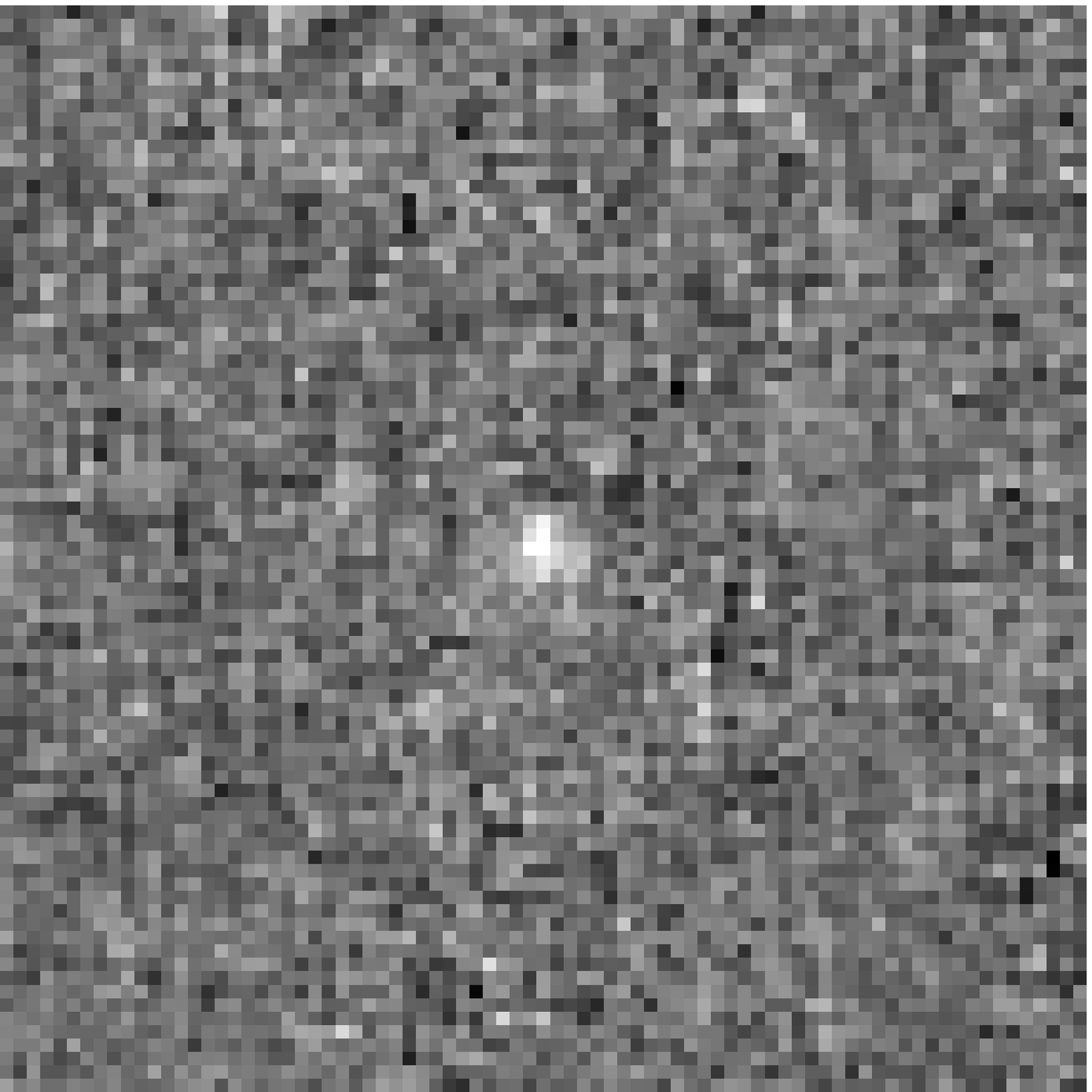}\\

\caption{Radial profiles and associated stacked images when stacking as a function of the UV continuum slope. Redshift and UV continuum bins are indicated in each case (\emph{Cont})
.              }
\label{images2}
\end{figure}

\begin{figure}
\centering
\includegraphics[width=0.19\textwidth]{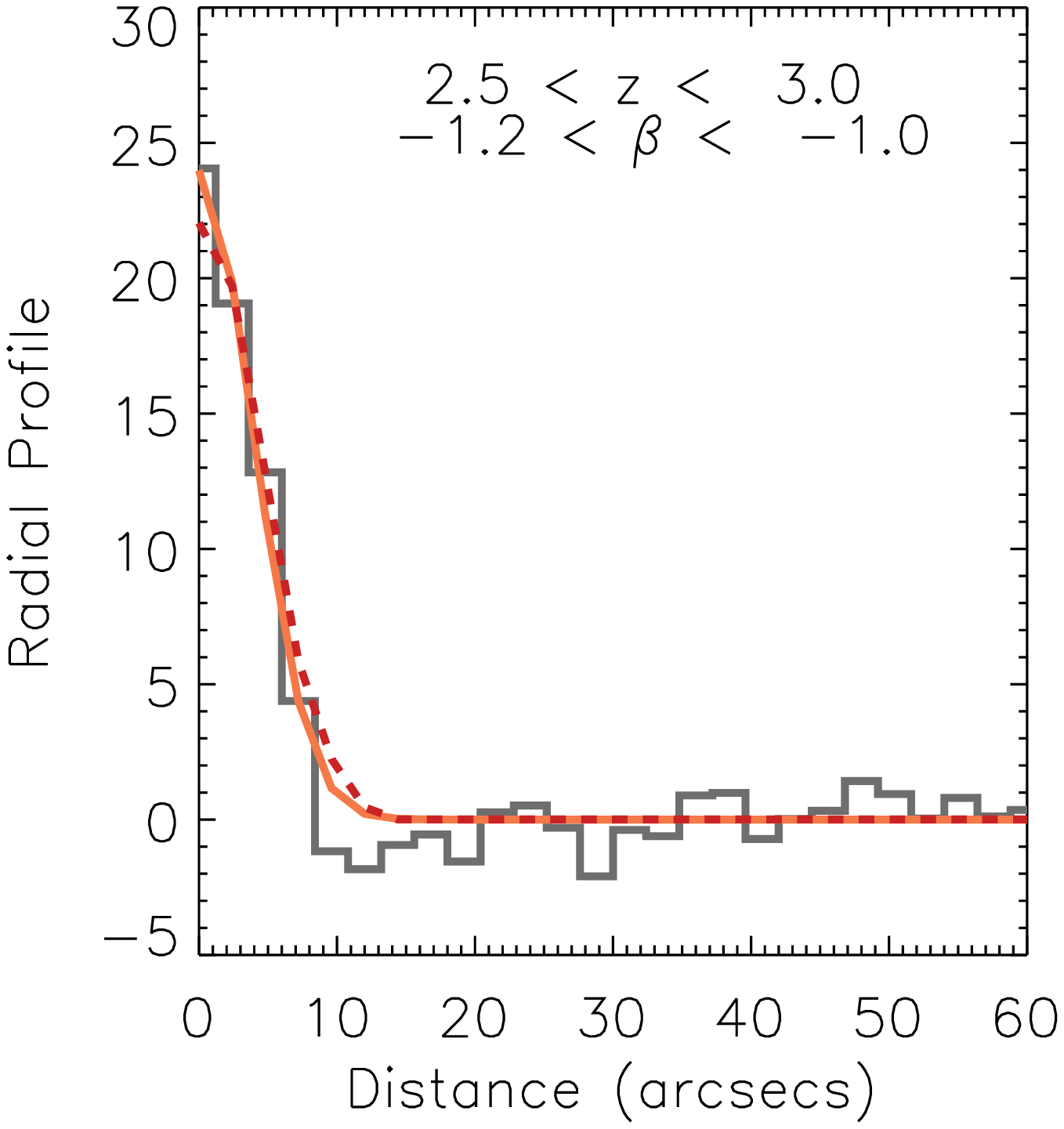}
\includegraphics[width=0.19\textwidth]{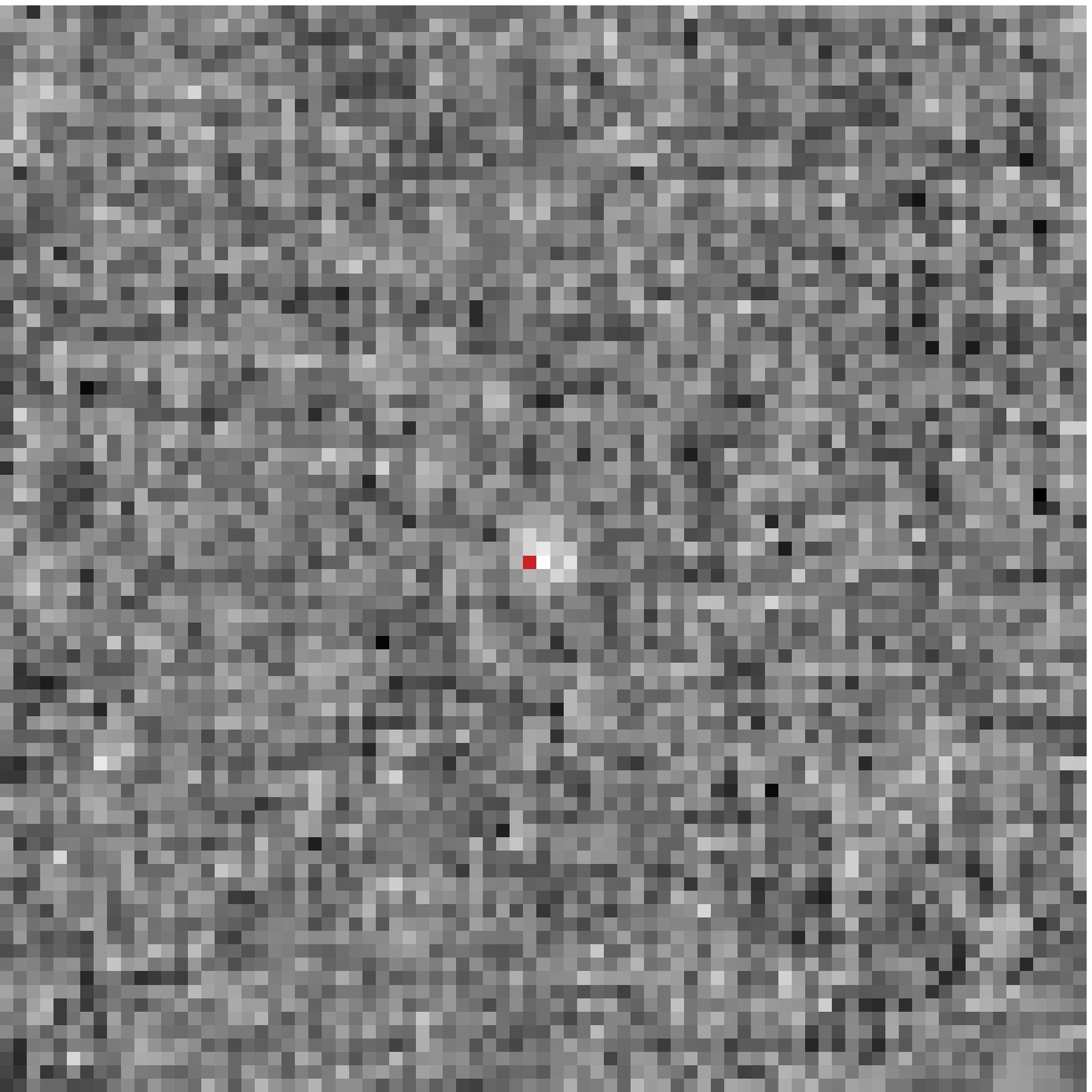}
\includegraphics[width=0.19\textwidth]{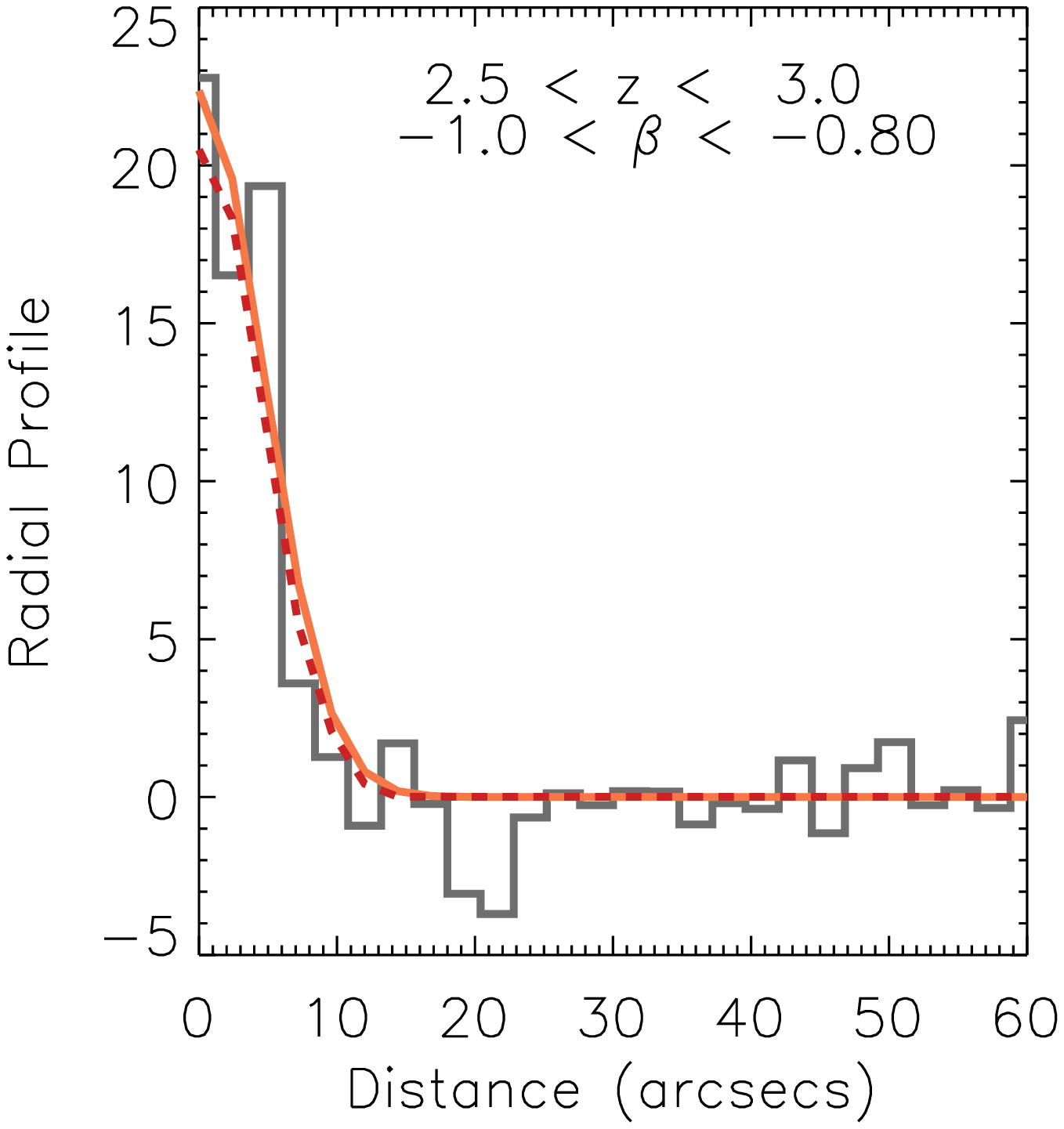}
\includegraphics[width=0.19\textwidth]{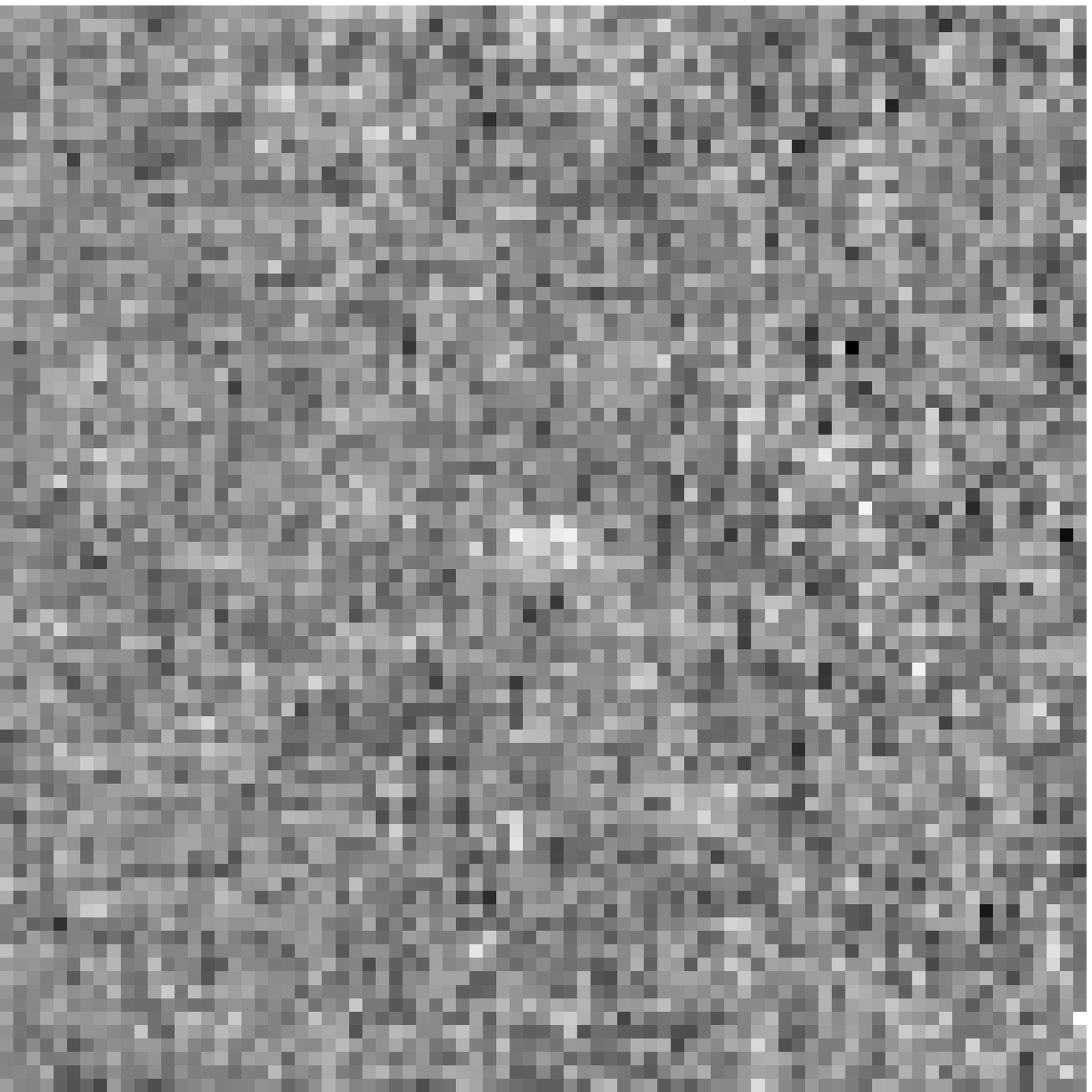}\\
\includegraphics[width=0.19\textwidth]{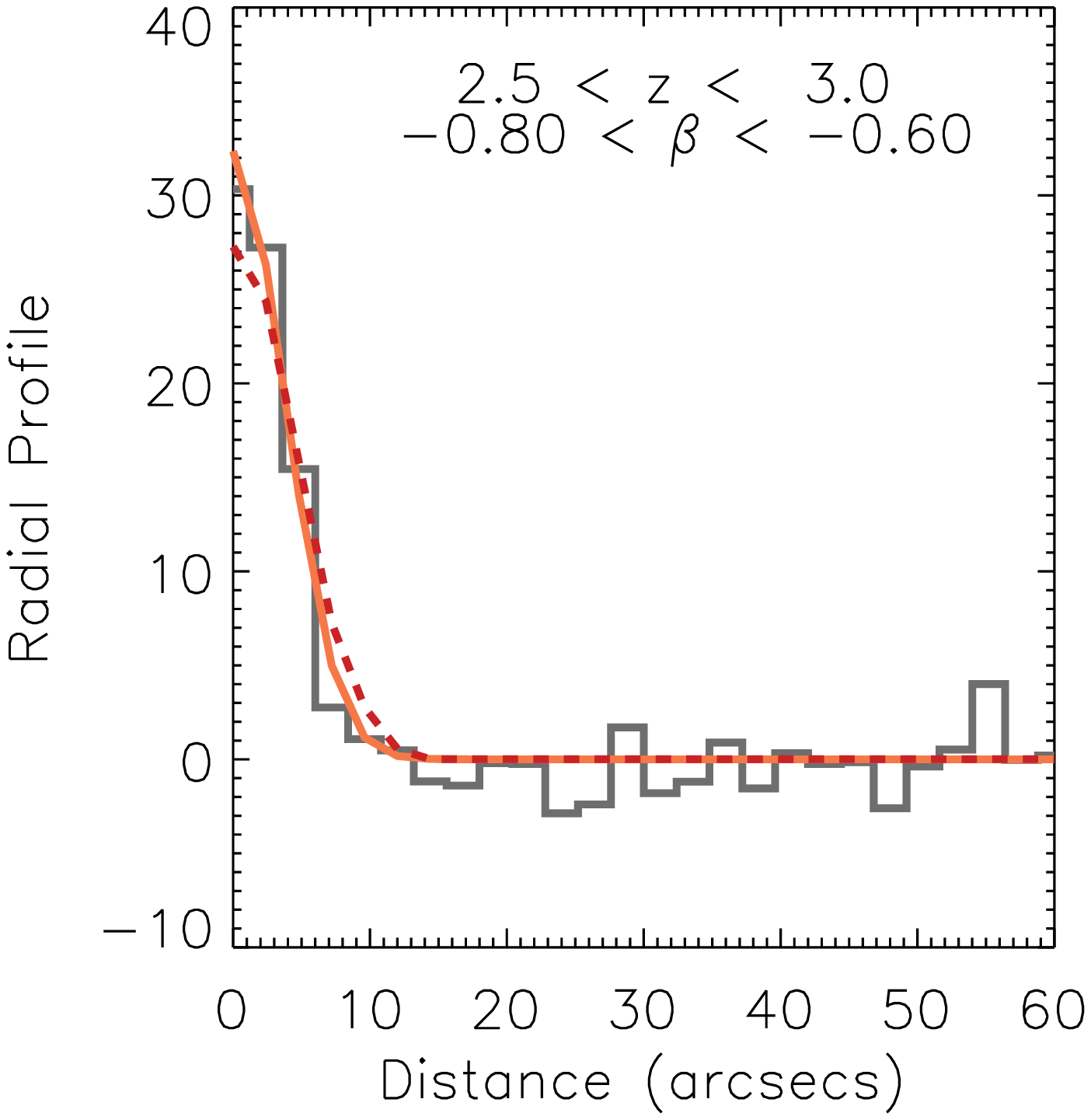}
\includegraphics[width=0.19\textwidth]{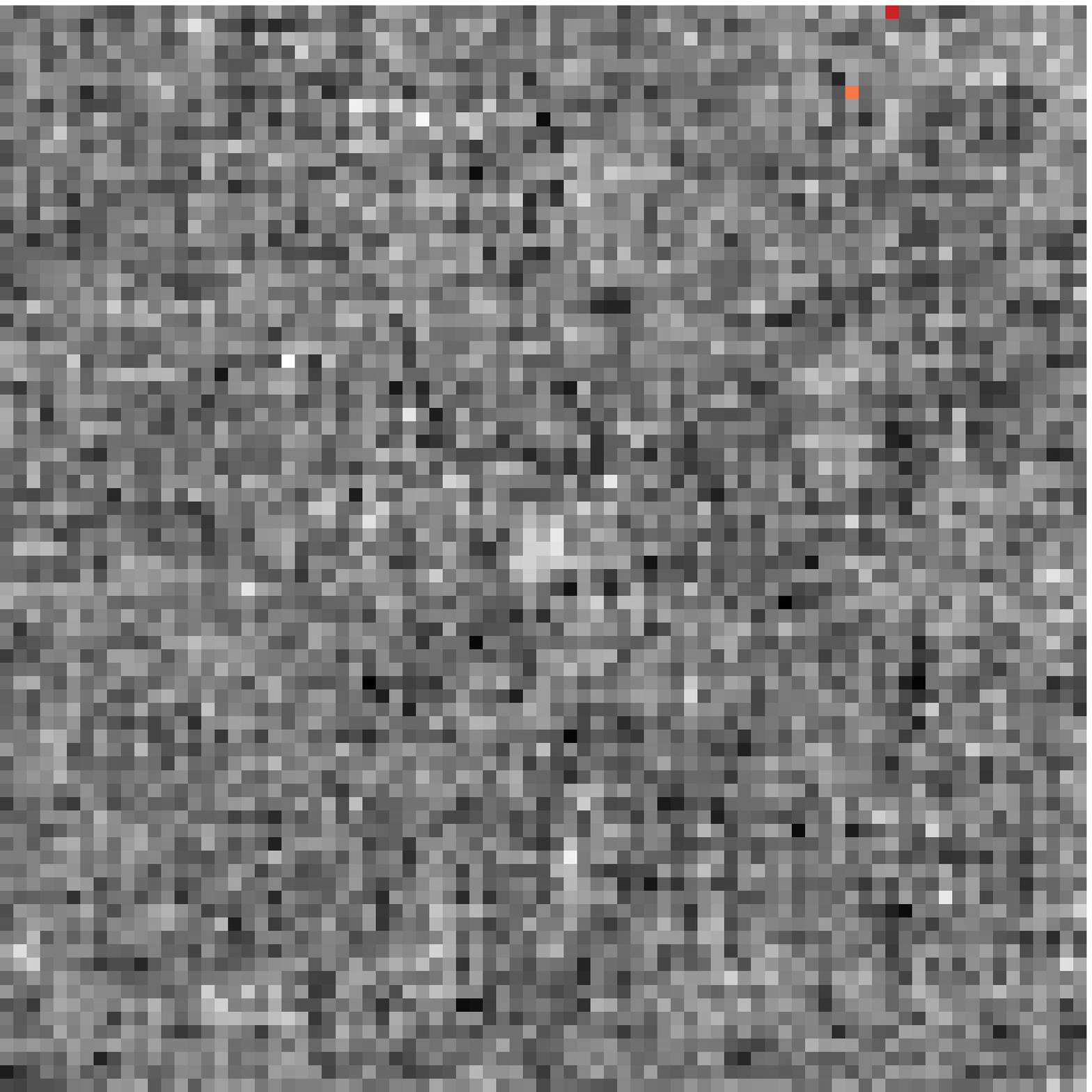}

\caption{Radial profiles and associated stacked images when stacking as a function of the UV continuum slope. Redshift and UV continuum bins are indicated in each case (\emph{Cont}).
              }
\label{images3}
\end{figure}

\begin{table*}\label{hola_tabla}
\begin{center}
\caption{Summary of stacked properties when stacking as a function of the UV continuum slope. The uncertainties in the rest-frame UV luminosity are related to the normalization of the templates to the observed photometry and they are typically lower tan 0.05 dex. The uncertainties in the total IR luminosities are obtained as the difference in the total IR luminosity when considering $f_{\rm 160 \mu m} \pm \Delta f_{\rm 160 \mu m}$. $N$ indicates the number of galaxies in each bin.}
\begin{tabular}{c c c c c c }
\hline
Redshift range & UV slope range & $f_{\rm 160 \mu m}$ [mJy] & $\log{L_{\rm UV}}$ & $\log{L_{\rm IR}}$ & $N$\\ 
\hline
\hline
     0.02$\leq z \leq$0.5 	&    	-1.4$\leq \beta \leq$-1.2	&	1.77$\pm$0.19		&     9.61		&	9.98$\pm$0.09		&    	194		\\
     0.02$\leq z \leq$0.5 	&    	-1.2$\leq \beta \leq$-1.0 	&     	2.73$\pm$0.29 		&     9.61		&    	10.16$\pm$0.09	&	121		\\
     0.02$\leq z \leq$0.5  	&   	-1.0$\leq \beta \leq$-0.8	&      3.25$\pm$0.46		&      9.58		&     	10.34$\pm$0.17	&	41		\\
     0.5$\leq z \leq$1.0  	& 	-1.4$\leq \beta \leq$-1.2	&      0.49$\pm$0.04 		&     9.72		&    	10.22$\pm$0.07	&	3125		\\
     0.5 $\leq z \leq$1.0	&     	-1.2$\leq \beta \leq$-1.0	&      0.82$\pm$0.06 		&     9.72		&    	10.44$\pm$0.06	&	2064		\\
     0.5$\leq z \leq$1.0  	&   	-1.0$\leq \beta \leq$-0.8	&      1.18$\pm$0.08		&      9.73		&     	10.60$\pm$0.05	&	1244		\\
     0.5$\leq z \leq$1.0  	&   	-0.8$\leq \beta \leq$-0.6	&      1.59$\pm$0.10		&      9.70		&     	10.72$\pm$0.05	&	963		\\
     1.0$\leq z \leq$1.5   	&  	-1.4$\leq \beta \leq$-1.2	&      0.18$\pm$0.02		&      9.86		&     	10.28$\pm$0.09	&	8136		\\
     1.0$\leq z \leq$1.5    	& 	-1.2$\leq \beta \leq$-1.0	&      0.41$\pm$0.03		&      9.84 		&    	10.61$\pm$0.06	&	5084		\\
     1.0$\leq z \leq$1.5     	&	-1.0$\leq \beta \leq$-0.8	&      0.65$\pm$0.03		&      9.84		&     	10.81$\pm$0.04	&	4187		\\
     1.0$\leq z \leq$1.5	&     	-0.8$\leq \beta \leq$-0.6	&      0.97$\pm$0.05		&      9.81		&     	10.98$\pm$0.05	&	2634		\\
     1.0$\leq z \leq$1.5	&     	-0.6$\leq \beta \leq$-0.4	&      1.16$\pm$0.06		&      9.80		&     	11.05$\pm$0.04	&	1628		\\
     1.0$\leq z \leq$1.5	&    	-0.4$\leq \beta \leq$-0.2	&      1.54$\pm$0.07		&      9.74		&     	11.15$\pm$0.04	&	1299		\\
     1.5$\leq z \leq$2.0 	&    	-1.2$\leq \beta \leq$-1.0	&      0.42$\pm$0.03		&      9.98		&     	10.92$\pm$0.05	&	4359		\\
     1.5$\leq z \leq$2.0  	&   	-1.0$\leq \beta \leq$-0.8	&      0.56$\pm$0.04		&      9.95		&     	11.05$\pm$0.06	&	3662		\\
     1.5$\leq z \leq$2.0	&     	-0.8$\leq \beta \leq$-0.6	&      0.77$\pm$0.04		&      9.88		&     	11.18$\pm$0.05	&	2501		\\
     1.5$\leq z \leq$2.0	&     	-0.6$\leq \beta \leq$-0.4	&      1.01$\pm$0.06		&      9.83		&     	11.31$\pm$0.05	&	1691		\\
     1.5$\leq z \leq$2.0	&     	-0.4$\leq \beta \leq$-0.2	&      1.08$\pm$0.09		&      9.77		&     	11.34$\pm$0.07	&	1061		\\
     1.5$\leq z \leq$2.0	&     	-0.2$\leq \beta \leq$-0.0	&      1.11$\pm$0.09		&      9.76		&     	11.33$\pm$0.06	&	825		\\
     2.0$\leq z \leq$2.5	&     	-1.2$\leq \beta \leq$-1.0	&      0.31$\pm$0.05		&     10.21		&     	11.13$\pm$0.15	&	2041		\\
     2.0$\leq z \leq$2.5	&     	-1.0$\leq \beta \leq$-0.8	&      0.62$\pm$0.06		&     10.13 	&    	11.43$\pm$0.09	&	1470		\\
     2.0$\leq z \leq$2.5	&     	-0.8$\leq \beta \leq$-0.6	&      0.82$\pm$0.09		&     10.07		&     	11.53$\pm$0.10	&	857		\\
     2.0$\leq z \leq$2.5	&     	-0.6$\leq \beta \leq$-0.4	&      0.73$\pm$0.09		&      9.91		&    	11.49$\pm$0.10	&	776		\\
     2.0$\leq z \leq$2.5	&     	-0.4$\leq \beta \leq$-0.2	&      1.14$\pm$0.12		&      9.86 		&   	11.69$\pm$0.08	&	526		\\
     2.5$\leq z \leq$3.0	&     	-1.2$\leq \beta \leq$-1.0	&      0.47$\pm$0.07		&     10.40		&     	11.56$\pm$0.15	&	1001		\\
     2.5$\leq z \leq$3.0	&     	-1.0$\leq \beta \leq$-0.8	&      0.58$\pm$0.08		&     10.32		&     	11.67$\pm$0.13	&	721		\\
     2.5$\leq z \leq$3.0	&     	-0.8$\leq \beta \leq$-0.6	&      0.63$\pm$0.13		&     10.22		&     	11.69$\pm$0.19	&	376		\\
\hline
\end{tabular}
\end{center}
\end{table*}

\begin{figure}
\centering
\includegraphics[width=0.19\textwidth]{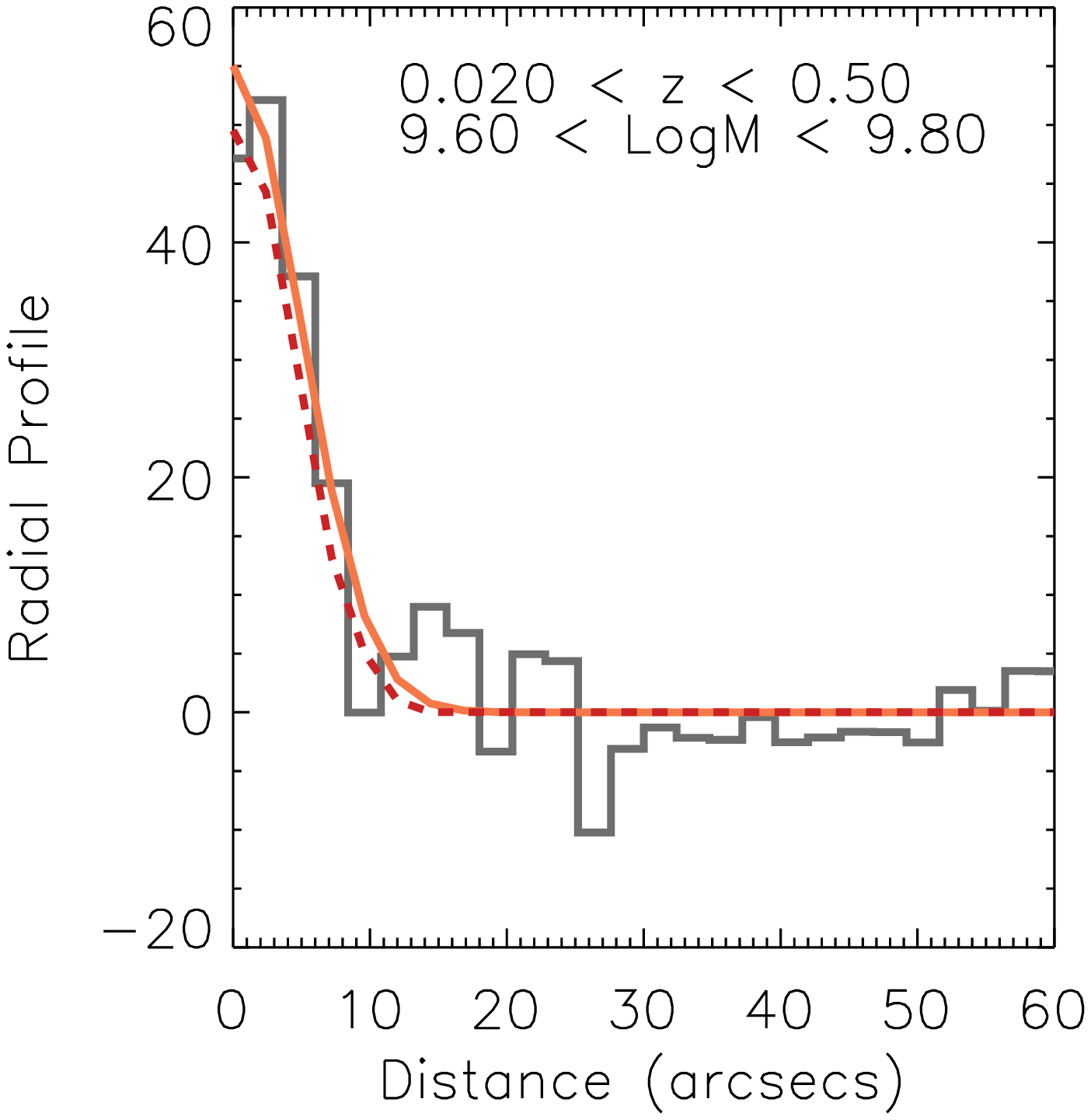}
\includegraphics[width=0.19\textwidth]{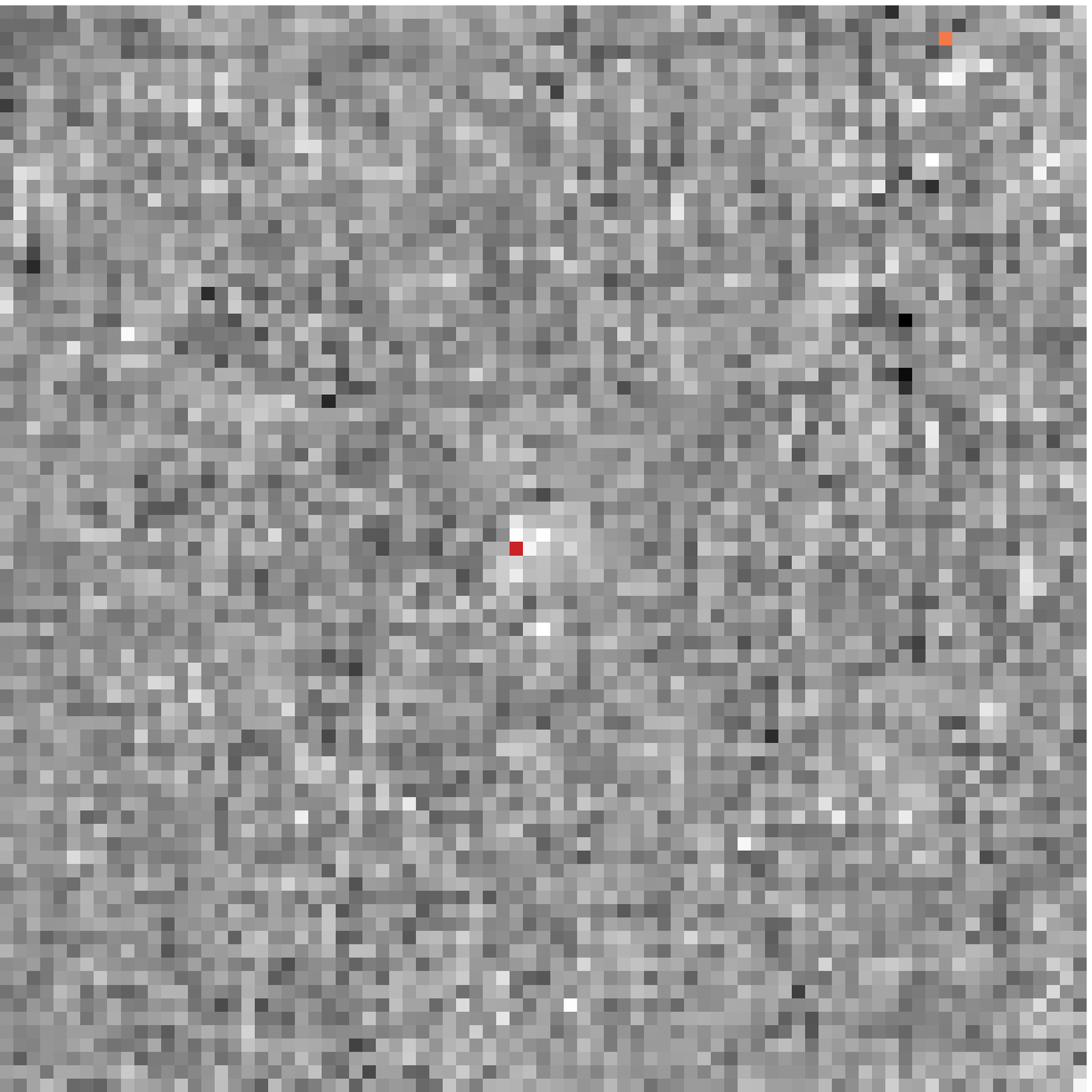}
\includegraphics[width=0.19\textwidth]{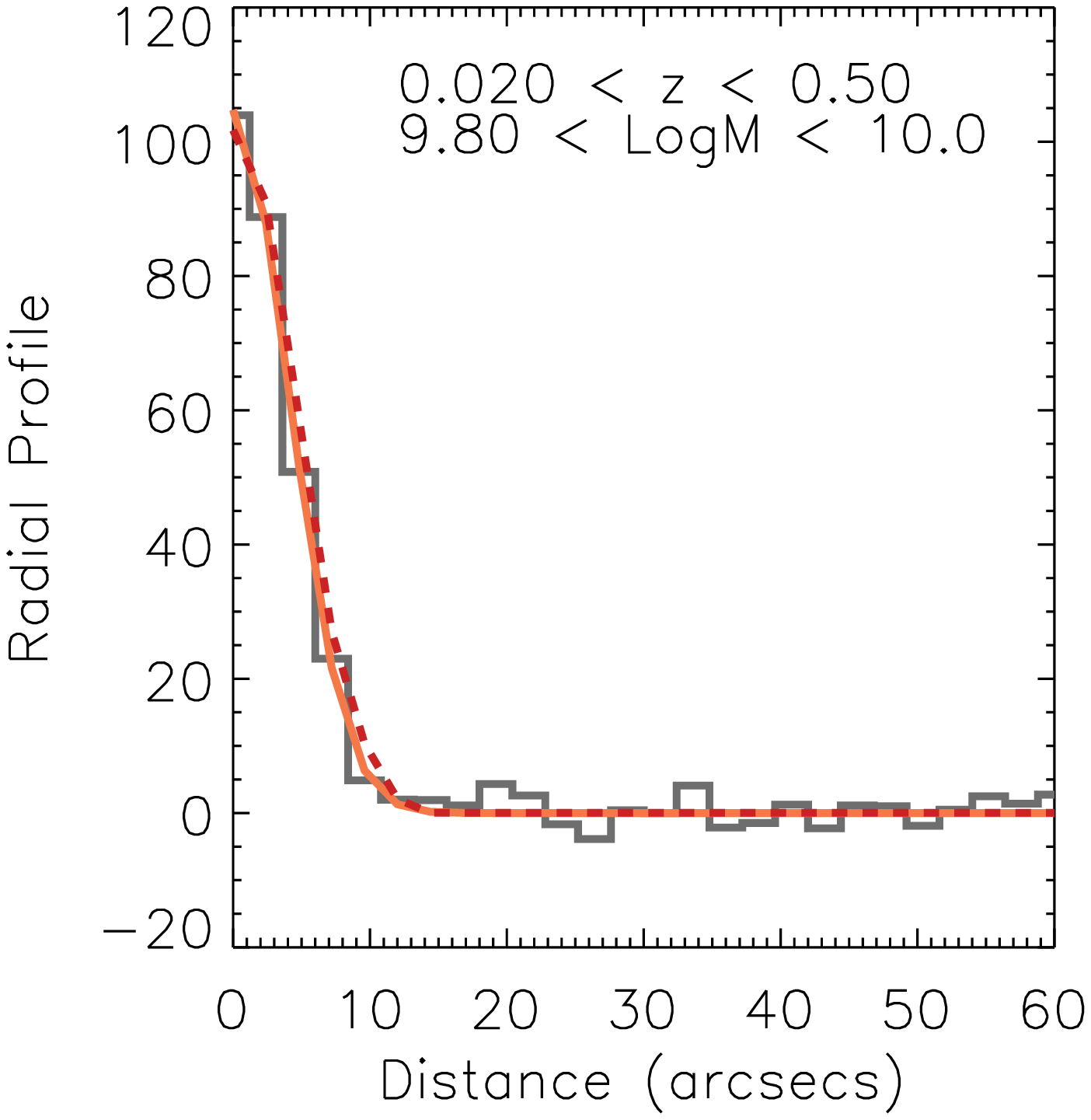}
\includegraphics[width=0.19\textwidth]{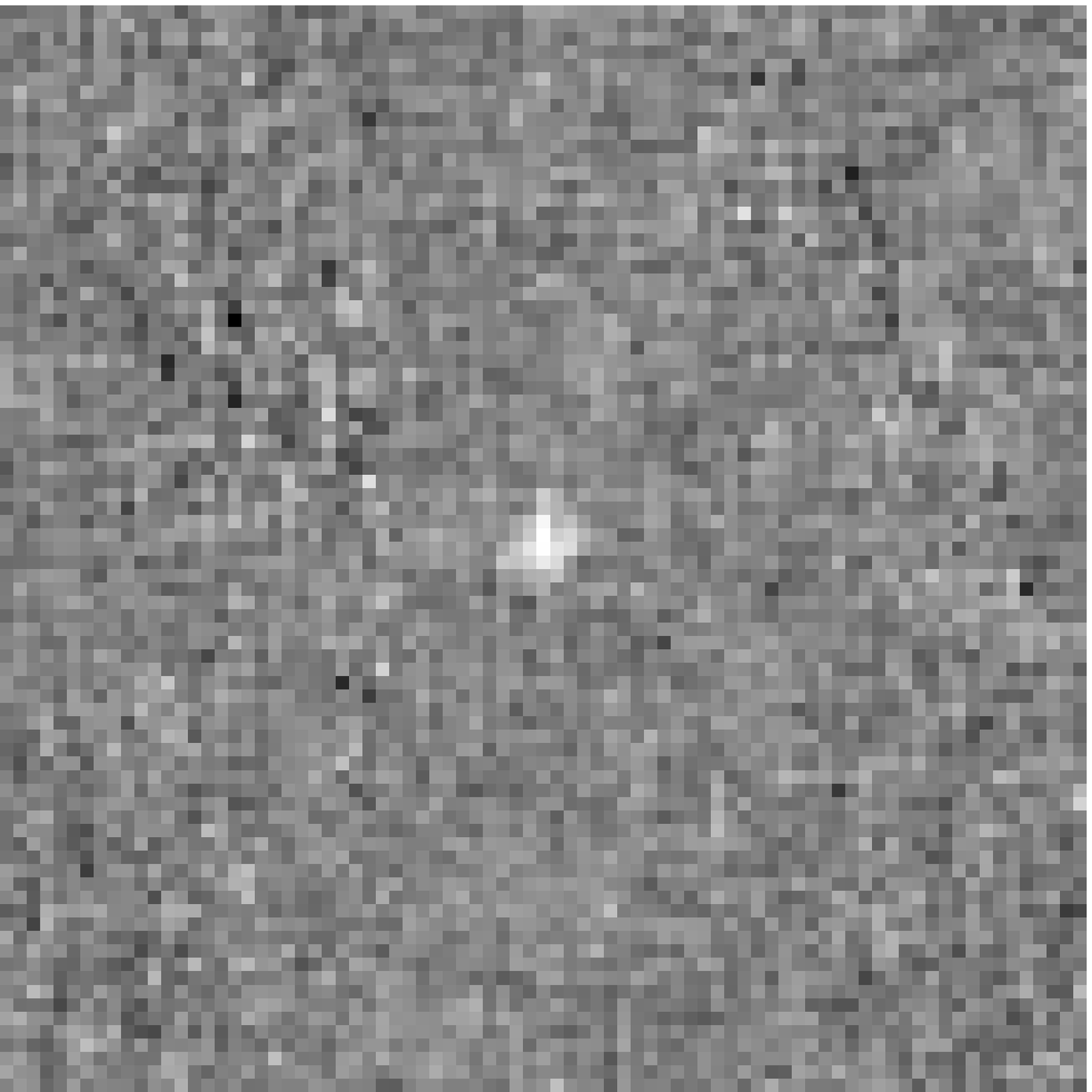}\\
\includegraphics[width=0.19\textwidth]{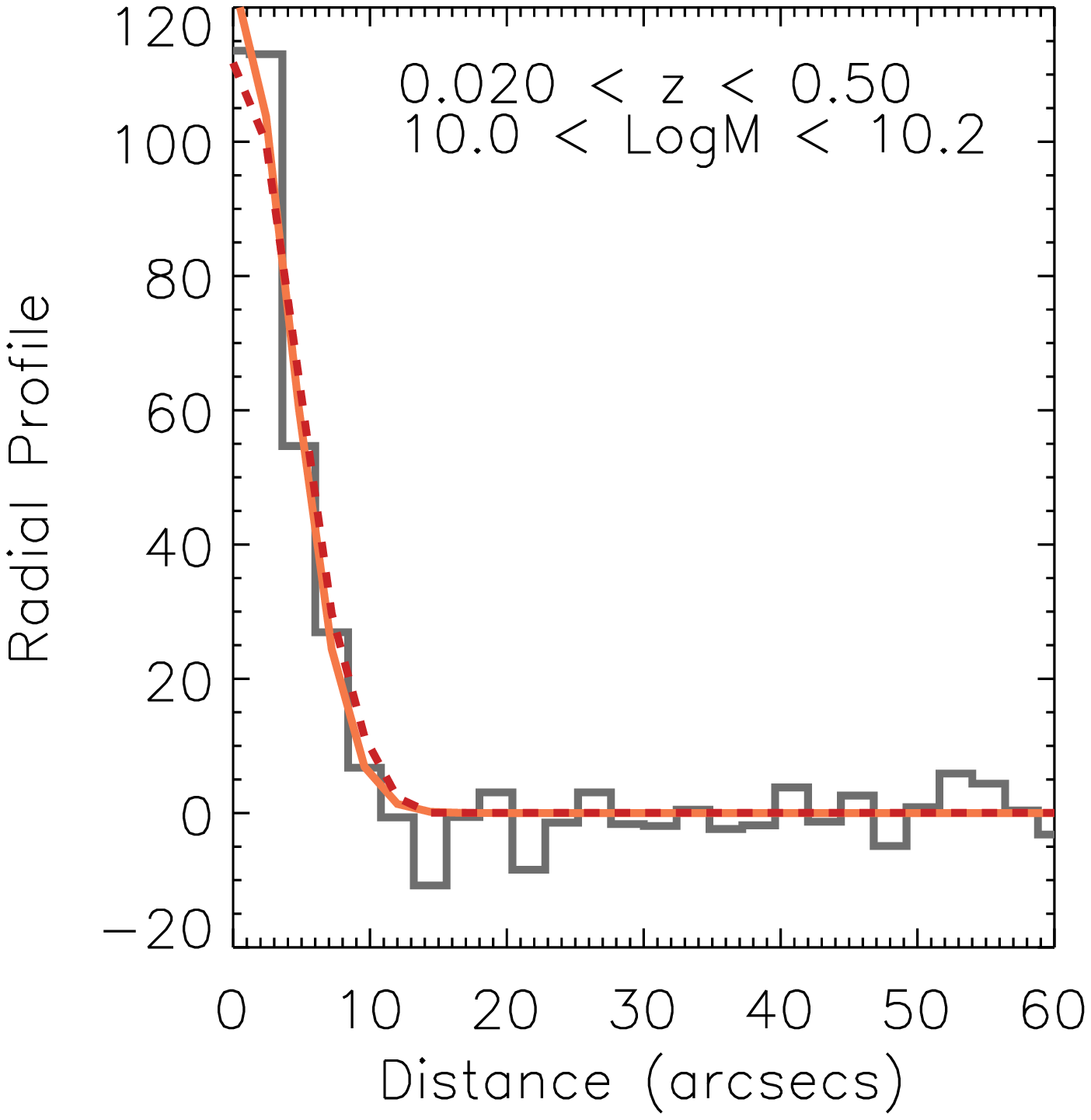}
\includegraphics[width=0.19\textwidth]{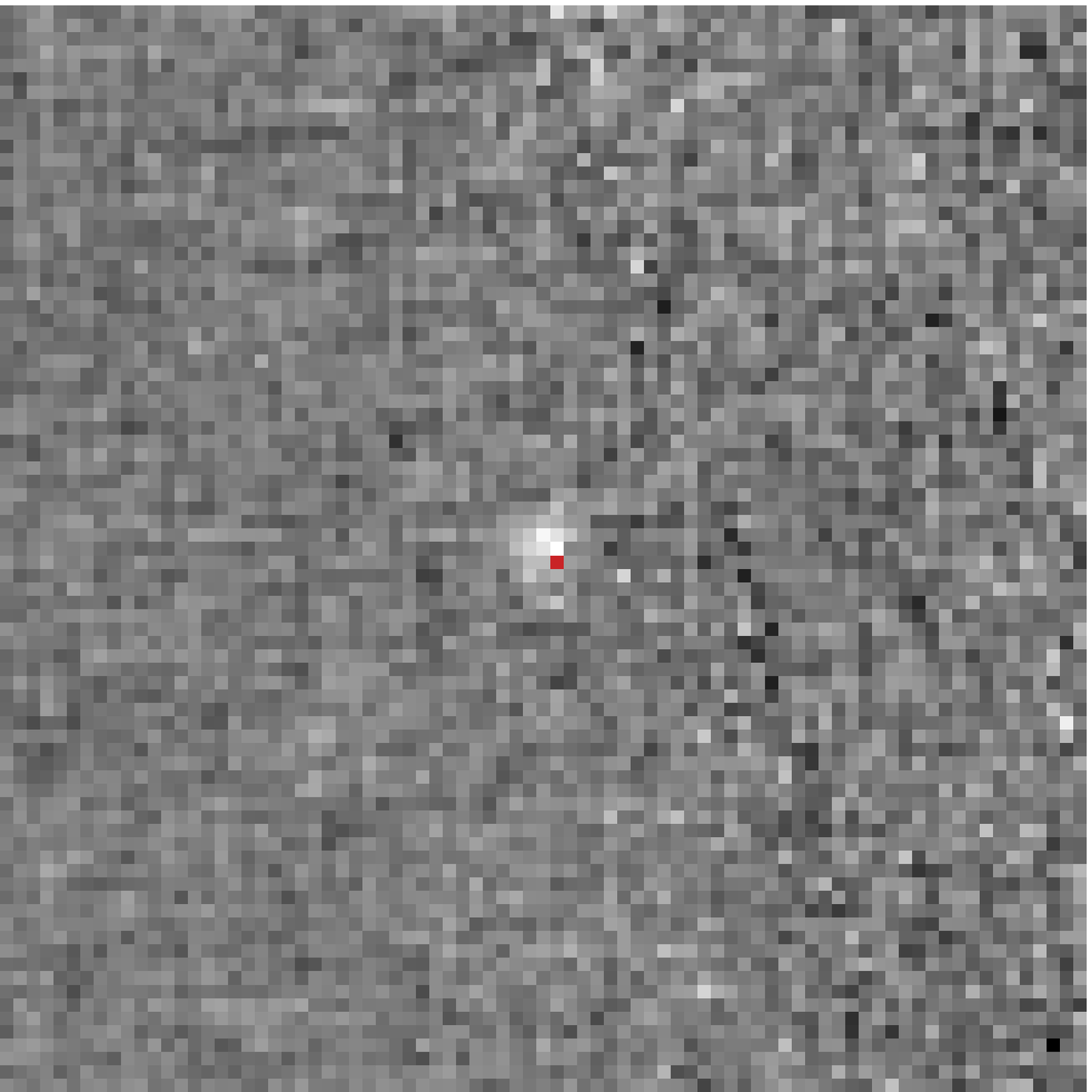}
\includegraphics[width=0.19\textwidth]{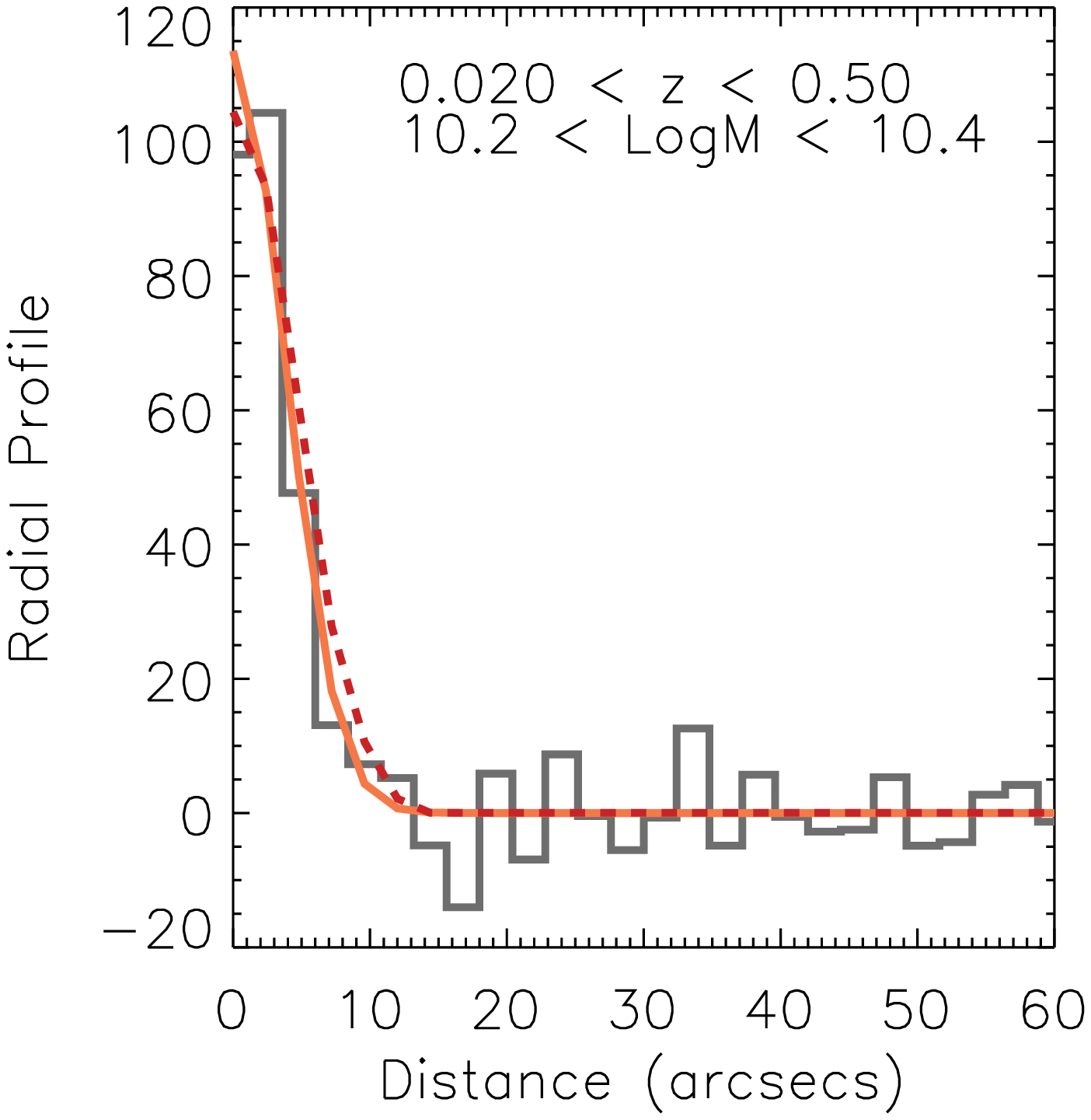}
\includegraphics[width=0.19\textwidth]{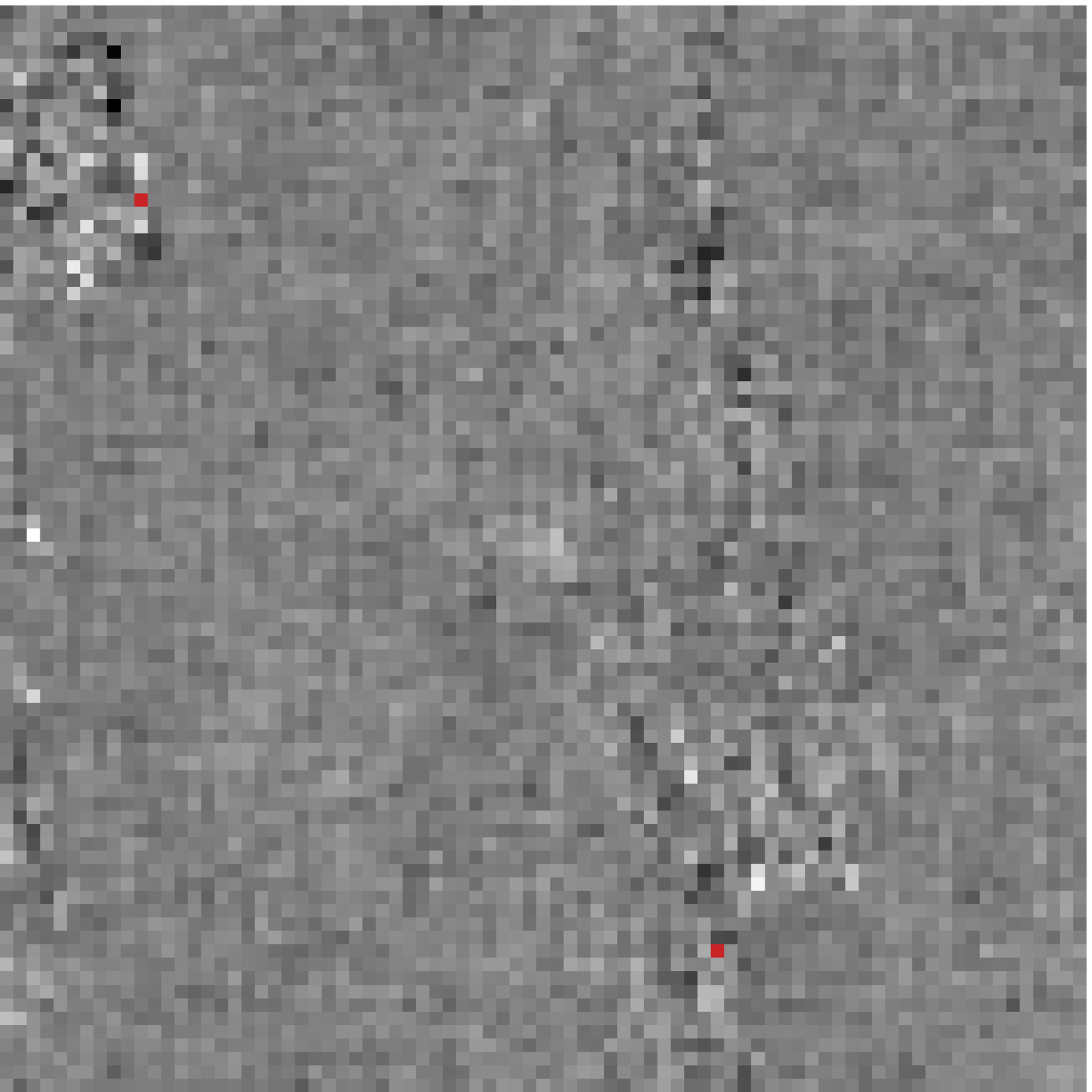}\\
\includegraphics[width=0.19\textwidth]{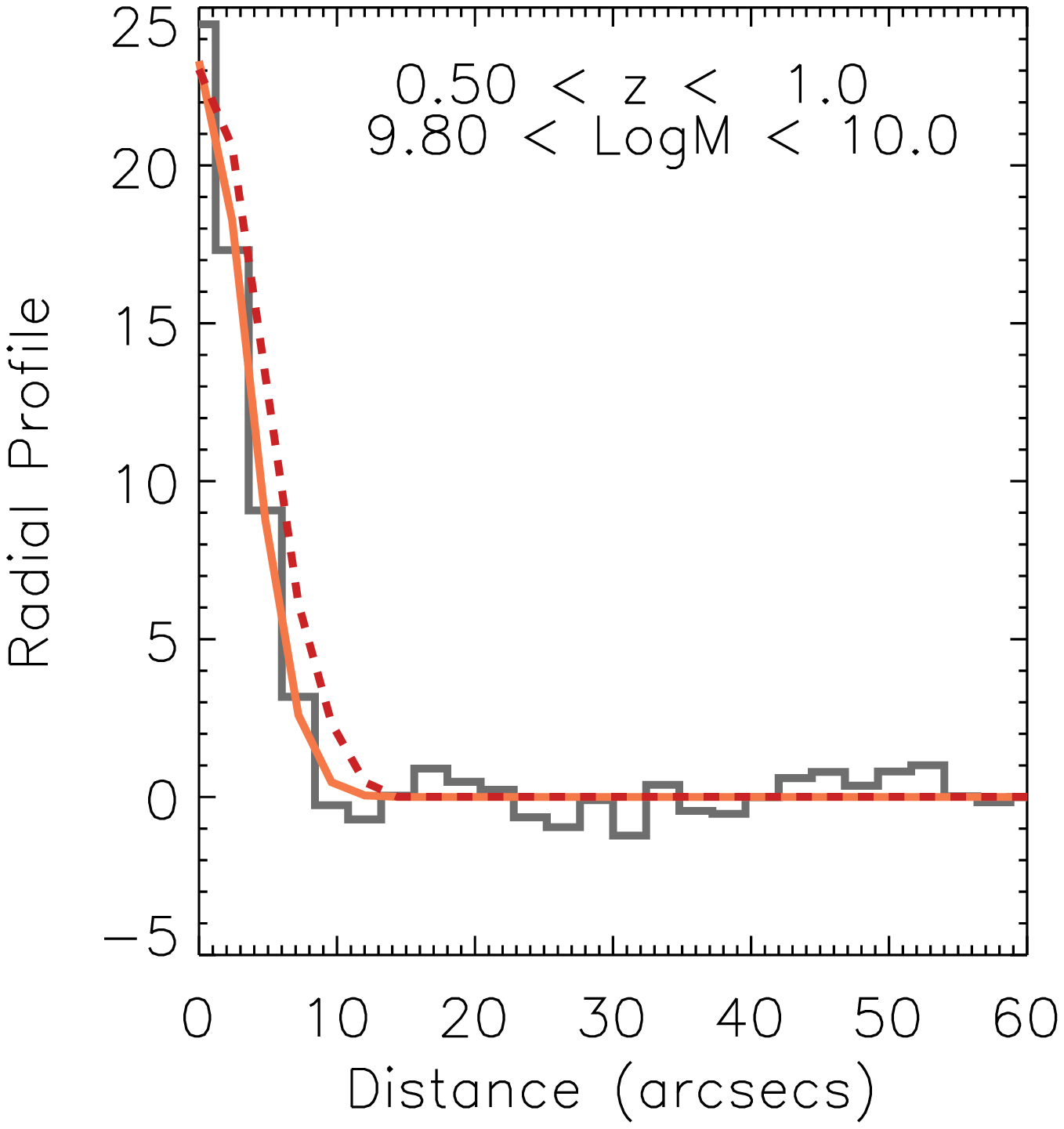}
\includegraphics[width=0.19\textwidth]{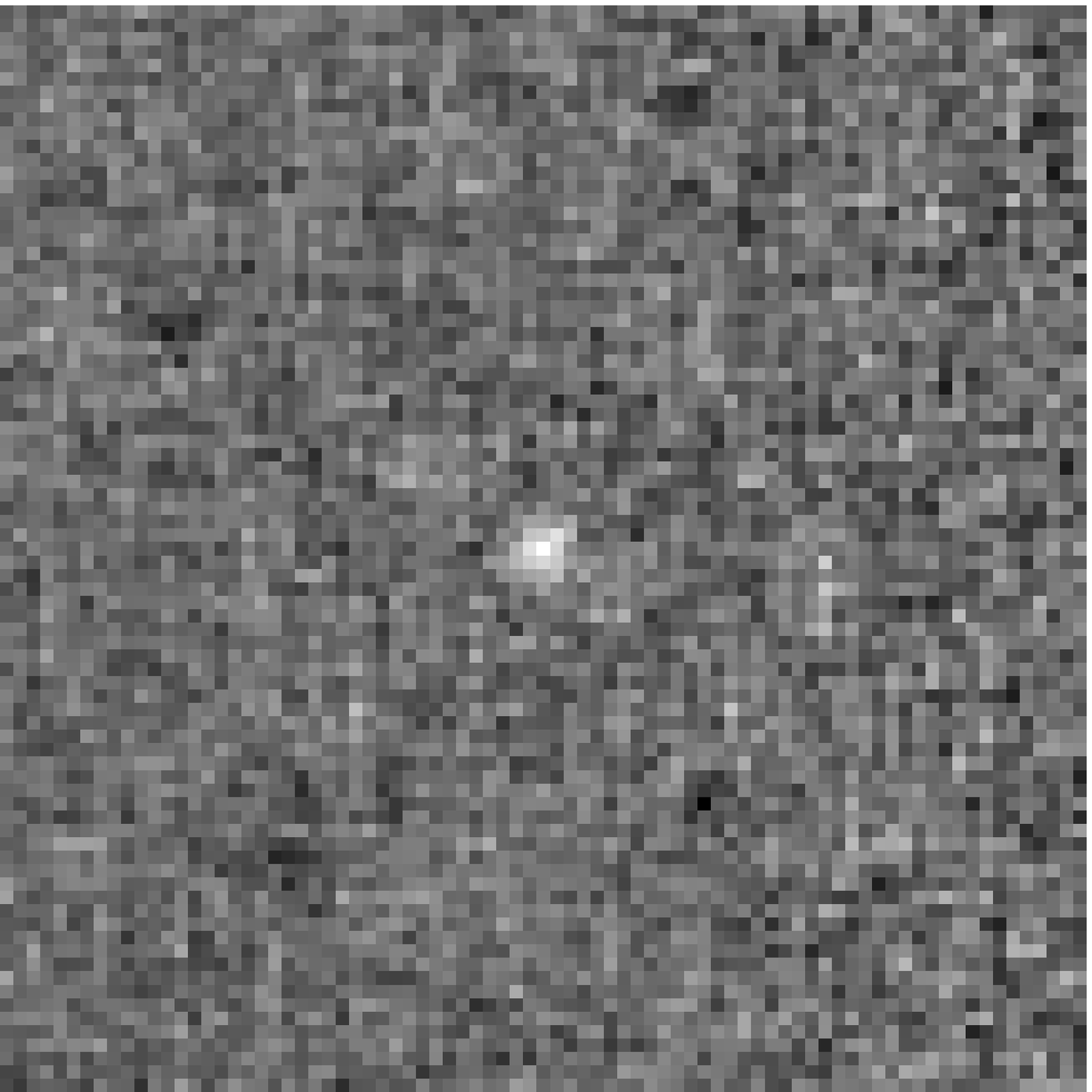}
\includegraphics[width=0.19\textwidth]{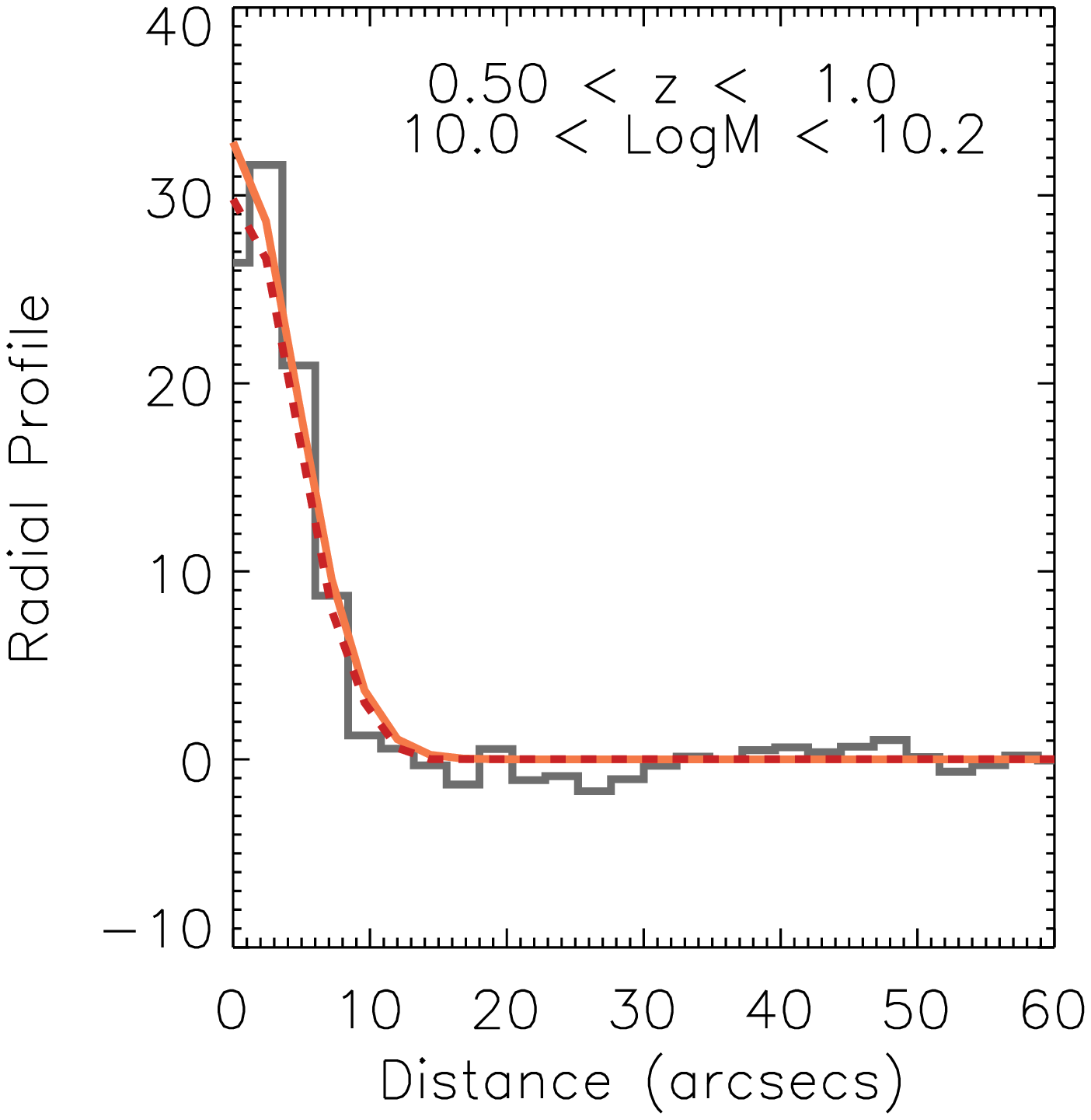}
\includegraphics[width=0.19\textwidth]{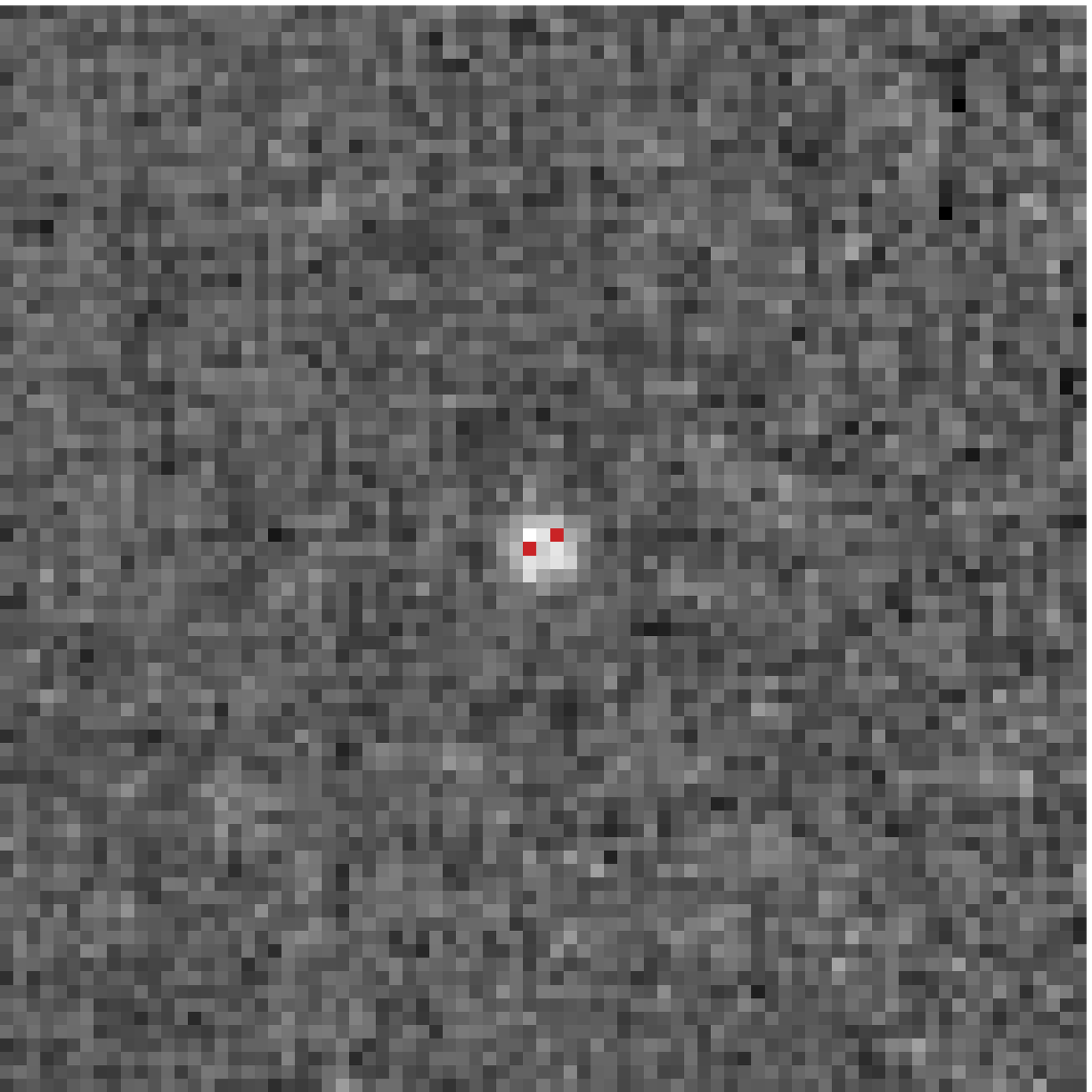}\\
\includegraphics[width=0.19\textwidth]{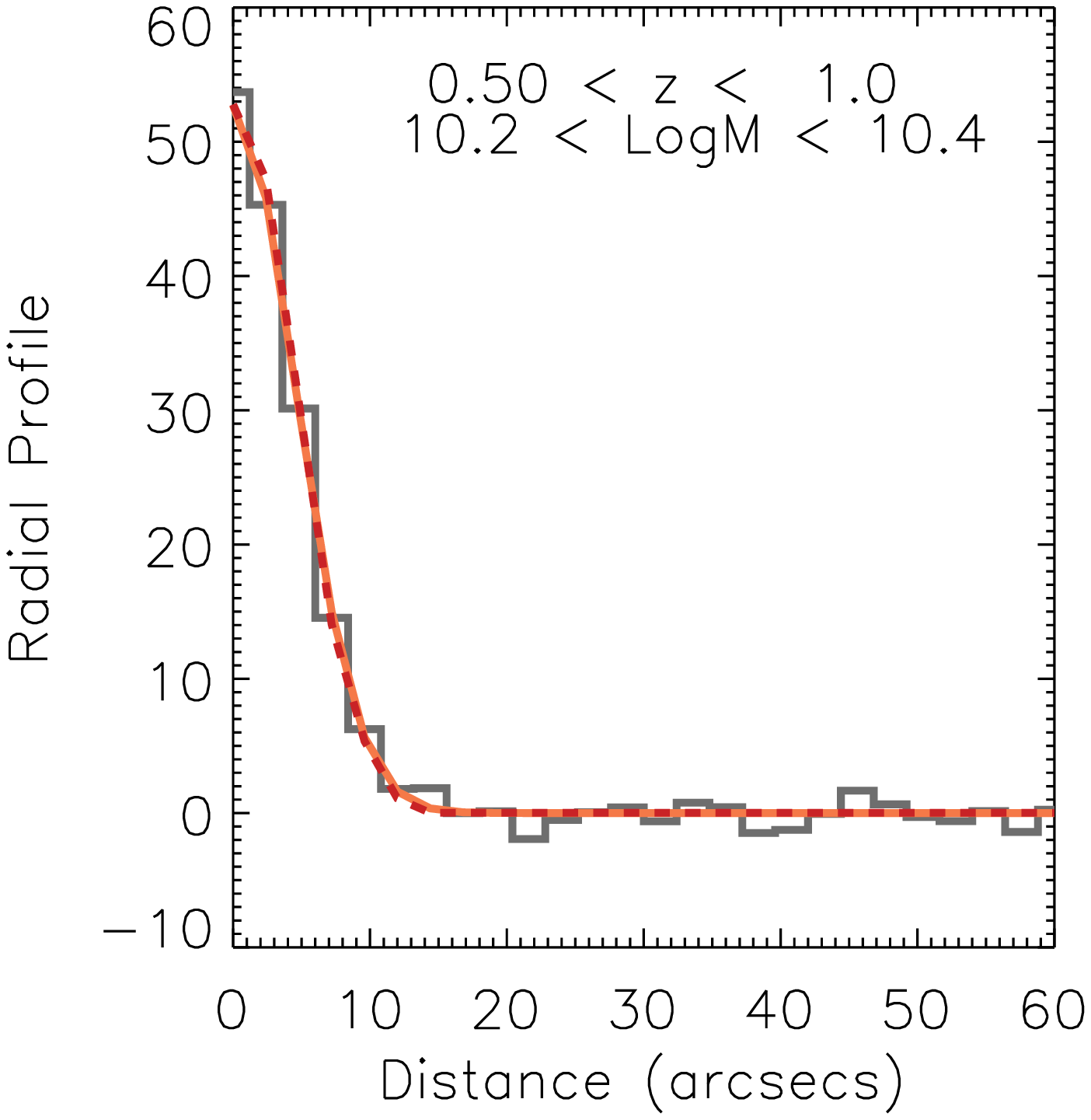}
\includegraphics[width=0.19\textwidth]{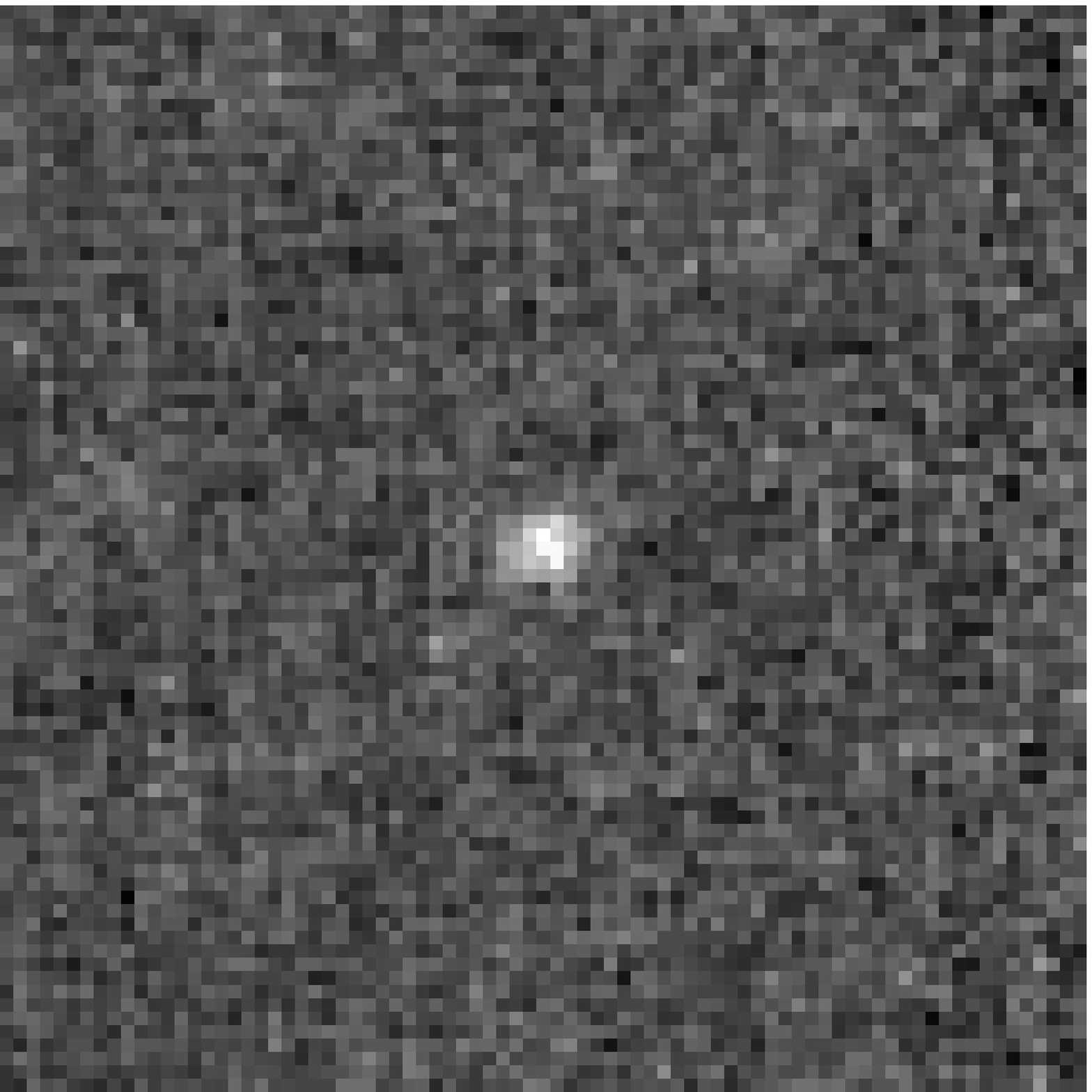}
\includegraphics[width=0.19\textwidth]{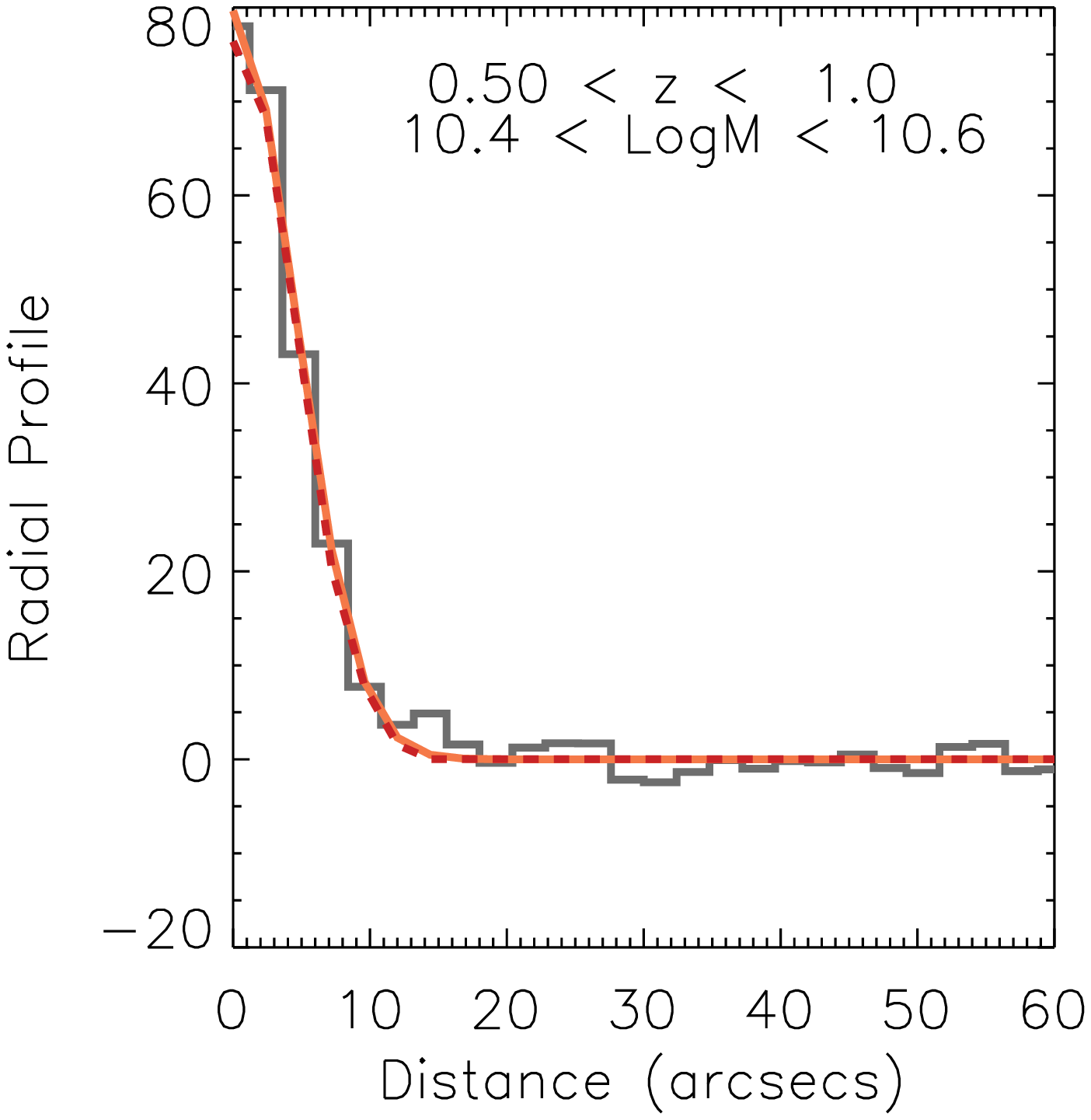}
\includegraphics[width=0.19\textwidth]{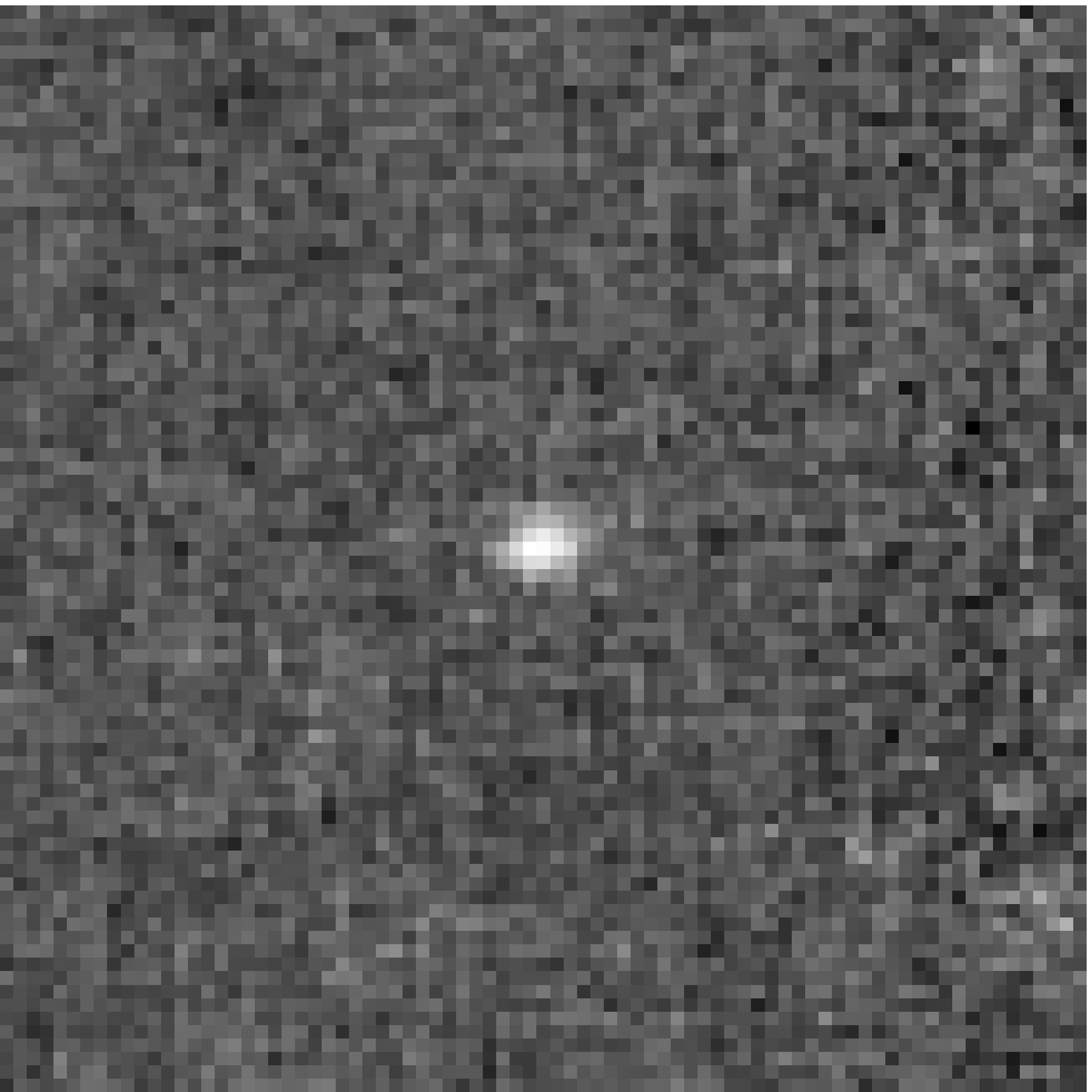}\\
\includegraphics[width=0.19\textwidth]{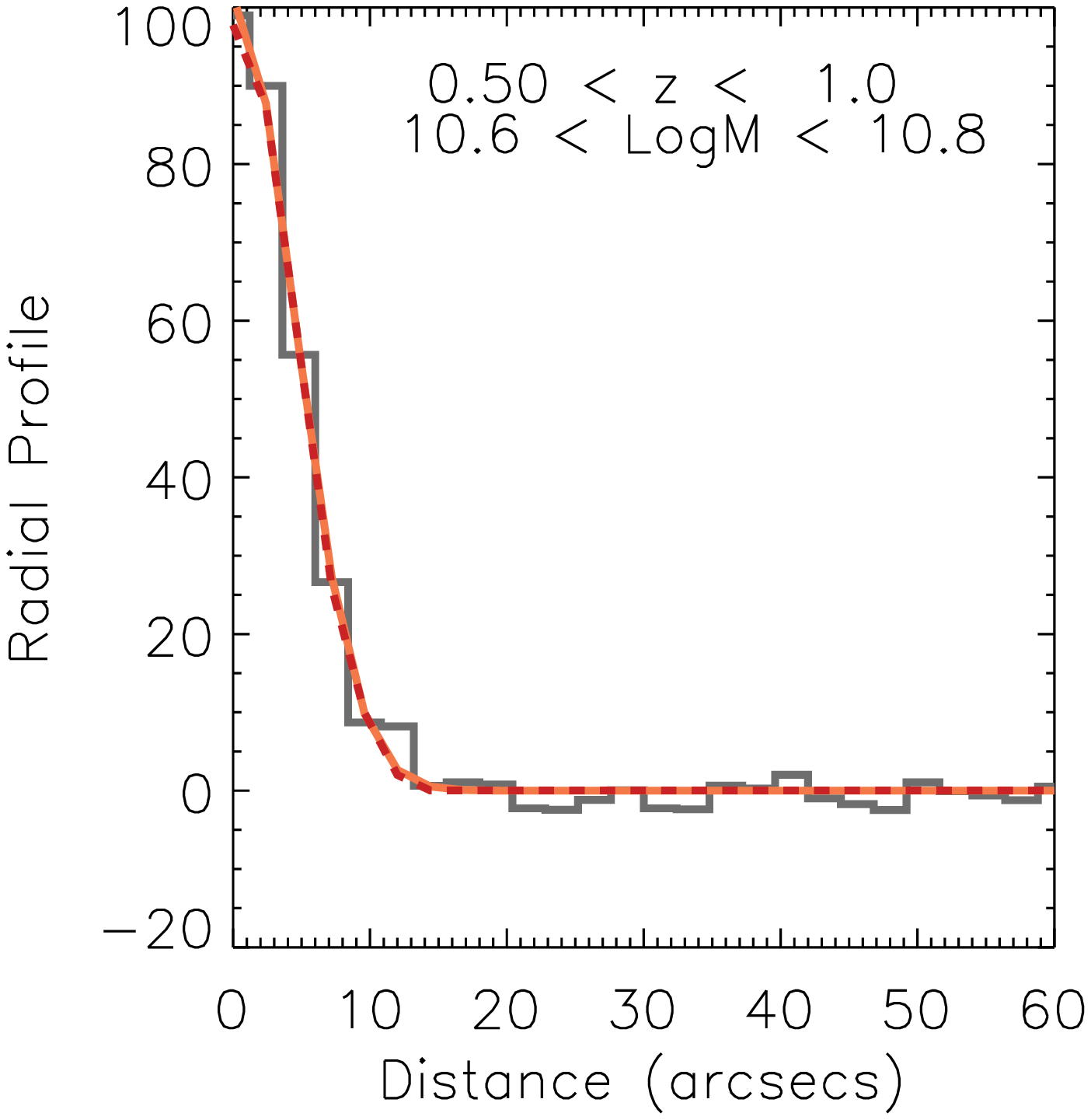}
\includegraphics[width=0.19\textwidth]{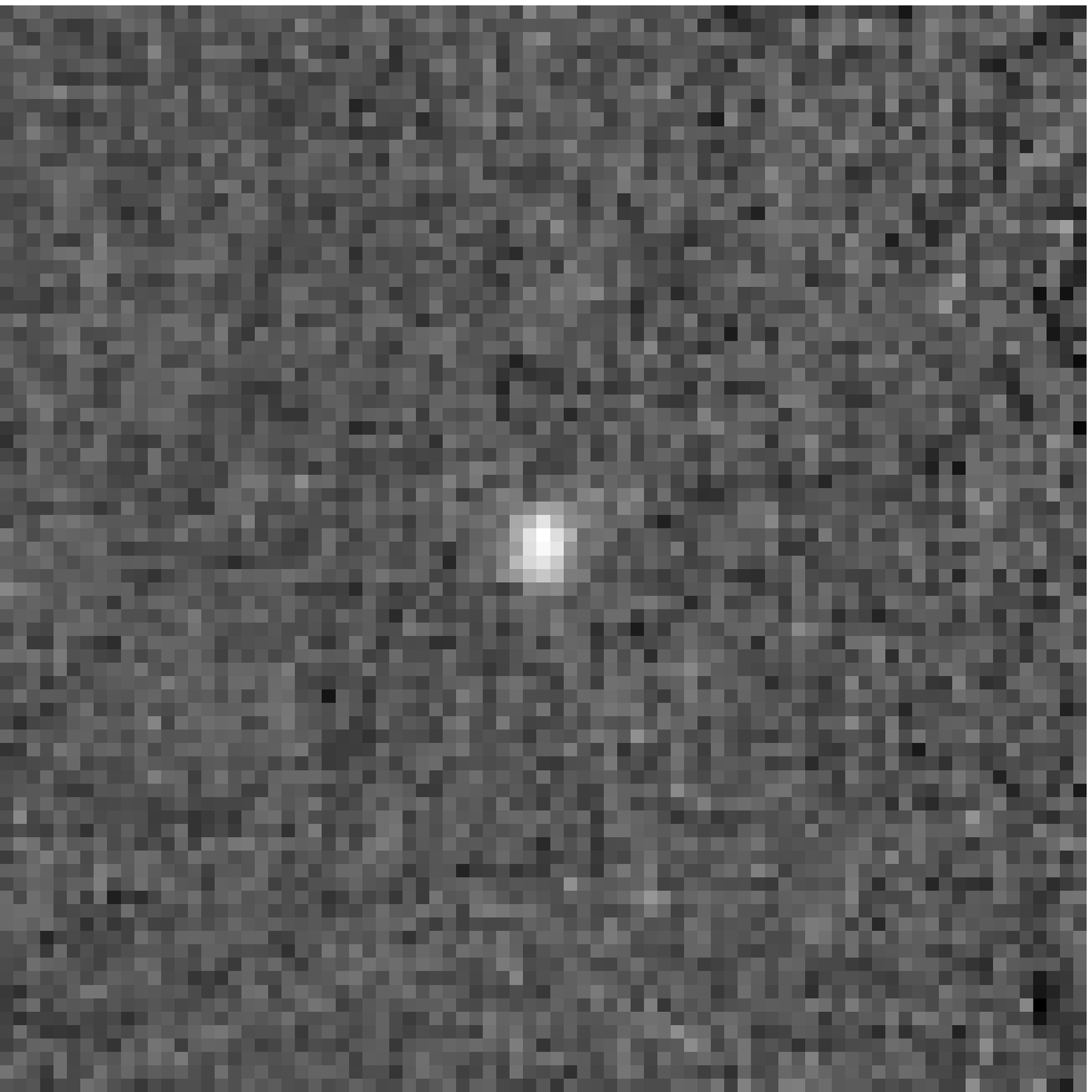}
\includegraphics[width=0.19\textwidth]{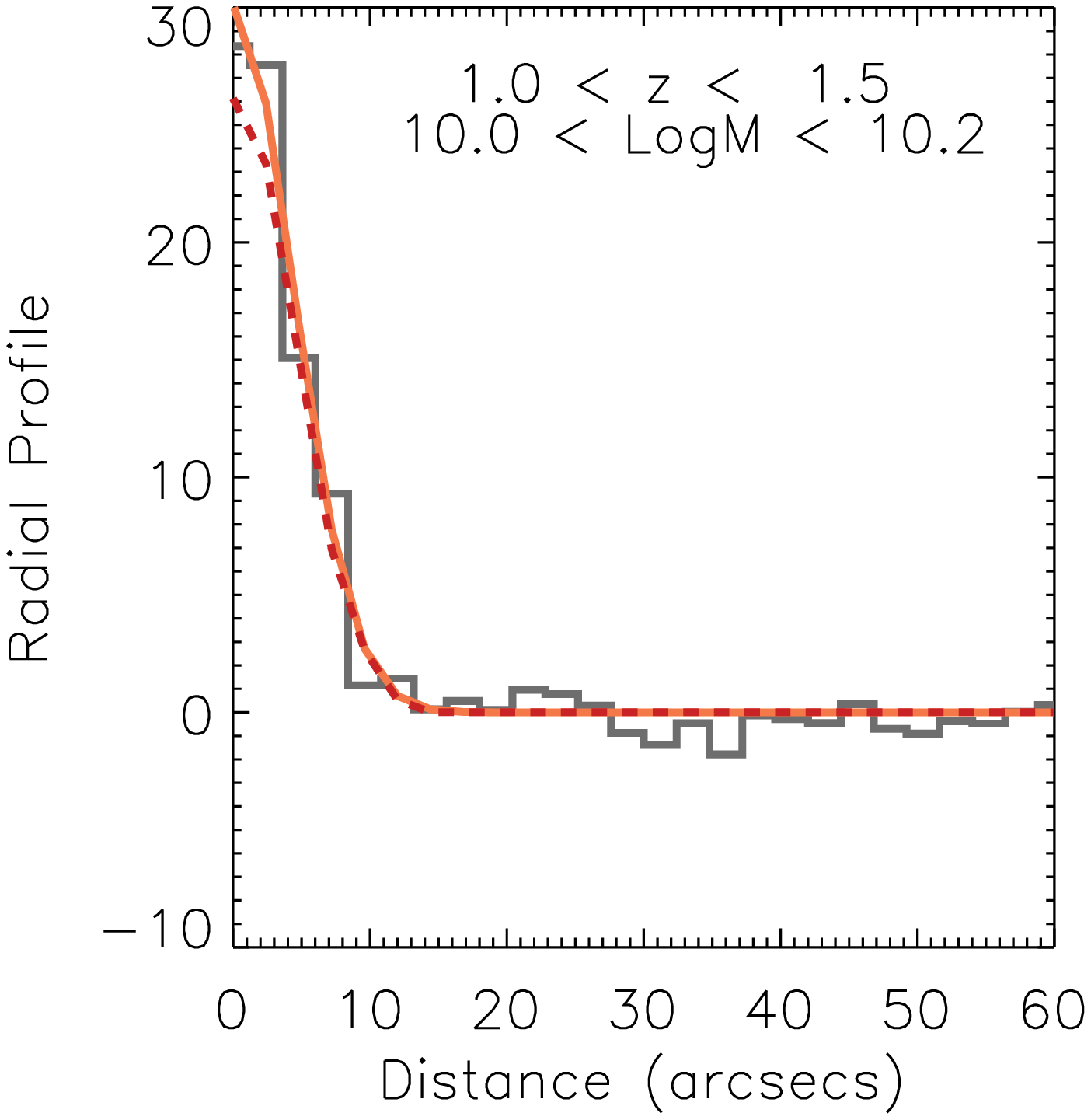}
\includegraphics[width=0.19\textwidth]{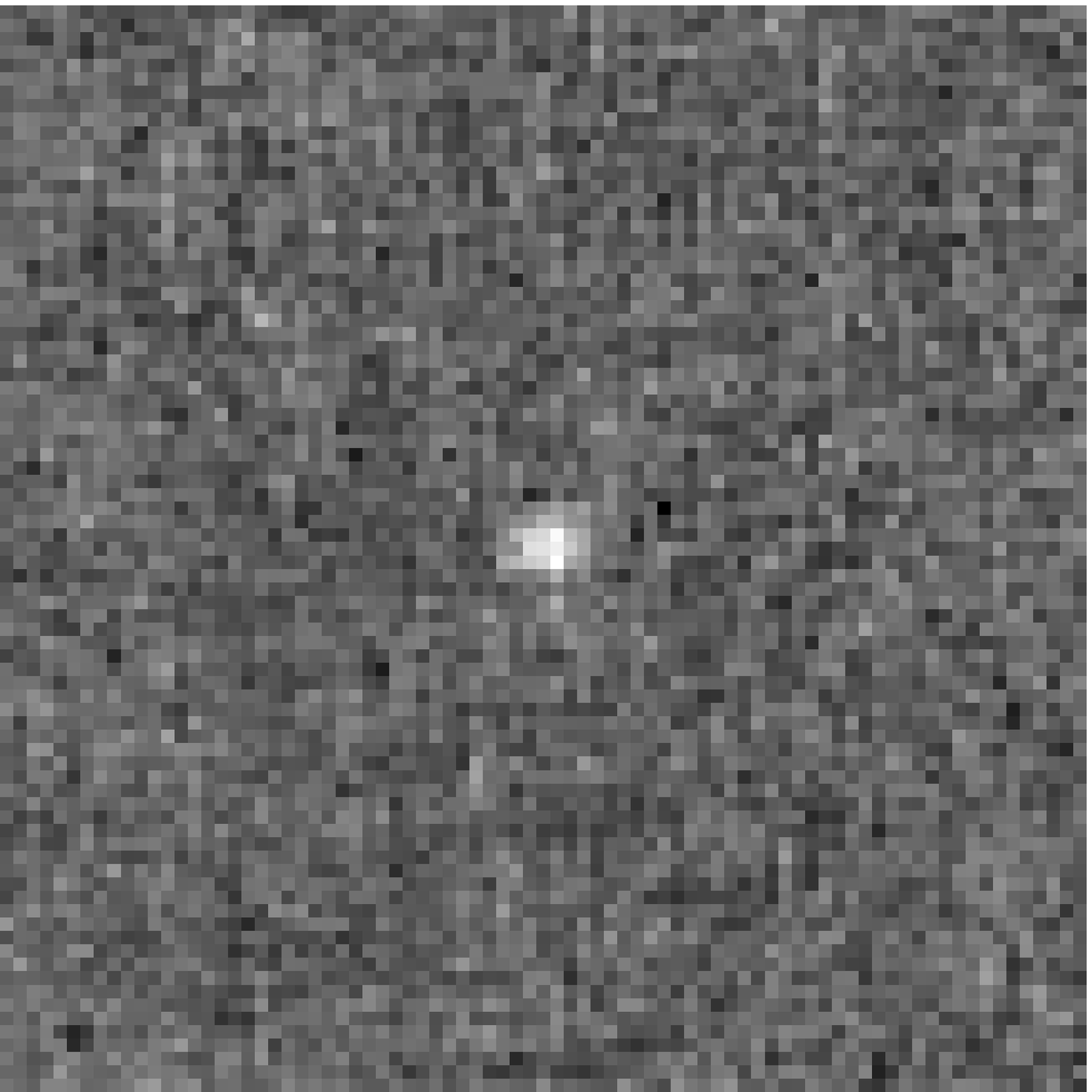}\\
\includegraphics[width=0.19\textwidth]{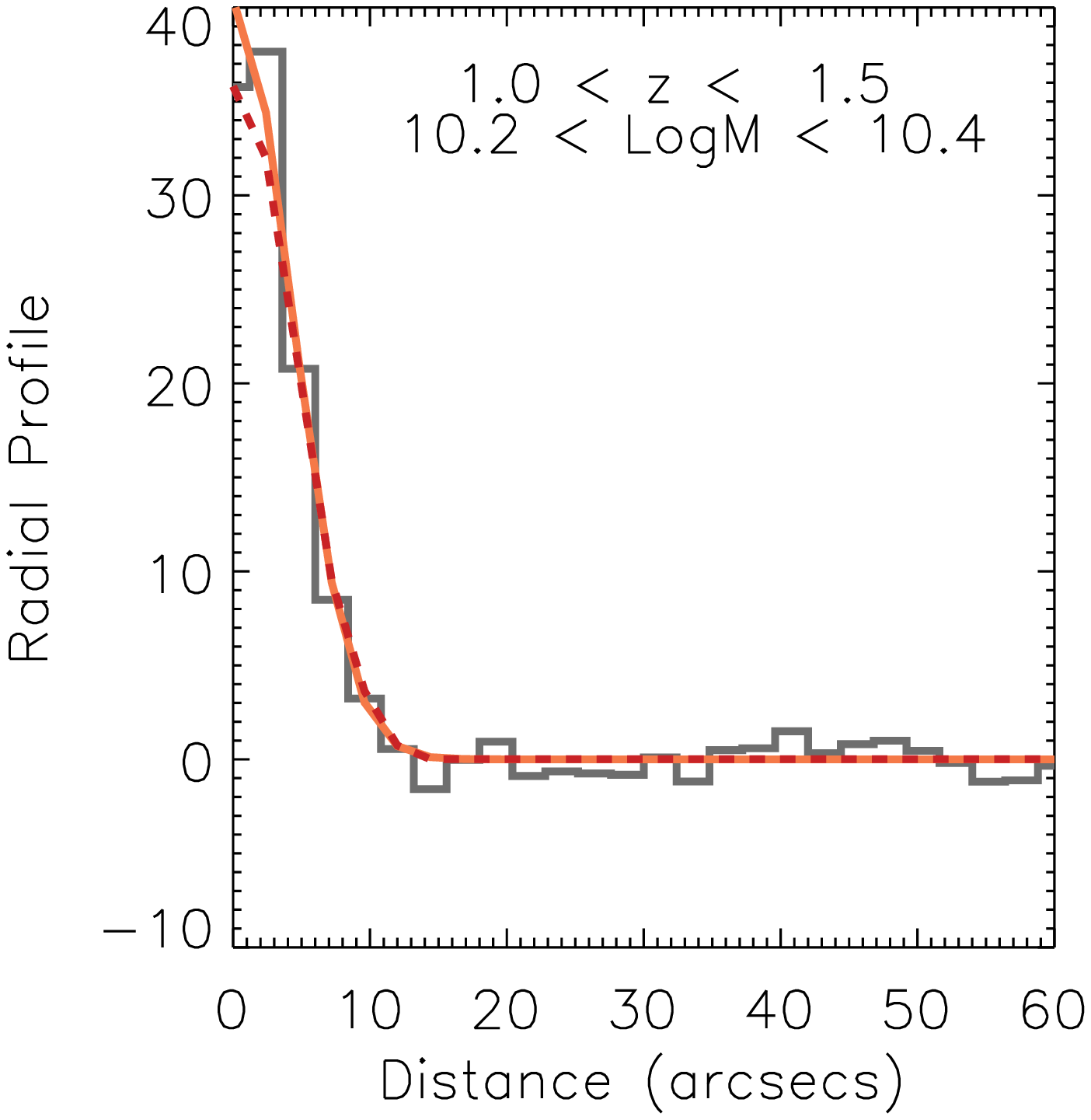}
\includegraphics[width=0.19\textwidth]{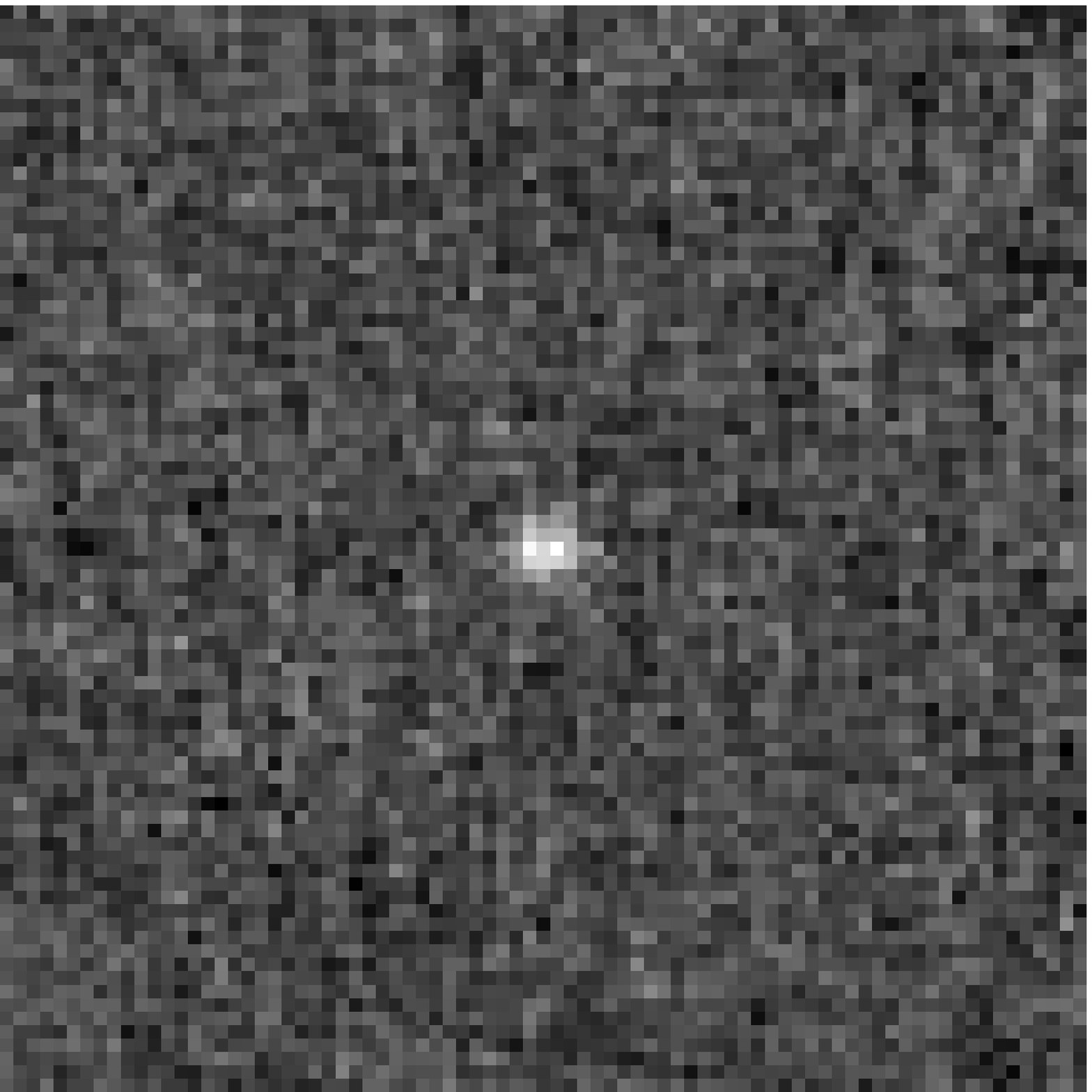}
\includegraphics[width=0.19\textwidth]{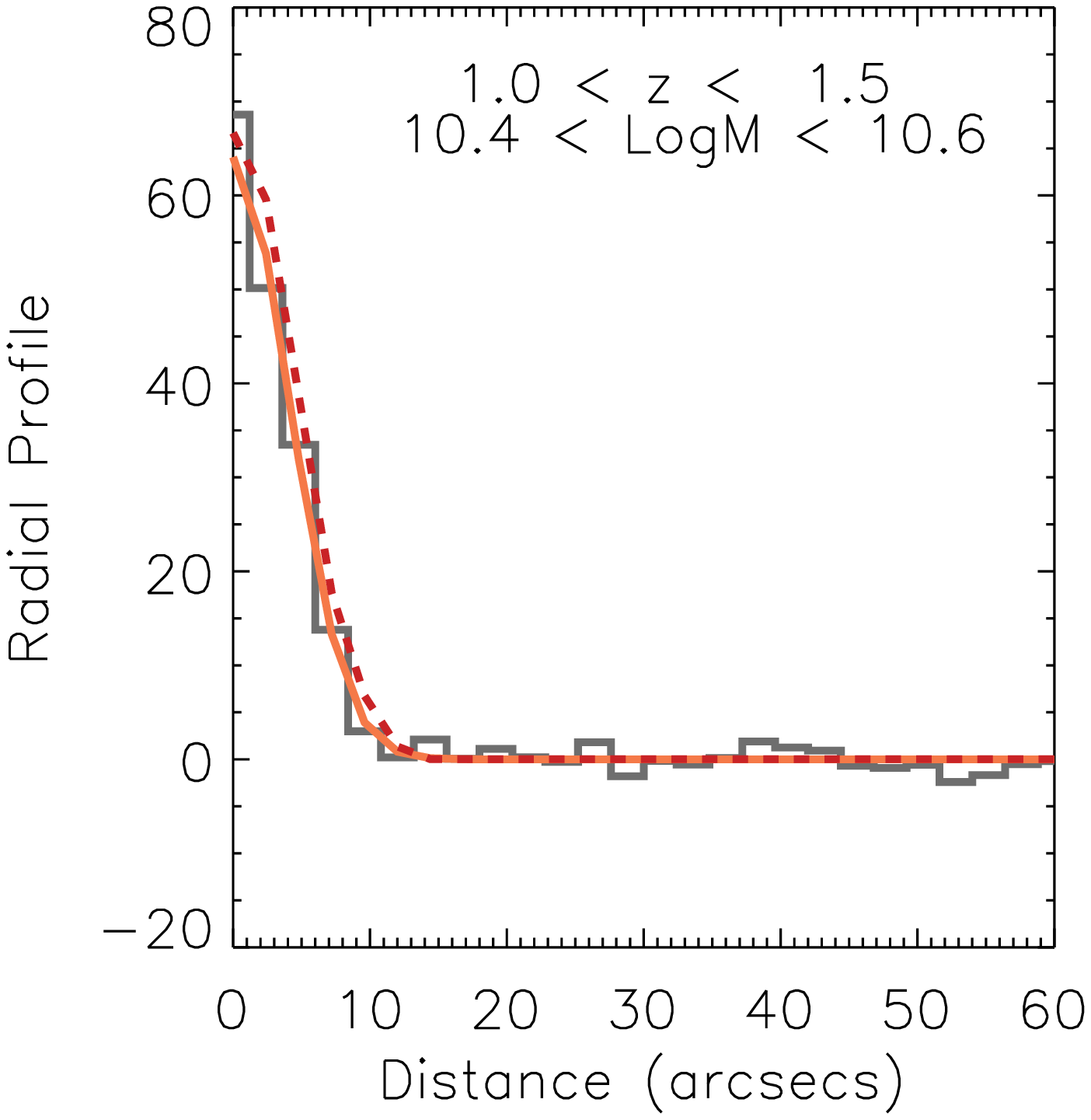}
\includegraphics[width=0.19\textwidth]{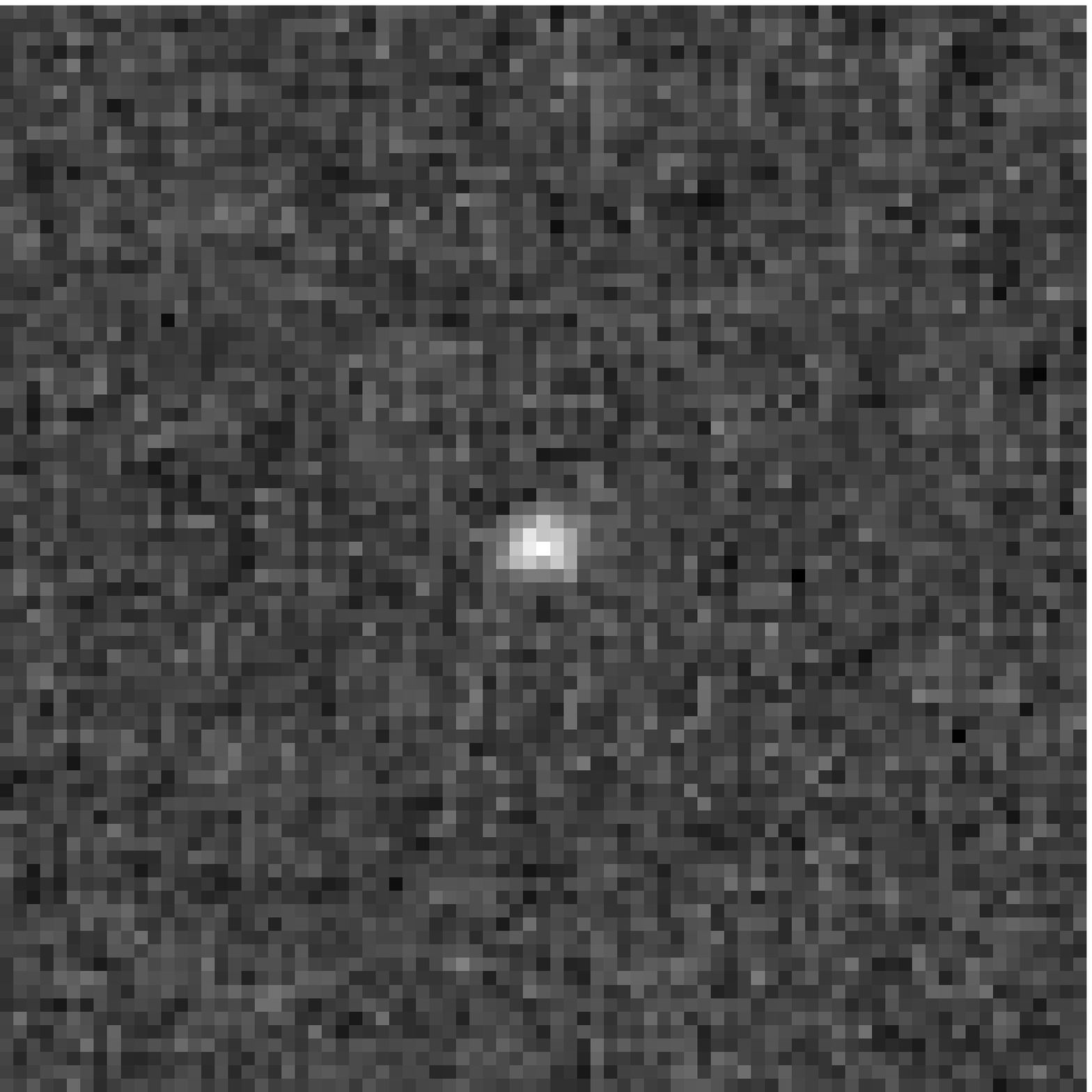}\\
\caption{Radial profiles and associated stacked images when stacking as a function of the stellar mass. Redshift and stellar mass bins are indicated in each case. Orange curves are the best-fitted gaussian to the radial profile, while red dashed curves represent the shape of the PACS-160 $\mu$m PSF.
              }
\label{masa1}
\end{figure}

\begin{figure}
\centering
\includegraphics[width=0.19\textwidth]{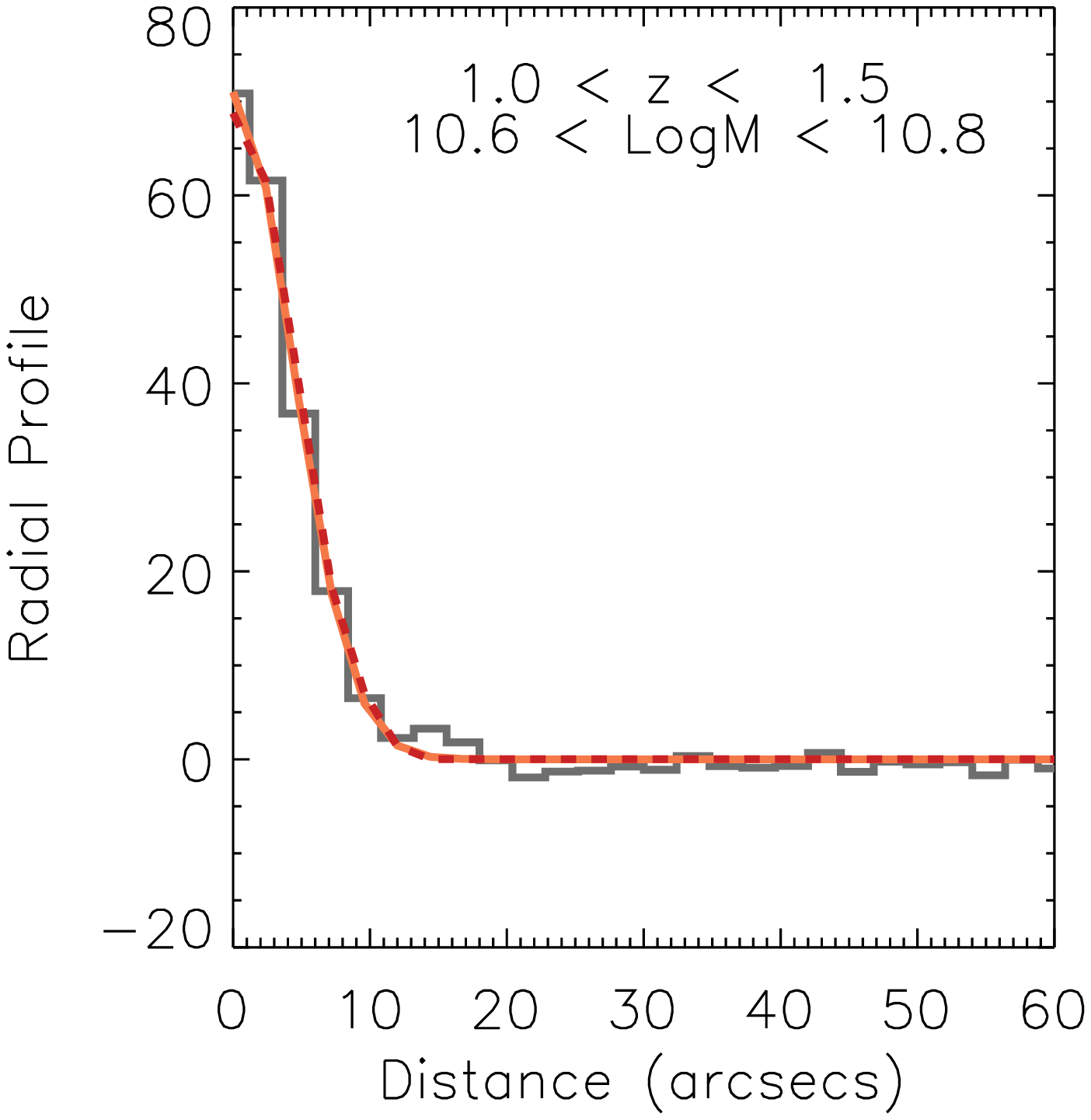}
\includegraphics[width=0.19\textwidth]{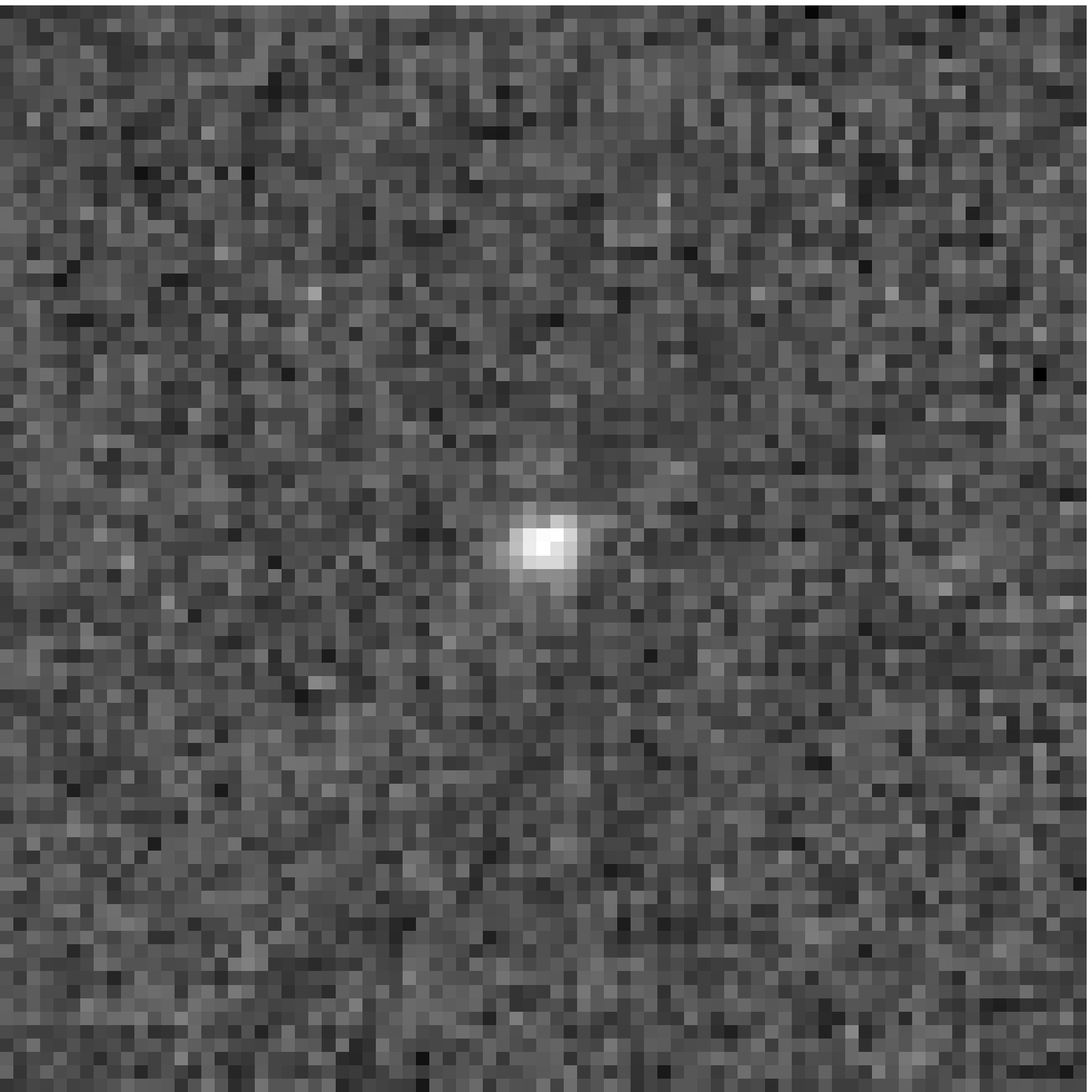}
\includegraphics[width=0.19\textwidth]{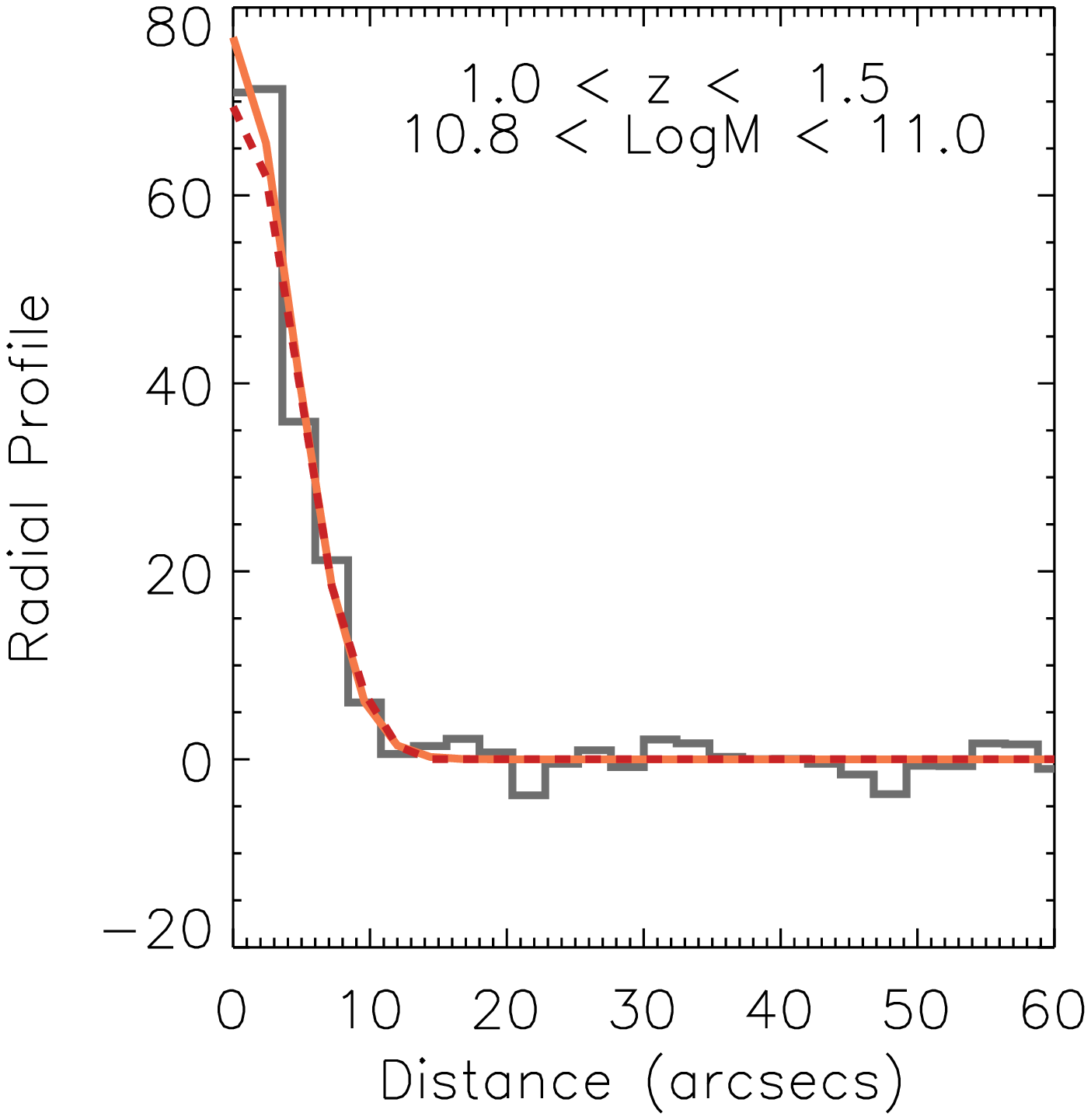}
\includegraphics[width=0.19\textwidth]{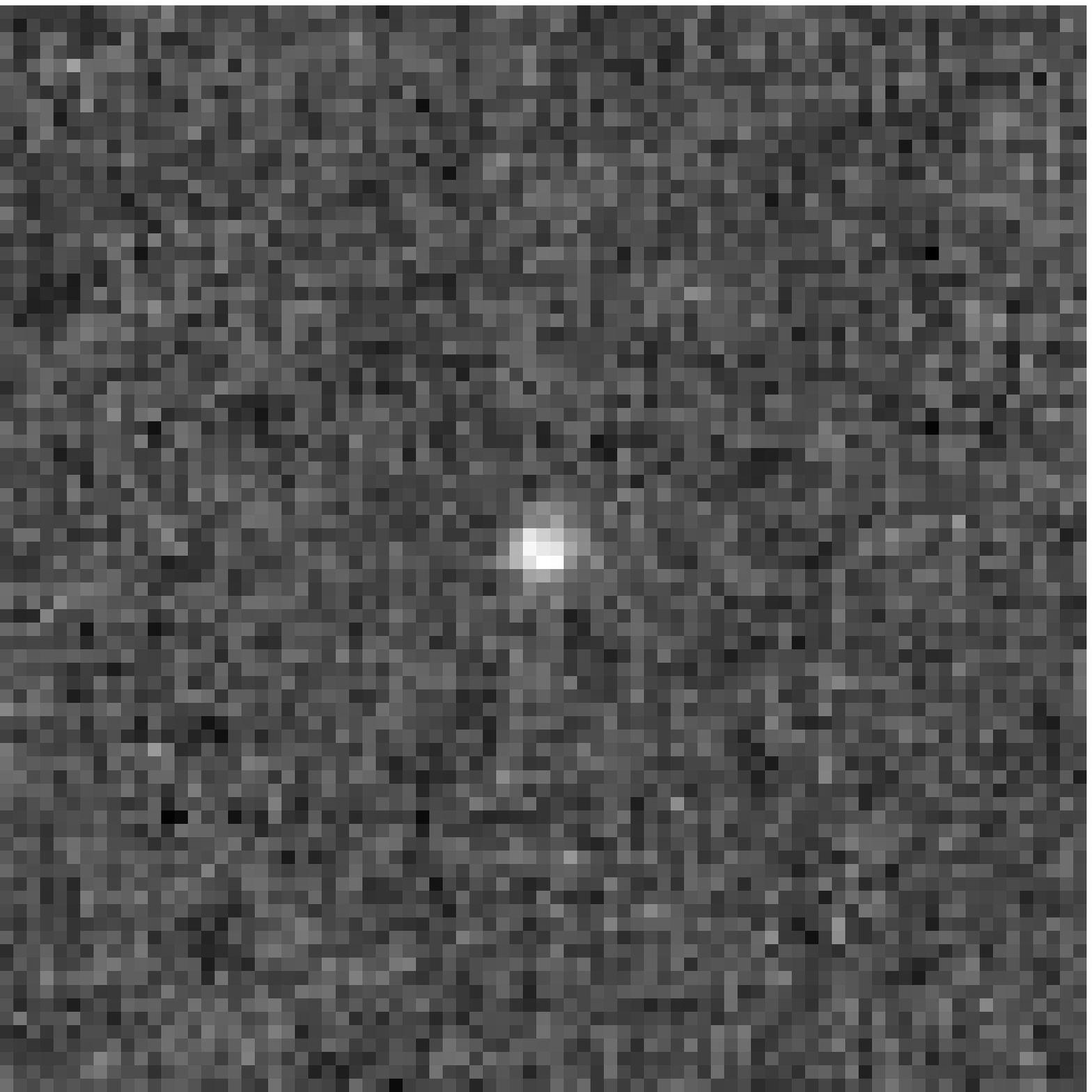}\\
\includegraphics[width=0.19\textwidth]{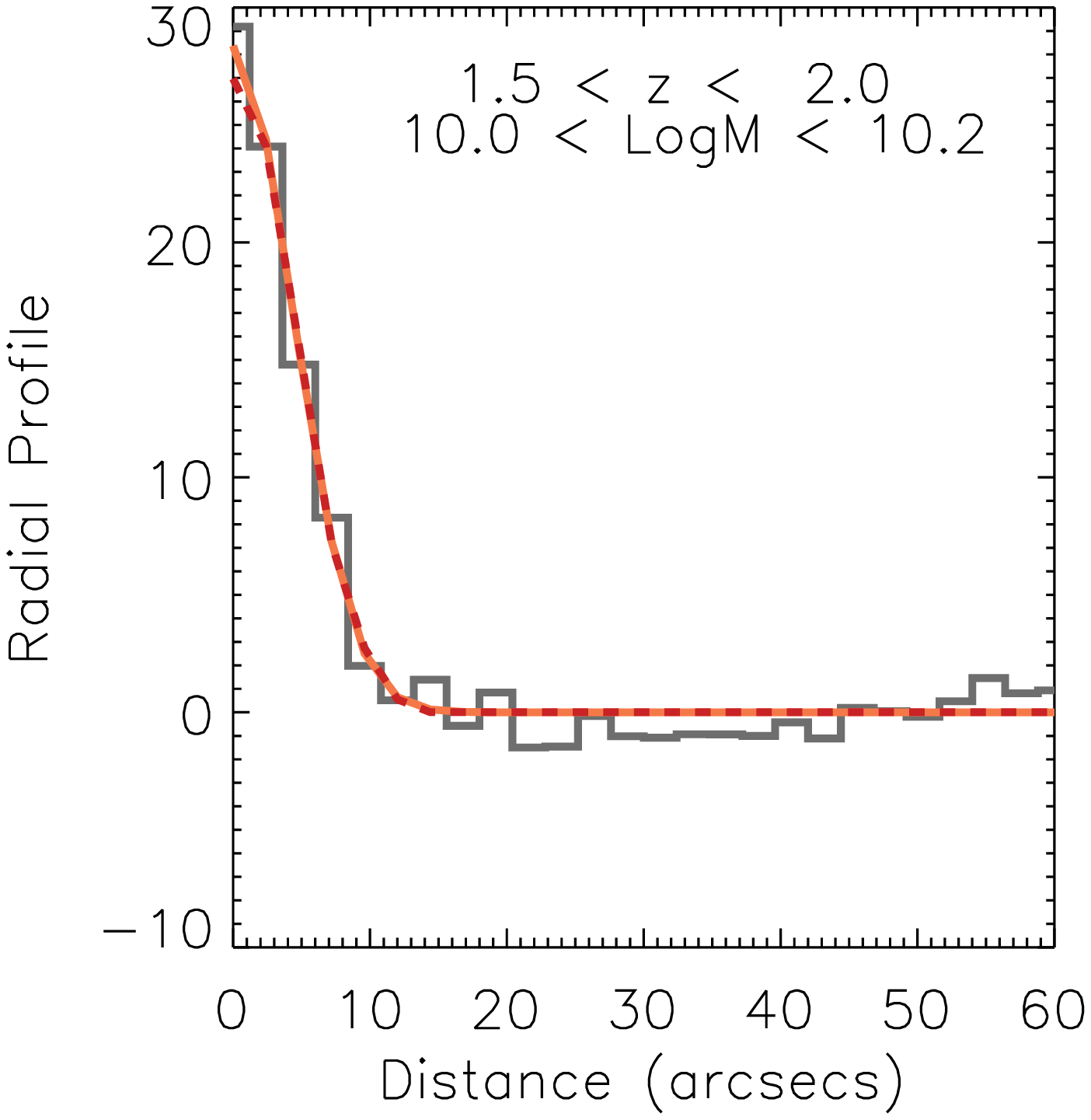}
\includegraphics[width=0.19\textwidth]{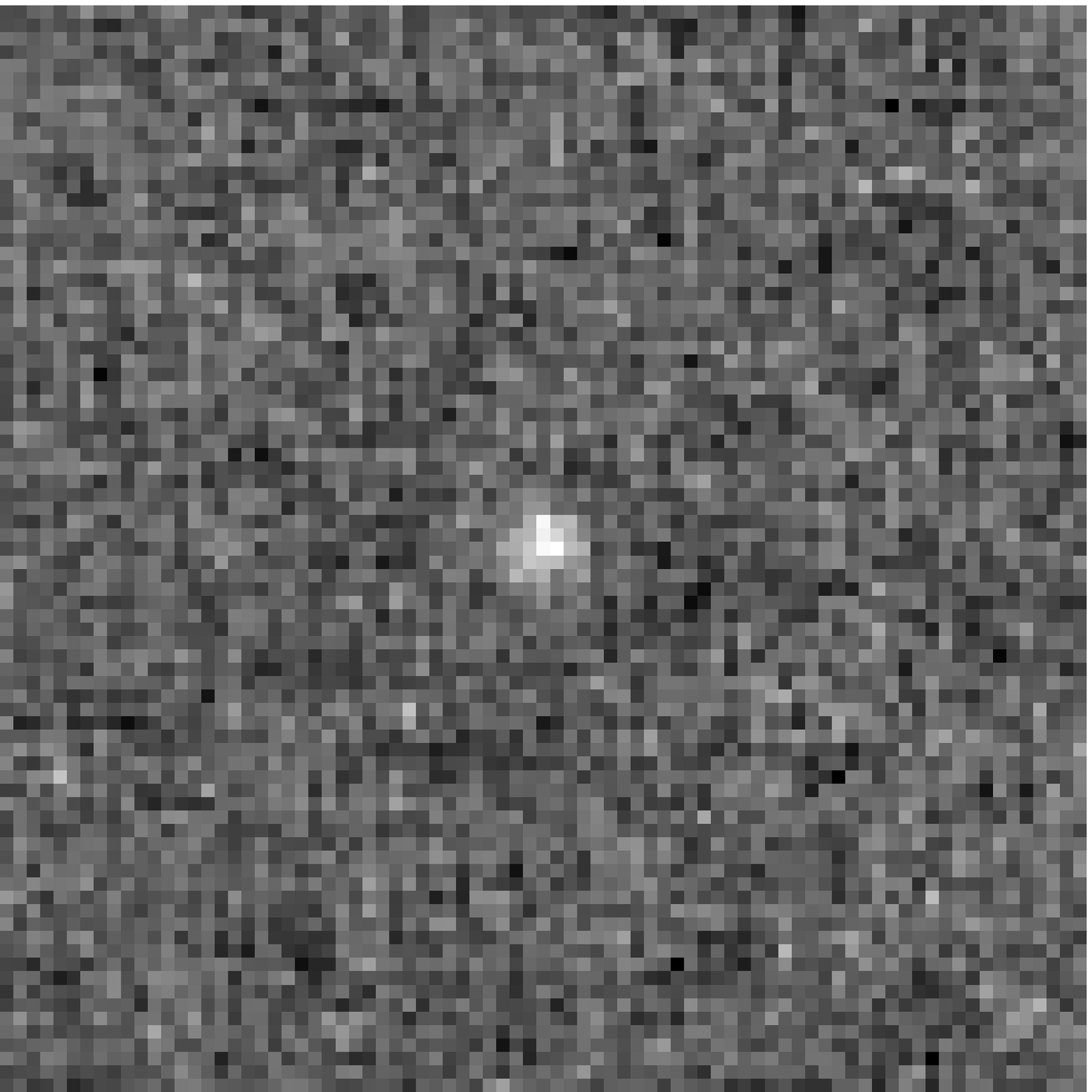}
\includegraphics[width=0.19\textwidth]{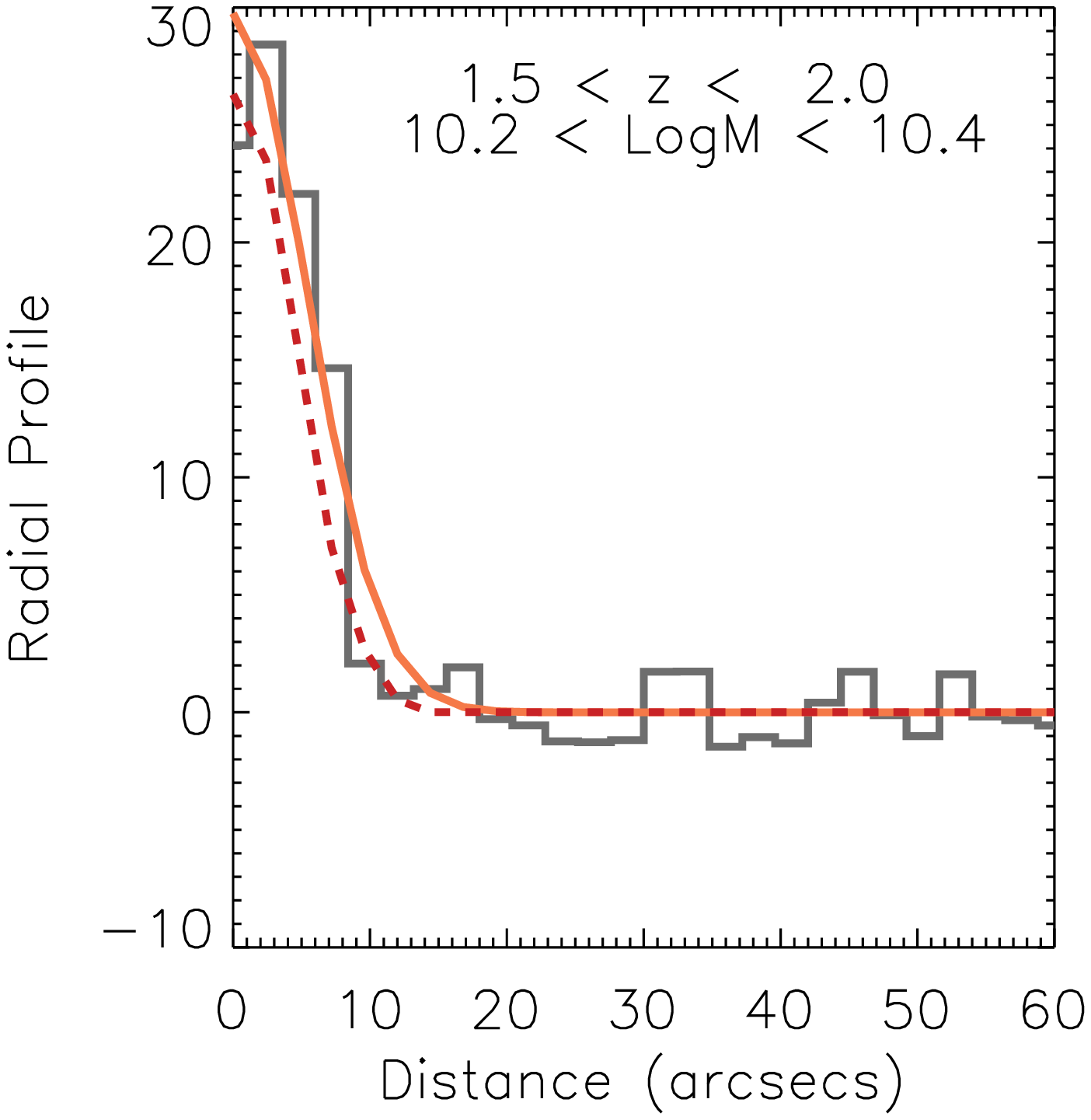}
\includegraphics[width=0.19\textwidth]{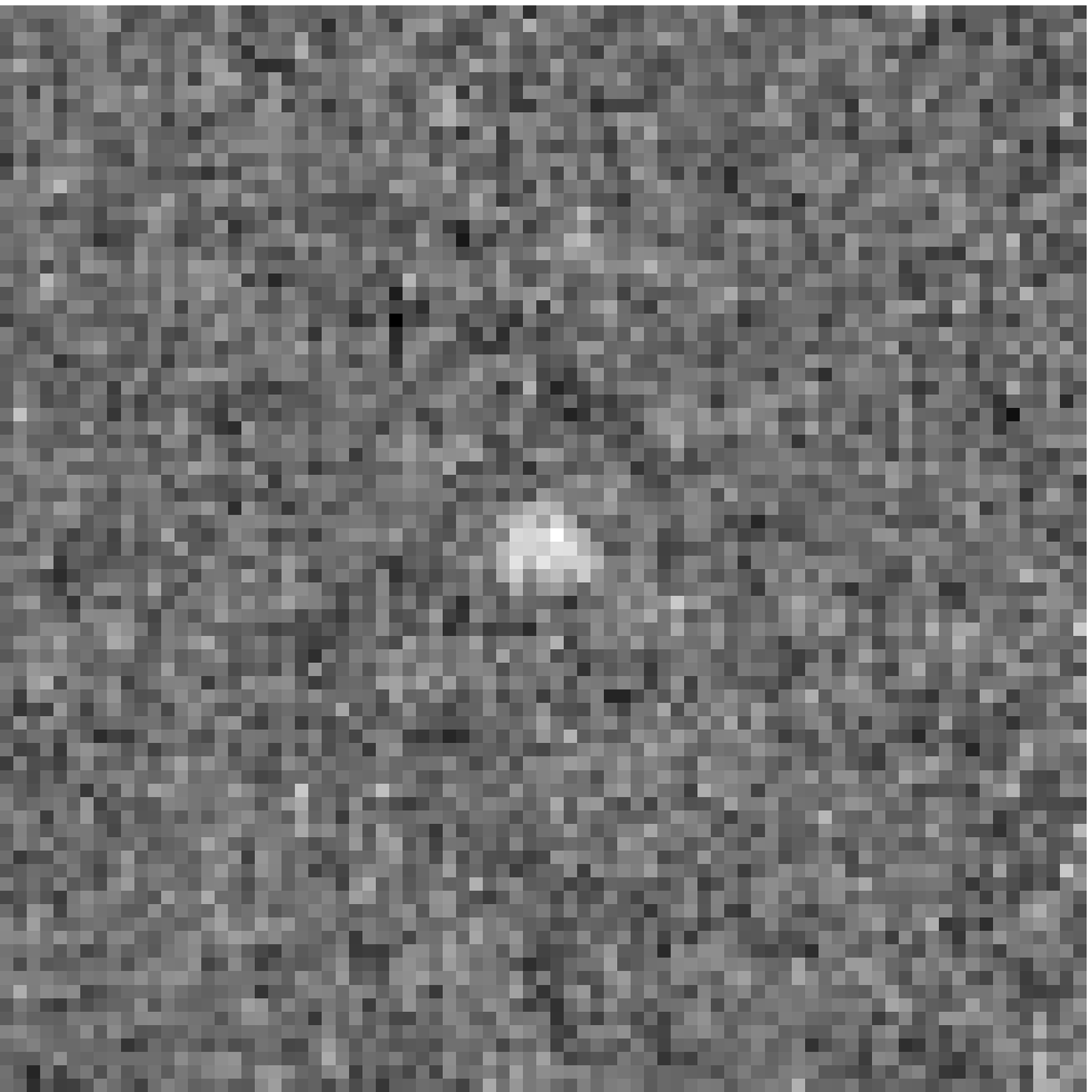}\\
\includegraphics[width=0.19\textwidth]{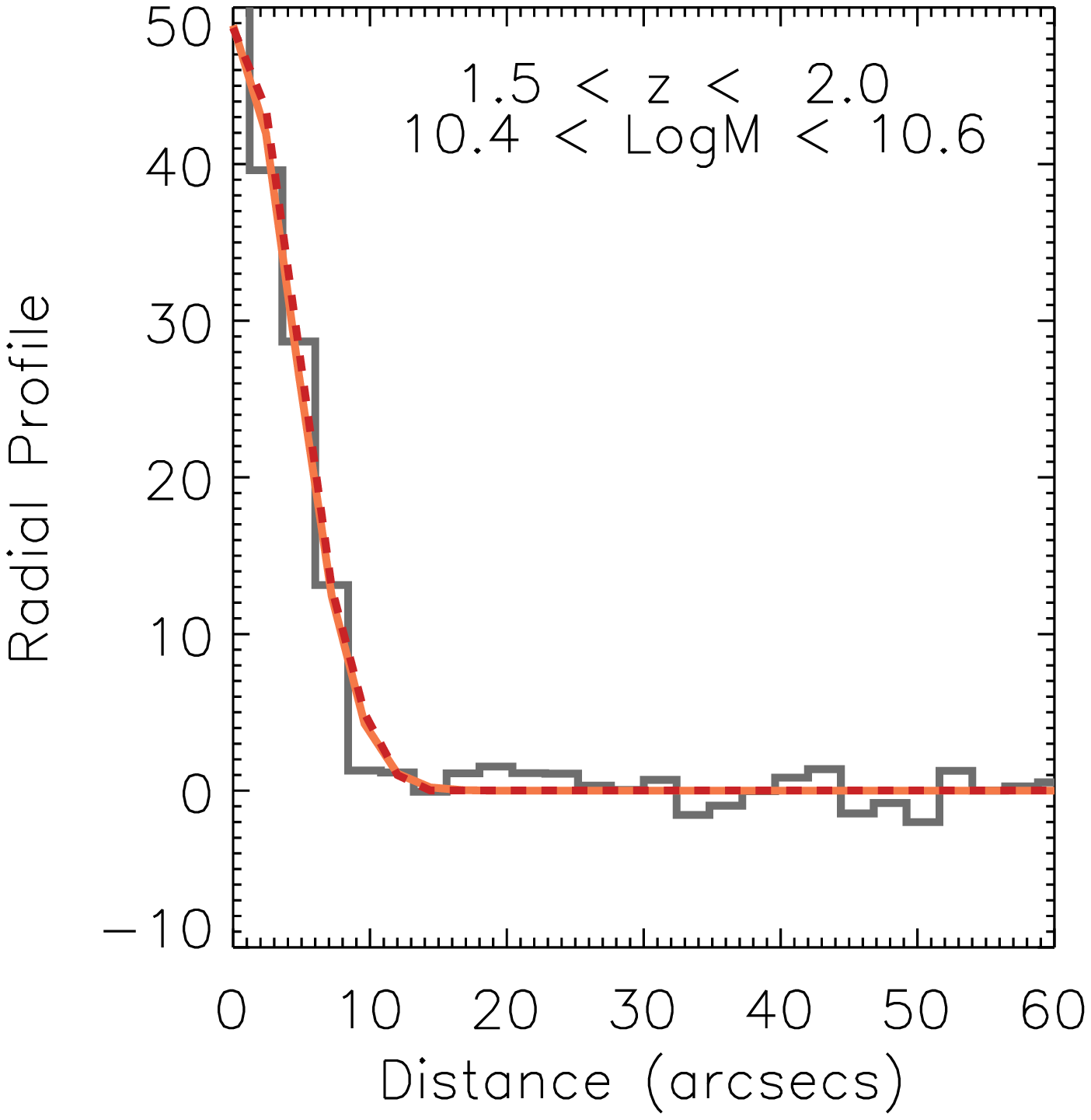}
\includegraphics[width=0.19\textwidth]{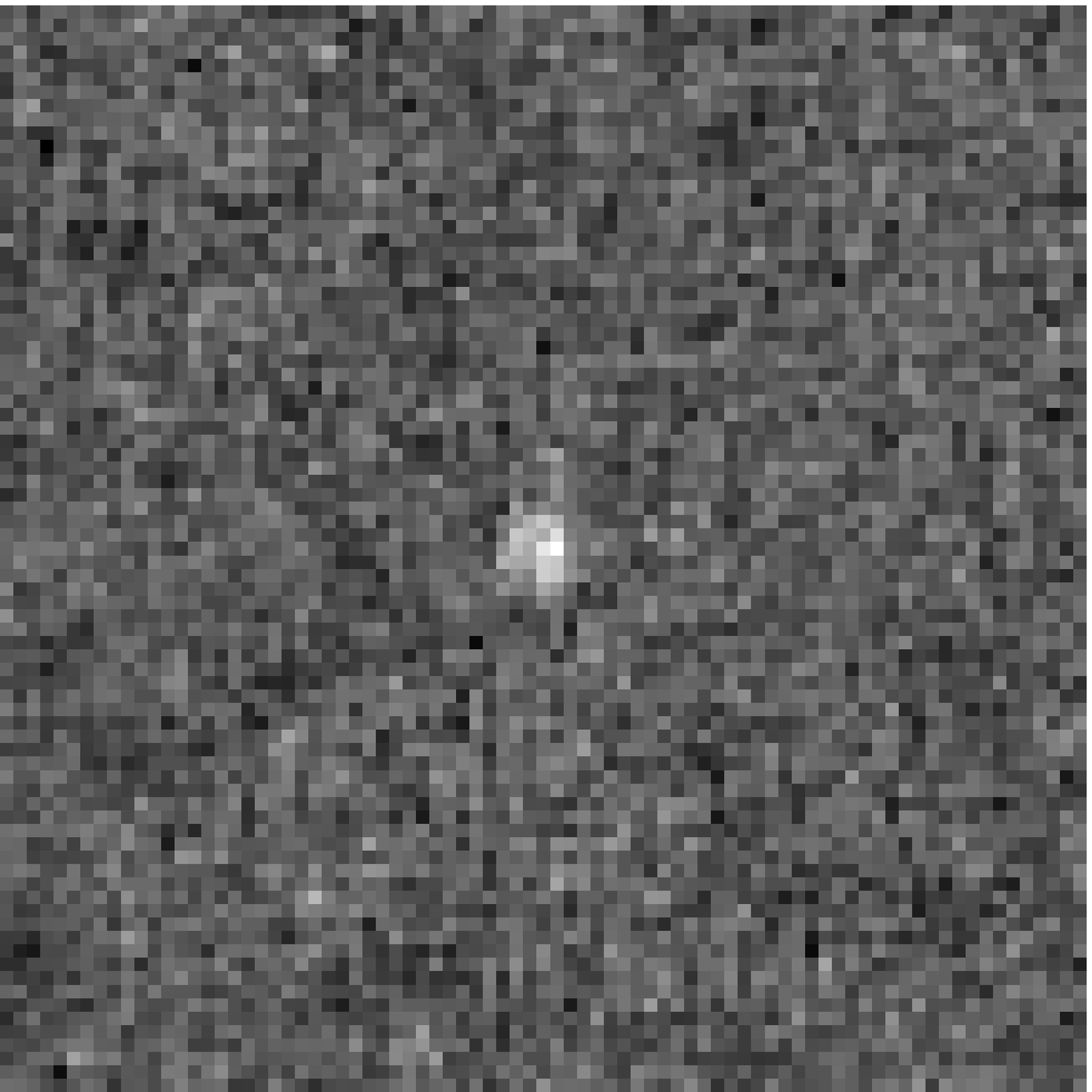}
\includegraphics[width=0.19\textwidth]{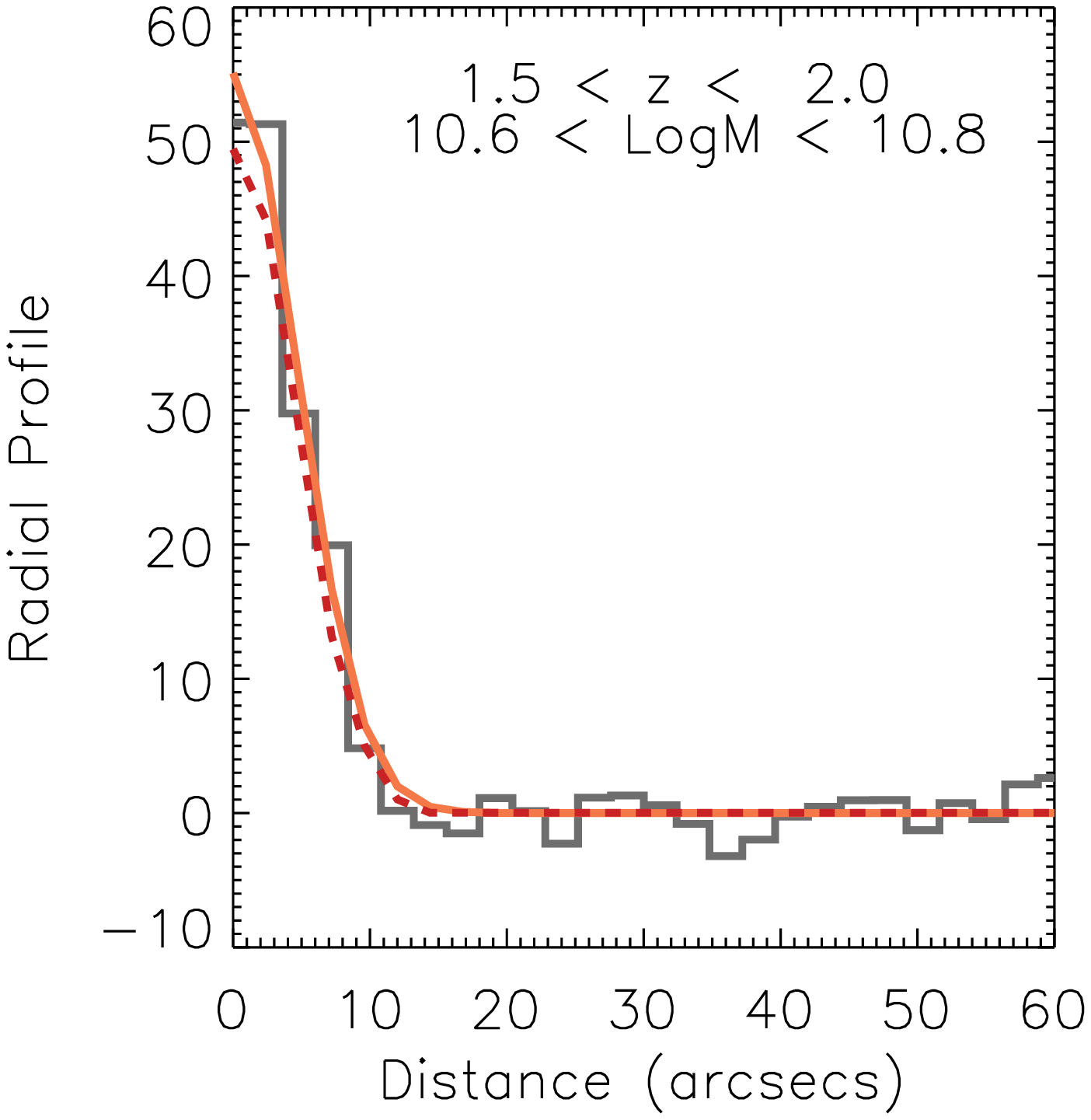}
\includegraphics[width=0.19\textwidth]{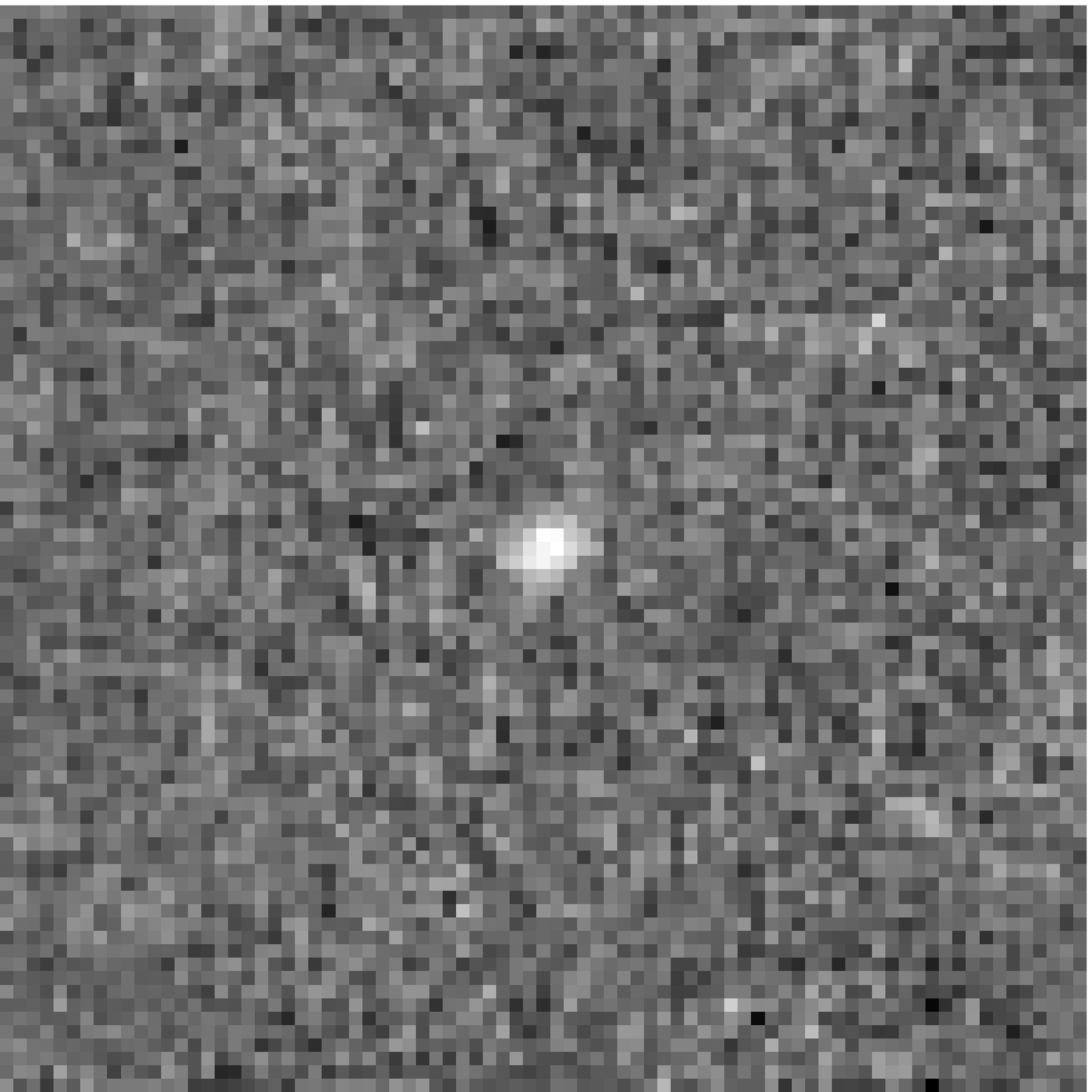}\\
\includegraphics[width=0.19\textwidth]{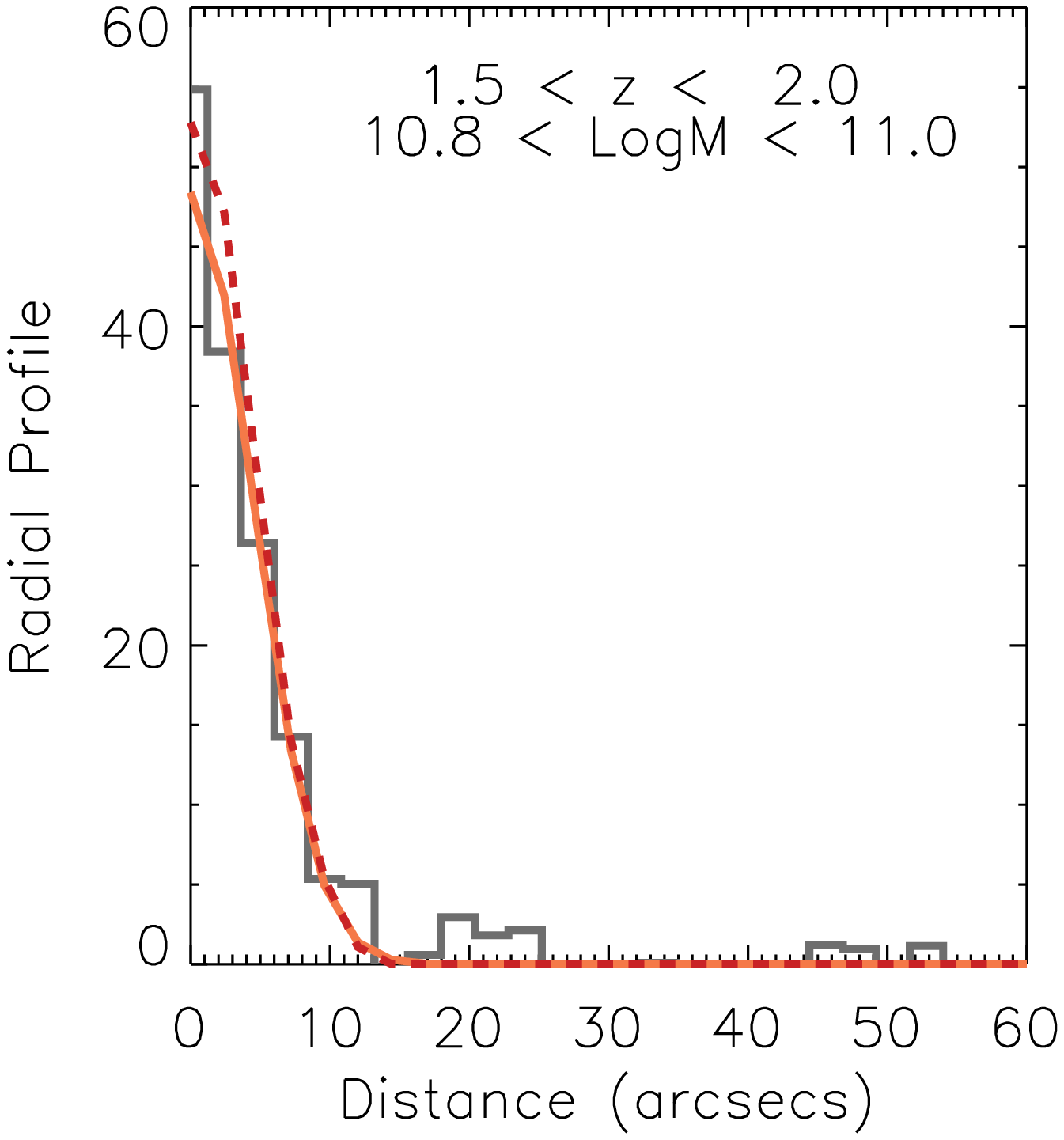}
\includegraphics[width=0.19\textwidth]{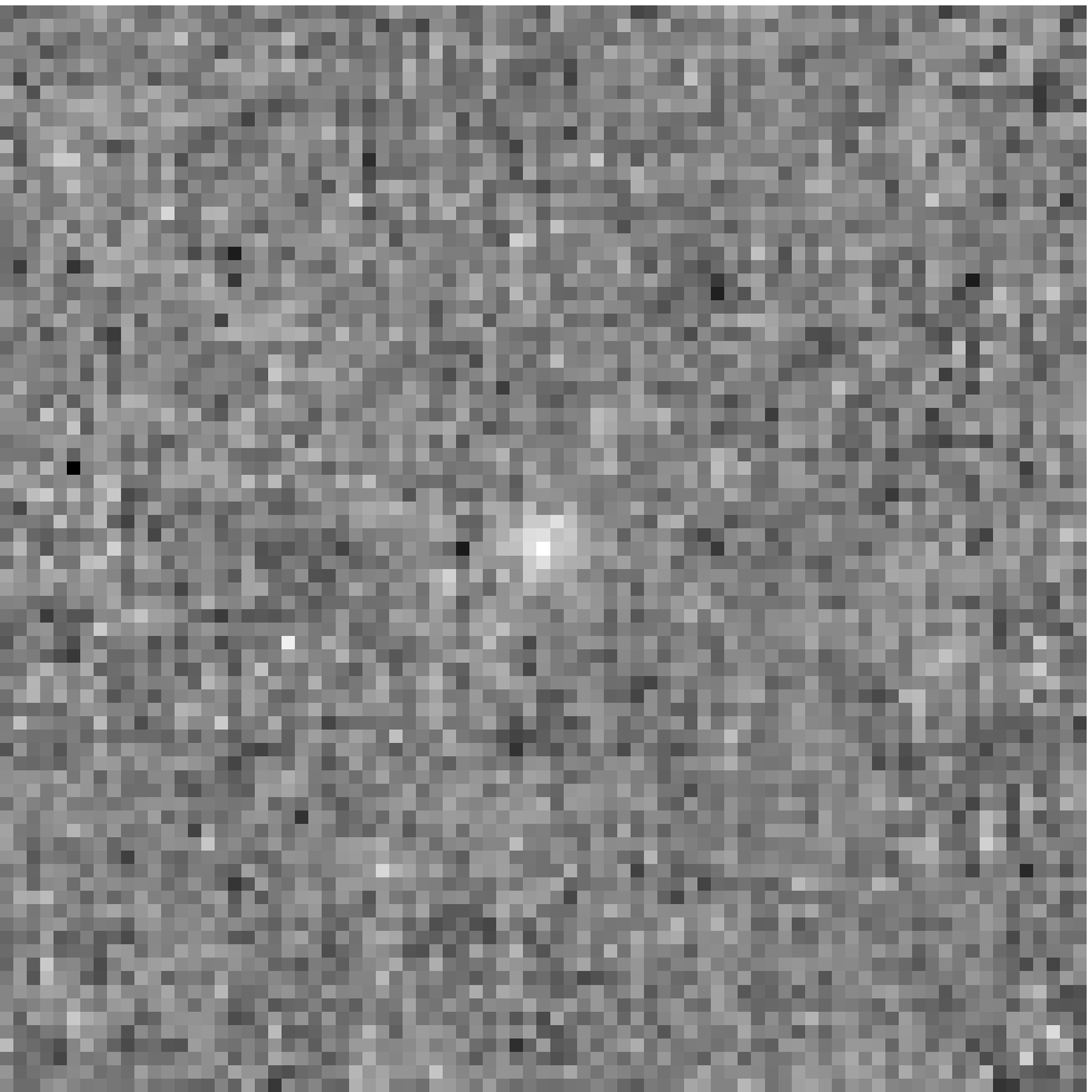}
\includegraphics[width=0.19\textwidth]{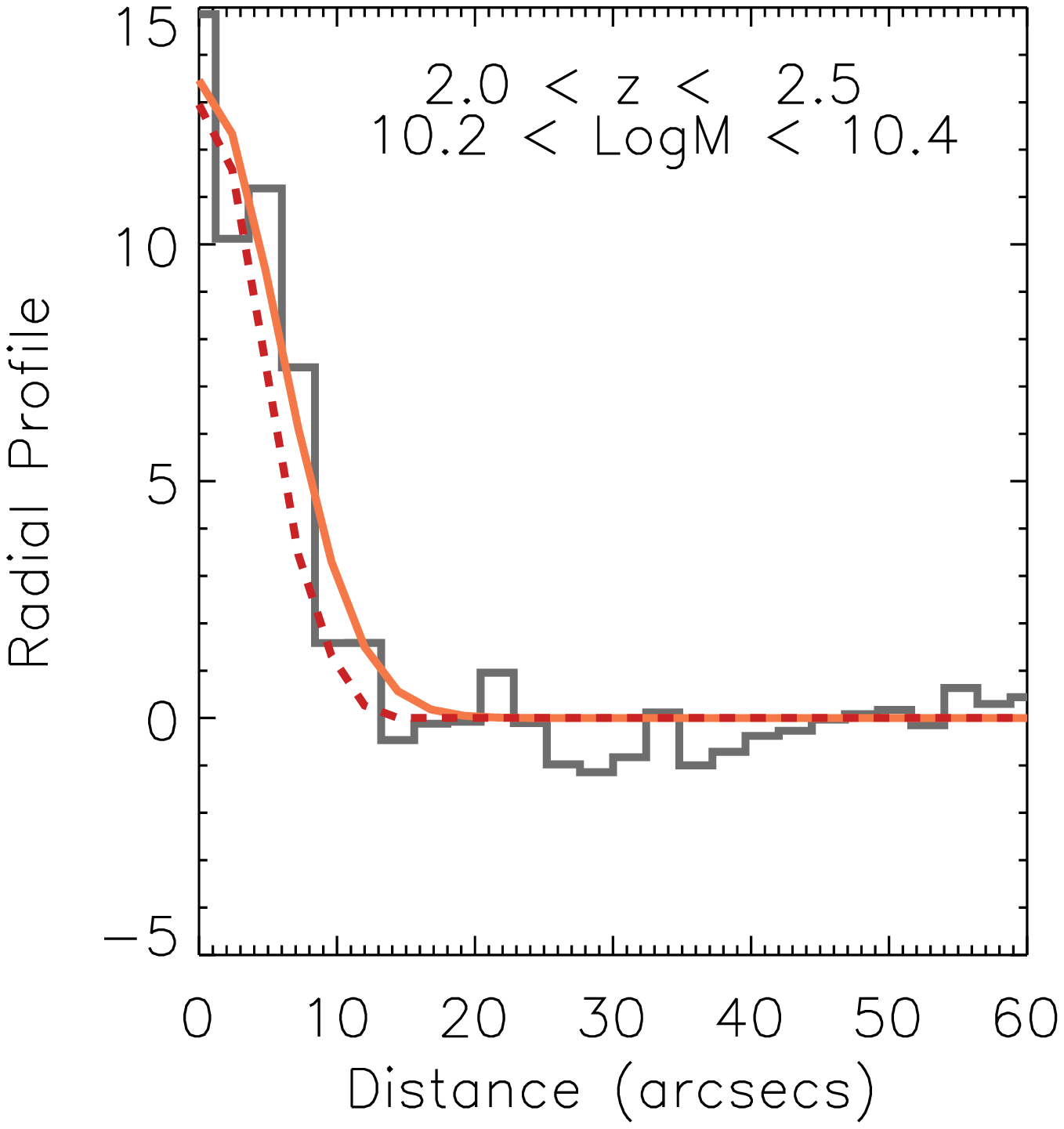}
\includegraphics[width=0.19\textwidth]{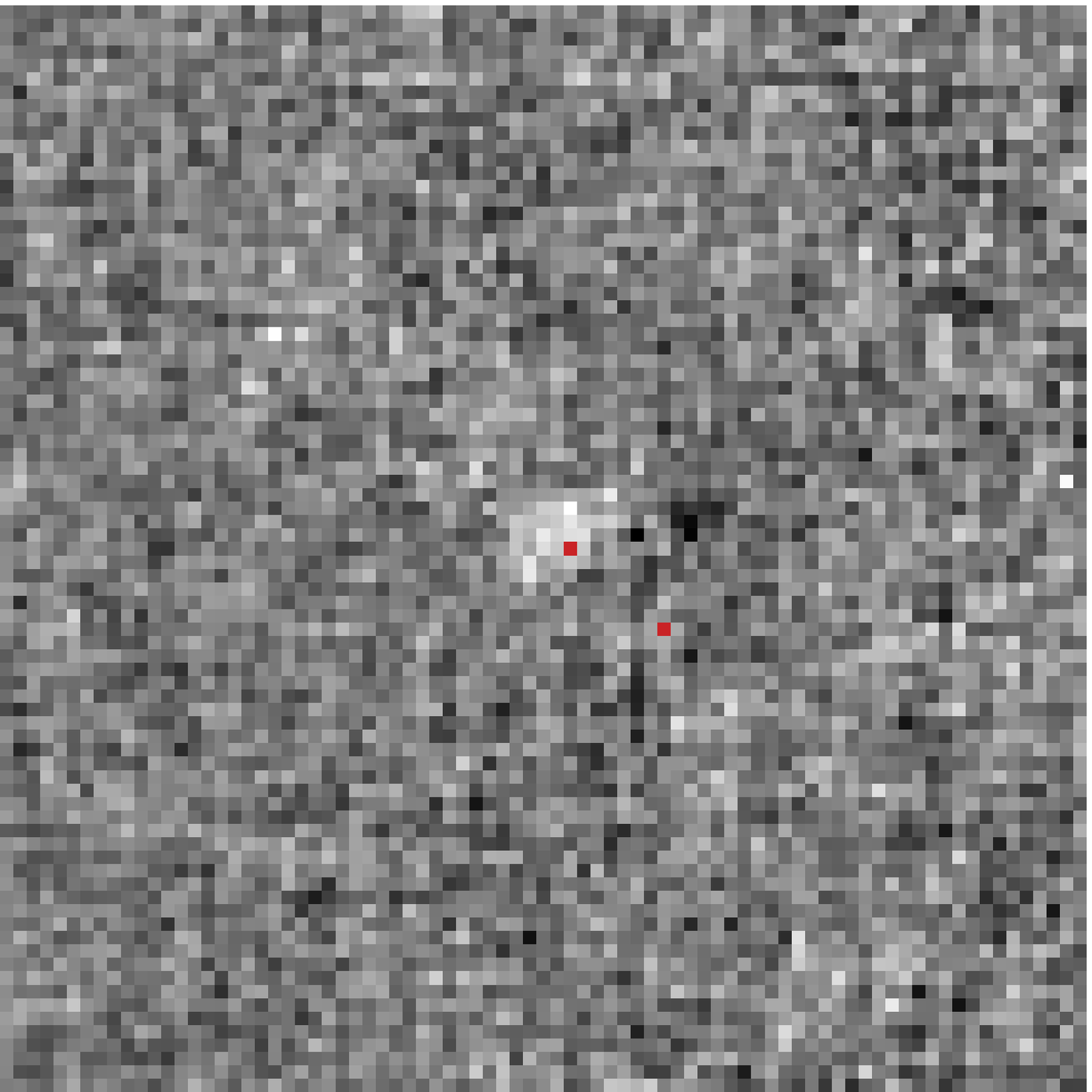}\\
\includegraphics[width=0.19\textwidth]{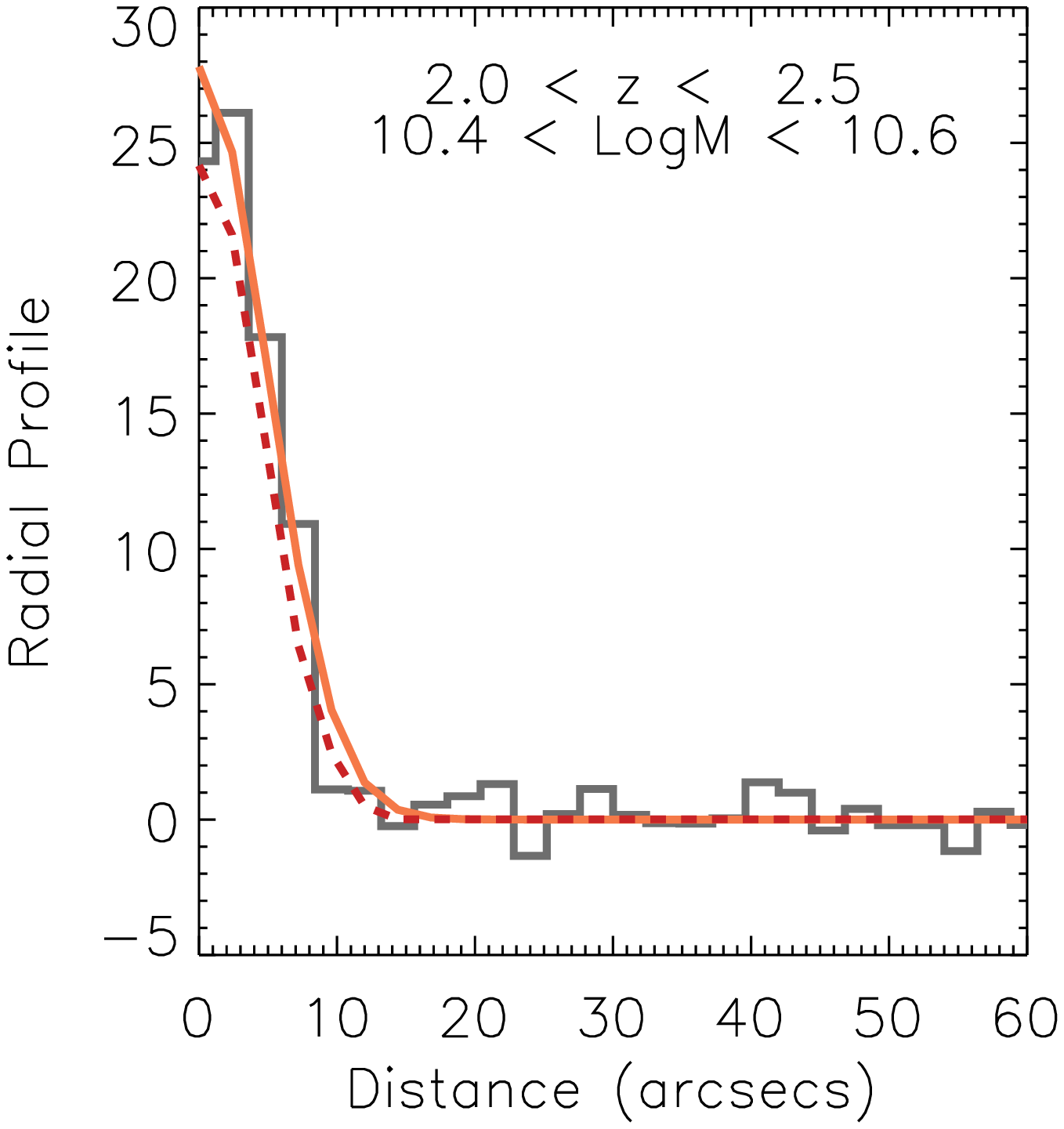}
\includegraphics[width=0.19\textwidth]{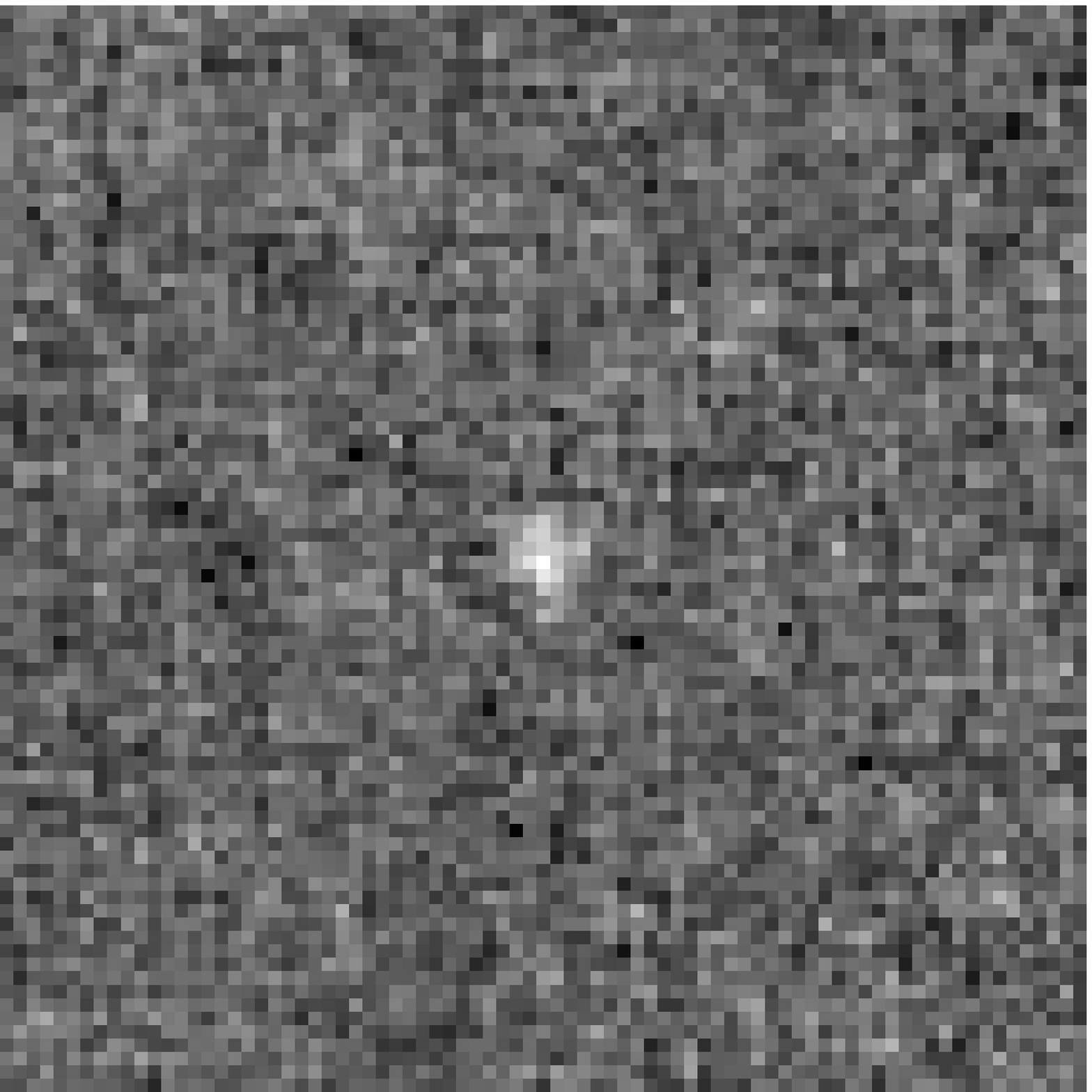}
\includegraphics[width=0.19\textwidth]{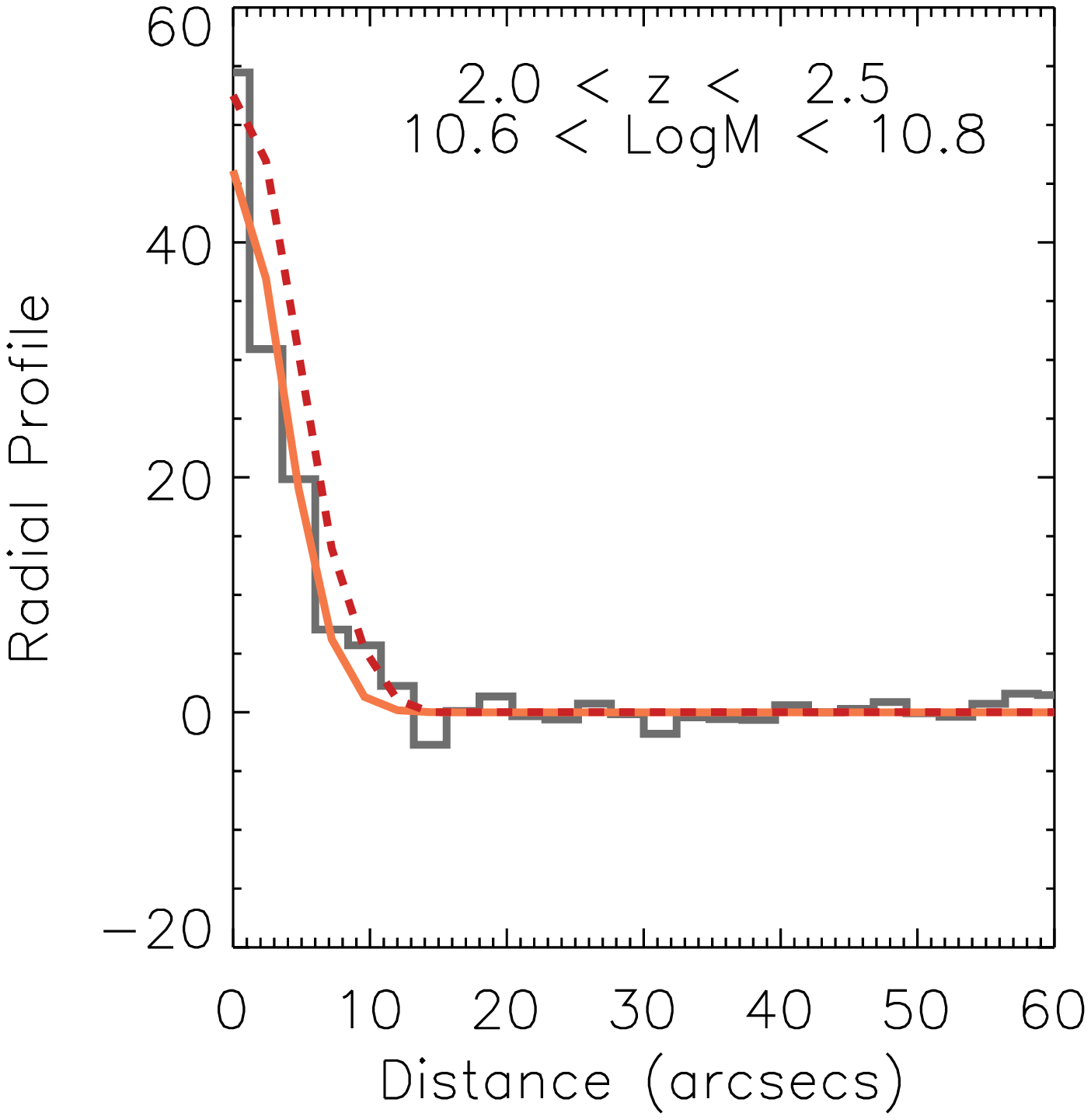}
\includegraphics[width=0.19\textwidth]{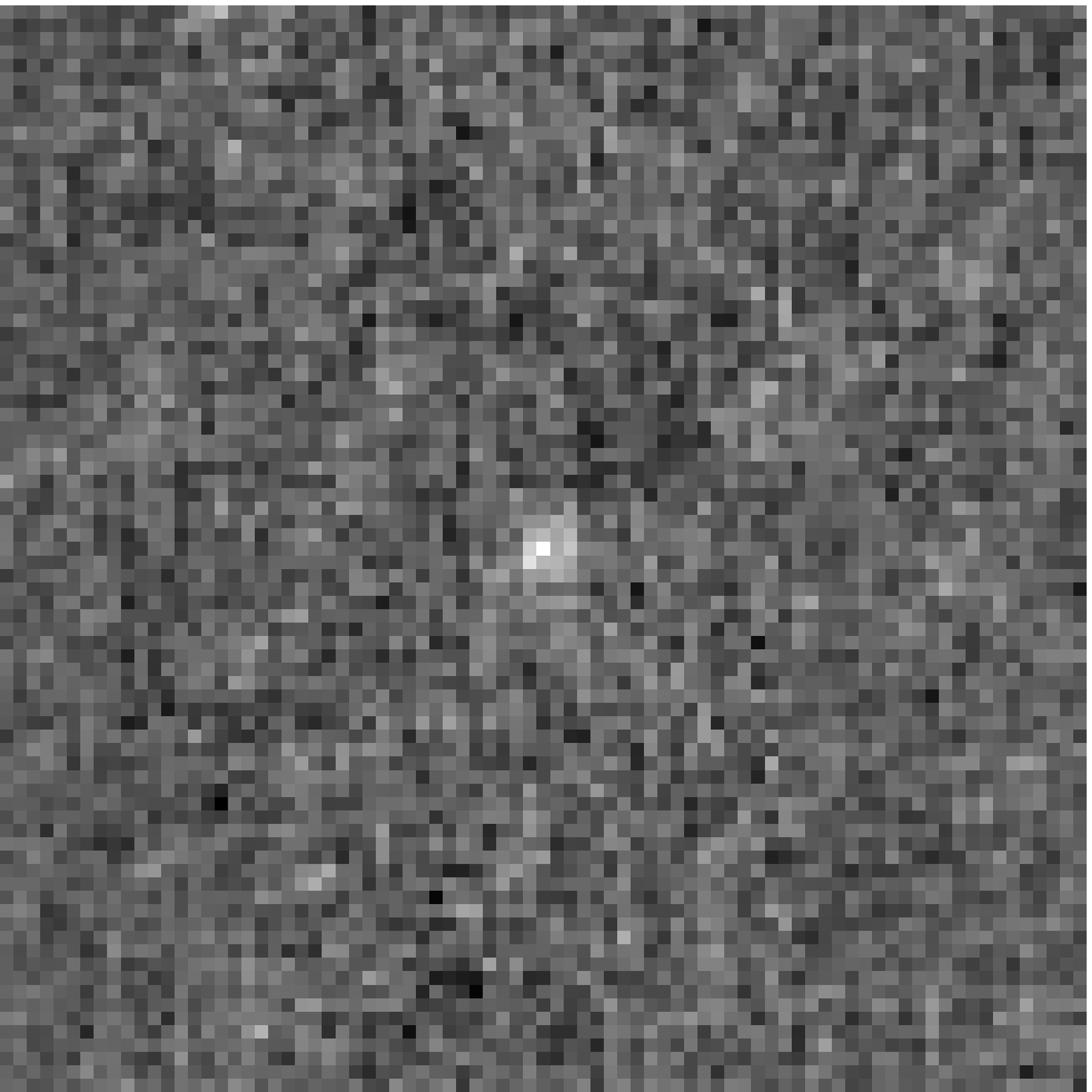}\\
\includegraphics[width=0.19\textwidth]{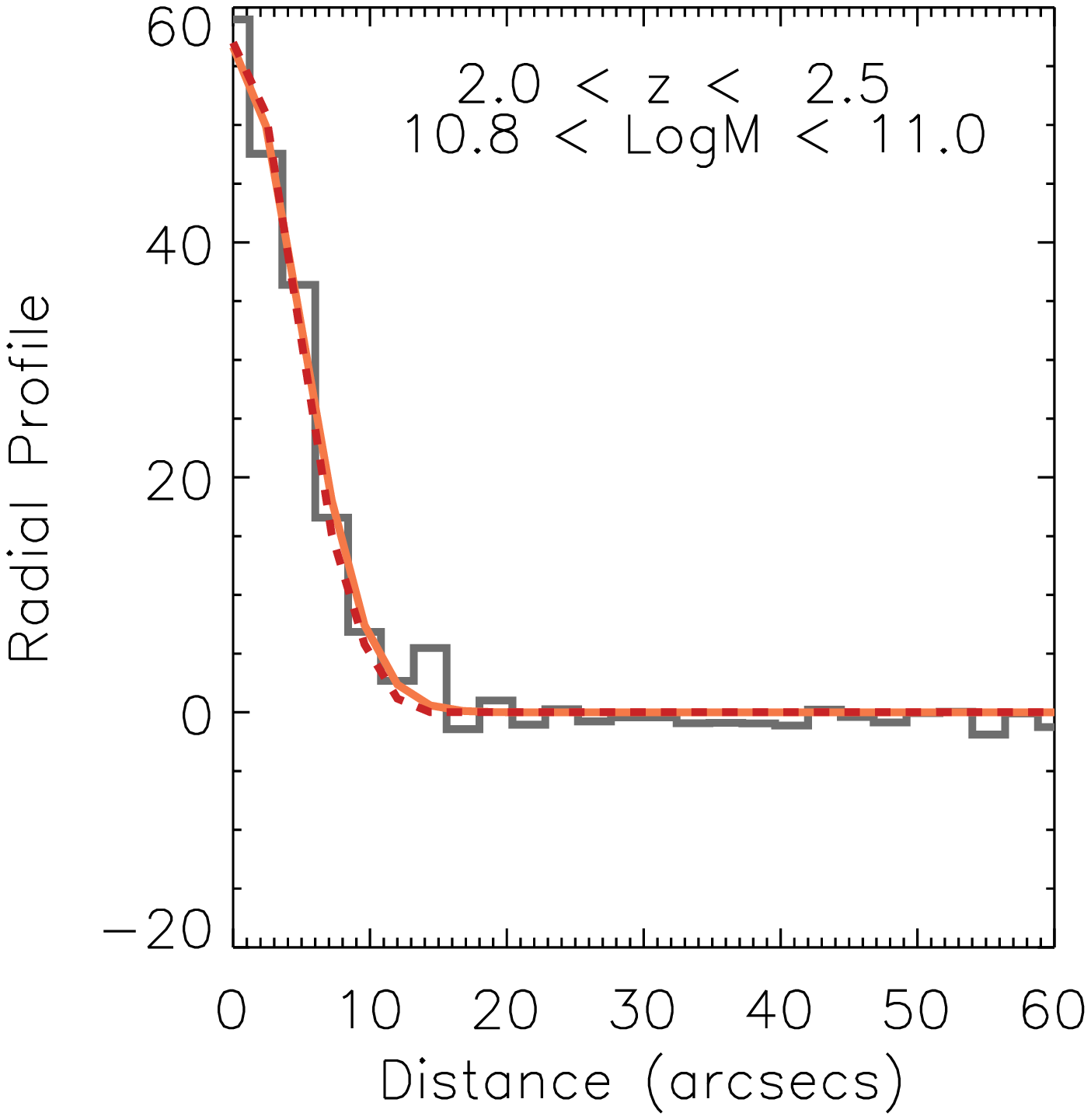}
\includegraphics[width=0.19\textwidth]{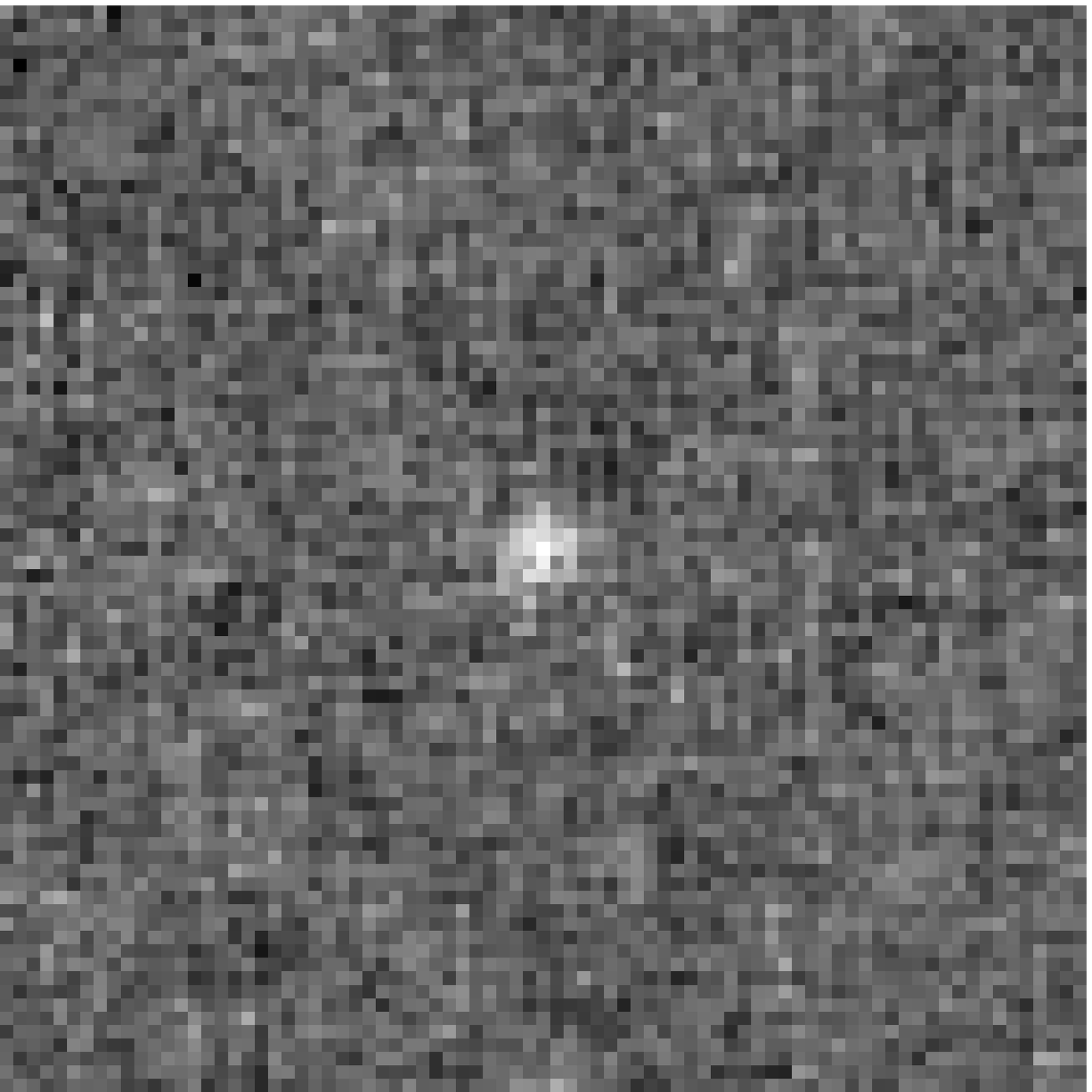}
\includegraphics[width=0.19\textwidth]{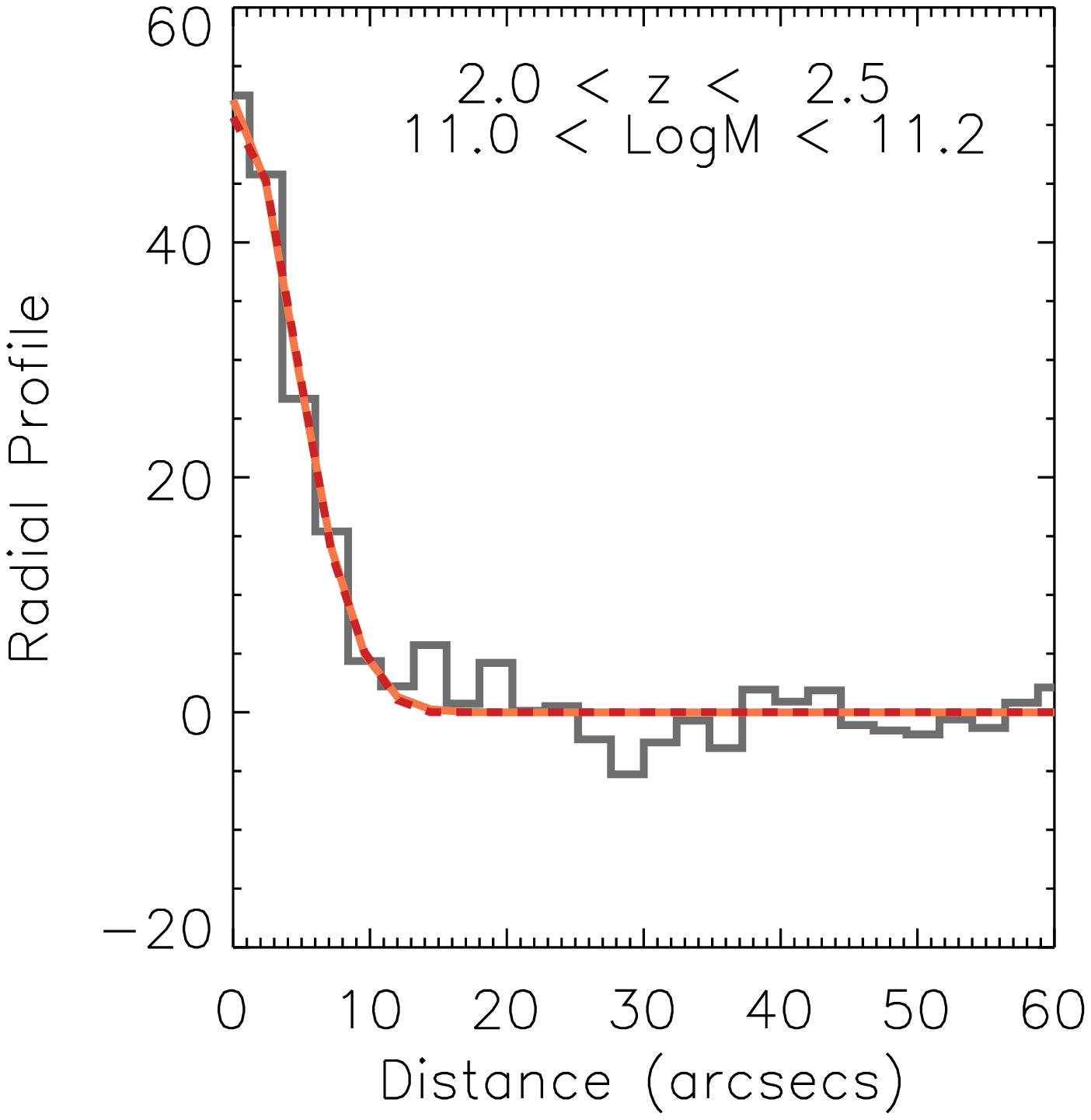}
\includegraphics[width=0.19\textwidth]{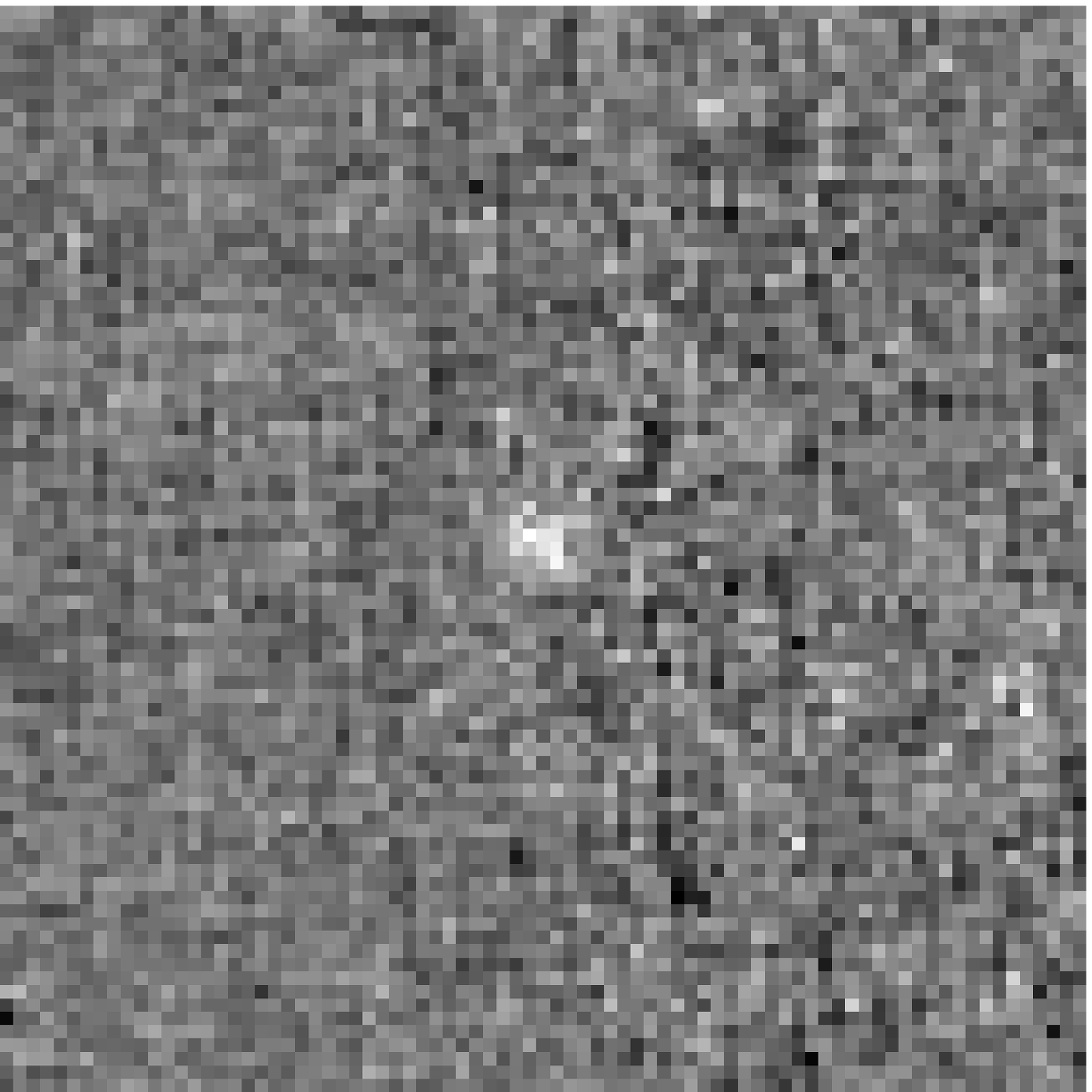}\\
\caption{Radial profiles and associated stacked images when stacking as a function of the stellar mass. Redshift and stellar mass bins are indicated in each case. Orange curves are the best-fitted gaussian to the radial profile, while red dashed curves represent the shape of the PACS-160 $\mu$m PSF (\emph{Cont}).
              }
\label{masa2}
\end{figure}

\begin{figure}
\centering
\includegraphics[width=0.19\textwidth]{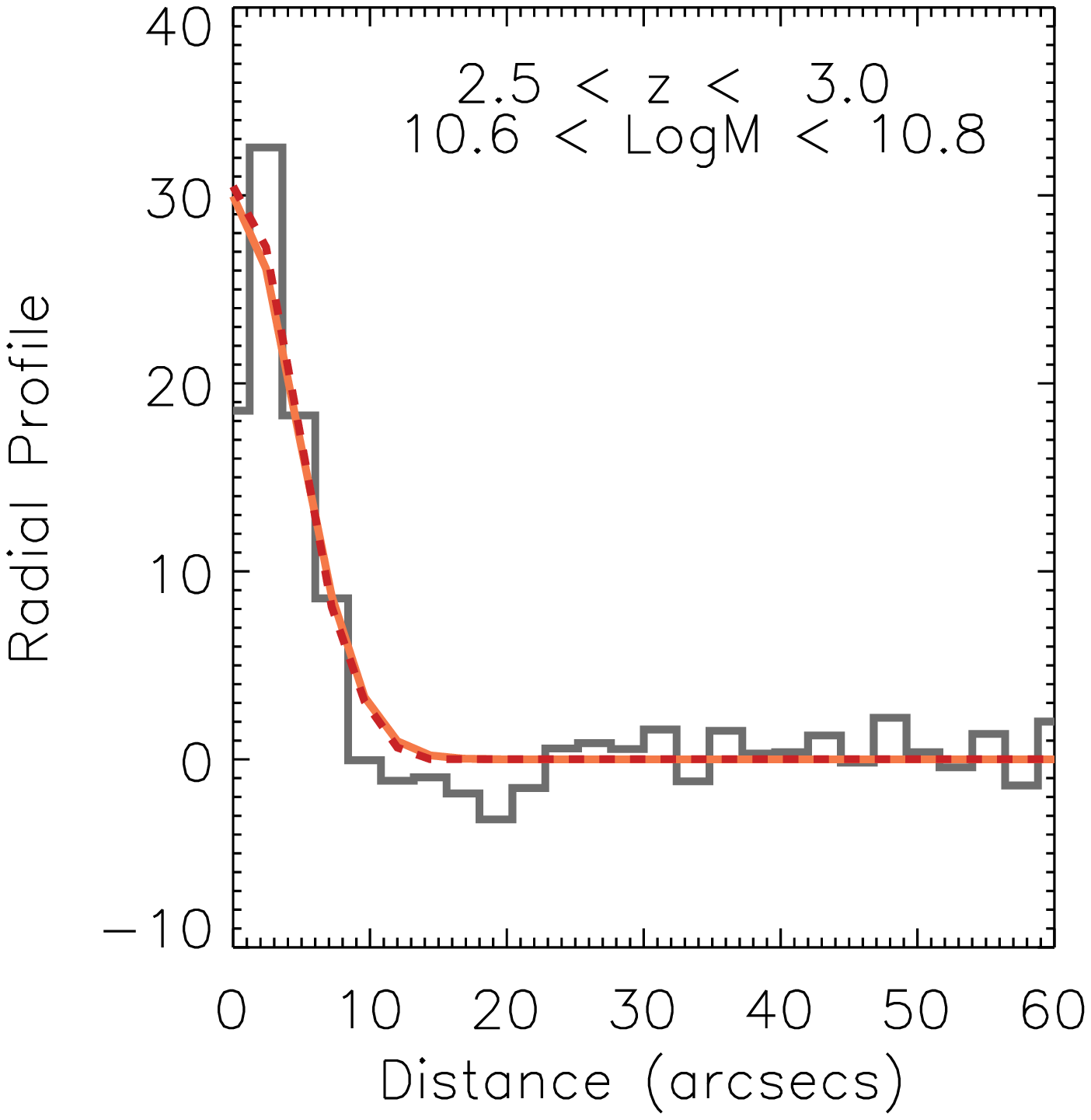}
\includegraphics[width=0.19\textwidth]{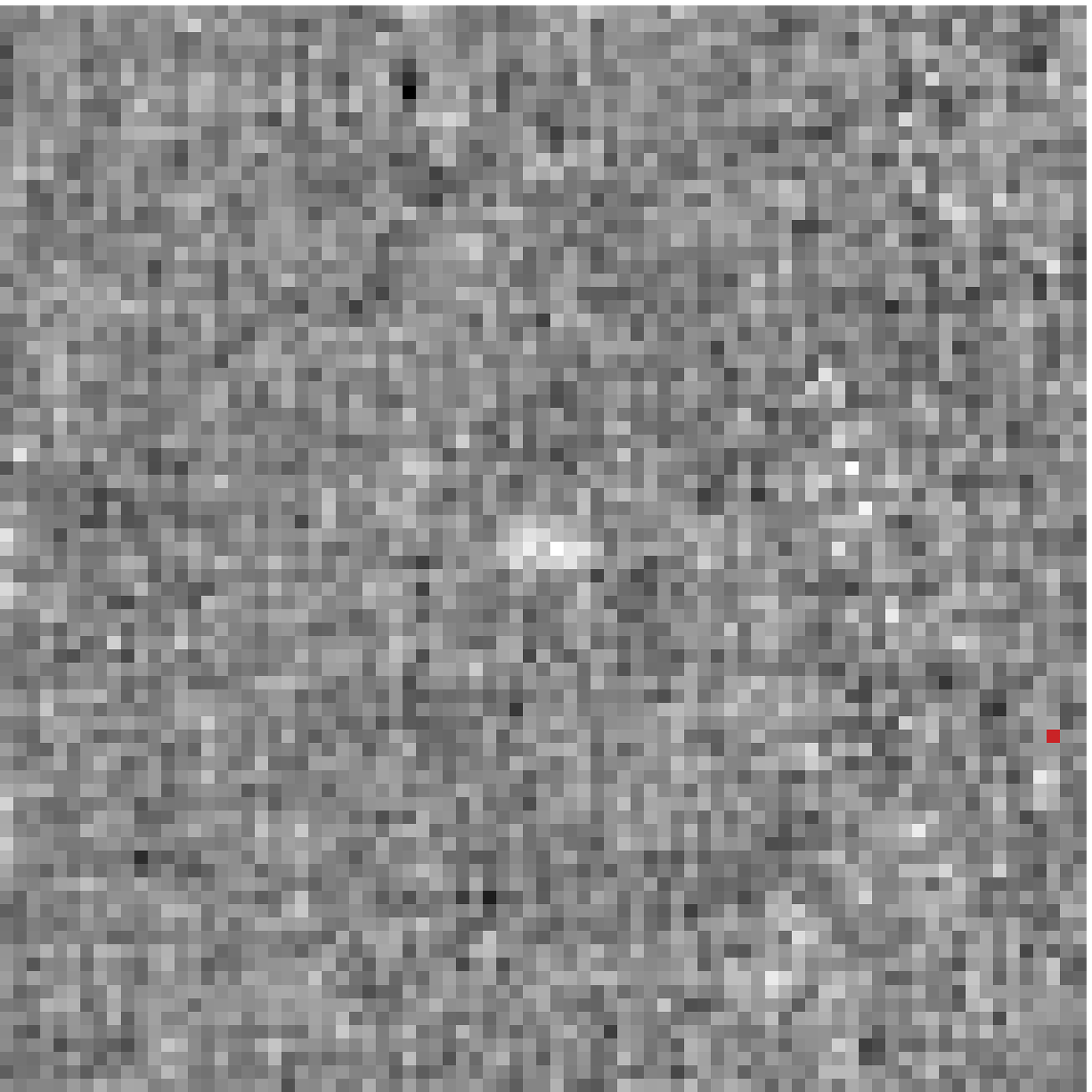}
\includegraphics[width=0.19\textwidth]{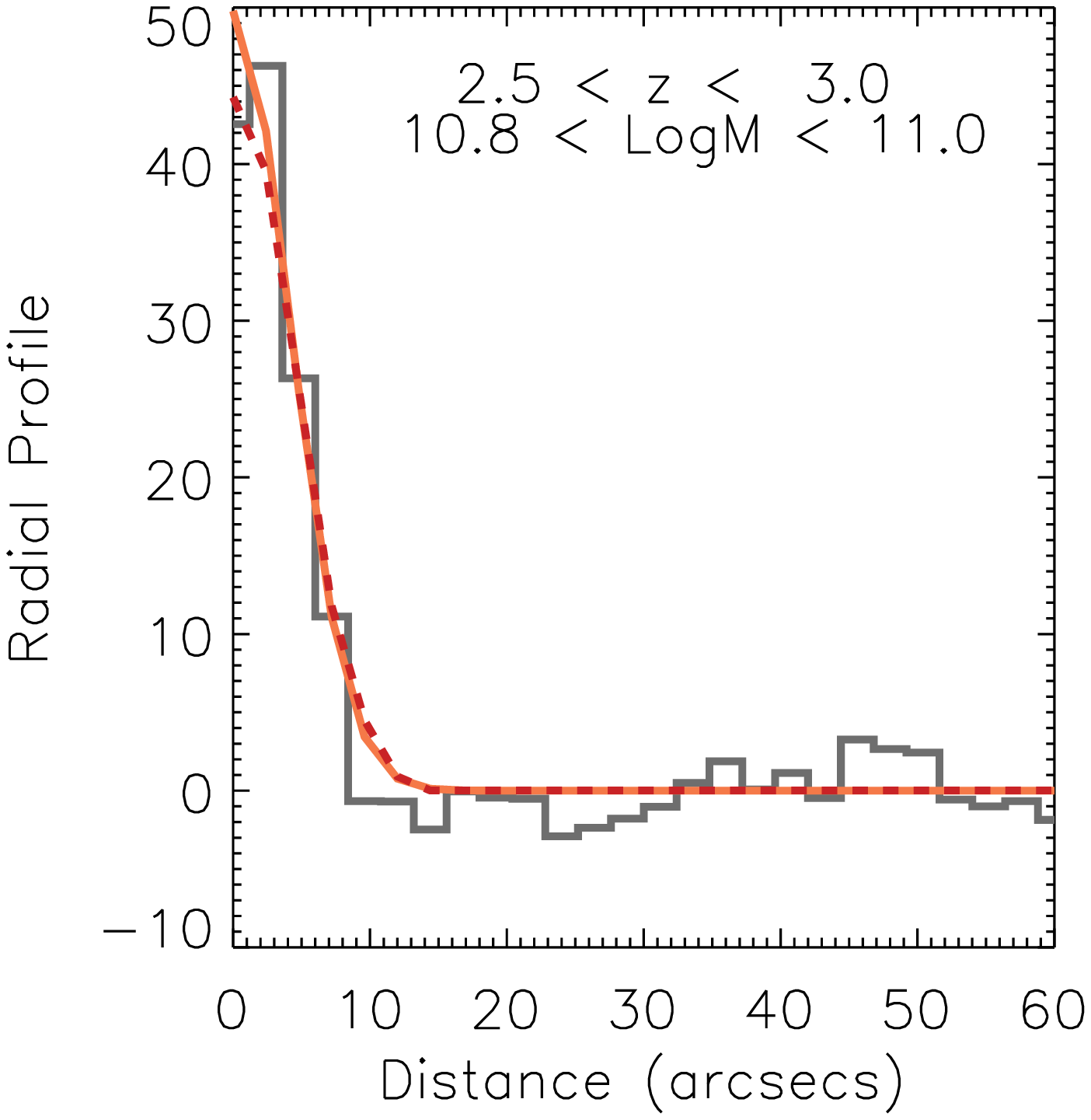}
\includegraphics[width=0.19\textwidth]{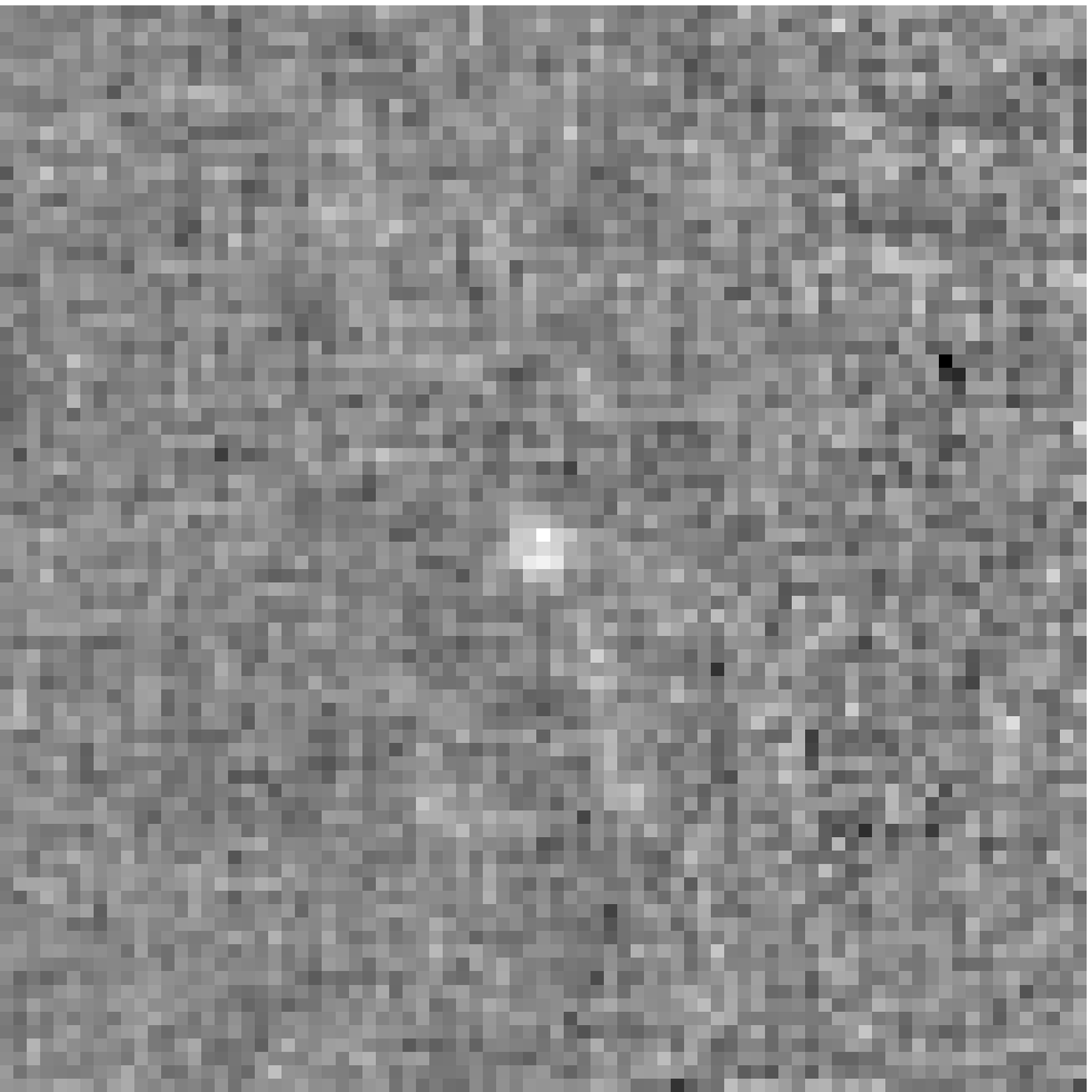}\\
\includegraphics[width=0.19\textwidth]{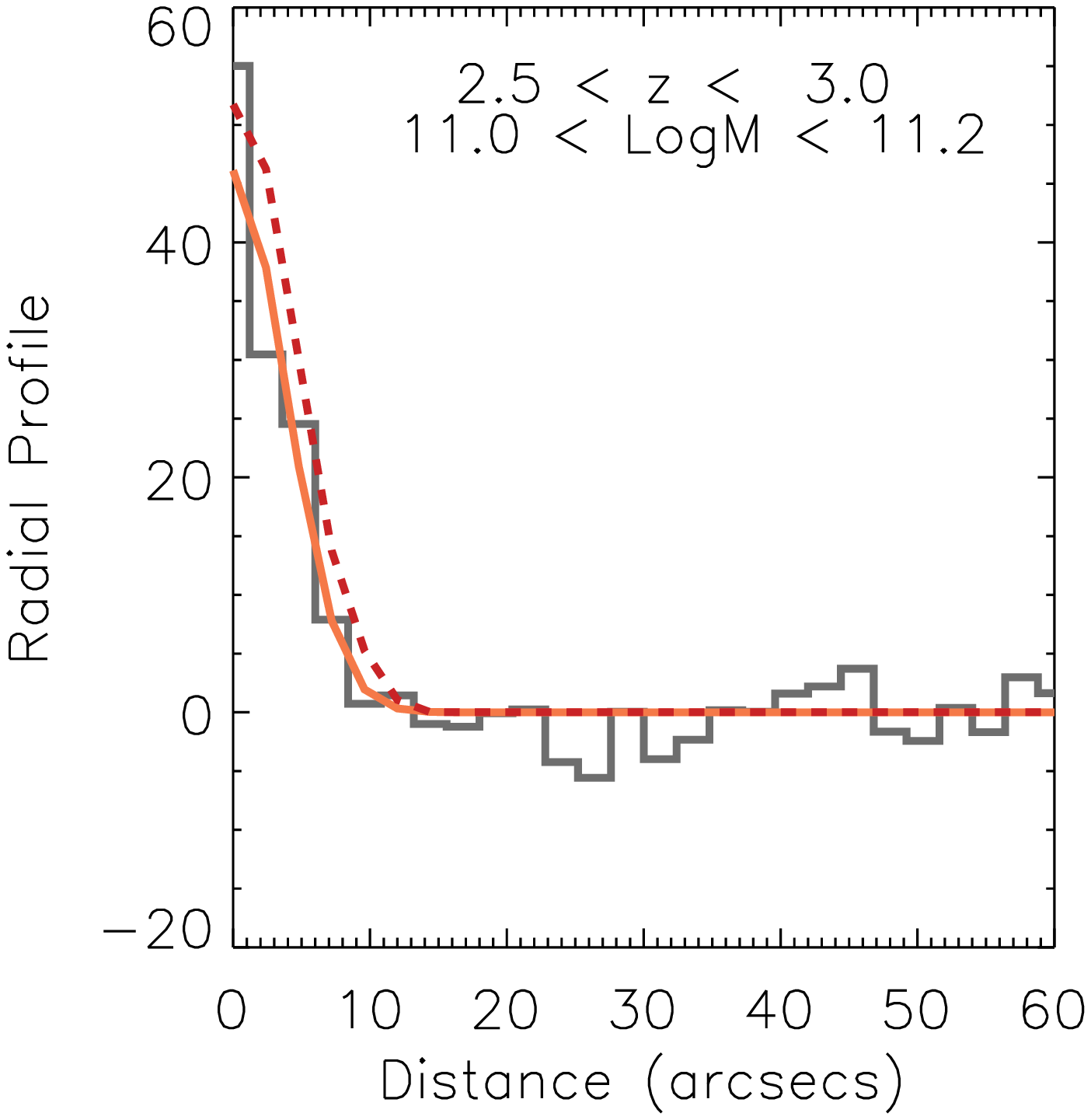}
\includegraphics[width=0.19\textwidth]{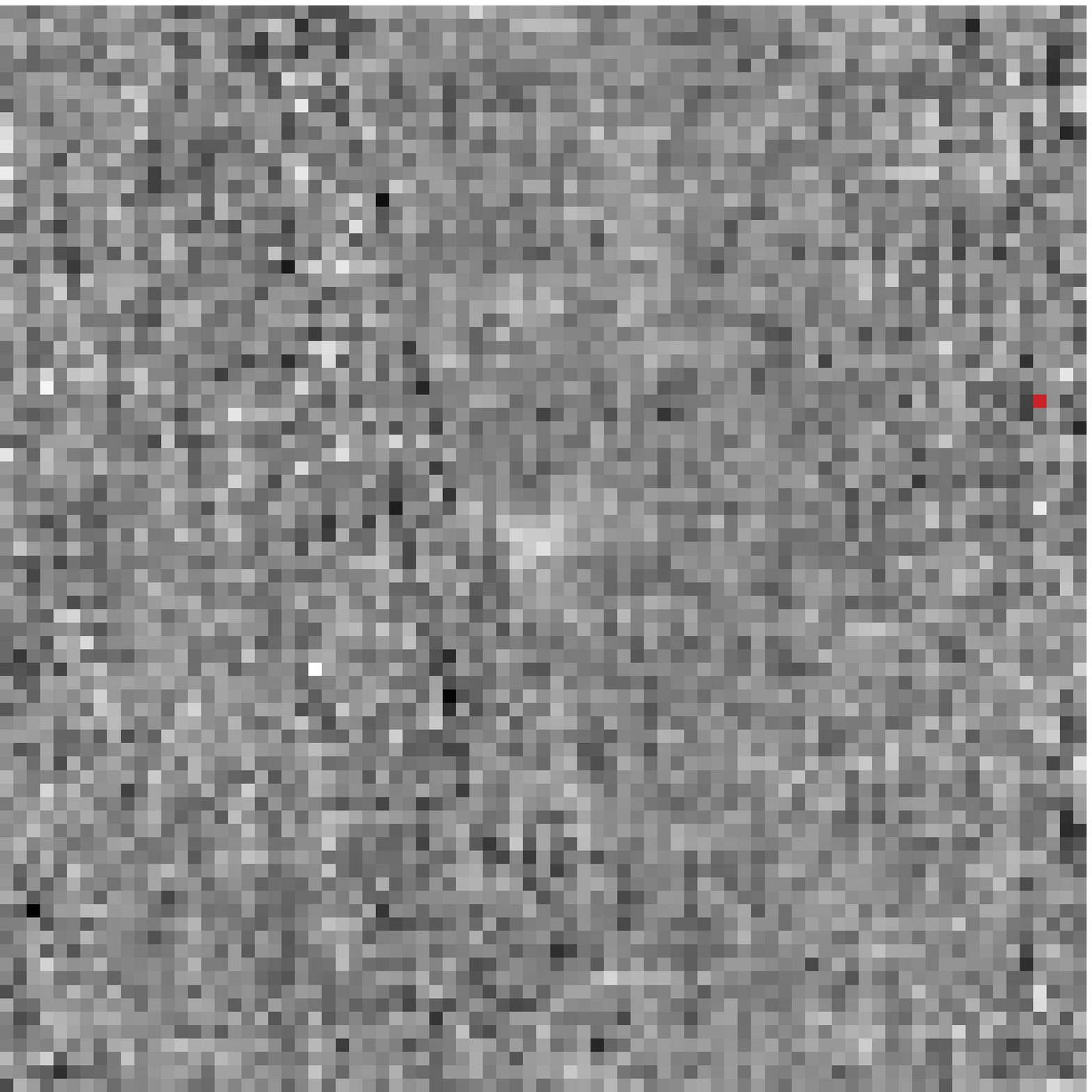}

\caption{Radial profiles and associated stacked images when stacking as a function of the stellar mass. Redshift and stellar mass bins are indicated in each case. Orange curves are the best-fitted gaussian to the radial profile, while red dashed curves represent the shape of the PACS-160 $\mu$m PSF (\emph{Cont}).
              }
\label{masa3}
\end{figure}

\begin{table*}\label{hola_tabla_masa}
\begin{center}
\caption{Summary of stacked properties when stacking as a function of stellar mass. The uncertainties in the rest-frame UV luminosity are related to the normalization of the templates to the observed photometry and they are typically lower tan 0.05 dex. The uncertainties in the total IR luminosities are obtained as the difference in the total IR luminosity when considering $f_{\rm 160 \mu m} \pm \Delta f_{\rm 160 \mu m}$. $N$ indicates the number of galaxies in each bin.}
\begin{tabular}{c c c c c c}
\hline
Redshift range & Stellar mass range & $f_{\rm 160 \mu m}$ [mJy]& $\log{L_{\rm UV}}$ & $\log{L_{\rm IR}}$ & $N$\\ 
\hline
\hline
     0.02$\leq z \leq$0.50&      9.60$\leq \log{\left( M_* / M_\odot \right)} \leq$   9.80&      1.39$\pm$0.16&9.64&     9.90$\pm$0.10&      184\\
     0.02$\leq z \leq$0.50&      9.80$\leq \log{\left( M_* / M_\odot \right)} \leq$ 10.00&      2.18$\pm$0.23&9.69&     10.12$\pm$0.09&    170\\
     0.02$\leq z \leq$0.50&     10.00$\leq \log{\left( M_* / M_\odot \right)} \leq$10.20&      2.55$\pm$0.23&9.67&     10.19$\pm$0.07&    111\\
     0.02$\leq z \leq$0.50&     10.20$\leq \log{\left( M_* / M_\odot \right)} \leq$10.40&      2.34$\pm$0.46&9.64&     10.11$\pm$0.17&     36\\
     
     0.50$\leq z \leq$1.00&      9.80$\leq \log{\left( M_* / M_\odot \right)} \leq$ 10.00&      0.42$\pm$0.04&9.78&     10.13$\pm$0.10&   2896\\
     0.50$\leq z \leq$1.00&     10.00$\leq \log{\left( M_* / M_\odot \right)} \leq$10.20&      0.83$\pm$0.05&9.83&     10.46$\pm$0.05&   2246\\
     0.50$\leq z \leq$1.00&     10.20$\leq \log{\left( M_* / M_\odot \right)} \leq$10.40&      1.30$\pm$0.06&9.82&     10.64$\pm$0.04&   1499\\
     0.50$\leq z \leq$1.00&     10.40$\leq \log{\left( M_* / M_\odot \right)} \leq$10.60&      1.85$\pm$0.09&9.79&     10.79$\pm$0.05&    865\\
     0.50$\leq z \leq$1.00&     10.60$\leq \log{\left( M_* / M_\odot \right)} \leq$10.80&      2.38$\pm$0.11&9.77&     10.98$\pm$0.04&    576\\
     
     1.00$\leq z \leq$1.50&     10.00$\leq \log{\left( M_* / M_\odot \right)} \leq$10.20&      0.70$\pm$0.04&9.95&     10.81$\pm$0.05&   2877\\
     1.00$\leq z \leq$1.50&     10.20$\leq \log{\left( M_* / M_\odot \right)} \leq$10.40&      0.90$\pm$0.05&9.97&     10.90$\pm$0.04&   2101\\
     1.00$\leq z \leq$1.50&     10.40$\leq \log{\left( M_* / M_\odot \right)} \leq$10.60&      1.41$\pm$0.06&9.93&     11.10$\pm$0.03&   1541\\
     1.00$\leq z \leq$1.50&     10.60$\leq \log{\left( M_* / M_\odot \right)} \leq$10.80&      1.64$\pm$0.10&9.87&     11.15$\pm$0.05&   1152\\
     1.00$\leq z \leq$1.50&     10.80$\leq \log{\left( M_* / M_\odot \right)} \leq$11.00 &     1.71$\pm$0.10&9.83 &    11.18$\pm$0.05&    869\\
     
     1.50$\leq z \leq$2.00&     10.00$\leq \log{\left( M_* / M_\odot \right)} \leq$10.20&      0.62$\pm$0.05&10.04&     11.11$\pm$0.06&   2688\\
     1.50$\leq z \leq$2.00&     10.20$\leq \log{\left( M_* / M_\odot \right)} \leq$10.40&      0.85$\pm$0.06& 9.98&     11.25$\pm$0.06&   1682\\
     1.50$\leq z \leq$2.00&     10.40$\leq \log{\left( M_* / M_\odot \right)} \leq$10.60&      1.13$\pm$0.08& 9.93&     11.37$\pm$0.06&   1158\\
     1.50$\leq z \leq$2.00&     10.60$\leq \log{\left( M_* / M_\odot \right)} \leq$10.80&      1.33$\pm$0.10& 9.89&     11.46$\pm$0.06&    851\\
     1.50$\leq z \leq$2.00&     10.80$\leq \log{\left( M_* / M_\odot \right)} \leq$11.00 &     1.08$\pm$0.12 & 9.83 &    11.35$\pm$0.09&    656\\
     
     2.00$\leq z \leq$2.50&     10.20$\leq \log{\left( M_* / M_\odot \right)} \leq$10.40&      0.40$\pm$0.05&10.27&     11.26$\pm$0.12&   2083\\
     2.00$\leq z \leq$2.50&     10.40$\leq \log{\left( M_* / M_\odot \right)} \leq$10.60&      0.70$\pm$0.07&10.21&     11.50$\pm$0.09&   1514\\
     2.00$\leq z \leq$2.50&     10.60$\leq \log{\left( M_* / M_\odot \right)} \leq$10.80&      0.90$\pm$0.08& 10.06&     11.60$\pm$0.07&    984\\
     2.00$\leq z \leq$2.50&     10.80$\leq \log{\left( M_* / M_\odot \right)} \leq$11.00&      1.32$\pm$0.12&9.976&     11.76$\pm$0.07 &   616\\
     2.00$\leq z \leq$2.50&     11.00$\leq \log{\left( M_* / M_\odot \right)} \leq$ 11.20&      1.30$\pm$0.15&9.90&     11.76$\pm$0.09&    389\\
    
     2.50$\leq z \leq$3.00&     10.60$\leq \log{\left( M_* / M_\odot \right)} \leq$10.80&      0.69$\pm$0.10&10.38&     11.75$\pm$0.12&    645\\
     2.50$\leq z \leq$3.00&     10.80$\leq \log{\left( M_* / M_\odot \right)} \leq$11.00&      1.05$\pm$0.13&10.16&     11.93$\pm$0.11&    418\\
     2.50$\leq z \leq$3.00&     11.00$\leq \log{\left( M_* / M_\odot \right)} \leq$11.20&      0.90$\pm$0.17&10.04&     11.87$\pm$0.16&    197\\
\hline
\end{tabular}
\end{center}
\end{table*}

\end{document}